\documentclass[pagebackref,showinstructions,showlabels,coverfontpercent=100,final]{adsphd2}

\usepackage[english,dutch]{babel}
\usepackage{stmaryrd}
\usepackage{amsfonts}
\usepackage{psfrag}
\usepackage{verbatim}
\usepackage{amssymb}
\usepackage{graphicx,times,amsmath} % Add all your packages here
\usepackage{array}
\usepackage{morefloats}
\usepackage{multirow}
\usepackage{url}
\usepackage{caption}
\usepackage[lined,ruled]{algorithm2e}
\usepackage{minipage}
\usepackage{nomencl}   % For nomenclature
\usepackage[style=long,number=none]{glossary} % For list of abbreviations
\usepackage[savepos]{zref}

\title{Clustering evolving data using kernel-based methods}

\author{Rocco}{Langone}

\promotor{Prof.~dr.~ir.~Johan A. K. ~Suykens}{}

\president{Em. prof.~dr.~ir.~Paul Van Houtte}
\jury{Em. prof.~dr.~ir.~Joos ~Vandewalle\\
      Prof.~dr.~ir.~Marc ~Van ~Barel\\
      Dr.~ir.~Bart ~De ~Ketelaere}
\externaljurymember{Prof.~dr.~Renaud ~Lambiotte}{Universit\'e de Namur}

\phddegree{Engineering} % "Doctor of ..."
\faculty{Faculty of Engineering}
\department{Department of Electrical Engineering (ESAT)}
\researchgroup{Stadius Center for Dynamical Systems, Signal Processing and Data Analytics}
\address{Kasteelpark Arenberg 10}
\email{rocco.langone@esat.kuleuven.be} % Leave empty to hide
\website{http://www.esat.kuleuven.be} % Leave empty to hide

\date{July 2014}
\depot{2014/7515/68}     % Leave out the initial D/ (it is added
\isbn{978-94-6018-844-2}

\renewcommand{\nomname}{List of Symbols}
\makeatletter
\let\@printnomenclatureorig\@printnomenclature
\def\@printnomenclature[#1]{%
  \cleardoublepage%
  \chaptermark{\nomname}
  \@printnomenclatureorig[#1]
}
\makeatother
\makenomenclature

\newcommand{\glossname}{Abbreviations}
\makeglossary

\let\printglossaryorig\printglossary
\renewcommand{\printglossary}{%
  \renewcommand{\glossaryname}{\glossname}
  \cleardoublepage%
  \printglossaryorig\chaptermark{\glossname}}

\bibliography{bib_phdthesis}

\bibstyle{abbrv}
\bibstyle{plain}

\InputIfFileExists{defs} % defs.tex, contains own preamble settings

\begin{document}

%%%%%%%%%%%%%%%%%%%%%%%%%%%%%%%%%%%%%%%%%%%%%%%%%%%%%%%%%%%%%%%%%%%%%%

%\makefrontcover
\makefrontcoverXII

\maketitle

\frontmatter % to get \pagenumbering{roman}

\vspace*{5cm} 
\textit{A mio pap\'a, che mi ha insegnato l'importanza della memoria storica per capire il presente ed immaginare il futuro}. \textit{A mia mamma, la cui disarmante semplicit\'a mi ricorda che i modelli pi\'u semplici ed eleganti vanno preferiti a quelli contorti e complicati}. \textit{Alle mie sorelle Luisa e Laura, studiose dell' ''intelligenza naturale''}. \textit{A mio fratello Rosario, ''ossessionato'' dai rankings.}

\includepreface{preface}
\includeabstract{abstract}

% To create a list of abbreviations, there are 2 options
% 1. manual creation and inclusion of this file
    \includeabbreviations{abbreviations}
% 2. automatic generation via the glossary package
%    \usepackage{glossary}
%    \makeglossary
%    \glossary{name=MD,description=molecular dynamics}
%    \printglossary
\printglossary

% To create a list of symbols, there are 2 options
% 1. include a manually created nomenclature as a chapter
    \includenomenclature{nomenclaturechapter}
% 2. automatic generation via the nomencl package
%    \usepackage{nomencl}
%    \makenomenclature
%    \nomenclature[cB]{$c_B(\vec{x})$}{Characteristic function of $B$}
%    \printnomenclature[3cm]
\printnomenclature[1.5cm]

\tableofcontents

%%%%%%%%%%%%%%%%%%%%%%%%%%%%%%%%%%%%%%%%%%%%%%%%%%%%%%%%%%%%%%%%%%%%%%

\mainmatter % to get \pagenumbering{arabic}

% Show instructions on a separate page
\instructionschapters\cleardoublepage

% Insert here your own chapters
% Chapters are expected to be in a tex-file with the given name dot
% tex and in a directory with the given name in the chapters
% directory.
%
  \graphicspath{{\chaptersdir/introduction/image/}}%
  %\cleardoublepage%
  \chapter{Introduction}
\label{ch:introduction}

%\ldots
%\instructionsintroduction
% Some dummy code to make sure bibtex does not complain.

%Illustration of how to include citations \cite{Meert2011PhD} and \cite{VandenBroeck2011IJCAI}. Lorem ipsum dolor sit amet, consetetur sadipscing elitr, sed diam nonumy eirmod tempor invidunt ut labore et dolore magna aliquyam erat, sed diam voluptua. At vero eos et accusam et justo duo dolores et ea rebum. Stet clita kasd gubergren, no sea takimata sanctus est Lorem ipsum dolor sit amet.

%And yet another citation \cite{FrRo2010Diffusion}.

% Some dummy code to get at least 1 entry in the nomenclature.
%Introducing some symbol: $\Theta$.

%\nomenclature{$\Theta$}{A nice symbol}

% Some dummy code to get at least 1 entry in the list of
% abbreviations.
%\glossary{name=MD,description=molecular dynamics}

% Some dummy code show how to include images.
%\begin{figure}
%  \centering
%  \medskip
%  \includegraphics[width=.9\textwidth]{sine}
%  \caption{Illustration of how to include a figure. }
%  \label{fig:sine}
%\end{figure}

\section{Background}
We live in the Information Age. The recent development of Information Technologies (computers, internet, smart phones, sensors etc.) has a big impact on science and society. In principle, the large amount of available data can help to grasp the complexity of many phenomena of interest, in order to make new scientific discoveries, designing optimal business strategies, optimizing industrial processes etc.   
     
Recognition of complex patterns in the data is of crucial importance to extract useful knowledge. In this context, clustering is a fundamental mode of understanding and learning \cite{50years_beyondkmeans,dataclu_review}. It refers to the task of organizing the data into meaningful groupings based only on the similarity between the data elements, and therefore is exploratory in its essence. Since no target or desired patterns are known a priori, it belongs to the family of unsupervised learning techniques \cite{Bishop2006}.             

Unveiling the underlying structure of the data through the cluster analysis is just one side of the coin. Other important elements are related to the dynamic version of the problem, i.e. monitoring the evolution of the clusters. Understanding how the behaviour of the system under study changes in time represents a key issue in many domains \cite{Mucha14052010,Guha_2003}. From this point of view dynamic clustering would be a useful tool to investigate how clusters form, evolve and disappear.      

The topic of this thesis is related to the design and application of kernel-based methods to perform dynamic clustering. Kernel methods are a class of machine learning techniques where two main modelling phases are present. First a mapping of the data into a high dimensional feature space is performed. Then, the design of learning algorithms in that space allows to discover complex and non-linear relations in the original input space \cite{KMforPA}.  
A major role in this work is played by Least Squares Support Vector Machine (LS-SVM) \cite{lssvm_book}, which is a class of Support Vector Machine (SVM) \cite{cortessvm} based on a constrained optimization framework with the presence of the $L_2$ loss function in the objective and equality instead of inequality constraints. By modifying and extending the objective and/or the constraints of the core formulation, it is possible to develop models tailored for a given application, with a systematic model selection procedure and high generalization abilities.

\section{Challenges}
\label{ch:challenges}
The main issues tackled in this thesis can be summarized as follows:

\begin{itemize}
\item \textbf{Community detection via kernel methods}: A network is a collection of nodes or vertices joined by edges and represents the patterns of connections between the components of complex systems \cite{newman_commdet}. Usually real-life networks display a high level of order and organization. For example, the distribution of edges is characterized by high concentrations of edges within special groups of vertices, and low concentrations between these groups. The problem of identifying such clusters of nodes is called community detection \cite{fortunatocommdet}. The main challenges posed by the usage of kernel methods for community detection are related to the choice of the kernel function, the model selection, the out-of-sample extension, the scalability to large datasets. 
\item \textbf{Analysis of dynamic communities}: Community detection of evolving networks aims to understand how the community structure of a complex network changes over time \cite{Bota:2011:ActaCybernetica,Mucha14052010}. A desirable feature of a clustering model which has to capture the evolution of communities is the temporal smoothness between clusters in successive time-steps. Providing a consistent clustering at each time results in a smooth view of the changes and a greater robustness against noise \cite{evol_spec_clustering,evol_clustering_chakrabarti}. 
\item \textbf{Fault detection}: With the development of information and sensor technology many process variables in a power plant like temperature, pressure etc. can be monitored. These measurements give an information on the current status of a machine and can be used to predict the faults and plan an optimal maintenance strategy \cite{choi2003,Kourti19953}. A useful model in this case must be able to catch, in an online fashion, the degradation process affecting the machine, to avoid future failures of the components and unplanned downtimes.
\item \textbf{Clustering in a non-stationary environment}: In many real-life applications non-stationary data are generated according to some distribution models which change over time. Therefore, a proper cluster analysis can be useful to detect important change points and in general to better understand the dynamics of the system under investigation. In this case a clustering algorithm is required to continuously adapt in response to new data and to be computationally efficient for real-time applications \cite{SAKM_boubacar}.   
\end{itemize} 

\section{Objectives}
In this thesis the following objectives can be outlined:
\begin{itemize} 
\item \textbf{to envisage a whole kernel-based framework for community detection}. In this context many issues arise. First of all it is important to choose a proper kernel function to describe the similarity between the nodes of the network under investigation. Then a key point is represented by the model selection, i.e. finding the natural number of communities which are present in the network and eventually tuning the kernel hyper-parameters. The kernel-based model must also be able to accurately predict the membership of new nodes joining the network, without performing the clustering from scratch. Moreover, since many real-world networks contain millions of nodes and edges, the network data have to be processed in a reasonable time. Finally, the research carried out for solving the static community detection problem paves the way for the development of models for the analysis of evolving networks.              
\item \textbf{to design a model for community detection in a changing scenario}. An evolving network can be described as a sequence of snapshot graphs, where each snapshot represents the configuration of the network at a particular time instant. When community detection is performed at time $t$, the clustering should be similar to the clustering at the previous time-step $t-1$, and should accurately incorporate the actual data. In this way, if the data at time $t$ does not deviate from
historical expectations, the clustering should be similar to that from time $t-1$, while if the structure of the data changes significantly, the clustering must be modified to account for the new structure. Thus, a good clustering algorithm must trade-off the benefit of maintaining a consistent clustering over time with the cost of deviating from an accurate
representation of the current data.    
\item \textbf{to conceive modelling strategies for clustering stationary and non-stationary data streams}. Data streams are a sequence of data records stamped and ordered by time \cite{Guha_2003}. Clustering data streams in real time is an ambitious problem with ample applications. If the data distribution is stationary, emphasis should be given to the off-line construction of the model. Once properly designed, such a model can be used to cluster the data stream in an online fashion by means of the out-of-sample extension. However, if the data distribution is non-stationary, the initial model soon becomes obsolete and must be quickly updated. Therefore the development of a fast, adaptive and accurate model is an important objective.    
\end{itemize}

\section{Chapter by Chapter Overview}
\begin{figure}
  \centering
  \medskip
  \includegraphics[width=\textwidth]{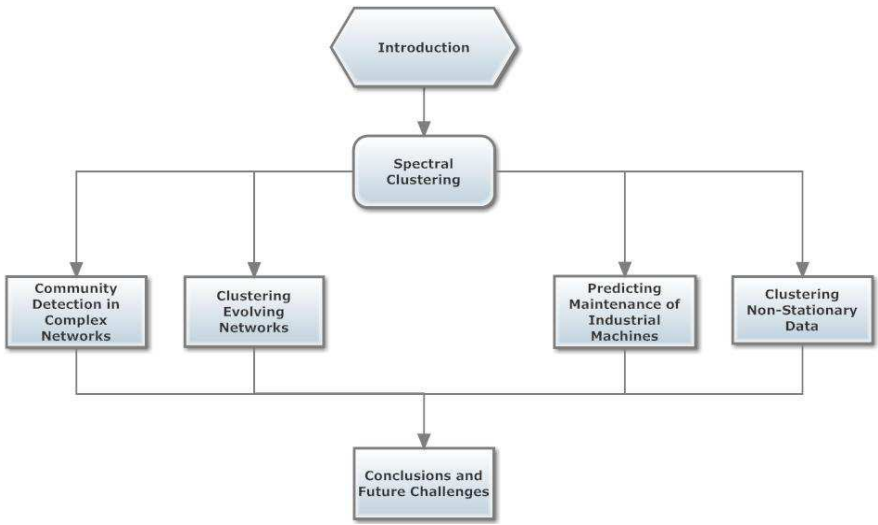}
  \caption{\textbf{Thesis overview}.}
  \label{Fig_chapter_overview}
\end{figure}

The general structure of this thesis is sketched in figure \ref{Fig_chapter_overview} and can be described as follows:   
\begin{itemize}
\item \textbf{Chapter $2$} contains four sections. First of all a general introduction to spectral clustering, one of the most successful clustering algorithms, is given. Then kernel spectral clustering (KSC) is reviewed. KSC is a spectral clustering  algorithm formulated in the LS-SVM optimization framework, with the possibility to extend the clustering model to out-of-sample data for predictive purposes. Moreover a model selection criterion called Balanced Line Fit (BLF) is also present. For these reasons it represents the starting point to face the challenges described in section \ref{ch:challenges}. Later on the soft kernel spectral clustering (SKSC) algorithm is introduced. Instead of using the hard assignment rule present in KSC, a fuzzy assignment based on the cosine distance from the cluster prototypes in the projections space is suggested. We also introduce a related model selection technique, called Average Membership Strength criterion (AMS), which solves the major drawbacks of BLF. Finally, we show that SKSC can improve the interpretability of the results and the clustering performance with respect to KSC, mainly in cases of large overlap between the clusters. 
\item \textbf{Chapter $3$} is dedicated to the community detection problem. After reviewing the important literature in the field we illustrate our methodology, which is composed by four main cornerstones. First, it is crucial to extract from the  given network a small sub-graph representative of its community structure, which is a challenging problem. This sub-graph can then be used to train a KSC model in a computationally efficient way and forms the basis for a good generalization. Second, the correct tuning of the kernel hyper-parameters (if any) and the number of communities is another important issue, which is solved by proposing a new model selection criterion based on the Modularity statistics. Third, the kernel functions used to properly describe the similarity between the nodes are presented. Finally, the out-of-sample extension allows not only to accurately predict the community affiliation of new nodes, but also can make the algorithm cluster millions of data in a short time on a desktop computer.                            
\item \textbf{Chapter $4$} introduces a novel model called kernel spectral clustering with memory effect (MKSC). This method is designed to cluster evolving networks (described as a sequence of snapshot graphs), aiming to track the long-term drift of the communities by ignoring meanwhile the short-term fluctuations. In this new formulation the desired temporal smoothness is incorporated in the objective function of the primal problem through the maximization of the correlation between the actual and the previous models. Moreover, new measures are presented in order to judge the quality of a partitioning produced at a given time. The new measures are the weighted sum of the snapshot quality and the temporal quality. The former only measures the quality of the current clustering with respect to the current data, while the latter measures the temporal smoothness in terms of the ability of the actual model to cluster the historic data. These new measures can be also used to perform the model selection. 
\item \textbf{Chapter $5$} discusses the application of KSC to an industrial case. Here we assume stationarity, i.e. the regimes experienced by the system under analysis do not change over time. Vibration data are collected from a packing machine to monitor its conditions. In order to describe the ongoing degradation process due to the dirt accumulation in the sealing jaws we first apply a windowing operation on the data, accounting for historical values of the sealing quality. The size of the window, together with the bandwidth of the Radial Basis Function (RBF) kernel and the number of clusters are tuned using the BLF criterion. Then an optimal kernel spectral clustering model is trained offline to identify two main regimes, that we can interpret as normal behaviour and critical conditions (need of maintenance). Thanks to the out-of-sample extension property, this model is used online to predict in advance when the machine needs maintenance. In principle, this implies the maximization of the production capacity and the minimization of downtimes.       
\item \textbf{Chapter $6$} deals with clustering non-stationary data. A new adaptive method named incremental kernel spectral clustering (IKSC) is devised. In a first phase a KSC model is constructed, then it is updated online according to the new points belonging to the data stream. The central idea behind the proposed technique concerns expressing the clustering model in terms of prototypes in the eigenspace, which are continuously adapted through the out-of-sample eigenvectors calculation. Moreover, the training set is formed only by the cluster centers in the input space, which are also updated in response to new data. This compact representation of the model and the training set in terms of cluster centroids makes the method computationally efficient and allows to properly track the evolution of complex patterns over time.            
\item \textbf{Chapter $7$} concludes the thesis and proposes future research directions. 
\end{itemize}
In general, in all the experiments discussed in each chapter the values reported to assess the cluster quality are average measures over $10$ runs of the algorithm under investigation. Although a fully statistical significance analysis has not always been performed, the mean values give good indications about the performances of the different methods.

\section{Main Contributions}

In what follows the main contributions of this thesis are resumed:
\begin{itemize}
\item \textbf{Community detection via KSC}. We conceived a complete methodology to cluster network data. The whole procedure can be summarized in three main stages: extract from the network a small sub-graph which retains the community structure of the entire graph, train an optimal KSC model using the selected sub-graph as training set, use the out-of-sample extension to assign the memberships to the remaining nodes. The precise tuning of the hyper-parameters and the choice of an appropriate kernel function are also essential. Furthermore, an algorithm to provide moderated outputs named SKSC and a related model selection criterion called AMS are proposed. Finally we mention how the technique can be used for large scale applications. The relevant papers are \cite{rocco_IJCNN2011,rocco_IJCNN2012,rocco_IJCNN2013,raghvendra_bigdata,raghvendra_sclara,FURS_paper,raghvendra_plosone,raghvendra_esann}.           
\item \textbf{KSC with memory effect}. A new model is designed to handle evolving networks. At the level of the primal optimization problem typical of LS-SVM, we introduce a memory effect in the objective function to account for the temporal smoothness of the clustering results over time. Also new model selection criteria specific for the given application are introduced. The related publications are \cite{Langone_physicA,rocco_ICMSQUARE2012,diego_IJCNN2013,paper_KDD2014}. 
\item \textbf{KSC for predictive maintenance}. We have successfully applied KSC to a complex industrial case. We developed a clustering model able to infer the degradation process affecting a packing machine from the vibration signals registered by accelerometers placed on the sealing jaws. A critical modelling phase concerned the windowing of the data in order to describe the history of the sealing quality. Moreover the model selection stage was also crucial. Finally, to improve the interpretability of the results a probabilistic output has been provided. This contribution is reported in \cite{rocco_singapore}. %and \cite{journal_POM}. %POM journal         
\item \textbf{Incremental KSC (IKSC)}. We presented a new algorithm to perform online clustering in a non-stationary environment. IKSC exploits the out-of-sample extension property of KSC to continuously adapt the initial model. In this way it is able to catch the dynamics of the clusters evolving over time. The IKSC method can model merging, splitting, appearance, death, expansion and shrinking of clusters, in a fast and accurate way \cite{IKSC_paper}.
\end{itemize}

%%%%%%%%%%%%%%%%%%%%%%%%%%%%%%%%%%%%%%%%%%%%%%%%%%
% Keep the following \cleardoublepage at the end of this file, 
% otherwise \includeonly includes empty pages.
\cleardoublepage

% vim: tw=70 nocindent expandtab foldmethod=marker foldmarker={{{}{,}{}}}
%

%
  \graphicspath{{\chaptersdir/chapter2/image/}}%
  %\cleardoublepage%
  \chapter{Spectral clustering}
\label{ch:spectralclustering}

Spectral clustering methods have been reported to often outperform the traditional approaches such as $K$-means and hierarchical clustering in many real-life problems. We start this chapter with a description of the basic concepts behind spectral partitioning. We discuss advantages, like the ability of detecting complex clustering boundaries, and disadvantages, mainly related to the absence of a model selection scheme and the out-of-sample extension to unseen data.

Then we summarize the kernel spectral clustering (KSC) model, which is formulated as a weighted kernel PCA problem in the primal-dual optimization framework typical of Least Squares Support Vector Machines (LS-SVMs). Thanks to this representation, KSC solves the above mentioned drawbacks of spectral clustering, since it can be trained, validated by means of a tuning criterion called Balanced Line Fit (BLF), and tested in an unsupervised learning procedure. 

Finally we propose an algorithm for soft (or fuzzy) clustering named soft kernel spectral clustering (SKSC). Basically, instead of using the hard assignment method present in KSC, we suggest a fuzzy assignment based on the cosine distance from the cluster prototypes in the space of the projections. We also introduce a related model selection technique, called average membership strength (AMS) criterion, which solves the main difficulties of BLF. Roughly speaking, SKSC is observed to improve the cluster performance upon KSC mainly when the clusters overlap to a large extent.

\section{Classical Spectral Clustering}
\label{sc:SC}

\subsection{Introduction}
Spectral clustering (SC) represents an elegant and effective solution to the graph partitioning problem. It makes use of the spectral properties of a matrix representation of the graph called Laplacian to divide it into weakly connected sub-graphs. It can be applied directly to network data to divide the vertices into several non-overlapping groups, or it can be used to cluster any kind of data. In this case the matrix of pairwise similarities between the data points serves as the network to partition. In \cite{chung} an explanation of SC by the point of view of graph theory is given. The authors of \cite{ngjordan} present a SC algorithm which successfully deals with a number of challenging clustering problems. Moreover, an analysis of the algorithm by means of matrix perturbation theory gives conditions under which a good performance is expected. In \cite{lsc} a new cost function for SC based on a measure of error between a given partition and a solution of the spectral relaxation of a minimum normalized cut problem is derived. The authors of \cite{Kannan00onclusterings} analyse the SC technique by means of a bi-criteria measure to assess the quality of a clustering result. An exhaustive tutorial on SC has been presented in \cite{luxburg_tutorial}. In what follows we only depict the basic idea behind spectral partitioning, which originated by studying the graph partitioning problem in graph theory. The interested reader can refer to the aforementioned works for a deeper discussion.   

\subsection{The Graph Partitioning Problem}
A graph (or network) $\mathcal{G = (V,E)}$ is a mathematical structure used to model pairwise relations between certain objects. It refers to a set of $N$ vertices or nodes $\mathcal{V}= \{v_i\}_{i=1}^N$ and a collection of edges $\mathcal{E}$ that connect pairs of vertices. If the edges are provided with weights the corresponding graph is weighted, otherwise it is referred as an unweighted graph. The topology of a graph is described by the similarity (or affinity) matrix, which is an $N \times N$ matrix $S$, where $S_{ij}$ indicates the link between the vertices $i$ and $j$. Associated to the similarity matrix there is the degree matrix $D = \textrm{diag}(d)$, with $d = [d_1,\hdots,d_N]^T = S1_N$ and $1_N$ indicating the $N \times 1$ vector of ones. Basically the degree $d_i$ of node $i$ is the sum of all the edges (or weights) connecting node $i$ with the other vertices: $d_i = \sum_{j=1}^{N}S_{ij}$.   

In the most basic formulation of the graph partitioning problem one is given an unweighted graph and asked to split it into $k$ non-overlapping groups $\mathcal{A}_1,\ldots,\mathcal{A}_k$ in order to minimize the cut size, which is the number of edges running between the groups\footnote{If the graph is weighted, the objective is to find a partition of the graph such that the edges between different groups have very low weights.}. In order to favour balanced clusters, we can consider the normalized cut $\textrm{NC}$, defined as:     
\begin{align}
	\label{mincut_def}
	\textrm{NC}(\mathcal{A}_1,\ldots,\mathcal{A}_k) = k - \textrm{tr}(G^TL_nG)   
\end{align}
where:
\begin{itemize} 
\item $L_n=I-D^{- \frac{1}{2}}SD^{-\frac{1}{2}}$ is called the normalized Laplacian
\item $G = [g_1,\ldots,g_k]$ is the matrix containing the normalized cluster indicator vectors $g_l = \frac{D^{\frac{1}{2}}f_l}{||D^{\frac{1}{2}}f_l||_2}$
\item $f_l$, with $l=1,\ldots,k$, is the cluster indicator vector for the $l$-th cluster. It has a $1$ in the entries corresponding to the nodes in the $l$-th cluster and $0$ otherwise. Moreover, the cluster indicator matrix can be defined as $F = [f_1,\ldots,f_k] \in \{0,1\}^{N \times k}$.      
\end{itemize} 
The $\textrm{NC}$ optimization problem is stated as follows:
\begin{equation}
\label{mincut_opt}
\begin{aligned}
& \underset{G}{\text{min}}
& & k - \textrm{tr}(G^TL_nG) \\
& \text{subject to}
& & G^TG=I
\end{aligned}
\end{equation}
with $I$ denoting the identity matrix.
Unfortunately this is a NP-hard problem. However we can find good approximate solutions in polynomial time by using a relaxation method, i.e. allowing $G$ to take continuous values:
\begin{equation}
\label{mincut_opt2}
\begin{aligned}
& \underset{\hat{G}}{\text{min}}
& & k - \textrm{tr}(\hat{G}^TL_n\hat{G})\\
& \text{subject to}
& & \hat{G}^T\hat{G}=I.
\end{aligned}
\end{equation}
with $\hat{G} \in \mathbb{R}^{N \times k}$.     
In this case it can be shown that solving problem (\ref{mincut_opt2}) is equivalent to finding the solution to the following eigenvalue problem:
\begin{align}
	\label{mincut_eig}
	L_ng = \lambda g.   
\end{align}
Basically, the relaxed clustering information is contained in the eigenvectors corresponding to the $k$ smallest eigenvalues of the normalized Laplacian $L_n$. 
In addition to the normalized Laplacian, other Laplacians can be defined, like the unnormalized Laplacian $L = D-S$ and the  random walk Laplacian $L_{rw} = D^{-1}S$. The latter is appealing for its suggestive interpretation in terms of a Markov random walk.        

\subsection{Link with Markov Chains}
A well known relationship between graphs and Markov chains exists: any graph has an associated random walk in which the probability of leaving a vertex is distributed among the outgoing edges according to their weight. For a given graph with $N$ nodes and $m$ edges the probability vector can be defined as $p_{t+1} = Pp_t$, where  $P = D^{-1}S$ indicates the transition matrix with the $ij$-th entry representing the probability of moving from node $i$ to node $j$ in one step. Under these assumptions we have an ergodic and reversible Markov chain with stationary distribution vector $\pi$ with components $\pi_i= \frac{d_i}{2m}$. It can be shown that this distribution describes the situation in which the random walker remains most of the time in the same cluster with rare jumps to the other clusters \cite{meila_random}. Moreover $L_{rw} = I-P$ and the eigenvectors corresponding to the smallest eigenvalues of $L_{rw}$ are the same as the eigenvectors related to the largest eigenvalues of $P$:
\begin{align}
	\label{rw_eig}
	L_ng = \lambda g \Rightarrow (I-P)g = \lambda g \Rightarrow g-Pg = \lambda g \Rightarrow Pg = (I-\lambda I)g.       
\end{align}
For the reader interested in having a deeper insight into this topic, we advice to explore \cite{meila_random,meila_learning,JeanCharlesPNAS2010}.     

\subsection{Basic Algorithm}
As mentioned before, the classical spectral clustering algorithm can be applied to partition any kind of data, not only networks. Indeed, if for the networks we are immediately provided with the affinity matrix $S$, in case of data points we have to construct $S$ starting from some similarity function. The basic steps are described in algorithm \ref{alg:SC}.                 
\begin{algorithm}[t]
\small
\LinesNumbered
\KwData{positive (semi-) definite similarity matrix $S \in \mathbb{R}^{N \times N}$, number $k$ of clusters to construct.} 
\KwResult{clusters $\mathcal{A}_1,\ldots,\mathcal{A}_k$.}
compute the graph Laplacian ($L$, $L_n$ or $L_{rw}$)\\
compute the eigenvectors $\hat{g_1},\ldots,\hat{g_k}$ corresponding to the smallest $k$ eigenvalues\\
let $\hat{G} \in \mathbb{R}^{N \times k}$ be the matrix containing the vectors $\hat{g_1},\ldots,\hat{g_k}$ as columns\\
for $i=1,\ldots,N$ let $u_i \in \mathbb{R}^{k}$ be the vector corresponding to the $i$-th row of $G$\\
cluster the points $u_i$ into clusters $\mathcal{C}_1,\ldots,\mathcal{C}_k$\\
compute the final partitioning $\mathcal{A}_1,\ldots,\mathcal{A}_k$, with $A_i = \{j|u_j \in C_i \}$.         
\caption{SC algorithm \cite{luxburg_graph}}
\label{alg:SC}
\end{algorithm}
Thanks to the mapping of the original input data in the eigenspace, SC is able to unfold the manifold the data are embedded in and to detect complex clustering boundaries. On the other hand, it has some clear disadvantages:
\begin{itemize}
\item it is not clear how to properly construct the similarity matrix $S$ and the number of clusters must be provided beforehand. In \cite{Zelnik-manor04self-tuningspectral} the authors proposed a solution to this model selection issue by introducing a parameter-free SC.      
\item there is no clear way as how test data points should be assigned to the initial clusters, since the embedding eigenvectors are only defined for the full dataset. In \cite{nystrom} and \cite{sc_nystrom} the authors employed the Nystr\"om method to find approximate eigenvectors for out-of-sample data and reduce the computational load for large scale applications. In \cite{vanbarel_ICD} the authors proposed a sparse spectral clustering method based on the incomplete Cholesky decomposition (ICD), which constructs an approximation of the Laplacian in terms of capturing the structure of the matrix. By using only the information related to the pivots selected by the ICD, a method to compute cluster memberships for out-of-sample points is also introduced.     
\end{itemize}     

\section{Kernel Spectral Clustering}
\label{sc:KSC}

\subsection{Generalities}
Kernel Spectral Clustering (KSC) represents a spectral clustering formulation in the LS-SVM optimization framework with primal and dual representations. The dual problem is an eigenvalue problem, related to spectral clustering. KSC has two main advantages with respect to classical spectral clustering:
\begin{itemize}
\item a precise model selection scheme to tune the hyper-parameters 
\item the out-of-sample extension to test points by means of an underlying model.
\end{itemize}
After finding the optimal hyper-parameters, the clustering model can be trained on a subset of the full data and readily applied to unseen test points in a learning framework. 

\subsection{Least Squares Support Vector Machine}
The Support Vector Machine (SVM) is a state-of-the-art classification method. It performs linear classification in a high-dimensional kernel-induced feature space, which corresponds to a non-linear decision boundary in the original input space. LS-SVM differs from SVM because it uses a $L_2$ loss function in the primal problem and equality instead of inequality constraints. This typically leads to eigenvalue problems or linear systems at the dual level, in the context of principal component analysis \cite{suykens_KPCA} and classification or regression \cite{suykens_svmclass}, respectively.

Given a training data set $\mathcal{D}_{\textrm{Tr}} = \{(x_i,y_i)\}_{i=1}^{N_{\textrm{Tr}}}$, where $x_i \in \mathbb{R}^{d}$ are the training points and $y_i \in \{-1,1\}$ are the related labels, the primal problem of the LS-SVM binary classifier\footnote{Multi-class classification
problems are decomposed into multiple binary classification tasks with the possibility to use several coding-decoding schemes \cite{VanGestel_multiclass}.} can be stated as \cite{suykens_svmclass}:
\begin{equation}
\label{primalLSSVM}
\begin{aligned}
& \underset{w,e_i,b}{\text{min}}
& & \frac{1}{2}w^Tw + \gamma \frac{1}{2} \sum_{i=1}^{N_{\textrm{Tr}}}e_i^2\\
& \text{subject to}
& & y_i(w^T \varphi(x_i)+b)=1-e_i, i=1,\hdots,N_{\textrm{Tr}}.
\end{aligned}
\end{equation}
The expression $\hat{y}=w^T\varphi(x)+b$ indicates the model in the primal space. It is linear with respect to the parameter vector $w$ but the relationship between $x$ and $y$ can be non-linear if the feature map $\varphi(\cdot)$ is a non-linear function. With $\gamma$ we indicate the regularization parameter which controls the trade-off between the model complexity and the minimization of the training error. If we construct the Lagrangian we have:
\begin{multline}
	\label{lagrangianLSSVM}
	\mathcal{L}(w,e_i,b,\alpha_i) = \frac{1}{2}w^Tw+\gamma\frac{1}{2}\sum_{i=1}^{N_{\textrm{Tr}}}e_i^2 -\sum_{i=1}^{N_{\textrm{Tr}}}\alpha_i(y_i(w^T\varphi(x_i)+b)-1+e_i)
\end{multline} 
where $\alpha_i$ are the Lagrange multipliers. The KKT optimality conditions are:\\\\
$\frac{\partial \mathcal{L}}{\partial w} = 0 \rightarrow w = \sum_{i=1}^{N_{\textrm{Tr}}}\alpha_iy_i\varphi(x_i)$,\\ $\frac{\partial \mathcal{L}}{\partial e_i} = 0 \rightarrow \alpha_i = \gamma e_i$,\\ $\frac{\partial \mathcal{L}}{\partial b} = 0 \rightarrow \sum_{i=1}^{N_{\textrm{Tr}}}\alpha_iy_i = 0$,\\ $\frac{\partial \mathcal{L}}{\partial \alpha_i} = 0 \rightarrow y_i(w^T\varphi(x_i)+b)-1+e_i = 0$.\\\\
Eliminating the primal variables $e_i$ and $w$ leads to the following linear system in the dual problem:\\
\begin{equation} 
\label{dualLSSVM}
\left[
\begin{array}{c|c}
%\mathrm{\Omega + I_{N_{\textrm{Tr}}}/\gamma} & \mathrm{y}\\
\Omega + I_{N_{\textrm{Tr}}}/\gamma & y\\
\hline
%\mathrm{y^T} & 0
y^T & 0
\end{array}
\right]
\left[
\begin{array}{c}
\alpha\\
\hline
b
\end{array}
\right]
=
\left[
\begin{array}{c}
1_{N_{\textrm{Tr}}}\\
\hline
0
\end{array}
\right]
\end{equation}\\
where $y=[y_1;\hdots;y_{N_{\textrm{Tr}}}]$, $1_{N_{\textrm{Tr}}}=[1;\hdots;1]$, $\alpha=[\alpha_1;\hdots;\alpha_{N_{\textrm{Tr}}}]$. The term $\Omega$ means the kernel matrix with entries $\Omega_{ij} = \varphi(x_i)^T\varphi(x_j) = K(x_i,x_j)$. With $K: \mathbb{R}^{d} \times \mathbb{R}^{d} \rightarrow \mathbb{R}$ we denote the kernel function which maps the input points into the high dimensional feature space $\varphi(\cdot)$. For example, by using a Radial Basis Function (RBF) kernel expressed by $K(x_i,x_j) = \exp(-||x_i-x_j||_2^2/\sigma^2)$, one is able to construct a model of arbitrary complexity. Finally, after solving the previous linear system, the LS-SVM classification model in the dual representation becomes:
\begin{equation}
y(x) = \textrm{sign}(\sum_{i=1}^{N_{\textrm{Tr}}}\alpha_i y_i K(x,x_i) + b).
\end{equation}   
The constrained optimization framework with explicit use of regularization explained above represents the core model not only for classification, but also regression and unsupervised learning, as we will see in the remainder of this dissertation.

\subsection{Primal-Dual Formulation}
\label{KSC_intro}
\begin{figure}[htbp]
\centering
\begin{tabular}{c}
\begin{psfrags}
\psfrag{x}[t][t]{{}}
\psfrag{y}[b][b]{{}}
\psfrag{t}[b][b]{{}}
\includegraphics[width=3.5in]{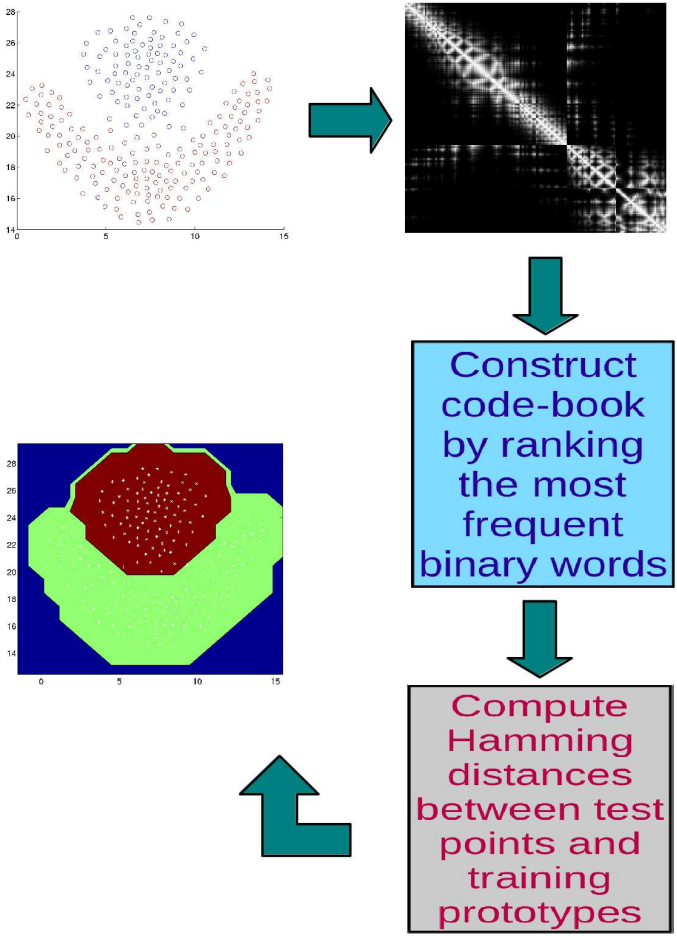}
\end{psfrags}
\end{tabular}
\caption{\textbf{KSC algorithm}. The dataset consists of a set $\mathcal{D} = \{x_i\}_{i=1}^{N}$ where $x_i \in \mathbb{R}^2$, and relates to a binary clustering problem with nonlinear boundary. After binarizing the matrix containing the eigenvectors of the Laplacian as columns, a code-book with the most frequent binary words representing the training cluster prototypes is formed. The test points are mapped into the training eigenspace through the out-of-sample extension. These projections are then binarized and the points are assigned to the closest prototype in terms of Hamming distance, by means of an ECOC decoding procedure.}\label{KSC_figure}
\end{figure}

Given a training data set $\mathcal{D}_{\textrm{Tr}} = \{x_i\}_{i=1}^{N_{\textrm{Tr}}}$, the multi-cluster KSC model \cite{multiway_pami} in the LS-SVM framework is formulated as a weighted kernel PCA problem \cite{mikakernelpca} decomposed in $l = k-1$ binary problems, where $k$ is the number of clusters to find:  
\begin{equation}
\label{primalKSC}
\begin{aligned}
& \underset{w^{(l)},e^{(l)},b_l}{\text{min}}
& & \frac{1}{2}\sum_{l=1}^{k-1}w^{(l)^T}w^{(l)}-\frac{1}{2N_{\textrm{Tr}}}\sum_{l=1}^{k-1}\gamma_l e^{(l)^T}Ve^{(l)}\\
& \text{subject to}
& & e^{(l)}=\Phi w^{(l)}+b_l1_{N_{\textrm{Tr}}}.
\end{aligned}
\end{equation}
The $e^{(l)}=[e^{(l)}_1,\hdots,e^{(l)}_{N_{\textrm{Tr}}}]^T$ are the projections of the data points $\{x_i\}_{i=1}^{N_{\textrm{Tr}}}$ mapped in the feature space along the direction $w^{(l)}$, also called score variables. The optimization problem (\ref{primalKSC}) can then be interpreted as the maximization of the weighted variances $C_l = e^{(l)^T}Ve^{(l)}$ and the contextual minimization of the squared norm of the vector $w^{(l)}$, $\forall l$. Through the regularization constants $\gamma_l\in\mathbb{R}^+$ we trade-off the model complexity expressed by $w^{(l)}$ with the correct representation of the training data. $V \in \mathbb{R}^{N_{\textrm{Tr}} \times N_{\textrm{Tr}}}$ is the weighting matrix and $\Phi$ is the $N_{\textrm{Tr}}\times d_h$ feature matrix $\Phi=[\varphi(x_1)^T;\hdots;\varphi(x_{N_{\textrm{Tr}}})^T]$. The clustering model is expressed by:
 \begin{equation}
	\label{primalModel}
	e_i^{(l)}=w^{(l)^T}\varphi(x_i)+b_l, i = 1,\hdots,N_{\textrm{Tr}}
\end{equation}
where as usual $\varphi:\mathbb{R}^d\rightarrow\mathbb{R}^{d_h}$ indicates the mapping to a high-dimensional feature space, $b_l$ are bias terms, with $l=1,\hdots,k-1$. The projections $e_i^{(l)}$ represent also the latent variables of the $k-1$ binary clustering indicators given by $\textrm{sign}(e_i^{(l)})$. 
The set of binary indicators form a code-book $\mathcal{CB} = \{c_p\}_{p = 1}^k$, where each code-word is a binary word of length $k-1$ representing a cluster. The Lagrangian associated with the primal problem is:
\begin{equation}
	\label{lagrangianKSC}
    \begin{split}	
	\mathcal{L}(w^{(l)},e^{(l)},b_l,\alpha^{(l)})= & \frac{1}{2}\sum_{l=1}^{k-1}w^{(l)^T}w^{(l)}-\frac{1}{2N_{\textrm{Tr}}}\sum_{l=1}^{k-1}\gamma_l e^{(l)^T}Ve^{(l)}-\\ 
    &\sum_{l=1}^{k-1}\alpha^{(l)^T}(e^{(l)}-\Phi w^{(l)}-b_l1_{N_{\textrm{Tr}}})
    \end{split}
\end{equation} 
where $\alpha^{(l)}$ are the Lagrange multipliers. The KKT optimality conditions are:\\\\
$\frac{\partial \mathcal{L}}{\partial w^{(l)}} = 0 \rightarrow w^{(l)} = \Phi^T \alpha^{(l)}$,\\ $\frac{\partial \mathcal{L}}{\partial e^{(l)}} = 0 \rightarrow \alpha^{(l)} = \frac{\gamma_l}{N_{\textrm{Tr}}}Ve^{(l)} $,\\ $\frac{\partial \mathcal{L}}{\partial b_l} = 0 \rightarrow 1_{N_{\textrm{Tr}}}^T \alpha^{(l)} = 0$,\\ $\frac{\partial \mathcal{L}}{\partial \alpha^{(l)}} = 0 \rightarrow e^{(l)}-\Phi w^{(l)}-b_l1_N = 0$.\\\\ 
Once we have solved the KKT conditions for optimality, if we set $V = D^{-1}$, we can derive the following dual problem:
\begin{equation}
\label{dual_KSC}
D^{-1}M_D\Omega\alpha^{(l)}=\lambda{_l}\alpha^{(l)} 
\end{equation}
where $\Omega$ is the kernel matrix with $ij$-th entry $\Omega_{ij}=K(x_i,x_j) = \varphi(x_i)^T \varphi(x_j)$. $D$ is the graph degree matrix which is diagonal with positive elements $D_{ii}=\sum_j\Omega_{ij}$, $M_D$ is a centering matrix defined as $M_D = I_{N_{\textrm{Tr}}} - \frac{1}{1_{N_{\textrm{Tr}}}^TD^{-1}1_{N_{\textrm{Tr}}}}1_{N_{\textrm{Tr}}}1_{N_{\textrm{Tr}}}^TD^{-1}$, the $\alpha^{(l)}$ are dual variables, $\lambda{_l} = \frac{N_{\textrm{Tr}}}{\gamma_l}$, $K: \mathbb{R}^d \times \mathbb{R}^d \rightarrow \mathbb{R}$ is the kernel function. The dual representation of the model becomes:
 \begin{equation}
	\label{dualModel}
	e_i^{(l)}=\sum_{j=1}^{N_{\textrm{Tr}}} K(x_j,x_i)\alpha_j^{(l)} +b_l, j = 1,\hdots,N_{\textrm{Tr}}.
\end{equation}
Moreover, by observing problem (\ref{dual_KSC}) one can realize that, apart from the presence of the centering matrix $M_D$, it is similar to problem (\ref{rw_eig}). In fact the kernel matrix plays the role of the similarity matrix associated to the graph $\mathcal{G = (V,E)}$ with $v_i \in \mathcal{V}$ equals to $x_i$. This is also the reason behind the choice of the weighting matrix in the primal problem as the inverse of the degree matrix $D$ related to the kernel matrix $\Omega$. Finally, the KSC method is sketched in algorithm \ref{alg:KSC}. 
\begin{algorithm}[t]
\small
\LinesNumbered
\KwData{Training set $\mathcal{D}_{\textrm{Tr}} = \{x_i\}_{i=1}^{N_{\textrm{Tr}}}$, test set $\mathcal{D}_{\textrm{test}} = \{x_m^{\textrm{test}}\}_{m=1}^{N_{\textrm{test}}}$ kernel function $K: \mathbb{R}^d \times \mathbb{R}^d \rightarrow \mathbb{R}$ positive definite and localized ($K(x_i,x_j) \rightarrow 0$ if $x_i$ and $x_j$ belong to different clusters), kernel parameters (if any), number of clusters $k$.} 
\KwResult{Clusters $\{\mathcal{A}_1,\hdots,\mathcal{A}_k\}$, codebook $\mathcal{CB} = \{c_p\}_{p = 1}^k$ with $\{c_p\} \in \{-1,1\}^{k-1}$.}
compute the training eigenvectors $\alpha^{(l)}$, $l = 1,\hdots,k-1$, corresponding to the $k-1$ largest eigenvalues of problem (\ref{dual_KSC})\\
let $A \in \mathbb{R}^{N_{\textrm{Tr}} \times k-1}$ be the matrix containing the vectors $\alpha^{(1)},\ldots,\alpha^{(k-1)}$ as columns\\
binarize $A$ and let the code-book $\mathcal{CB} = \{c_p\}_{p = 1}^k$ be composed by the $k$ encodings of $Q = \textrm{sign}(A)$ with the most occurrences\\ 
$\forall i$, $i=1,\hdots,N_{\textrm{Tr}}$, assign $x_i$ to $A_{p^*}$ where $p^* = \textrm{argmin}_pd_H(\textrm{sign}(\alpha_i),c_p)$ and $d_H(.,.)$ is the Hamming distance\\
binarize the test data projections $\textrm{sign}(e_m^{(l)})$, $m = 1,\hdots,N_{\textrm{test}}$, and let $\textrm{sign}(e_m) \in \{-1,1\}^{k-1}$ be the encoding vector of $x_m^{\textrm{test}}$\\ 
$\forall m$, assign $x_m^{\textrm{test}}$ to $A_{p^*}$, where $p^* = \textrm{argmin}_pd_H(\textrm{sign}(e_m),c_p)$.  
\caption{KSC algorithm \cite{multiway_pami}}
\label{alg:KSC}
\end{algorithm}
A visual representation of the KSC technique is also illustrated in figure \ref{KSC_figure}.

\subsection{Model Selection}
\label{sc:BLF}
In general the kernel hyper-parameters should be chosen carefully in order to ensure good generalization. This is particularly important for very flexible kernels such as the RBF kernel where too small values of the bandwidth $\sigma$ will result in overfitting and too high values in a poor model. 

To deal with this crucial issue, the KSC algorithm is provided with a model selection procedure based on the Balanced Line Fit (BLF) criterion \cite{multiway_pami}. It can be shown that in the ideal situation of compact and well separated clusters, the points $[e_i^{(1)},\hdots,e_i^{(k-1)}]$, $i=1,\hdots,N_{\textrm{Tr}}$, form lines, one per each cluster. Then, by exploiting this shape of the points in the projections space, BLF can be used to select optimal clustering parameters such as the number of clusters and eventually the kernel hyper-parameters. In particular the BLF is defined in the following way:
\begin{equation}
\label{BLF}
\textrm{BLF}(\mathcal{D}^V,k)= \eta\textrm{linefit}(\mathcal{D}^V,k)+ (1-\eta)\textrm{balance}(\mathcal{D}^V,k) 
\end{equation}
where $\mathcal{D}^V$ represents the validation set and $k$ as usual indicates the number of clusters. The linefit index equals $0$ when the score variables are distributed spherically and equals $1$ when the score variables are collinear, representing points in the same cluster. The balance index equals $1$ when the clusters have the same number of elements and tends to $0$ in extremely unbalanced cases. The parameter $\eta$ controls the importance given to the linefit with respect to the balance index and takes values in the range $[0,1]$. Thus, for instance BLF reaches its maximum value $1$ in case of well distinct clusters of the same size if $\eta = 0.5$.   

Extensive experiments have shown the usefulness of BLF for model selection. However, some drawbacks have been observed:
\begin{itemize}
\item often the criterion is biased toward $k=2$ clusters
\item it is not clear how to choose $\eta$  
\item it fails in case of large overlap between the clusters
\item it is more suited for data points than network data.
\end{itemize}   
In sections \ref{sc:SKSC} and \ref{sc:MS_community_detection} we will show how two new model selection criteria introduced in \cite{rocco_IJCNN2013} and \cite{rocco_IJCNN2011} can solve these difficulties and then be used as a valid alternative to BLF.           

\subsection{Generalization}
\label{sc:OOSE_KSC}
Spectral clustering methods provide a clustering only for the given training data without a clear extension to test points, as discussed in \cite{sc_nystrom}. Moreover, the out-of-sample technique proposed therein consists of applying the Nystr\"om method \cite{nystrom_book} in order to give an embedding for the test points by approximating the underlying eigenfunction. In \cite{vanbarel_ICD} the information related to the pivots selected by the Incomplete Cholesky Decomposition (ICD) allows to compute cluster memberships for out-of-sample points. Other similar numerical techniques are used in \cite{Ning_PR} and \cite{Dhanjal_arXiv} as a solution to using spectral clustering for large scale applications. On the other hand, the extension proposed in KSC is model-based, in the sense that  the out-of-sample points are mapped onto the eigenvectors found in the training phase:
\begin{equation}
\label{OOSE}
e_{\textrm{test}}^{(l)} = \Omega_{\textrm{test}}\alpha^{(l)} + b_l1_{\textrm{Ntest}} 
\end{equation}
where $\Omega_\textrm{test}$ is the $N_\textrm{test}\times N$ kernel matrix evaluated using the test points with entries $\Omega_\textrm{test,ri} = K(x_r^\textrm{test},x_i)$, $r = 1,\hdots,N_\textrm{test}$, $i = 1,\hdots,N_{\textrm{Tr}}$. The cluster indicators can be obtained by binarizing the score variables. As for the training nodes, the memberships are assigned by comparing these indicators with the code-book and selecting the nearest prototype based on Hamming distance. This scheme corresponds to an ECOC (Error Correcting Output Codes) decoding procedure.
 
To conclude, the LS-SVM framework in which KSC has been designed allows to train, validate and test the clustering model in an unsupervised learning scheme.

\section{Soft Kernel Spectral Clustering}
\label{sc:SKSC}

\subsection{Overview}
Most of the clustering methods performs only hard clustering, where each item is assigned to only one group. However, this works fine when the clusters are compact and well separated, while the performance can decrease dramatically when they overlap. Since this is the case in many real-world scenarios, soft or fuzzy clustering is becoming popular in many fields \cite{MKL_fuzzy,softclu_graphs}. In soft clustering each object belongs to several groups at the same time, with a different degree of membership. This is desirable not only to cope in a more effective way with overlapping clusters, but the uncertainties on data-to-clusters assignments help also to improve the overall interpretability of the results. 

In what follows we describe a novel algorithm for fuzzy clustering named soft kernel spectral clustering (SKSC) \cite{rocco_IJCNN2013}. SKSC is characterized by the same core model as KSC, but it is provided with a different assignment rule allowing soft cluster memberships. A first attempt to provide a sort of probabilistic output in KSC was already done in \cite{highlysparse}. However, in the cited work the underlying assumption is that there is few overlap between the clusters. On the other hand SKSC can handle cases where a large amount of overlap between clusters is present. Moreover, SKSC uses a new method to tune the number of clusters and the kernel hyper-parameters based on the soft assignment. This model selection technique is called Average Membership Strength (AMS) criterion. The latter can solve the issues arising with BLF mentioned in section \ref{sc:BLF}. In fact, unlike BLF, AMS is not biased toward two clusters, does not have any parameter to tune and can be used in an effective way also with overlapping clusters. 

\subsection{Algorithm}
\label{sc:algSKSC}
The main idea behind soft kernel spectral clustering is to use KSC as an initialization step in order to find a first division of the training data into clusters. Then this grouping is refined by re-calculating the prototypes in the score variables space, and the cluster assignments are performed by means of the cosine distance between each point and the prototypes. This allows also to obtain highly sparse models as explained in \cite{mall_PREMI}, where a possible alternative to reduced set methods (see \cite{raghvendra_IJCNN2014}) is proposed.  

\begin{figure}[htbp]
\centering
\begin{tabular}{c}
\begin{psfrags}
\psfrag{x}[t][t]{{}}
\psfrag{y}[b][b]{{}}
\psfrag{t}[b][b]{{}}
\includegraphics[width=2in]{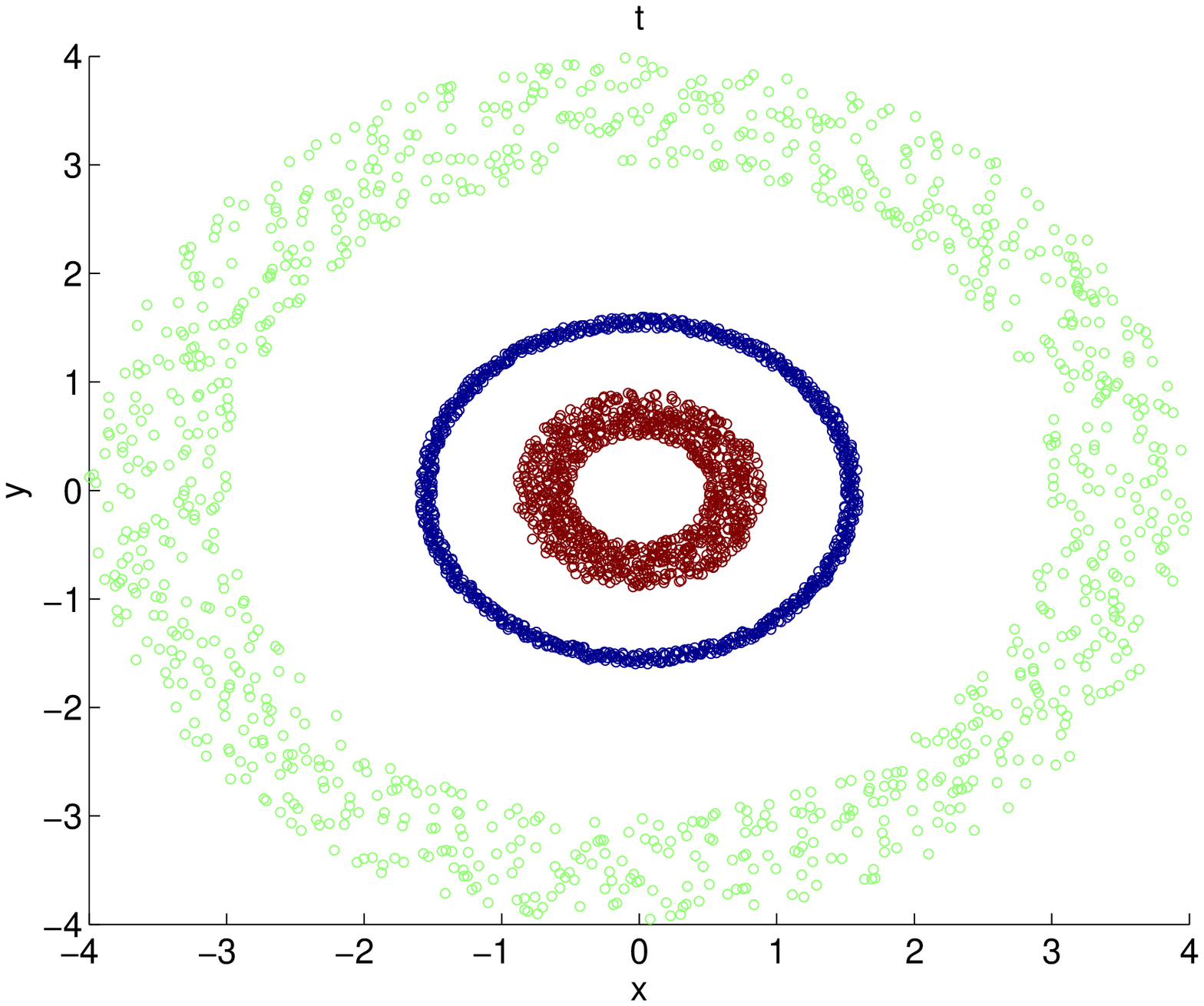}
\end{psfrags}\\
\begin{psfrags}
\psfrag{x}[t][t]{{$e^{(1)}$}}
\psfrag{y}[b][b]{{$e^{(2)}$}}
\psfrag{t}[b][b]{{}}
\includegraphics[width=2in]{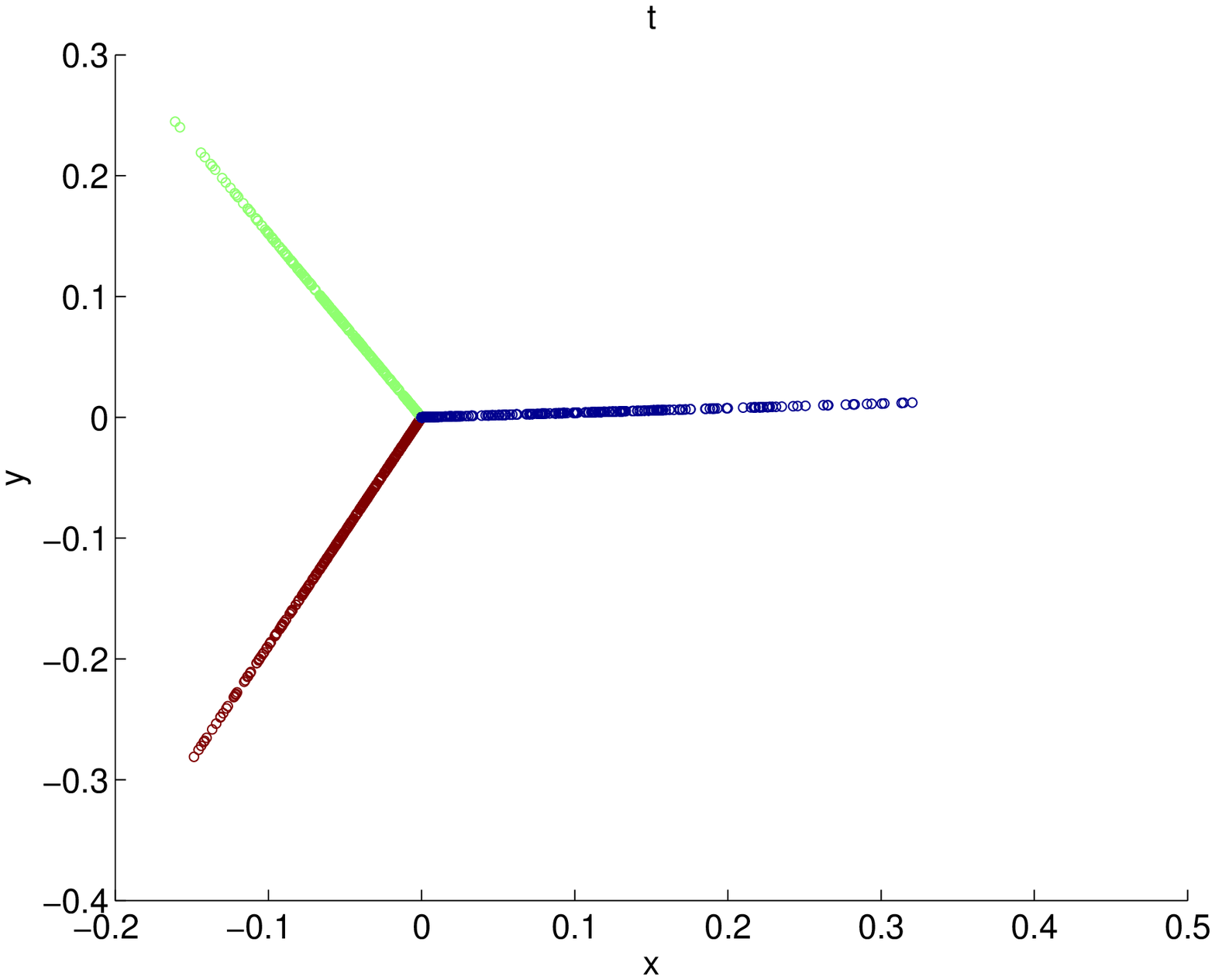}
\end{psfrags}\\
\begin{psfrags}
\psfrag{x}[t][t]{{$e^{(1)}$}}
\psfrag{y}[b][b]{{$e^{(2)}$}}
\psfrag{t}[b][b]{{}}
\includegraphics[width=2in]{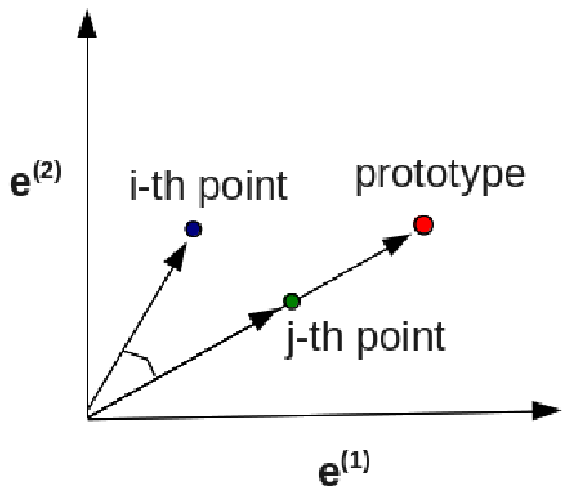}
\end{psfrags}
\end{tabular}
\caption{\textbf{Clusters as lines in the $e^{(l)}$ space}: example of how three distinct clusters ($3$ concentric rings on the top in a 2D space) appear in the projection space in the KSC formulation (center). On the bottom an illustration of the cosine distance in the score variable space is given: the $j$-th point is parallel to the prototype so its membership is $1$. This is not the case for the $i$-th point.}\label{fig_cluproj}
\end{figure}

As already pointed out in section \ref{sc:BLF}, in the projections/score variables space the points belonging to the same cluster appear aligned in the absence of overlap (see center of Figure \ref{fig_cluproj}). In this ideal situation of clear and well distinct groupings, any soft assignment should reduce to a hard assignment, where every point must belong to one cluster with membership $1$ and to the other clusters with membership $0$. In fact, the membership reflects the certainty with which we can assign a point to a cluster and it can be thought as a kind of subjective probability. In order to cope with this situation, the cosine distance from every point to the prototypes can be used as the basis for the soft assignment. In this way, in the perfect above-mentioned scenario, every point positioned along one line will be assigned to that cluster with membership or probability equal to $1$, since the cosine distance from the corresponding prototype is $0$, being the two vectors parallel (see bottom of Figure \ref{fig_cluproj}). %Moreover, next to the origin, 
Given the projections for the training points $e_i = [e_i^{(1)},\hdots,e_i^{(k-1)}]$, $i=1,\hdots,N_{\textrm{Tr}}$ and the corresponding hard assignments $q_i^p$ we can calculate for each cluster the new prototypes $s_1,\hdots,s_p,\hdots,s_k$, $s_p \in \mathbb{R}^{k-1}$ as:
\begin{equation}
\centering
\label{e_prototypes}
s_p = \frac{1}{n_p}\sum_{i=1}^{n_p}e_i 
\end{equation}             
where $n_p$ is the number of points assigned to cluster $p$ during the initialization step by KSC.
Then we can calculate the cosine distance between the $i$-th point in the score variables space and a prototype $s_p$ using the following formula:
\begin{equation}
\centering
\label{cosine_dist}
d^{\textrm{cos}}_{ip} = 1-e_i^Ts_p/(||e_i||_2||s_p||_2).
\end{equation}
The membership of point $i$ to cluster $q$ can be expressed as:
\begin{equation}
\centering
\label{soft_membership}
cm_i^{(q)} = \frac{\prod_{j\neq q}d^{\textrm{cos}}_{ij}}{\sum_{p=1}^k\prod_{j\neq p}d^{\textrm{cos}}_{ij}}
\end{equation}
with $\sum_{p=1}^{k}cm_i^{(p)} = 1$.             
As discussed in \cite{prob_clust_israel}, this membership is given as a subjective probability and it indicates the strength of belief in the clustering assignment.

\begin{figure}[htbp]
\centering
\begin{tabular}{c}
\begin{psfrags}
\psfrag{x}[t][t]{{}}
\psfrag{y}[b][b]{{}}
\psfrag{t}[b][b]{{}}
\includegraphics[width=3.5in]{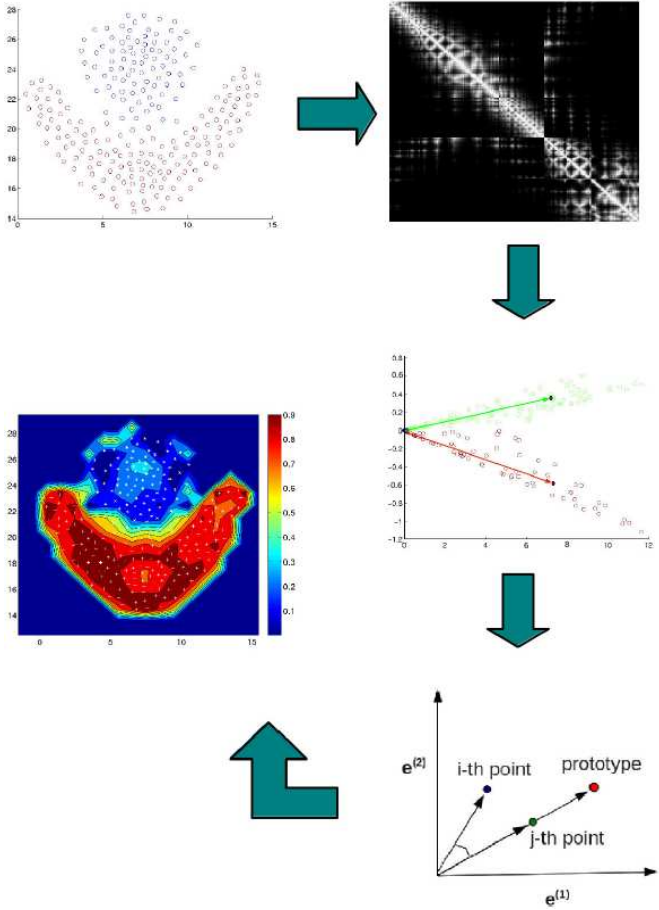}
\end{psfrags}
\end{tabular}
\caption{\textbf{SKSC algorithm}. The dataset consists of a set $\mathcal{D} = \{x_i\}_{i=1}^{N}$ where $x_i \in \mathbb{R}^2$, and relates to a binary clustering problem with nonlinear boundary. From the projections of the training points in the eigenspace of the Laplacian some cluster prototypes are computed. The projection of the test points are then assigned to the different clusters with a probability depending on the cosine distance from each of them.}\label{SKSC_Figure}
\end{figure}

The out-of-sample extension on unseen data consists of two steps:
\begin{enumerate}
\item project the test points onto the eigenspace spanned by $[\alpha^{(1)},\hdots,\alpha^{(k-1)}]$ using eq. (\ref{OOSE})
\item calculate the cosine distance between these projections and the training cluster prototypes, and then the corresponding soft assignment by means of eq. (\ref{soft_membership}).  
\end{enumerate}
The SKSC method is summarized in algorithm \ref{alg:SKSC}.
\begin{algorithm}[t]
\small
\LinesNumbered
\KwData{Training set $\mathcal{D}_{\textrm{Tr}} = \{x_i\}_{i=1}^{N_{\textrm{Tr}}}$ and test set $\mathcal{D}_{\textrm{test}} = \{x_m^{\textrm{test}}\}_{m=1}^{N_{\textrm{test}}}$, kernel function $K: \mathbb{R}^d \times \mathbb{R}^d \rightarrow \mathbb{R}$ positive definite and localized ($K(x_i,x_j) \rightarrow 0$ if $x_i$ and $x_j$ belong to different clusters), kernel parameters (if any), number of clusters $k$.}
\KwResult{Clusters $\{\mathcal{A}_1,\hdots,\mathcal{A}_p,\hdots,\mathcal{A}_k\}$, soft cluster memberships $cm^{(p)}, p =1,\hdots,k$, cluster prototypes $\mathcal{SP} = \{s_p\}_{p = 1}^k$, $s_p \in \mathbb{R}^{k-1}$.} 
Initialization by solving eq. (\ref{dualModel}).\\ 
Compute the new prototypes $s_1,\hdots,s_k$ (eq. (\ref{e_prototypes})).\\ 
Calculate the test data projections $e_m^{(l)}$, $m = 1,\hdots,N_{\textrm{test}}$, $l = 1,\hdots,k-1$.\\
Find the cosine distance between each projection and all the prototypes (eq. (\ref{cosine_dist}))  
$\forall m$, assign $x_m^{\textrm{test}}$ to cluster $A_{p}$ with membership $cm^{(p)}$ according to eq. (\ref{soft_membership}).  
\caption{SKSC algorithm \cite{rocco_IJCNN2013}}
\label{alg:SKSC}
\end{algorithm}
The main steps of this technique are depicted in figure \ref{SKSC_Figure}.

%\begin{comment}
%\end{comment}
 
\subsection{Model Selection}
\label{AMS_algorithm}
From the soft assignment technique explained in the previous section a new model selection method can be derived. In fact, we can calculate a kind of mean membership per cluster indicating the average degree of belonging of the points to that cluster. If we do the same for every cluster and we take the mean, we have what we call the Average Membership Strength (AMS) criterion\footnote{Future work can be related to experiment with a weighted version of the AMS criterion. The weights could be chosen, for instance, to emphasize the contribution of points with highest membership, in order to reduce the influence of the noise.}:
\begin{equation}
\centering
\label{ams}
\textrm{AMS} = \frac{1}{k}\sum_{p=1}^{k}\frac{1}{n_p}\sum_{i=1}^{n_p}cm_i^{(p)}.
\end{equation}             
For $k=2$ the Euclidean distance between the points and the prototypes is used\footnote{In this case the cosine distance is not discriminative because the points lie on a line.}, while for $k>2$ we consider the cosine distance. 
Unlike BLF, AMS does not have any parameter to tune, it is not biased toward $k=2$ and as we will show in the experimental section it can be used in an effective way also in case of overlapping clusters. The model selection procedure is described in algorithm \ref{alg:MS_AMS}. 
\begin{algorithm}[t]
\small
\LinesNumbered
\KwData{training and validation sets, kernel function $K: \mathbb{R}^d \times \mathbb{R}^d \rightarrow \mathbb{R}$ positive definite and localized.}
\KwResult{selected number of clusters $k$, kernel parameters (if any).} 
Define a grid of values for the parameters to select\\
Train the related SKSC model using the training set\\ 
Compute the soft cluster assignments for the points of the validation set\\
For every partition calculate the related $\textrm{AMS}$ score using eq. (\ref{ams})\\
Choose the model with the highest score.
\caption{AMS model selection algorithm for SKSC \cite{rocco_IJCNN2013}}
\label{alg:MS_AMS}
\end{algorithm}

\subsection{Toy Examples}
\label{sec_toy}
In this section some experimental examples with artificial data are presented to illustrate the proposed soft clustering technique. We compare SKSC with KSC in terms of Adjusted Rand Index (ARI) \cite{arindex}, mean Silhouette value (MSV) \cite{silhouette} and Davies-Bouldin index (DBI) \cite{davies_bouldin} (see appendix\footnote{A summary of the main cluster quality measures used in this thesis will be given in the appendix.}), and AMS with BLF. In general we observe that SKSC achieves better results than KSC and the AMS criterion appears to be more useful than BLF for model selection, mostly when a big amount of overlap between clusters is present.      

\begin{figure}[htbp]
\centering
\begin{tabular}{c}
\begin{psfrags}
\psfrag{x}[t][t]{{}}
\psfrag{y}[b][b]{{}}
\psfrag{t}[b][b]{{No Overlap}}
\includegraphics[width=2.5in]{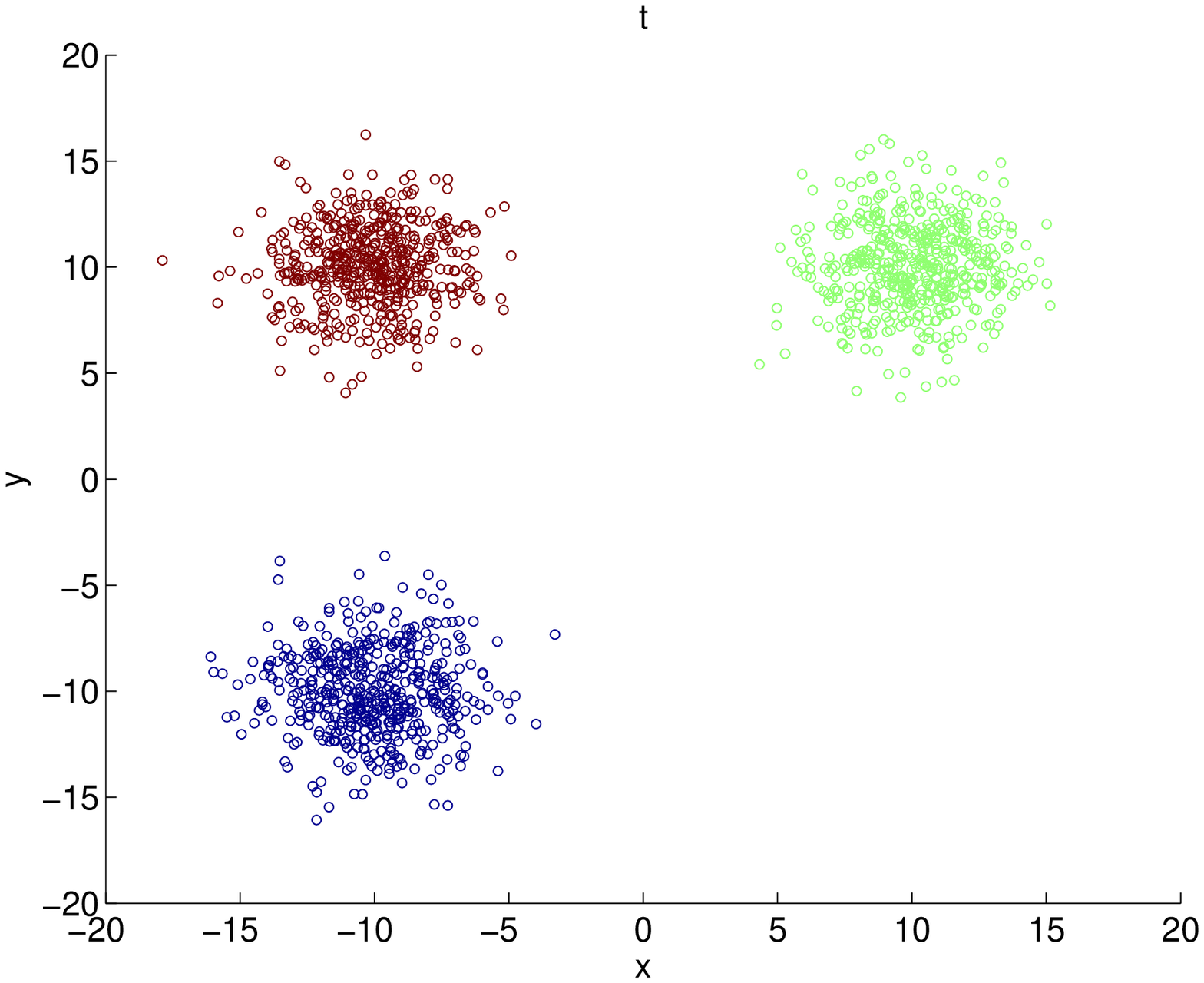}
\end{psfrags}\\
\begin{psfrags}
\psfrag{x}[t][t]{{}}
\psfrag{y}[b][b]{{}}
\psfrag{t}[b][b]{{Few Overlap}}
\includegraphics[width=2.5in]{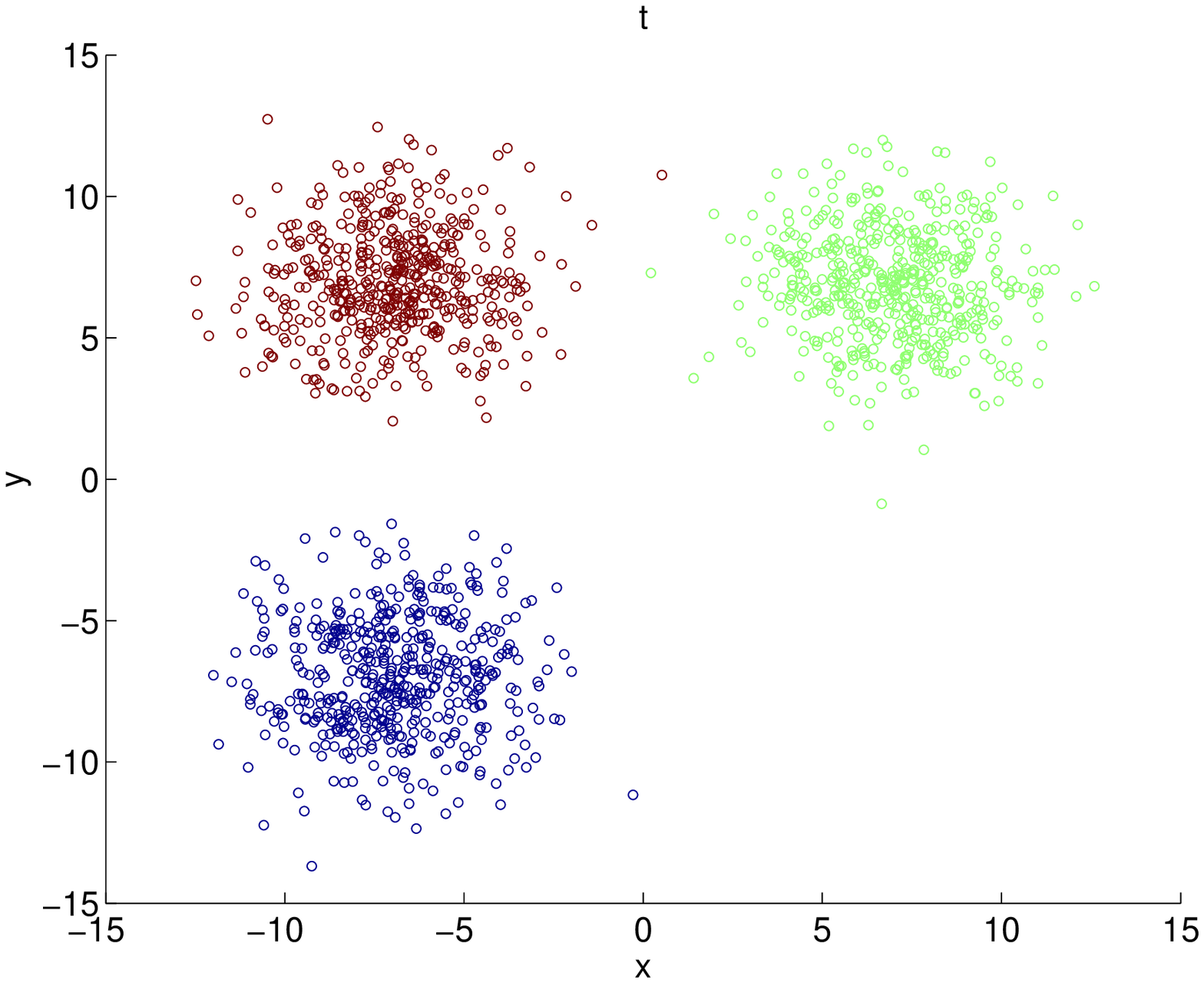}
\end{psfrags}\\
\begin{psfrags}
\psfrag{x}[t][t]{{}}
\psfrag{y}[b][b]{{}}
\psfrag{t}[b][b]{{Large Overlap}}
\includegraphics[width=2.5in]{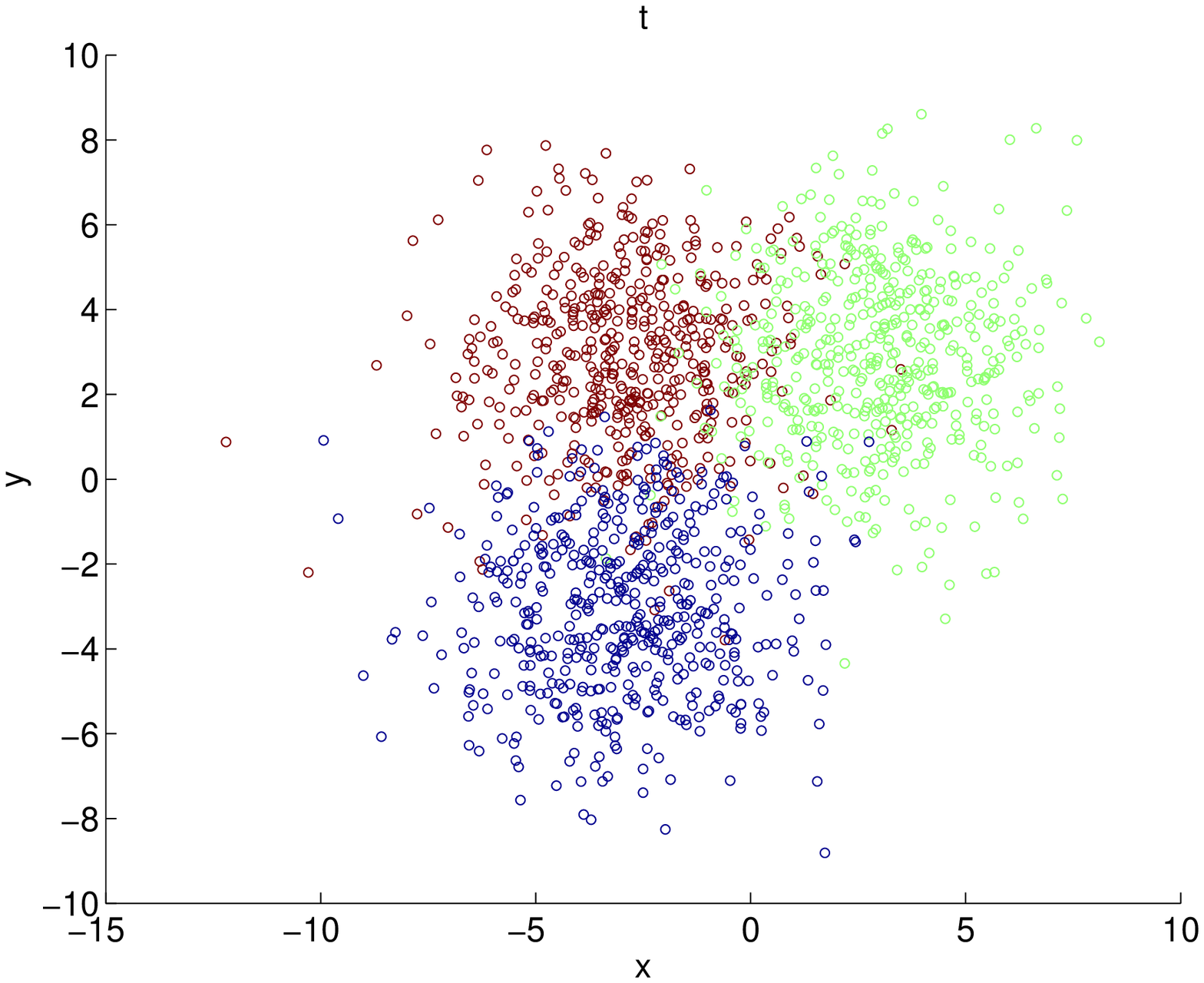}
\end{psfrags}
\end{tabular}
\caption{\textbf{Three Gaussians dataset}. The dataset consists of three sets of $N$ points in a 2D space ($x_i \in \mathbb{R}^2, i=1,\hdots,N$) representing Gaussian mixtures with different amount of overlap. }\label{fig_Gaussian}
\end{figure}

We have designed an experiment with the goal of understanding the role of the overlap rather than the nonlinearities to affect the clustering performances. Thus, synthetic data consisting of $1000$ data points have been generated from a mixture of three 2D Gaussians, with $500$ samples drawn from each component of the mixture. We then consider different amounts of overlap between these clouds of points, as depicted in figure \ref{fig_Gaussian} from top to bottom. 

\begin{figure}[htbp]
\centering
\begin{tabular}{c}
\begin{psfrags}
\psfrag{x}[t][t]{{Number of clusters k}}
\psfrag{y}[b][b]{{}}
\psfrag{t}[b][b]{{No Overlap}}
\includegraphics[width=2.5in]{no_overlap_MStogether.eps}
\end{psfrags}\\\\
\begin{psfrags}
\psfrag{x}[t][t]{{Number of clusters k}}
\psfrag{y}[b][b]{{}}
\psfrag{t}[b][b]{{Few Overlap}}
\includegraphics[width=2.5in]{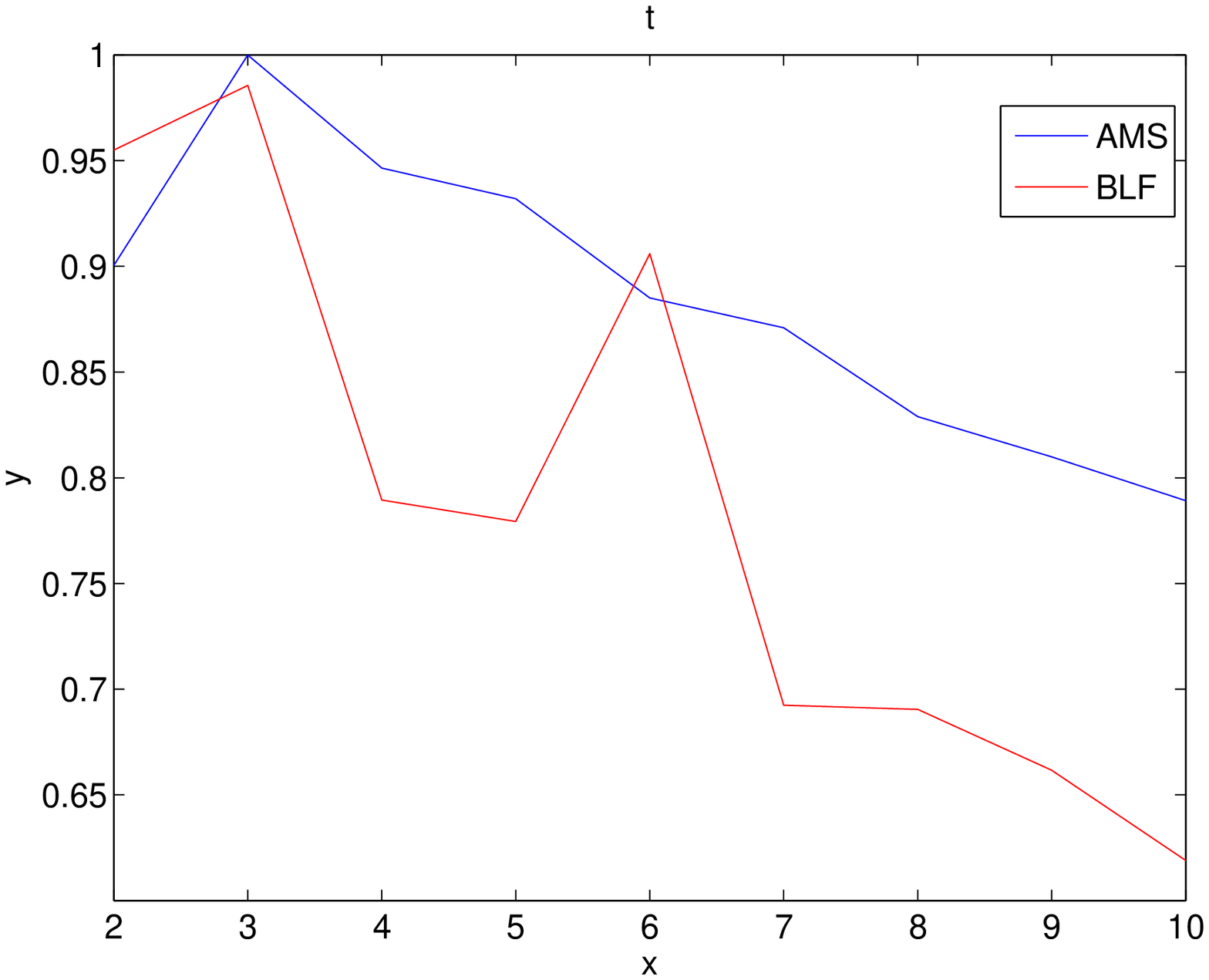}
\end{psfrags}\\\\
\begin{psfrags}
\psfrag{x}[t][t]{{Number of clusters k}}
\psfrag{y}[b][b]{{}}
\psfrag{t}[b][b]{{Large Overlap}}
\includegraphics[width=2.5in]{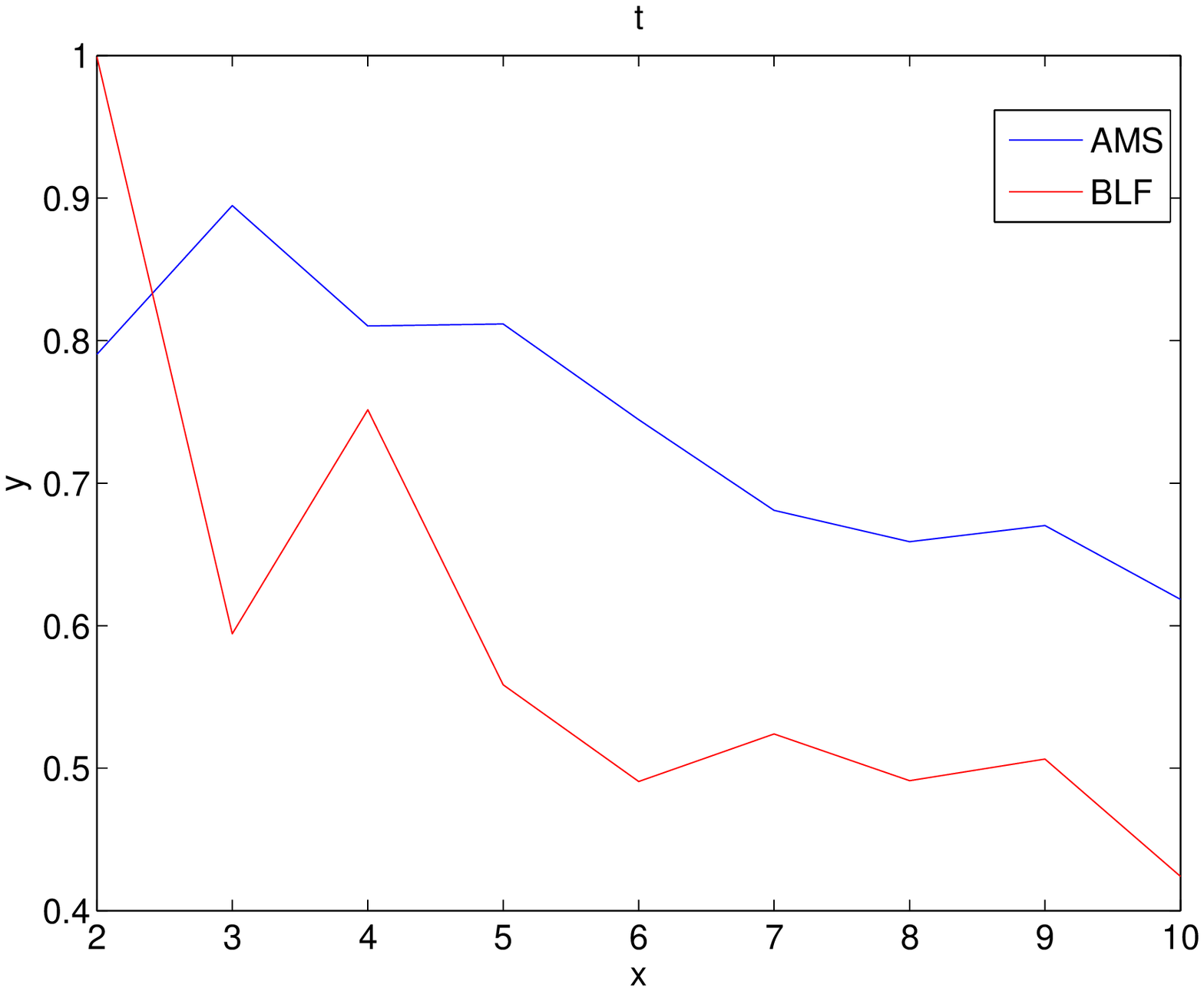}
\end{psfrags}
\end{tabular}
\caption{\textbf{Model selection for the three Gaussians dataset}. The true number of clusters is $k=3$. Two model selection criteria, namely AMS and BLF, are contrasted. We can notice how the former can better deal with overlapping clusters.}\label{fig_GaussianMS}
\end{figure}

\begin{figure}[htbp]
\centering
\begin{tabular}{c}
\begin{psfrags}
\psfrag{x}[t][t]{{}}
\psfrag{y}[b][b]{{}}
\psfrag{t}[b][b]{{$p_1$}}
\includegraphics[width=2.5in]{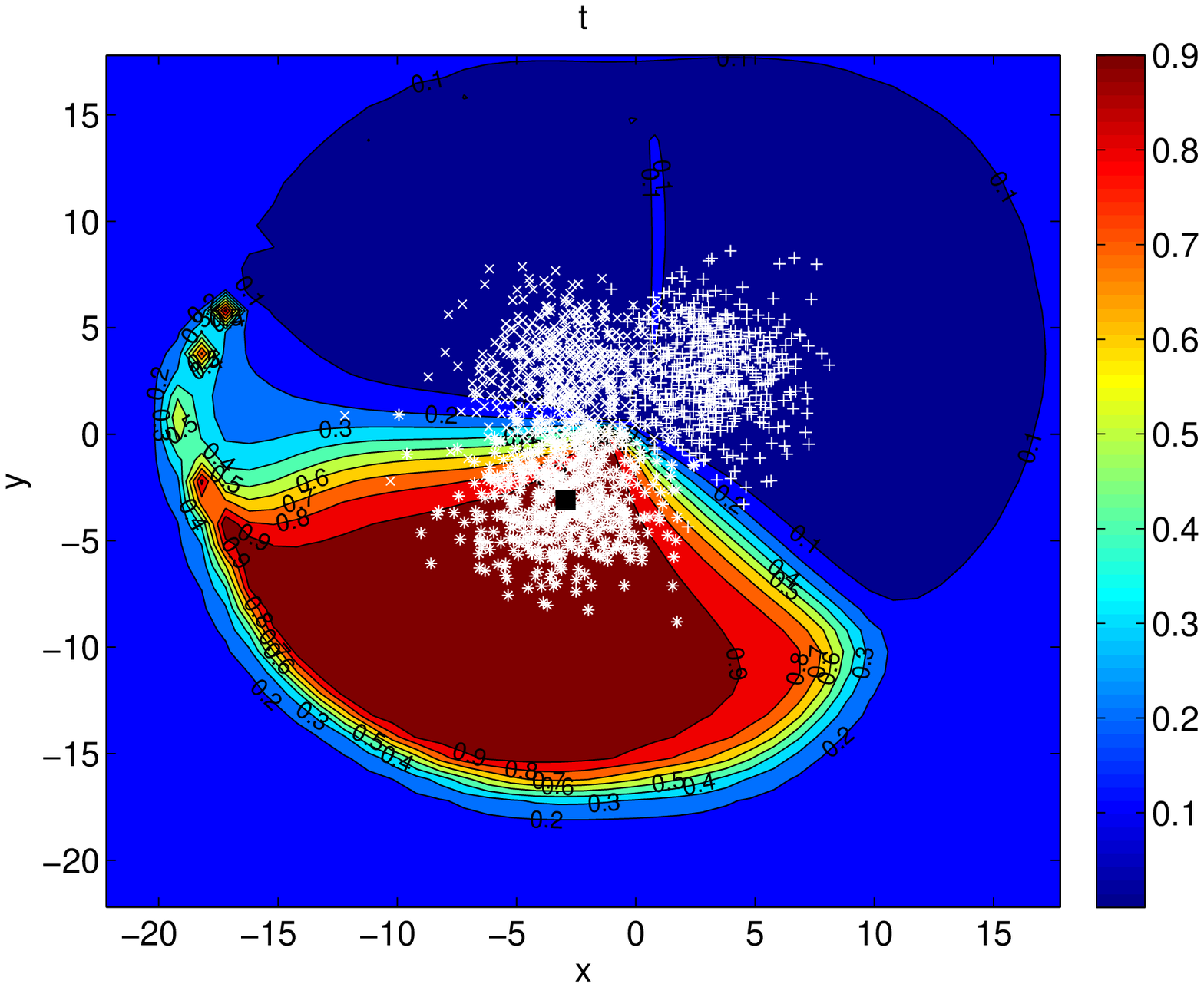}
\end{psfrags}\\
\begin{psfrags}
\psfrag{x}[t][t]{{}}
\psfrag{y}[b][b]{{}}
\psfrag{t}[b][b]{{$p_2$}}
\includegraphics[width=2.5in]{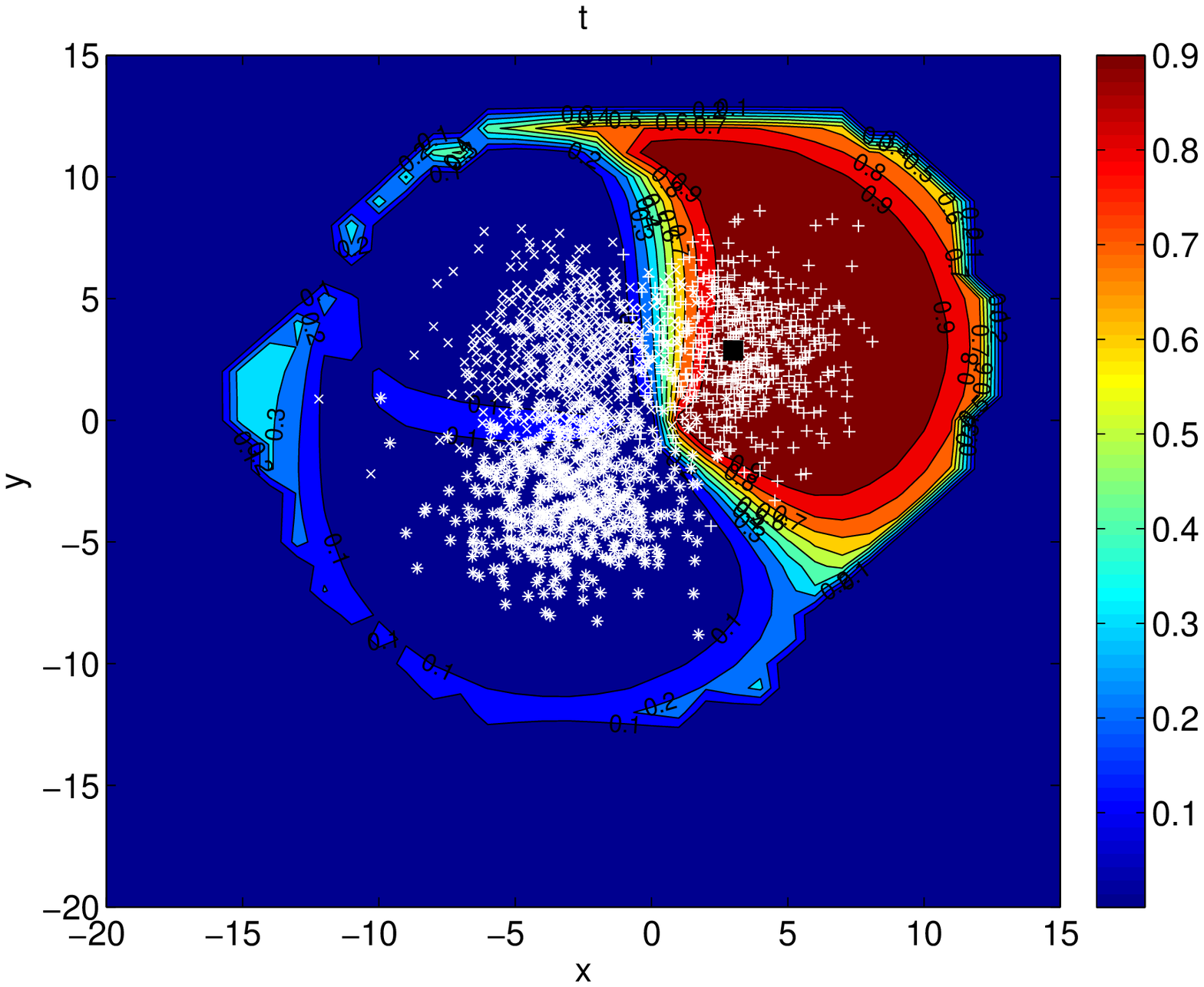}
\end{psfrags}\\
\begin{psfrags}
\psfrag{x}[t][t]{{}}
\psfrag{y}[b][b]{{}}
\psfrag{t}[b][b]{{$p_3$}}
\includegraphics[width=2.5in]{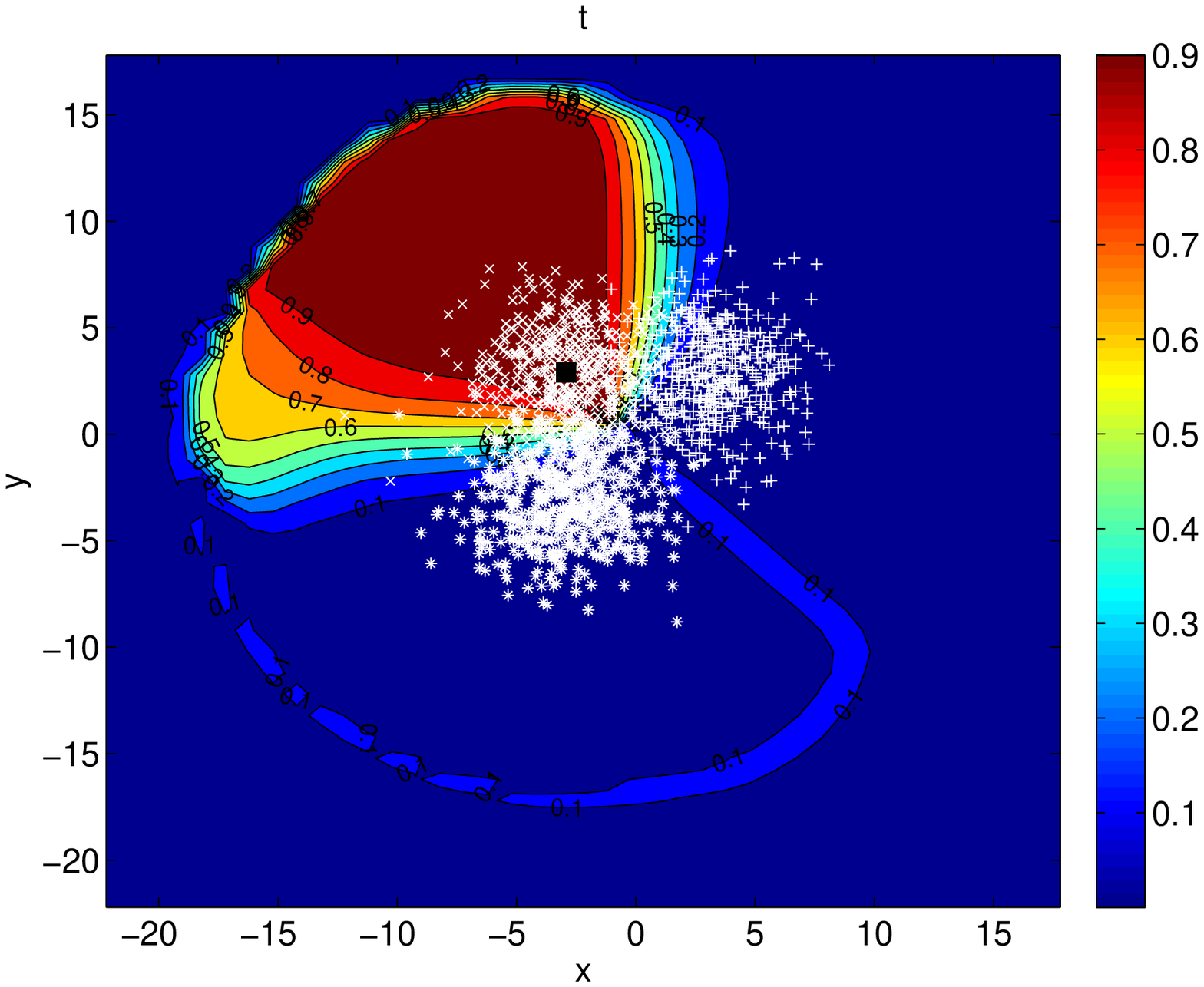}
\end{psfrags}
\end{tabular}
\caption{\textbf{Soft clustering results for three Gaussians dataset with large overlap}: probability to belong to cluster $1$ ($p_1$, top), to cluster $2$ ($p_2$, center) and to cluster $3$ ($p_3$, bottom) detected by SKSC for the large overlap case. We can notice that all the probabilities reach their maximum value around the centroids and then decrease in the overlapping region.}\label{fig_oos_Gaussian}
\end{figure}

In figure \ref{fig_GaussianMS} we show the model selection plots for these three cases. We can see that AMS, even in case of large overlap, is able to correctly identify the presence of the three distinct Gaussian clouds. This is not the case for the BLF criterion. In tables \ref{table_GaussianS}, \ref{table_GaussianM}, \ref{table_GaussianL} the clustering results of KSC and SKSC (when fed with correct parameters) are evaluated. We can notice that SKSC outperforms KSC when the overlap between the clusters increases. Finally, in figure \ref{fig_oos_Gaussian} the soft clustering results produced by SKSC for the large overlap case are depicted. As we would expect, the probability to belong to each of the $3$ clusters decreases in the overlapping regions.  

\begin{table}[htbp]
\begin{center}
\begin{tabular}{|c|c|c|} 
\hline 
\textbf{}  & \textbf{KSC} &  \textbf{SKSC} \\
\hline
\textbf{MSV} & $0.96$ & $0.96$\\
\hline
\textbf{DBI} & $0.24$ & $0.24$\\
\hline
\textbf{ARI} & $1$  & $1$ \\
\hline
\end{tabular}
\end{center}
\caption{\textbf{Three Gaussians, no overlap}: cluster quality evaluation of KSC and SKSC based on Mean Silhouette Value (MSV), Davies-Bouldin Index (DBI), and Adjusted Rand Index (ARI). A summary of these and others cluster quality criteria is given in the Appendix.}
\label{table_GaussianS}
\end{table}

\begin{table}[htbp]
\begin{center}
\begin{tabular}{|c|c|c|} 
\hline 
\textbf{}  & \textbf{KSC} &  \textbf{SKSC} \\
\hline
\textbf{MSV} & $0.90$ & $\mathbf{0.91}$\\
\hline
\textbf{DBI} & $0.34$ & $\mathbf{0.33}$ \\
\hline
\textbf{ARI} & $0.96$  & $\mathbf{0.99}$ \\
\hline
\end{tabular}
\end{center}
\caption{\textbf{Three Gaussians, few overlap}: clustering results of KSC and SKSC according to Mean Silhouette Value, Davies-Bouldin Index, and Adjusted Rand Index (see Appendix). The best performances are in bold.}
\label{table_GaussianM}
\end{table}          

\begin{table}[htbp]
\begin{center}
\begin{tabular}{|c|c|c|} 
\hline 
\textbf{}  & \textbf{KSC} &  \textbf{SKSC} \\
\hline
\textbf{MSV} & $0.59$ & $\mathbf{0.64}$ \\
\hline
\textbf{DBI} & $0.79$  & $\mathbf{0.76}$ \\
\hline
\textbf{ARI} & $0.65$   & $\mathbf{0.74}$  \\
\hline
\end{tabular}
\end{center}
\caption{\textbf{Three Gaussians, large overlap}: cluster quality measures (see Appendix) for KSC and SKSC, with the best performance in bold.}
\label{table_GaussianL}
\end{table}

\subsection{Application: Image Segmentation}
Here we show the usage of SKSC for the image segmentation task. In computer vision, image segmentation is the process of partitioning a digital image into multiple sets of pixels, such that pixels in the same group share certain visual characteristics. 

We randomly selected $10$ color images from the Berkeley image dataset \cite{berkeley_image}. Before clustering the images, a pre-processing of the data was necessary. In particular, we computed a local color histogram with a $5 \times 5$ pixels window around each pixel using minimum variance color quantization of $8$ levels. Then, in order to compare the similarity between two histograms $h^{(i)}$ and $h^{(j)}$, the positive definite $\chi^2$ kernel $K(h^{(i)},h^{(j)}) = \exp(-\dfrac{\chi_{ij}^2}{\sigma_{\chi}^2})$ has been adopted \cite{sc_nystrom}. The symbol $\chi_{ij}^2$ denotes the $\chi_{ij}^2$ statistical test used to compare two probability distributions \cite{chi2test}, $\sigma_{\chi}$ as usual indicates the bandwidth of the kernel.           

\begin{figure}[htbp]
\centering
\begin{tabular}{c}
\begin{psfrags}
\psfrag{x}[t][t]{{}}
\psfrag{y}[b][b]{{}}
\psfrag{t}[b][b]{{}}
\includegraphics[width=3in]{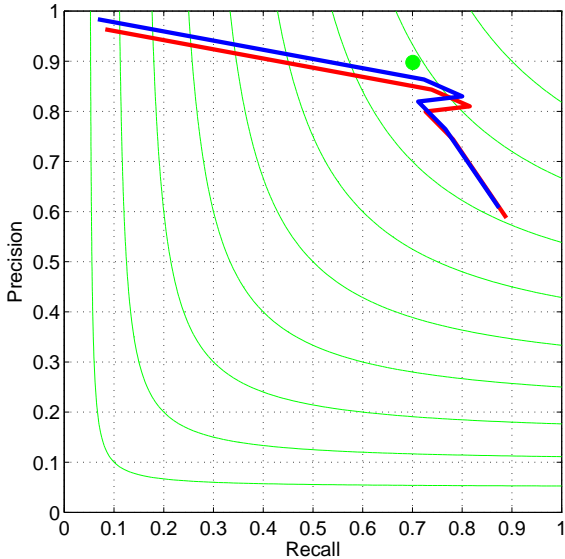}
\end{psfrags}
\end{tabular}
\caption{\textbf{F-measure plot}: evaluation of KSC (red) and SKSC (blue) algorithms with respect to human boundaries in the image segmentation task. Iso-F curves are shown in green. The green dot indicates the average agreement between the human subjects who performed the segmentations used as ground-truth.}\label{fig_F}
\end{figure}

\begin{figure}[htbp]
\centering
\begin{tabular}{cc}
\begin{psfrags}
\psfrag{x}[t][t]{{}}
\psfrag{y}[b][b]{{}}
\psfrag{t}[b][b]{{}}
\includegraphics[width=1.95in]{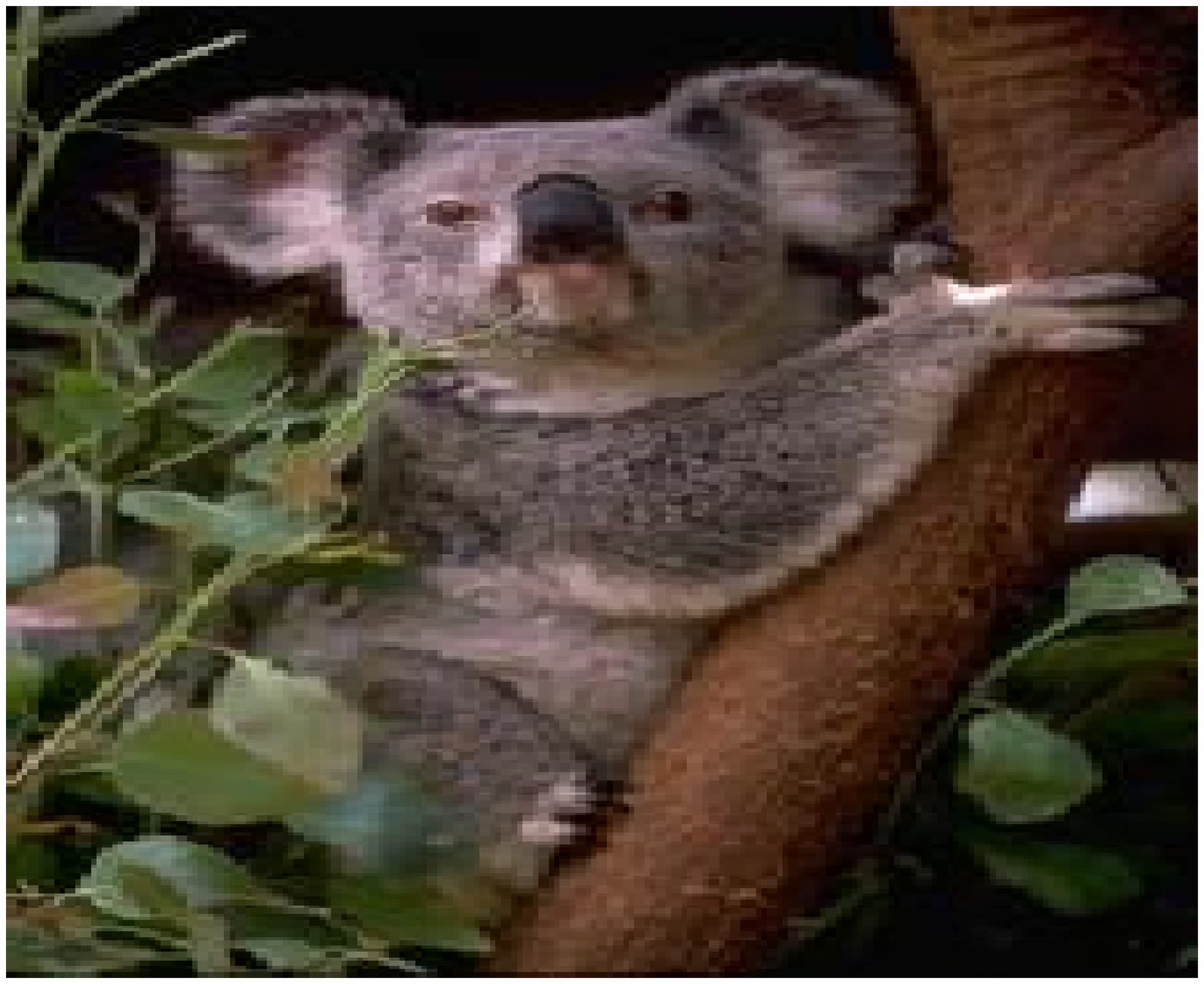}
\end{psfrags}&
\begin{psfrags}
\psfrag{x}[t][t]{{}}
\psfrag{y}[b][b]{{}}
\psfrag{t}[b][b]{{}}
\includegraphics[width=2in]{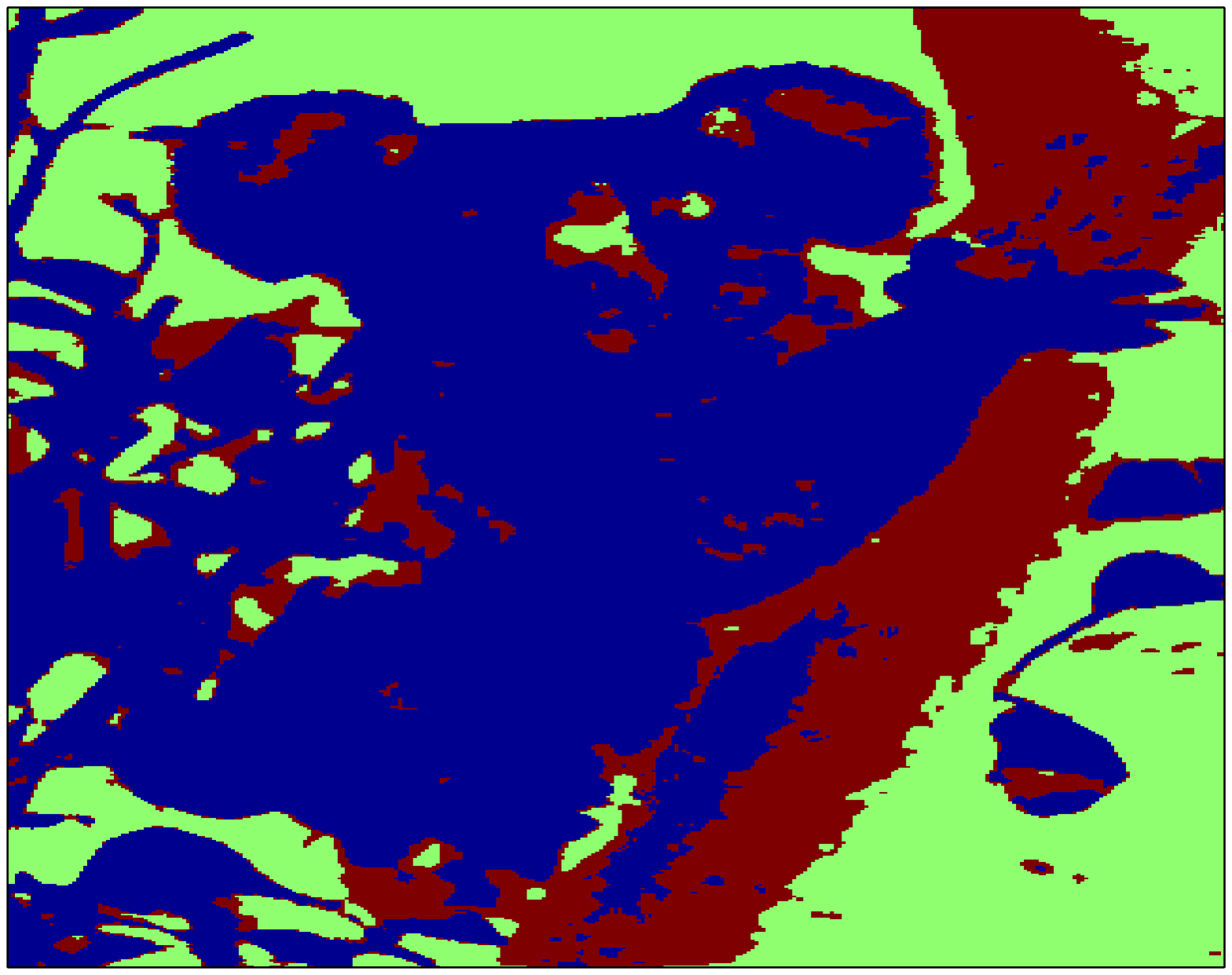}
\end{psfrags}\\
\begin{psfrags}
\psfrag{x}[t][t]{{}}
\psfrag{y}[b][b]{{}}
\psfrag{t}[b][b]{{}}
\includegraphics[width=1.9in]{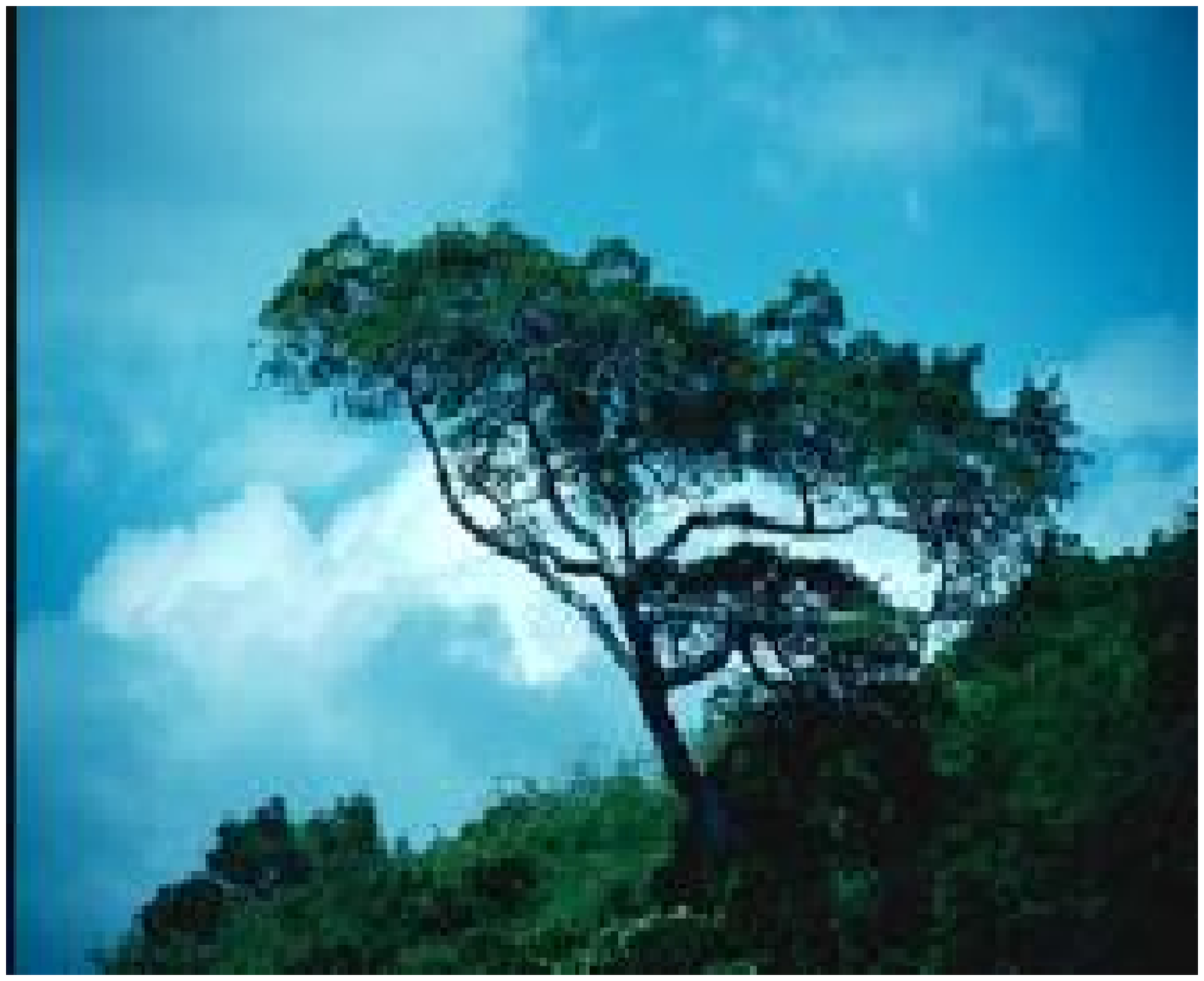}
\end{psfrags}&
\begin{psfrags}
\psfrag{x}[t][t]{{}}
\psfrag{y}[b][b]{{}}
\psfrag{t}[b][b]{{}}
\includegraphics[width=2in]{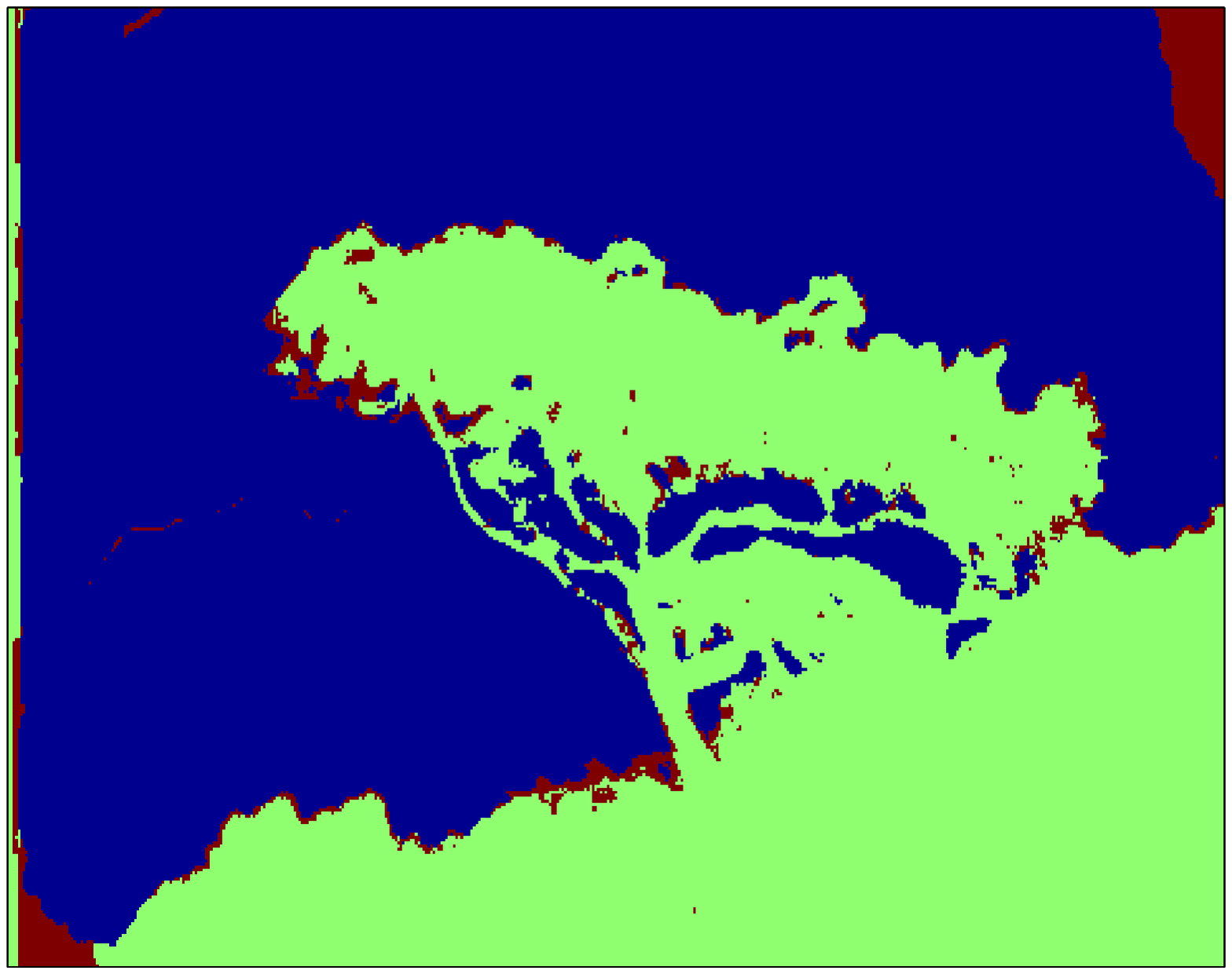}
\end{psfrags}\\
\begin{psfrags}
\psfrag{x}[t][t]{{}}
\psfrag{y}[b][b]{{}}
\psfrag{t}[b][b]{{}}
\includegraphics[width=1.95in]{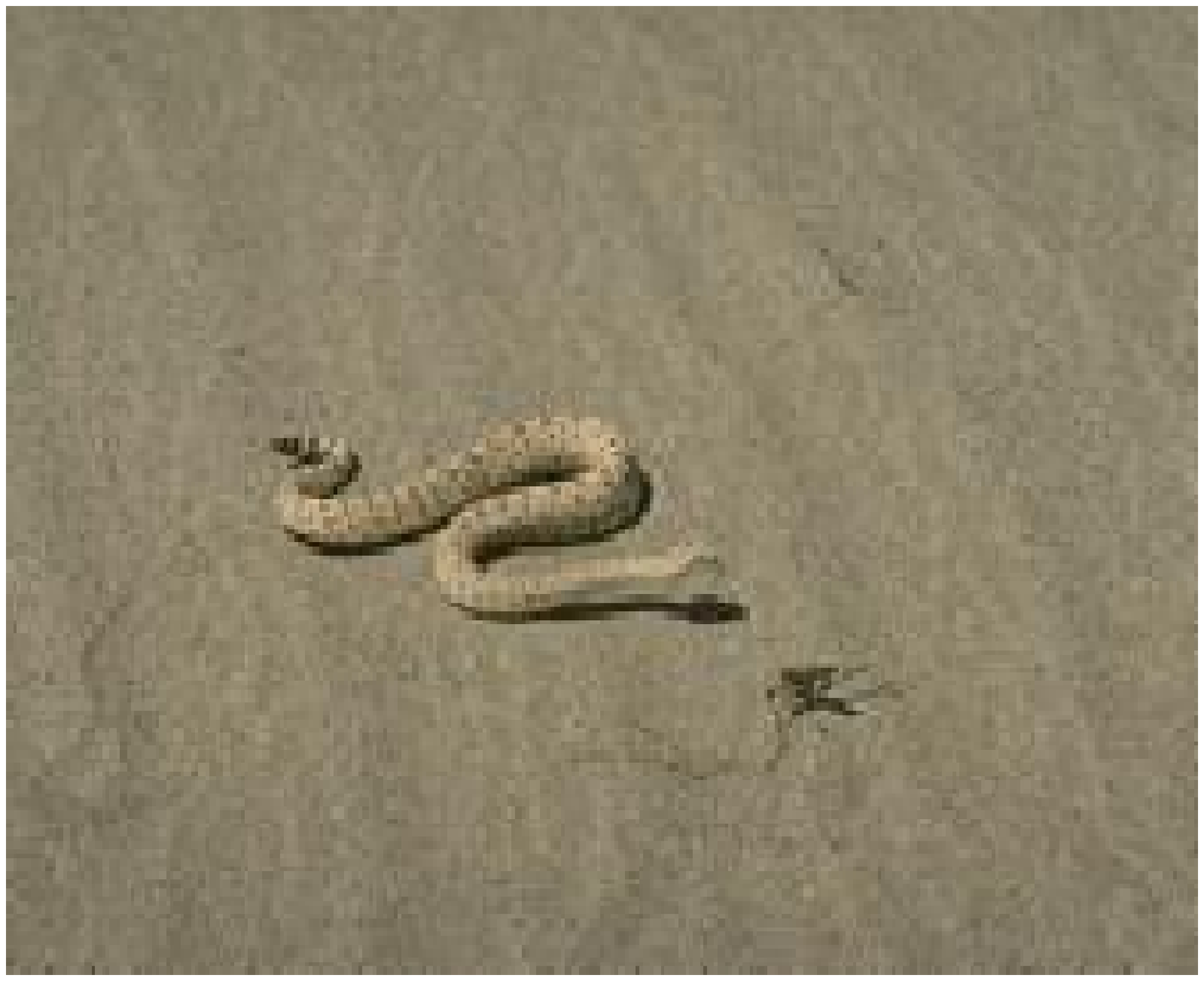}
\end{psfrags}&
\begin{psfrags}
\psfrag{x}[t][t]{{}}
\psfrag{y}[b][b]{{}}
\psfrag{t}[b][b]{{}}
\includegraphics[width=2in]{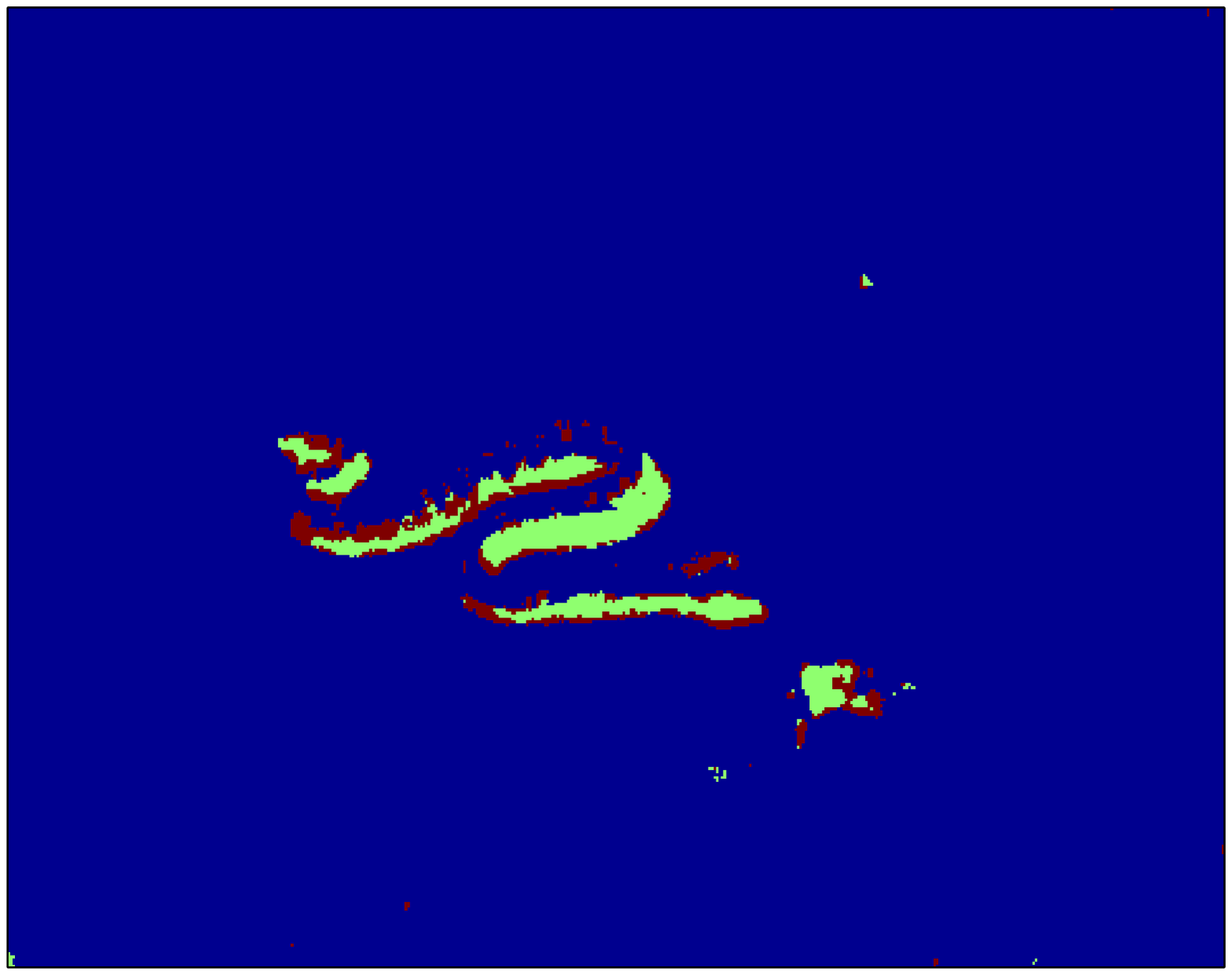}
\end{psfrags}
\end{tabular}
\caption{\textbf{SKSC segmentations}: original images (left) and SKSC segmentations (right) for image IDs $69015$, $147091$, $196073$. }\label{res_SKSC}
\end{figure}

A randomly selected subset of $1000$ pixels forms the training set and the whole image is the test set, in order to obtain the final segmentation. After properly selecting the number of clusters and the kernel bandwidth for each image, we compared the segmentation obtained by KSC and SKSC by using three evaluation measures \cite{amfm_pami2011}:
\begin{itemize}
\item F-measure ($\frac{2 \cdot Precision \cdot Recall}{Precision+Recall}$) with respect to human ground-truth boundaries
\item Variation of information (VI): it measures the distance between two segmentations in terms of their average conditional entropy. Low values indicate good match between the segmentations.
\item Probabilistic Rand Index (PRI): it operates by comparing the congruity of assignments between pairs of elements in the clusters found by an algorithm and multiple ground-truth segmentations.  
\end{itemize}     
The segmentation results are shown in figure \ref{fig_F} (F-measure) and tables \ref{table_VI} (VI) and \ref{table_PRI} (PRI). 
For comparison purposes, the soft partitioning obtained by SKSC is transformed in hard clustering results by assigning the pixels to the closest prototype. In this case the performances of SKSC and KSC are overall similar, most probably due to the fact that the images are characterized by clear segments (i.e. small overlap). Finally, in figure \ref{res_SKSC} the segmentations accomplished by SKSC on three selected images are depicted for illustration purpose.

\section{Conclusions}
In this chapter, after introducing the basics of spectral partitioning, we have discussed the LS-SVM modelling framework and we summarized the KSC algorithm. The latter is a spectral clustering model formulated as weighted kernel PCA in a primal-dual optimization setting. This formulation allows a simple extension of the model to out-of-sample points and a systematic model selection procedure based on the BLF criterion. Then, we presented the soft kernel spectral clustering (SKSC) technique, which is based on KSC but it uses a fuzzy assignment rule. The latter depends on the cosine distance of every point from the cluster prototypes in the projections space. A new model selection technique derived from this soft assignment, namely the average membership strength (AMS) criterion, is also proposed. We showed how AMS solves the main drawbacks of BLF and can be used more effectively in presence of overlap between clusters. We illustrated on toy data and an image segmentation problem, that SKSC outperforms KSC mainly in the most difficult tasks (i. e. when clusters overlap to a large extent). Moreover, SKSC produces more interpretable results due to its fuzzy nature. 

\begin{table}[htbp]
\begin{center}
\begin{tabular}{|c|c|c|} 
\hline 
\textbf{Image ID}  & \textbf{KSC} &  \textbf{SKSC} \\
\hline
$145086$ & $3.11$ & $\mathbf{3.10}$ \\
\hline
$42049$ & $3.42$  & $\mathbf{3.41}$ \\
\hline
$167062$ & $1.84$  & $\mathbf{1.69}$ \\
\hline
$147091$ & $1.10$  & $1.10$ \\
\hline
$196073$ & $0.18$  & $0.18$ \\
\hline
$62096$ & $0.35$  & $\mathbf{0.31}$ \\
\hline
$101085$ & $2.77$  & $\mathbf{2.69}$ \\
\hline
$69015$ & $2.36$  & $2.36$ \\
\hline
$119082$ & $0.90$  & $\mathbf{0.89}$ \\
\hline
$3096$ & $2.97$  & $2.97$ \\
\hline
\end{tabular}
\end{center}
\caption{\textbf{VI criterion}: comparison of KSC and SKSC segmentations in terms of variation of information (the lower the better). Best performance in bold. The algorithms have a similar performance.}
\label{table_VI}
\end{table}

\begin{table}[htbp]
\begin{center}
\begin{tabular}{|c|c|c|} 
\hline 
\textbf{Image ID}  & \textbf{KSC} &  \textbf{SKSC} \\
\hline
$145086$ & $0.59$ & $0.59$ \\
\hline
$42049$ & $0.70$  & $\mathbf{0.71}$ \\
\hline
$167062$ & $\mathbf{0.83}$  & $0.77$ \\
\hline
$147091$ & $0.83$  & $0.83$ \\
\hline
$196073$ & $0.98$  & $0.98$ \\
\hline
$62096$ & $\mathbf{0.95}$  & $0.94$ \\
\hline
$101085$ & $0.37$  & $\mathbf{0.38}$ \\
\hline
$69015$ & $0.58$  & $0.58$ \\
\hline
$119082$ & $0.87$  & $0.87$ \\
\hline
$3096$ & $0.19$  & $0.19$ \\
\hline
\end{tabular}
\end{center}
\caption{\textbf{PRI criterion}: comparison of KSC and SKSC segmentations in terms of probabilistic rand index (the higher the better). Best performance in bold. In general, the methods produce a similar outcome.}
\label{table_PRI}
\end{table}

%%%%%%%%%%%%%%%%%%%%%%%%%%%%%%%%%%%%%%%%%%%%%%%%%%
% Keep the following \cleardoublepage at the end of this file, 
% otherwise \includeonly includes empty pages.
\cleardoublepage

% vim: tw=70 nocindent expandtab foldmethod=marker foldmarker={{{}{,}{}}}
%

%
  \graphicspath{{\chaptersdir/chapter3/image/}}%
  %\cleardoublepage%
  \chapter{Community Detection in Complex Networks}
\label{ch:community_detection}

Community detection refers to the problem of partitioning a complex network into clusters of nodes with high density of edges, in order to understand its structure and function. A profusion of algorithms has been proposed in the recent past since many real-world applications arise in different fields, from biology to computational social sciences. In this chapter we propose a whole kernel-based framework for community detection. The main steps of our approach are the extraction of a sub-graph representative of the community structure of the entire network for training the clustering model, the validation stage by means of proper model selection criteria, and the membership assignments for the remainder of the nodes via out-of-sample extension. In contrast to most of the methods present in the literature which are rather specific, our technique is more flexible because in the model selection phase the user can provide the desired criterion, in order to obtain a final partitioning with certain characteristics. Moreover, the out-of-sample extension allows to readily assign the membership to new nodes joining the network without using heuristics. Also we can deal with weighted and unweighted networks by choosing the suitable kernel function, and graphs with overlapping communities can be analysed. Finally, by exploiting the high sparsity of the majority of the real graphs, the method can scale to large network data even on a desktop computer and can be easily parallelized.

\section{Related work}
In nature, science and technology complex systems are usually organized in networks of interactions between many components. Although every system has its peculiar properties, a common architecture can be recognized \cite{Simon62thearchitecture}. It seems that complex networks are organized in modules or communities, and this feature can be explained by considering two important aspects: resilience and adaptation. Many studies suggested that if a part of a complex system stops working properly, the modular organization allows to circumvent the problem, making it more resilient to failures. Moreover, the community structure helps such a system to adapt quickly to changes in the environment \cite{evol_origin_mod}.

A network (or graph) with $N$ nodes is characterized by a community structure when groups of nodes forming tightly connected units are only weakly linked to each other. Uncovering the natural communities present in a complex network has recently become a major topic in science and engineering with an interdisciplinary effort \cite{fortunatocommdet,newman_commdet,CNM,arenas_fernandez_gomez,Radicchi02032004}. This interest is due to the fact that revealing the modular structure of a network can be very useful for a number of reasons like visualizing large graphs composed of millions of nodes and edges, investigating the function of each module, discovering the role of the different nodes inside a  community. 

Many algorithms to handle the community detection problem have been proposed, with different characteristics. The most popular algorithm is that introduced in \cite{Girvan11062002}. The method is historically important because it popularized community detection among physicists and engineers and favoured the search of new algorithmic solutions. It is based on the removal of edges with high betweenness, which expresses the participation of an edge to the information flow on a graph. The connected components of the remaining network are the communities detected by the method. Even if quite appealing the algorithm is slow (it scales as $O(N^3)$) and cannot deal with overlapping communities, as each vertex is assigned to a single cluster. A large number of successful techniques is based on Modularity maximization, which  describes the strength of the community structure by comparing the actual graph with a random graph sharing the same degree distribution \cite{newmanmod2}. A well-known algorithm (with linear run-time $O(N)$) is the Louvain method \cite{louvainmethod}, based on a greedy optimization of the Modularity. The Modularity maxima found by the method are also better than those found with another well-known greedy technique used in \cite{CNM}. The most popular algorithm for discovering overlapping communities is the Clique Percolation Method (CPM) \cite{PallaCPM}, which is based on the idea that nodes inside a community are likely to form cliques, in contrast to inter-community edges. Another recent and very fast technique for overlapping community detection is \cite{bigclam}, built on a novel and counter-current observation that in many networks overlaps between communities are densely connected and can reveal the modules. In \cite{evans:2009} the authors use a partition of the links of a network (instead of clustering the nodes) to uncover its community structure. In this way their method can better detect nodes belonging to more than one community. A different class of methods are those based on information-theoretic ideas, such as the minimum-description-length method called Infomap introduced in \cite{infomap}. All the above-mentioned algorithms can discover clusters at a particular scale, and even if they produce hierarchical results there is no guarantee that the relevant partitions are not artefacts. On the other hand multi-resolution techniques able to identify the different levels of organization of vertices in clusters have been proposed. They are inspired by different principles. For example in \cite{multiRmod_Reichardt} and \cite{arenas_fernandez_gomez} the authors use different definitions of Modularity and the Laplacian matrix respectively, the algorithm introduced in \cite{JeanCharlesPNAS2010} is developed on the stability of clustering under random walk, in \cite{lancichinetti_OLAM} a fitness function which estimates the strength of a cluster and entails a resolution parameter is optimized. 

In this context, also spectral algorithms play an important role. Although most spectral methods have been focused on data clustering, many applications to network data for community detection exist \cite{spectral_donetti,jiangmod,spectral_Capocci2005}. As for data clustering some issues arise when using spectral clustering for community detection:
\begin{itemize}
\item the construction and storage of the $N \times N$ adjacency matrix describing the graph becomes infeasible for large networks 
\item the computation of the eigenvectors of the Laplacian is computationally expensive $O(N^3$), although some approximation techniques like the Lanczos method \cite{lanczos_iterationmethod}, the Nystr\"om algorithm \cite{sc_nystrom} and the power iteration method \cite{pic} can speed-up the process
\item it is not clear how to choose the number of communities into which to split the network at hand
\item it does not incorporate a clear extension to out-of-sample nodes. However, numerical techniques like the Nystr\"om method \cite{sc_nystrom}, or incremental algorithms like \cite{Ning_PR,Dhanjal_arXiv} provide a possible way to tackle this issue. Recently, also a sparse spectral clustering method based on the Incomplete Cholesky Decomposition (ICD) gives the possibility to perform out-of-sample extension by exploiting the pivots selected by the ICD \cite{vanbarel_ICD}. Besides spectral clustering, most of the algorithms for community detection work only at the training level. They produce a partition of the network into modules but cannot assign the membership to new nodes joining the network without running from the beginning or using heuristics. In the community detection field, to the best of our knowledge the only kind of algorithm providing a systematic out-of-sample extension to test nodes are those one based on statistical inference by means of block models \cite{block_snijders,block_nowicki,block_Airoldi}.                   
\end{itemize}     

In what follows we describe a methodology that casts the community detection problem in a machine learning algorithm. Our core model is KSC \cite{multiway_pami}, which has been summarized in the previous chapter. In this way we are able to handle the aforementioned issues in a unique framework, and to propose a competitive technique for the analysis of complex networks. Moreover, this work paves the way to the development of the MKSC model for the analysis of evolving networks discussed in chapter \ref{ch:evclustering}.               
    
\begin{comment}     
\begin{figure}[htbp]
\centering
\begin{tabular}{c}
\begin{psfrags}
\psfrag{x}[t][t]{{}}
\psfrag{y}[b][b]{{}}
\psfrag{t}[b][b]{{}}
\includegraphics[width=3in]{brain.eps}
\end{psfrags}\\
\begin{psfrags}
\psfrag{x}[t][t]{{}}
\psfrag{y}[b][b]{{}}
\psfrag{t}[b][b]{{}}
\includegraphics[width=3in]{socialnet.eps}
\end{psfrags}\\
\begin{psfrags}
\psfrag{x}[t][t]{{}}
\psfrag{y}[b][b]{{}}
\psfrag{t}[b][b]{{}}
\includegraphics[width=3in]{internet.eps}
\end{psfrags}
\end{tabular}
\caption{\textbf{Some example of networks.} From top to bottom, the human brain, a social network and the internet.}\label{net_example}
\end{figure}    
\end{comment}

\section{Methods}
The usage of KSC for community detection can be summarized in three main stages:
\begin{itemize}
\item subset extraction: in order to solve an affordable eigenvalue problem for training the model, we need to select a subset of the entire network that captures the inherent community structure\footnote{How to choose such a subgraph is an open problem. In general it is selected by experimental evaluation, where one wants to find a trade-off between a small size (to reduce the computational burden) and a good quality of the clustering results on the entire network (which depends on how well the subset represents the community structure of the whole graph).}    
\item model selection: the proper number of communities and the kernel hyper-parameters must be tuned carefully to ensure good results  
\item out-of-sample extension: once constructed an optimal model, the memberships for the remainder of the network can be assigned via eq. (\ref{OOSE}).  
\end{itemize}
Furthermore, the choice of an appropriate kernel function which depicts the similarity between the nodes is very important.   
  
\subsection{Representative Sub-graph Extraction}
\label{exp_factor}
Sampling a sub-graph representative of the community structure of the whole network under study is a crucial task in our framework, since it allows a meaningful out-of-sample extension to nodes not present in the training set. Simply taking a random sample of nodes can lead to very different results in several runs, since the quality of the selected sub-graph can have a huge variability. Also selecting a subset in such a way that it follows the same degree distribution or betweenness centrality distribution of the whole graph can produce samples that are not representative of the community structure of the larger network. Recently, a new quality function describing the representativeness of the sample with respect to the community structure of the whole graph has been introduced in \cite{samp_comm_struc}. This quality function is called Expansion Factor (EF) and is defined as $\frac{|N(\mathcal{A}_G)|}{|\mathcal{A}_G|}$, where $\mathcal{A}_G$ indicates a subset of the graph $\mathcal{G}$, $N(\mathcal{A}_G)$ its neighbourhood, i.e. the remaining part of the network to which $\mathcal{A}_G$ is connected, and $|\cdot|$ denotes the cardinality of a set. The idea is that by selecting a sub-graph for which the expansion factor is maximum, one samples a representative subset. Roughly speaking, by including in $\mathcal{A}_G$ nodes that best contribute to the expansion factor, we are taking nodes that are more connected to the rest of the network than to $\mathcal{A}_G$. These nodes are then very likely to belong to clusters not yet represented in the sub-graph, allowing us to produce a sample which is a condensed representation of the community structure of the whole network.          

In \cite{rocco_IJCNN2012} we proposed a greedy strategy for the optimization of EF, which can be summarized in algorithm \ref{alg:EF}. The selection of the active subset can take from a few to several minutes or hours depending on the size $N$ of the entire network and its sparsity, the chosen size $N_A$ for the active subset and the threshold $\epsilon$. Moreover, although the variability of the method is less compared to the random sampling, the technique is stochastic by nature. In \cite{FURS_paper} the Fast and Unique Representative Subset (FURS) selection technique has been proposed to cope with these issues. The method has been shown to be computationally less expensive and deterministic: it greedily selects nodes with high-degree centrality from most or all the modules present in the network, which are usually located at the center rather than the periphery and can better capture the community structure.
 
\begin{algorithm}[ht]
\small
\LinesNumbered
\KwData{network $\mathcal{G}=(\mathcal{V},\mathcal{E})$ composed of $N$ nodes $\mathcal{V} = \{n_i\}_{i = 1}^N$  (represented as the adjacency matrix $A \in \mathbb{R}^{N \times N}$), $N_A$ (size of sub-graph $\mathcal{A}_G$).}
\KwResult{active set of $N_A$ selected nodes belonging to the subset $\mathcal{A}_G = (\mathcal{V}_A,\mathcal{E}_A)$ 
.} 
select randomly an initial sub-graph $\mathcal{A}_G = \{n_j\}_{j = 1}^{N_A} \subset \mathcal{V}$\\   
\While{$\delta(EF) > \epsilon$}{
compute $EF(\mathcal{A}_G)$\\
randomly pick two nodes as $n_* \in \mathcal{V}_A$ and $n_+ \in \mathcal{V} \setminus \mathcal{V}_A$\\ 
let $\{\mathcal{W} = \mathcal{V}_A \setminus \{n_*\}\} \cup \{n_+\}$\\  
\eIf{$EF(\mathcal{W}) > EF(\mathcal{A}_G)$}{swap($\{n_+\}$,$\{n_*\}$)}
{
do not swap($\{n_+\}$,$\{n_*\}$)
}
}
\caption{EF-based subset selection \cite{rocco_IJCNN2012}}
\label{alg:EF}
\end{algorithm}

\subsection{Model Selection Criteria}
\label{sc:MS_community_detection}
Often people use heuristics to select the tuning parameters present in their models. Since model selection is a crucial
point, here we describe a systematic way to do it properly. Our method is based on a validation procedure. We train
the KSC model with different number of communities and (where needed) several values of the kernel hyper-parameters. In the validation step the obtained groupings are judged depending on some quality functions like BLF \cite{multiway_pami}, AMS \cite{rocco_IJCNN2013} (described in the previous chapter), Modularity \cite{rocco_IJCNN2011} (see Appendix): the one (or more) partition with the highest value of the chosen criterion is selected. 

\begin{figure}[htbp]
\centering
\begin{tabular}{cc}
\textbf{Stage I} & \textbf{Stage II}\\
\includegraphics[width=1.5in, height=1.5in]{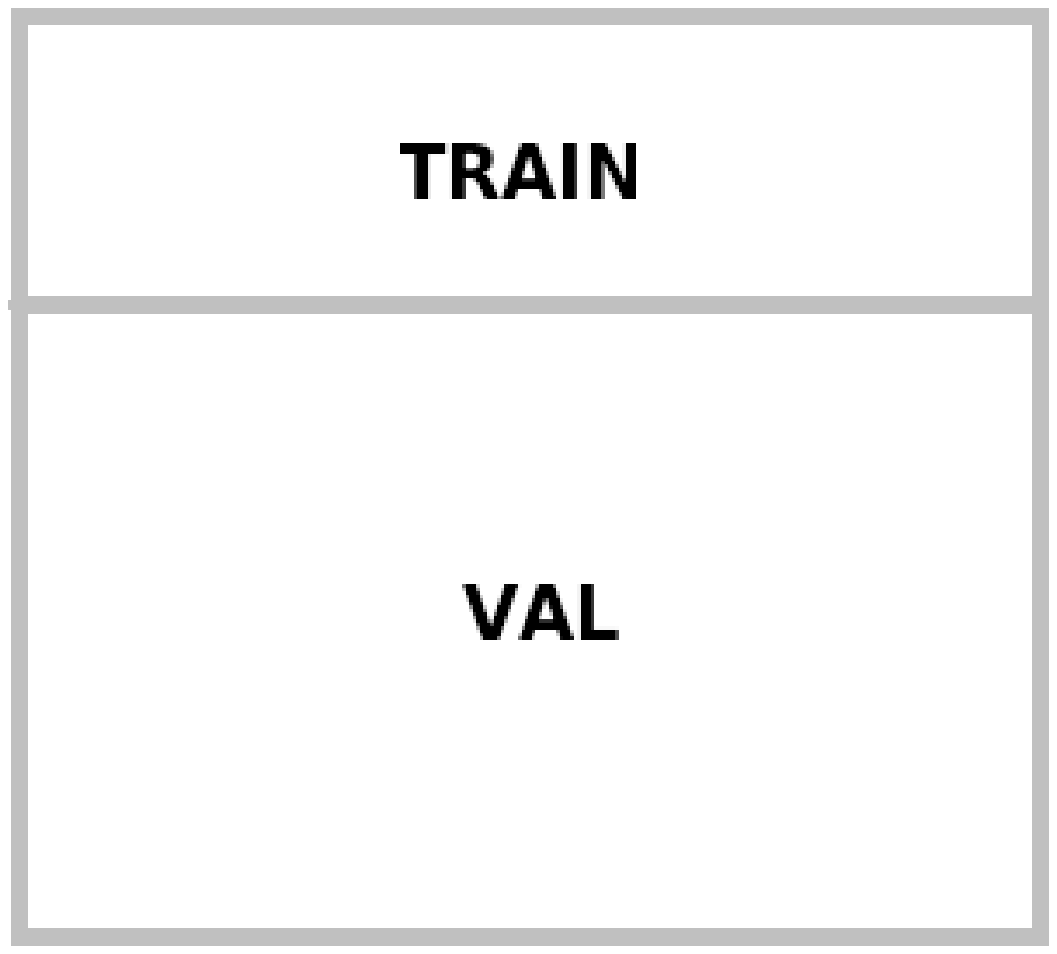} & \includegraphics[width=1.5in, height=1.5in]{prova3.eps}\\
\includegraphics[width=1.5in, height=1.5in]{prova3.eps} & \includegraphics[width=1.5in, height=1.5in]{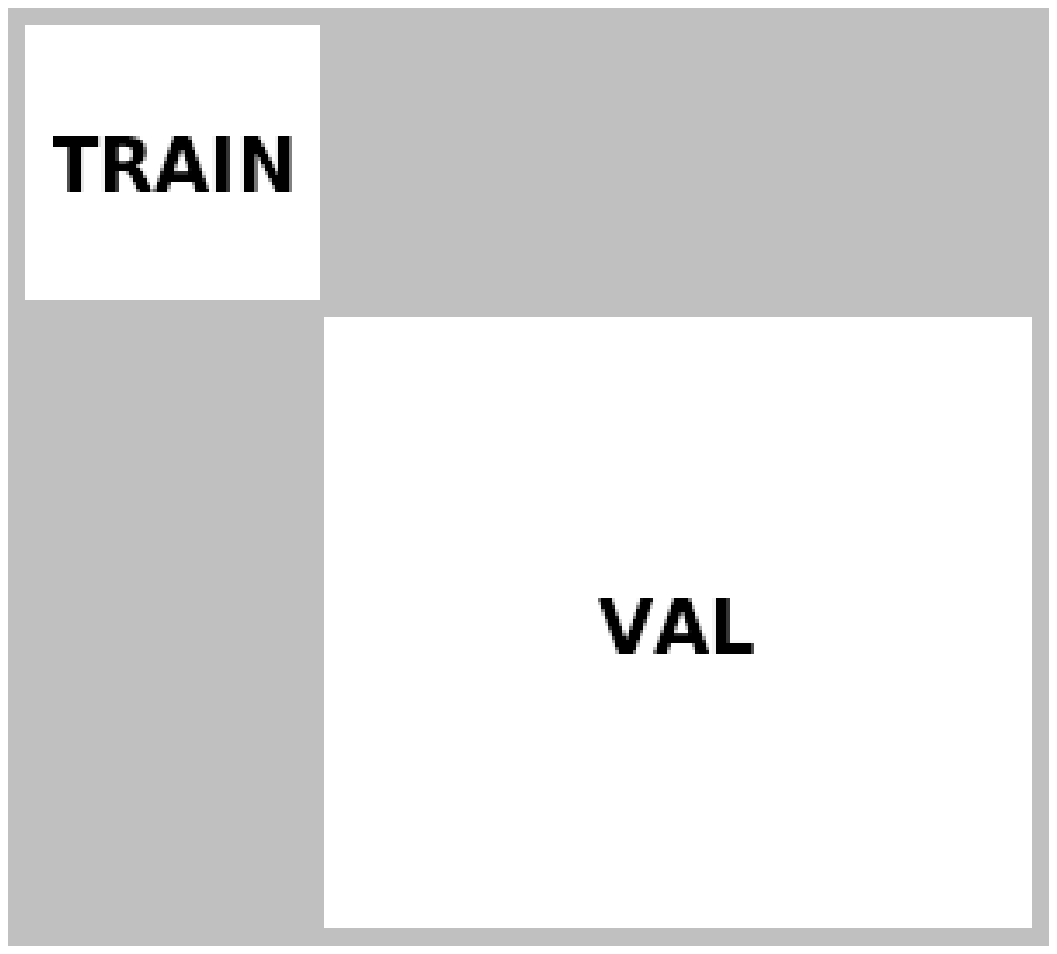}
\end{tabular}
\caption{\textbf{Training and validation sets}. Example showing the way the datasets for the study of a network via KSC are built up. In this specific case the first $25\%$ of the total nodes form the training set and the remaining $75\%$ the validation set. The first row refers to BLF or AMS, the second to Modularity. The first column represents the first stage of the validation process (prediction of memberships), the second column depicts the second stage. For Modularity, in the second stage of the validation process we change representation from adjacency list to adjacency matrix.}\label{fig_unweightsets}
\end{figure}

For the model selection algorithm using AMS the reader can refer to section \ref{AMS_algorithm}, while the tuning scheme performed by means of Modularity can briefly be expressed by algorithm \ref{alg:MS_mod}.
\begin{algorithm}[!ht]
\small
\LinesNumbered
\KwData{training set, validation set stage I, validation set stage II,
positive (semi-) definite kernel function $K(x_i,x_j)$}
\KwResult{selected number of clusters $k$ and (if any) kernel hyper-parameters.} 
compute cluster indicator matrix \begin{math} X \end{math} from the cluster results of the different models, obtained using the training set and the validation set I stage in the learning process\\
compute the Modularity matrix \begin{math} M = S - \frac{dd^T}{2m} \end{math}, where all the quantities refer to the validation set used in the II stage of the validation process\\
compute the Modularity $ \textrm{Mod} = \frac{1}{2m}\textrm{tr}(X^TMX) $\\
select the model (i.e. $k$ and the eventually the kernel hyper-parameters) corresponding to the partition(s) which gives the highest Modularity value.
\caption{Modularity-based model selection for KSC \cite{rocco_IJCNN2011}}
\label{alg:MS_mod}
\end{algorithm}
The training set, validation set and the two stages of the validation process have the following meaning. The training set is the matrix given as input to the kernel spectral clustering model during the training phase. The validation process can be divided into two stages:
\begin{enumerate}
\item stage I: the cluster memberships for the validation set (data not belonging to the training set) are predicted by the model based on eq. (\ref{OOSE})
\item stage II: the quality of the predicted memberships are judged by means of a certain criterion. 
\end{enumerate}
In these two stages the validation sets involve the same data (the nodes of the graph under study) but represented in different ways. In stage I some rows of the adjacency matrix are considered: this is called an adjacency list. In stage II, if we consider the Modularity criterion, the validation set is an adjacency matrix, because this is needed in order to calculate the related Modularity. For BLF and AMS this is not the case (see figure \ref{fig_unweightsets} for a further clarification).

Finally we can point out that the definition of the Modularity function is general because it does not make any assumption on the kind of the community structure of the network to detect. This, together with the fact that it is a quality function particularly suited for graphs, is one of the main advantages with respect to the BLF criterion, which as we already mentioned in section \ref{sc:BLF} is optimized to detect clusters that are well separated. Moreover, the Modularity-based model selection algorithm is feasible for large scale applications because a sparse representation of the validation adjacency matrix can be used.      

%\begin{comment}
%\end{comment}

\subsection{Choice of the Kernel Function}
\label{kernel_description}
Unlike classical spectral clustering, when using KSC or SKSC for community detection we need to build a graph over the initial network to represent its topology in the feature space. In this context, the choice of a proper kernel function is fundamental to be able to unveil the modular organization of the network.

In the analysis of the weighted networks we use the RBF kernel to capture the similarity within the nodes. We treat the $i$-th row of the adjacency list of the entire graph representing node $v_i$ as a point in an Euclidean space of dimension $N$, say $x_i$. Then the similarity between two nodes is given by $K(x_i,x_j) = \exp(-||x_i-x_j||_2^2/\sigma^2)$, where the hyper-parameter $\sigma$ denotes the kernel bandwidth.        
 
In dealing with unweighted networks a recently proposed kernel function named community kernel \cite{community_kernel} is used to build up the similarity matrix of the graph. This kernel function does not have any parameter to tune and the similarity $\Omega_{ij}$ between two nodes $v_i$ and $v_j$ is defined as the number of edges connecting the common neighbours of these two nodes: $ \Omega_{ij} = \sum_{k, l \in \mathcal{N}_{ij}} A_{kl}$. Here $\mathcal{N}_{ij}$ is the set of the common neighbours of nodes $v_i$ and $v_j$, $A$ indicates the adjacency matrix of the graph, $\Omega$ is the kernel matrix. As a consequence, even if two nodes are not directly connected to each other, if they share many common neighbours their similarity $\Omega_{ij}$ will be set to a large value. Moreover, in \cite{community_kernel} it is empirically observed that this kernel matrix is positive definite, a fundamental requirement in order to use the community kernel in our kernel-based framework.

The above-mentioned kernel functions are able to correctly capture the similarity between the nodes of weighted and unweighted networks. Unfortunately, they are not computationally efficient. In order to be able to perform in a reasonable time the computation of the kernel matrix for large graphs, it is advisable to use the cosine kernel, since it works with a sparse representation of the variables \cite{raghvendra_bigdata}.

\subsection{Computational Complexity}
\begin{figure}[htbp]
\centering
\begin{tabular}{c}
\begin{psfrags}
\psfrag{x}[t][t]{{N}}
\psfrag{y}[b][b][1][360]{{CPU time (s)}}
\psfrag{t}[b][b]{{}}
\includegraphics[width=3in]{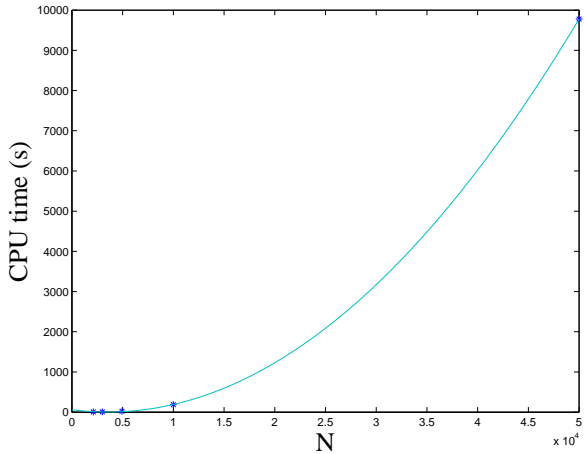}
\end{psfrags}
\end{tabular}
\caption{\textbf{Computational complexity KSC} Run-time required to obtain the partitioning of some computer generated networks with different size. The curve represents a second degree polynomial yielding the best fit to the simulation data. }\label{plot_cputime}
\end{figure}

Since we can use the Lanczos method \cite{lanczos_iterationmethod} or the Incomplete Cholesky Decomposition (ICD) \cite{sparsesc} to compute the top $k$ eigenvectors of eq. (\ref{dual_KSC}) in quadratic run-time\footnote{Quadratic with respect to the number of training points or the pivots of the ICD respectively.}, the computational complexity for the training phase is dominated by the time required to construct the kernel matrix $\Omega$. Since the networks are usually very sparse, the latter is given by $O(N_{\textrm{Tr}}^2)$. If we consider the whole network as test set, the run-time of the test phase concerns the computation of eq. (\ref{OOSE}) and equals $O(N_{\textrm{Tr}}N)$. Moreover, the out-of-sample extension can be easily scaled down of a big factor in a distributed environment. In fact, by chunking the test set into blocks with size proportional to the training set and parallelizing the computations, the overall complexity is reduced to $O(2N_{\textrm{Tr}}^2)$. Finally, in figure \ref{plot_cputime} the computational burden required by KSC is evaluated on some of the synthetic networks described in the next section. Here the complexity seems quadratic and not linear according to a least squares curve fitting: this can be explained by the fact that a non-sparse representation of the variables and an inefficient implementation of the kernel function have been used in the simulations.     
    
\section{Simulations on Synthetic Networks}
Methods to detect communities in graphs are required to be deeply tested. To do that, one needs benchmark graphs with a known community structure that the algorithms should identify. In \cite{fortunatobench,fortunatobench2} the authors proposed new classes of benchmark graphs referred as LFR where the distributions of node degree and community size are both power laws, which mimics an important feature of real networks. By means of the LFR benchmark it is possible to generate undirected, unweighted, directed, weighted and hierarchical graphs with overlapping nodes and different degree of mixing between communities. Thanks to its generality, it quickly became a standard tool for testing community detection methods, and for this reason we used it in all the experiments described in this dissertation.  

\begin{figure}[htbp]
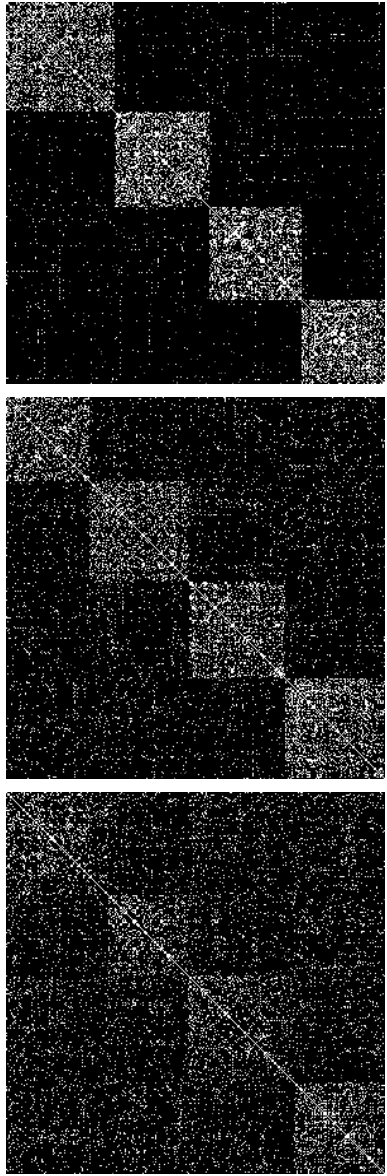

\centering
\begin{tabular}{c}
\begin{psfrags}
\psfrag{x}[t][t]{{}}
\psfrag{y}[b][b]{{}}
\psfrag{t}[b][b]{{Small Overlap}}
\includegraphics[width=2in]{kmatrix_net_small_overlap.eps}
\end{psfrags}\\
\begin{psfrags}
\psfrag{x}[t][t]{{}}
\psfrag{y}[b][b]{{}}
\psfrag{t}[b][b]{{Medium Overlap}}
\includegraphics[width=2in]{kmatrix_net_medium_overlap.eps}
\end{psfrags}\\
\begin{psfrags}
\psfrag{x}[t][t]{{}}
\psfrag{y}[b][b]{{}}
\psfrag{t}[b][b]{{Large Overlap}}
\includegraphics[width=2in]{kmatrix_net_big_overlap.eps}
\end{psfrags}
\end{tabular}
\caption{\textbf{Artificial networks Net\_{unw4C}}. Adjacency matrix of the unweighted synthetic networks composed by $4$ communities with  small overlap (top), medium overlap (center) and large overlap (bottom).}\label{fig_net}
\end{figure}

In table \ref{table_synthnet} a description of all the artificial graphs used to test KSC and SKSC is given. For instance \textit{Net\_{unw3C1000ov}} indicates an unweighted network formed by $3$ communities with $1000$ overlapping nodes. For illustration purposes, in figure \ref{fig_net} the synthetic networks with $1\,000$ nodes and $4$ communities with mixing parameters equal to $0.05$ (small overlap), $0.2$ (medium overlap), $0.35$ (large overlap) are illustrated by means of the corresponding adjacency matrix. We can recognize the block configuration, indicating the presence of an underlying community structure. 

\begin{figure}[htbp]
\centering
\begin{tabular}{c}
\begin{psfrags}
\psfrag{x}[t][t]{{\% of the whole network used as training set}}
\psfrag{y}[b][b][1][360]{{ARI}}
\psfrag{t}[b][b]{{Active selection of training sub-graph}}
\includegraphics[width=3in]{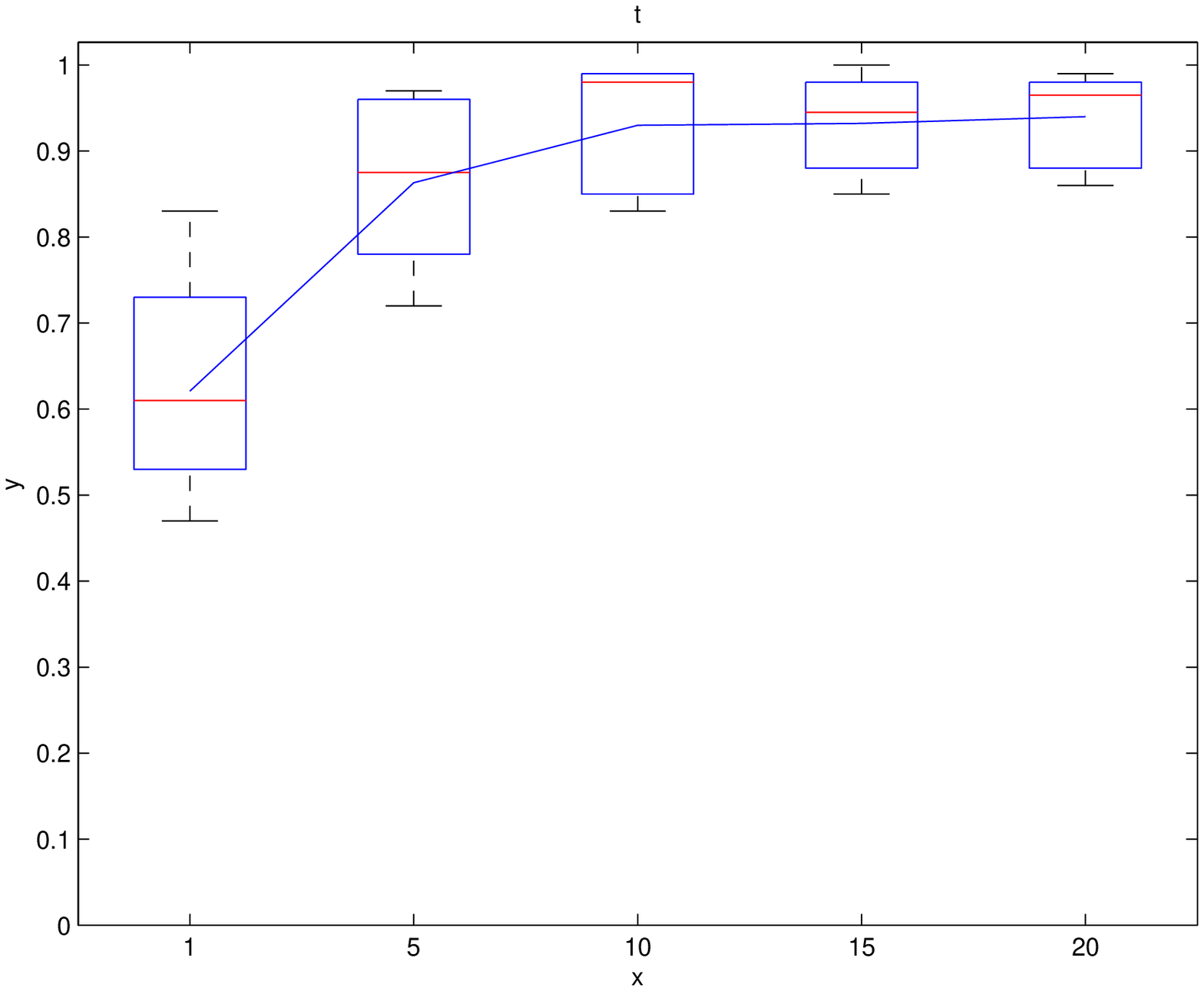}
\end{psfrags}\\\\
\begin{psfrags}
\psfrag{x}[t][t]{{\% of the whole network used as training set}}
\psfrag{y}[b][b][1][360]{{ARI}}
\psfrag{t}[b][b]{{Random selection of training sub-graph}}
\includegraphics[width=3in]{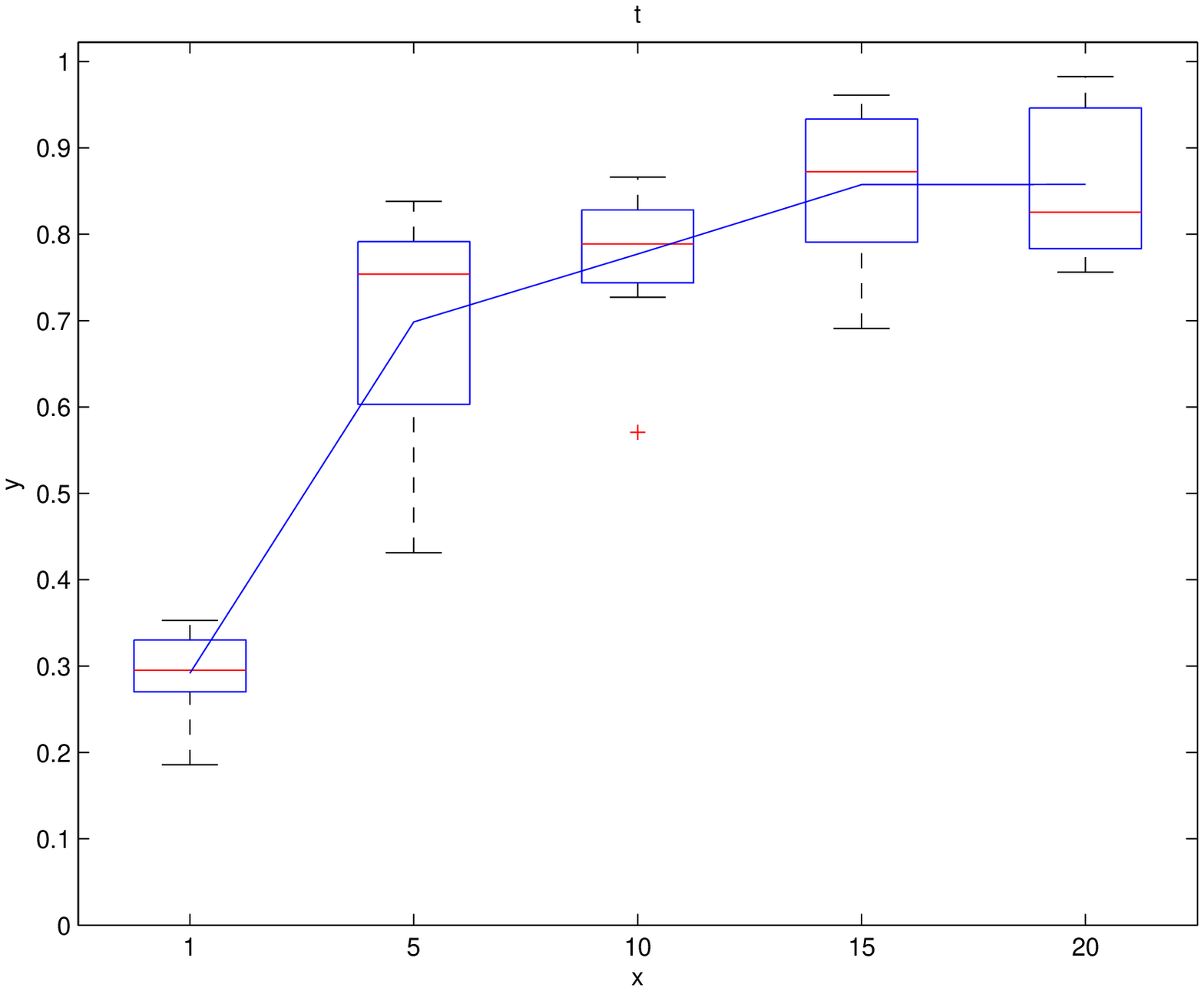}
\end{psfrags}
\end{tabular}
\caption{\textbf{Quality EF sub-graph}. The quality of the sub-graph extracted from the artificial graph \textit{Net\_{unw9C}} using the EF algorithm is assessed (top). The ARI between the predicted partition where the EF sub-graph is used as training set and the true partition is computed. A comparison with random sampling is performed (bottom). We can notice that, mostly when few nodes are selected, the EF sub-graph allows a more accurate out-of-sample extension. Moreover the variability is reduced when more nodes are considered.}\label{all_boxplot_M9C}
\end{figure}

In most of the experiments the training set is usually composed by the $15 \%$ of the nodes of the starting graph extracted by using algorithm \ref{alg:EF}, and it is randomized $10$ times. How to chose in practice the size of the maximal subgraph is an important aspect. On one hand one would like to select a very small subset in order to save computing time. On the other hand the subgraph should be large enough to accurately represent the community structure of the whole graph. By experimental evaluations (like the one depicted in figure \ref{all_boxplot_M9C}) we set the size of the training subgraph to $15 \%$ of the size of the entire network. Surprisingly, the same percentage is used in \cite{Leskovec_sampling}, where the optimal subgraph size is determined based on its representativeness in terms of graph parameters like degree distribution, clustering coefficient distribution etc.

We consider as test set the entire network. The model selection results are shown by means of boxplots or only the average values, and usually the validation set is formed by the $35 \%$ of the starting graph. Moreover, for big network data, the training and validation sets size are constrained by the maximum size of the kernel matrix that can be stored in the memory of our PC, which is $5000 \times 5000$.

When evaluating the quality of the final partitions, for the synthetic networks we compare KSC or SKSC with spectral clustering using the Nystr\"om method for out-of-sample extension. In case of the real graphs, the results obtained with KSC or SKSC depending on the data-sets are reported.

\begin{table}[htbp]
\begin{center}
\begin{tabular}{|c|c|c|c|c|} %tabella datasets
\hline 
\textbf{Network}  & \textbf{Nodes} &  \textbf{Edges} &  \textbf{Sparsity ($\%$)} & \textbf{Overlap ($\mu$)} \\
\hline
Net\_{unw9C} & $3\,000$ & $22\,904$ & $99.49$ & $0.1$ \\
\hline
Net\_{unw13C} & $10\,000$ & $76\,789$ & $99.85$ & $0.1$\\
\hline
Net\_{unw22C} & $50\,000$ & $383\,220$  & $99.97$ & $0.1$\\ 
\hline
Net\_{unw3C1000ov} & $3\,000$ & $149\,535$  & $98.34$ & $0.1$\\ 
\hline
Net\_{unw4CmuS} & $1\,000$ & $11\,867$  & $98.81$ & $0.05$\\ 
\hline
Net\_{unw4CmuM} & $1\,000$ & $13\,451$  & $98.65$ & $0.2$\\ 
\hline
Net\_{unw4CmuL} & $1\,000$ & $14\,223$  & $98.57$ & $0.35$\\ 
\hline
Net\_{w6C} & $3\,000$ & $148\,928$  & $99.35$ & $0.1$\\ 
\hline
Net\_{w4C1000ov} & $3\,000$ & $149\,033$  & $98.34$ & $0.1$\\ 
\hline
\end{tabular}
\end{center}
\caption{\textbf{Synthetic network data}. Summary of some properties of the computer generated graphs that have been analysed. In the name of the data-sets, the acronym \textit{unw} means unweighted, \textit{w} stands for weighted, \textit{ov} indicates the presence of overlapping nodes, \textit{C} means communities. Sparsity refers to the adjacency matrix associated to each graph and indicates the percentage of zero entries with respect the total number of elements of the matrix. The mixing parameter $\mu$ is related to the number of inter-community edges and describes the degree of overlap between the communities. The acronym \textit{muS}  means small overlap, \textit{muM} medium overlap, \textit{muL} large overlap.}
\label{table_synthnet}
\end{table}

\begin{figure}[htbp]
\centering
\begin{tabular}{c}
\begin{psfrags}
\psfrag{x}[t][t]{{Number of iterations}}
\psfrag{y}[b][b][1][360]{{EF}}
\psfrag{t}[b][b]{{Net\_{unw9C}}}
\includegraphics[width=3in]{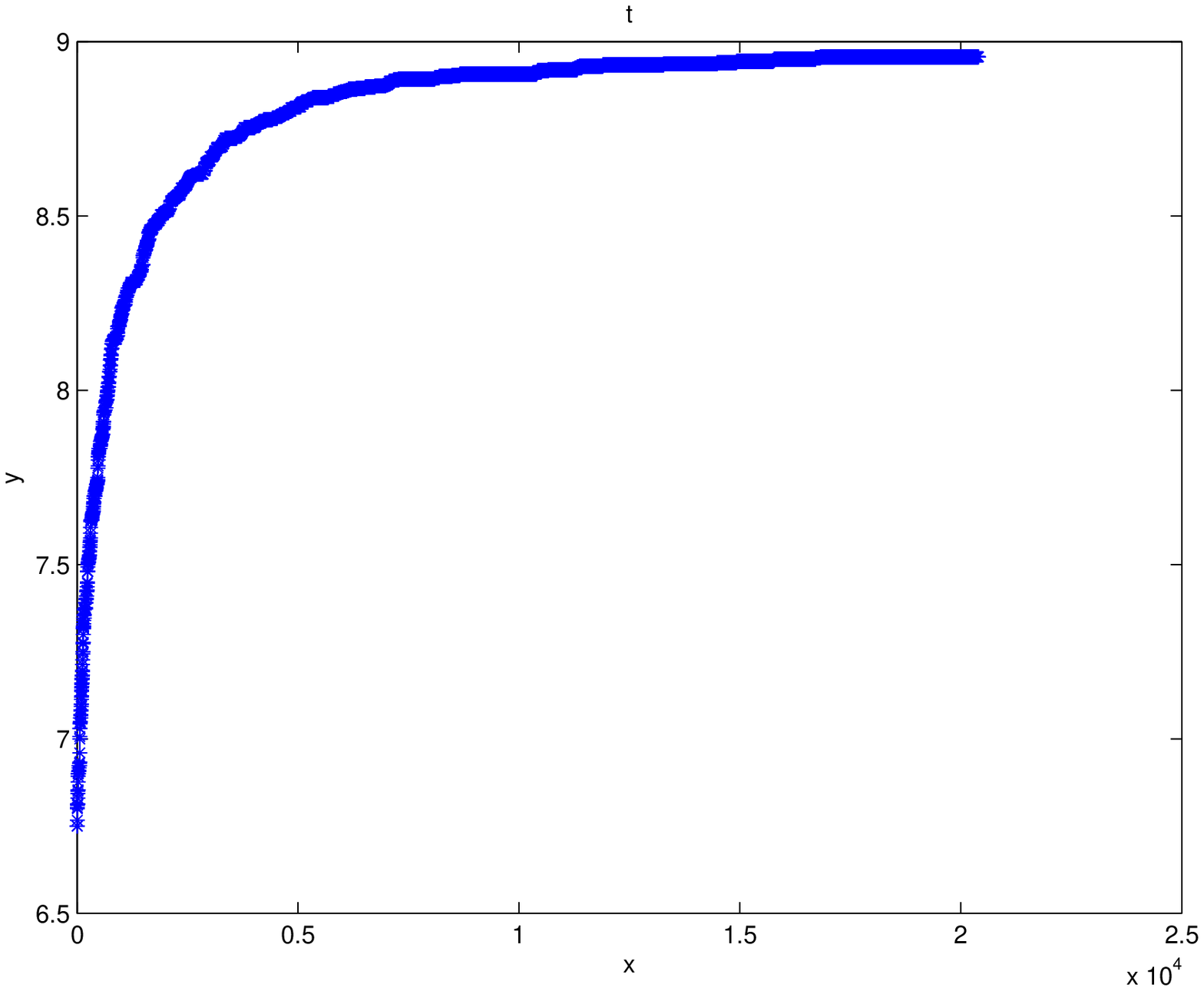}
\end{psfrags}
\end{tabular}
\caption{\textbf{Expansion factor optimization}. Example of the greedy optimization of the expansion factor (EF) for extracting a sub-graph from the artificial network \textit{Net\_{unw9C}}.}\label{expfactor_net9C}
\end{figure}

\begin{figure}[htbp]
\centering
\begin{tabular}{cc}
\begin{psfrags}
\psfrag{x}[t][t]{{}}
\psfrag{y}[b][b][1][360]{{}}
\psfrag{t}[b][b]{{}} %Degree distribution artificial graph with $9$ communities
\includegraphics[width=2in]{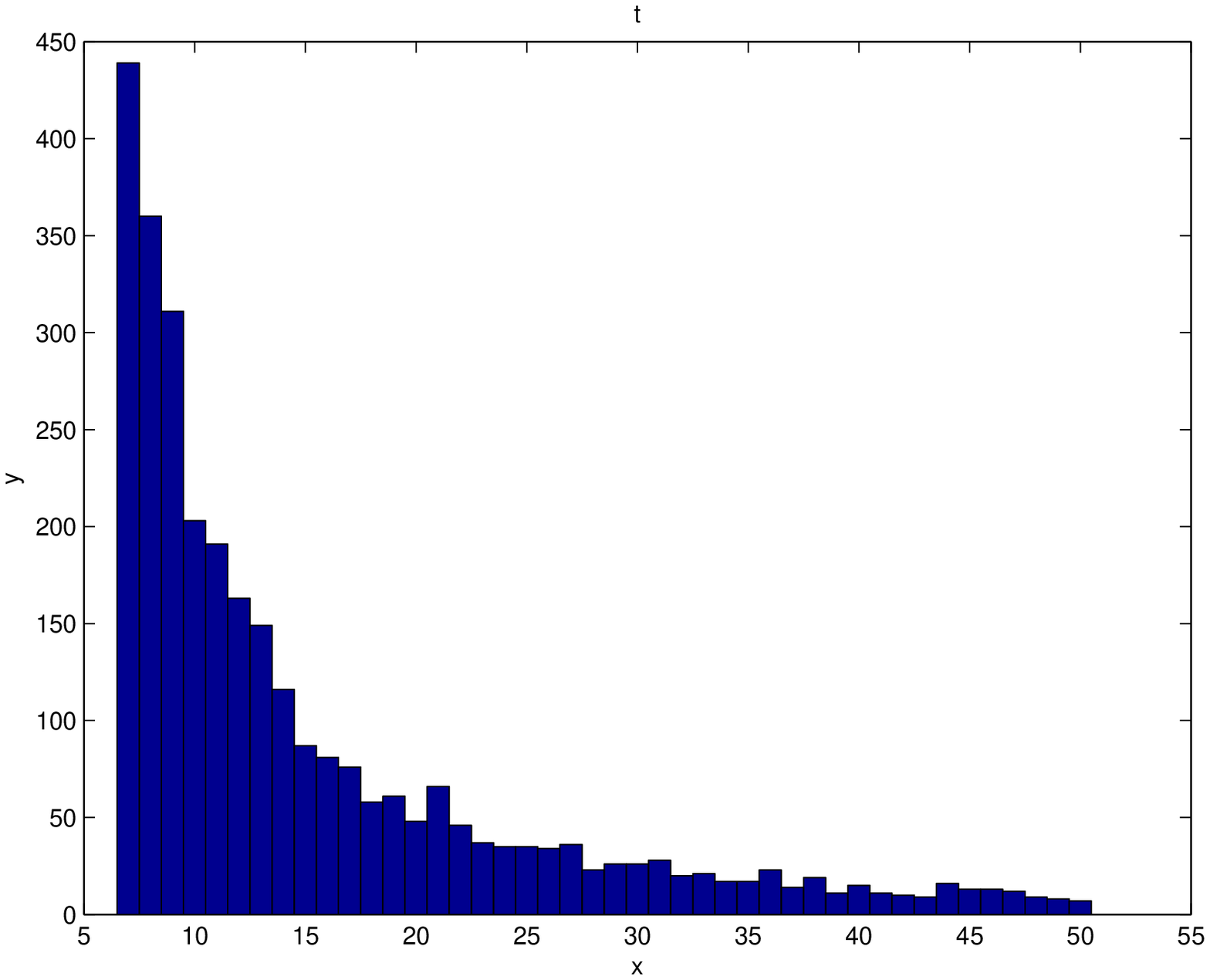}
\end{psfrags}&
\begin{psfrags}
\psfrag{x}[t][t]{{}}
\psfrag{y}[b][b][1][360]{{}}
\psfrag{t}[b][b]{{}}
\includegraphics[width=2in]{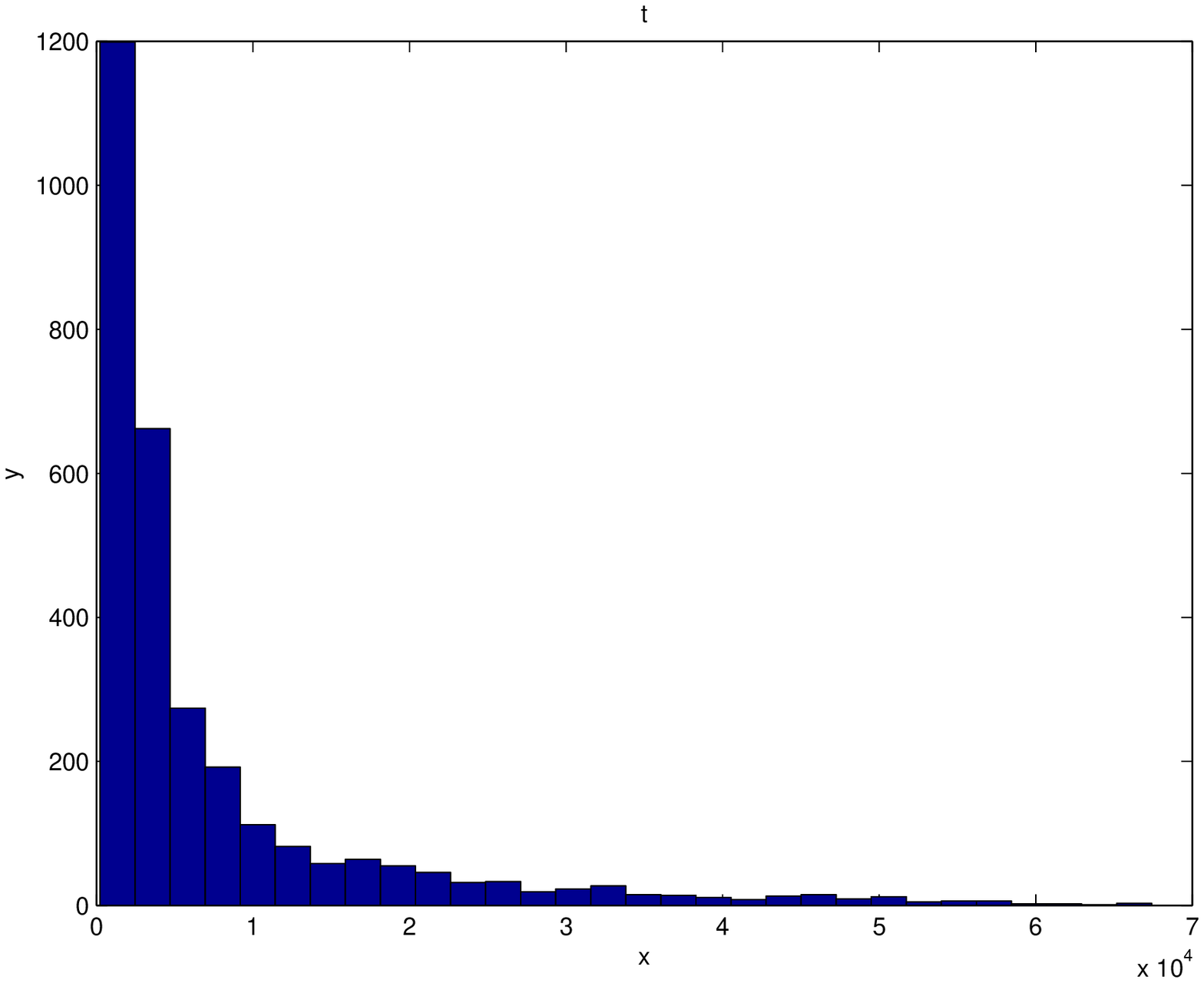}
\end{psfrags}\\
\begin{psfrags}
\psfrag{x}[t][t]{{}}
\psfrag{y}[b][b][1][360]{{}}
\psfrag{t}[b][b]{{}}%Artificial network with $9$ communities
\includegraphics[width=2in]{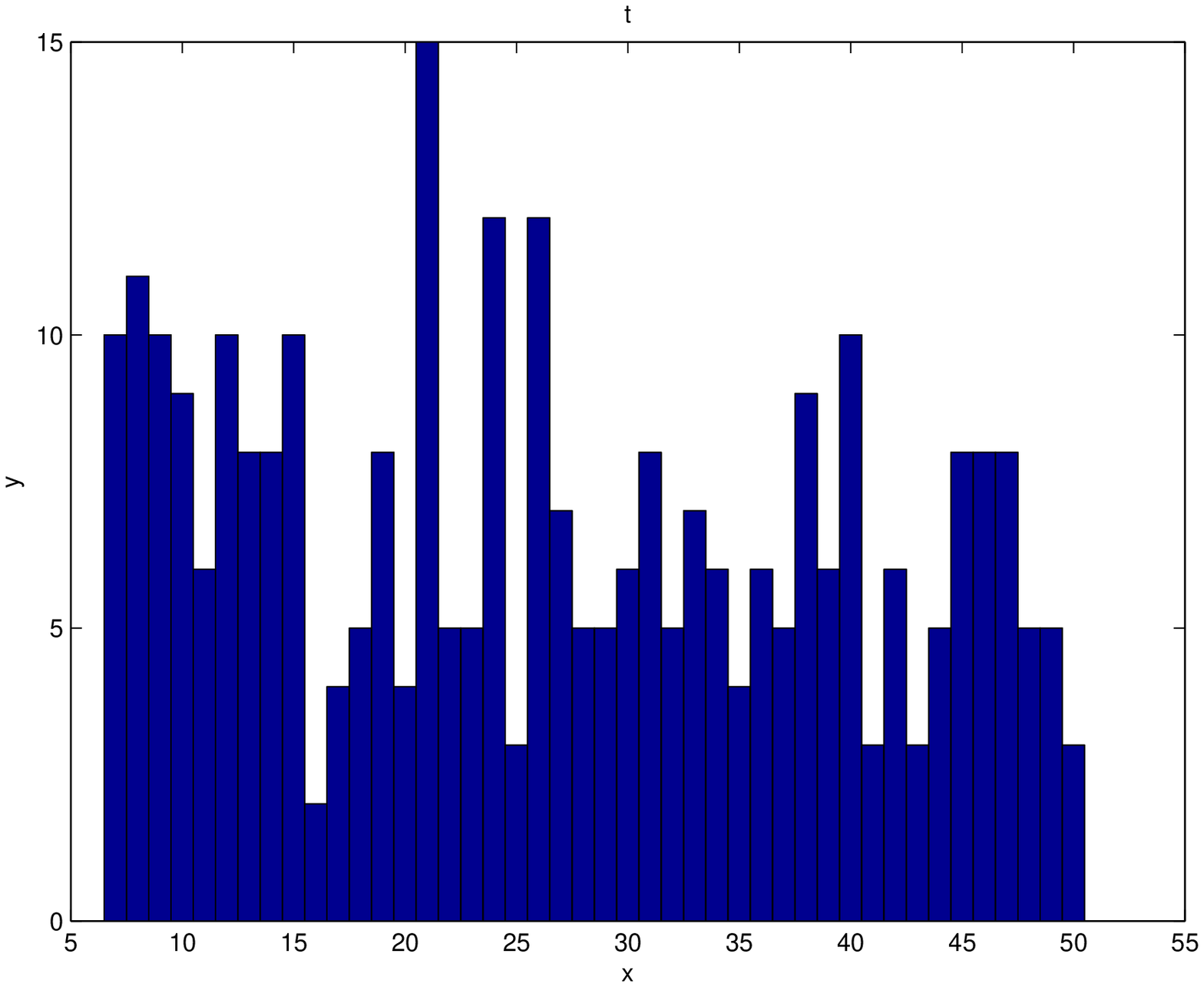}
\end{psfrags}&
\begin{psfrags}
\psfrag{x}[t][t]{{}}
\psfrag{y}[b][b][1][360]{{}}
\psfrag{t}[b][b]{{}}%Selected subgraph
\includegraphics[width=2in]{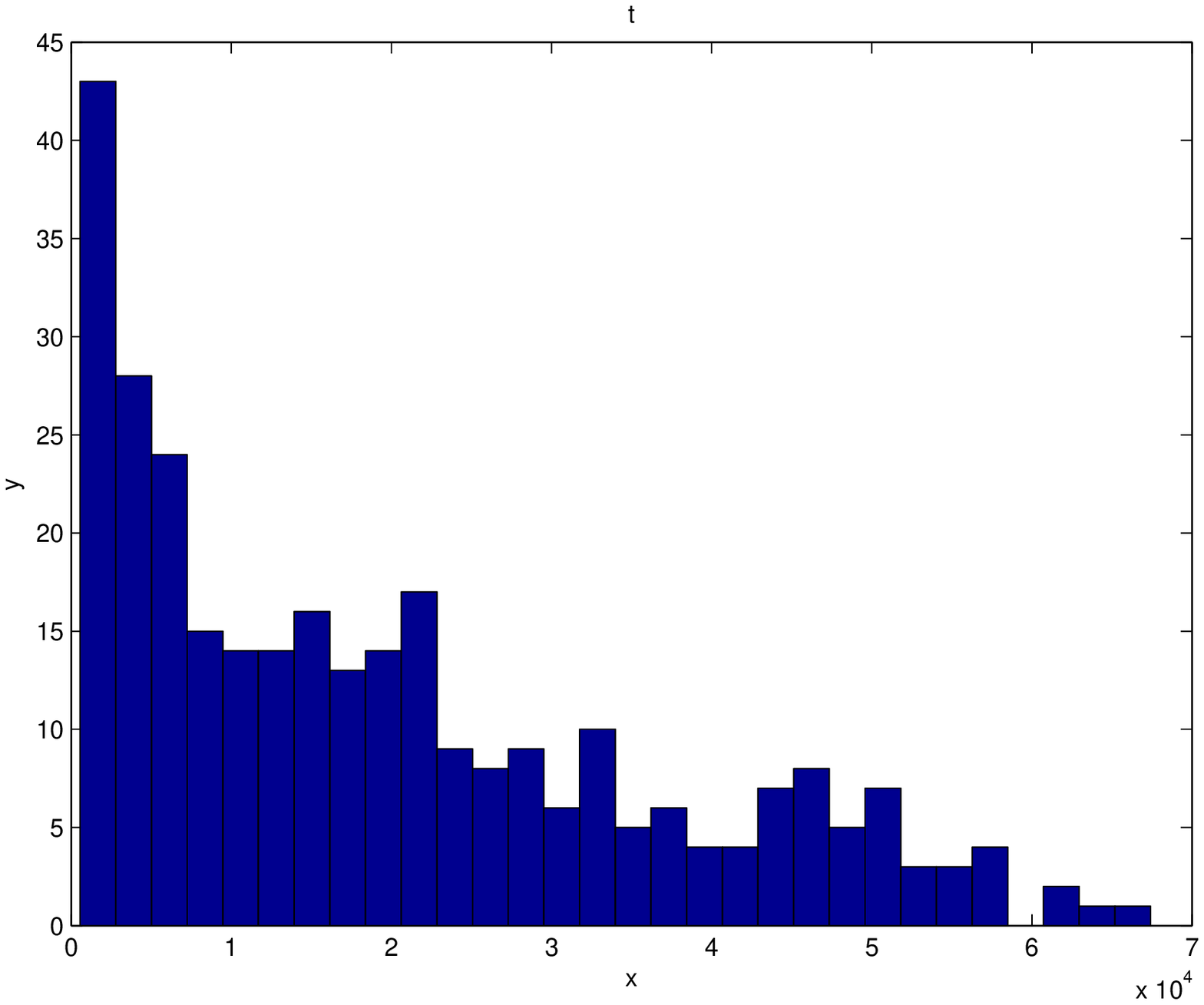}
\end{psfrags}\\
\scriptsize{    Bin division  } & \scriptsize{    Bin division}\\
\end{tabular}
\caption{\textbf{Properties EF sub-graph}. Degree and betweenness centrality distribution (number of nodes) of the synthetic network \textit{Net\_{unw9C}} (top left and top right) and of a typical active set selected from the EF algorithm (bottom left and bottom right). We can notice that the representativeness of the set in terms of community structure can not be related to its representativeness in terms of degree and betweenness centrality distribution.}\label{deg_bC_distr_M9C}
\end{figure}

\begin{figure}[htbp]
\centering
\begin{tabular}{c}
\begin{psfrags}
\psfrag{x}[t][t]{{}}
\psfrag{y}[b][b][1][360]{{}}
\psfrag{t}[b][b]{{Degree distribution of the entire kernel matrix $\Omega$}}
\includegraphics[width=3in]{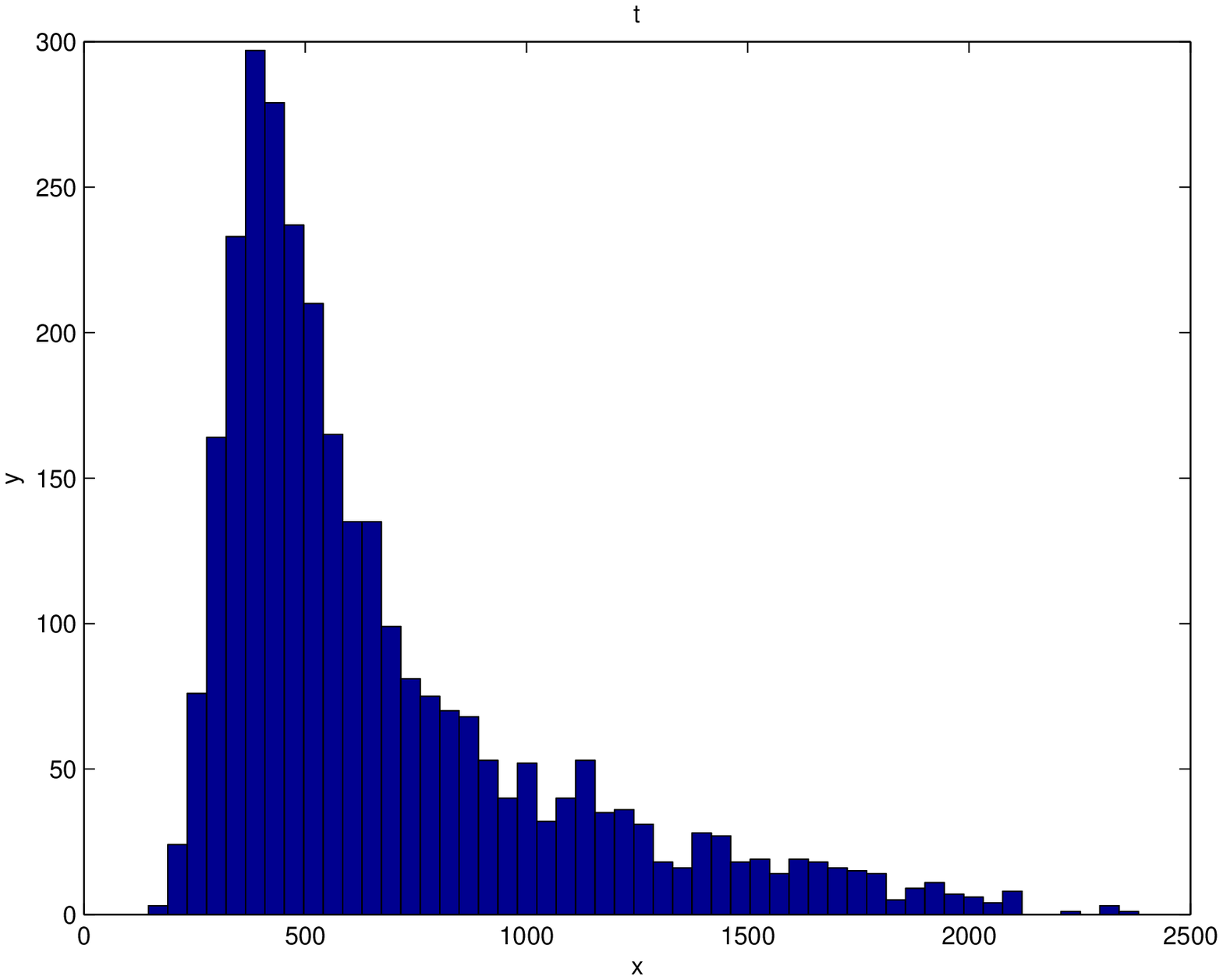}
\end{psfrags}\\
\begin{psfrags}
\psfrag{x}[t][t]{{}}
\psfrag{y}[b][b][1][360]{{}}
\psfrag{t}[b][b]{{Degree distribution of $\Omega_{test}$}}
\includegraphics[width=3in]{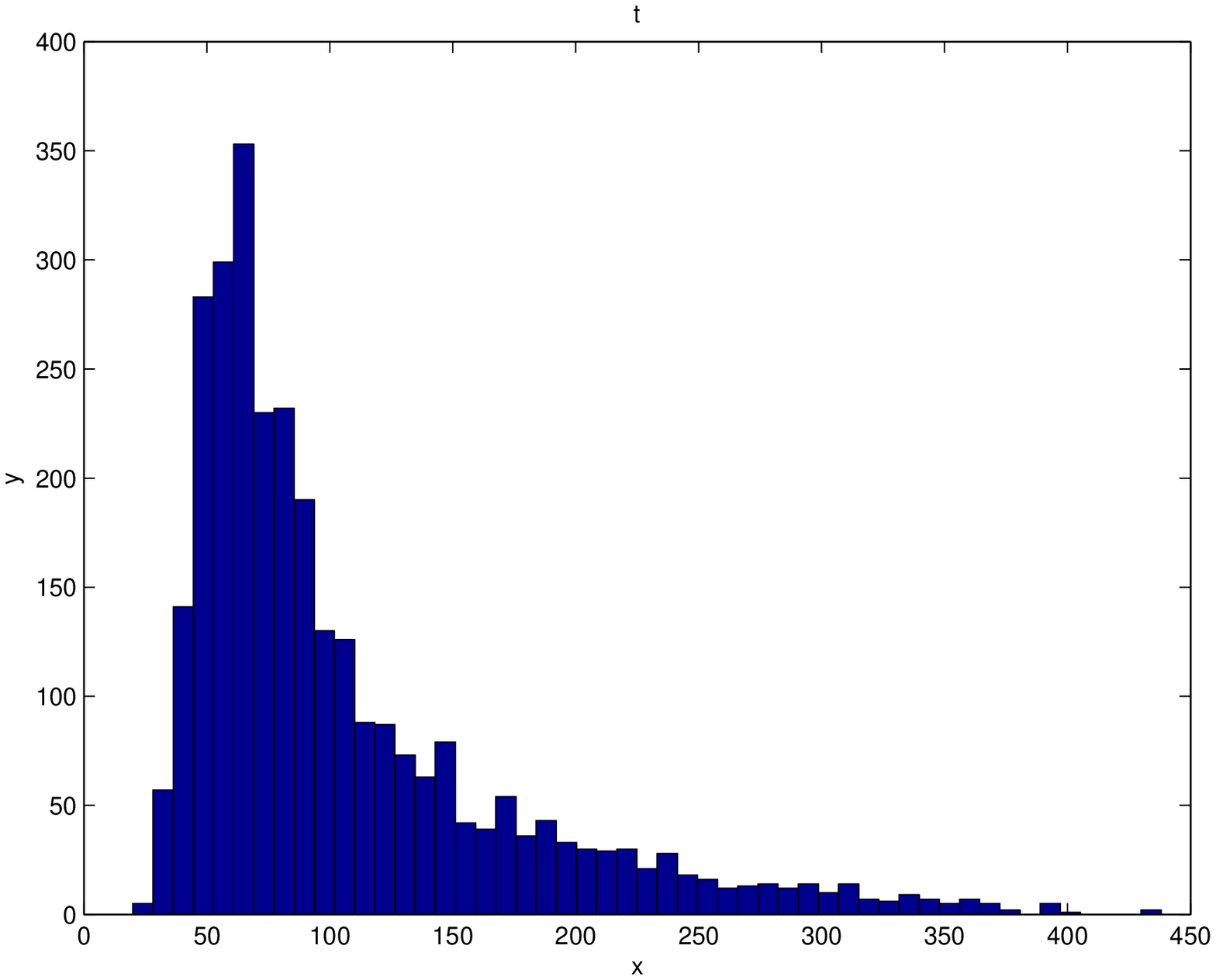}
\end{psfrags}
\end{tabular}
\caption{\textbf{Generalization with the EF sub-graph}. The degree distribution related to the full kernel matrix (of size $3\,000 \times 3\,000$) describing the similarity between the nodes of the network \textit{Net\_{unw9C}} is pictured in the first row. In the second row the degree distribution associated to the test kernel matrix (of size $3\,000 \times 300$) is shown. The two distributions look like quite similar, meaning that the KSC model trained on the active selected subset can generalize well in the test phase.}\label{out_degree}
\end{figure}

Figure \ref{expfactor_net9C} depicts a typical example of the greedy optimization of the EF performed by the algorithm explained in section \ref{exp_factor}, where the artificial network \textit{Net\_{unw9C}} is considered. Moreover, in figure \ref{all_boxplot_M9C} it is shown how the active sampling technique produces a better sample than random sampling. To see this, we compare the ARI index \cite{arindex} (see appendix) between the partitions predicted by the kernel spectral clustering model and the true memberships, by using the training set selected randomly or actively. In figure \ref{deg_bC_distr_M9C} we can notice that the degree and betweenness centrality distribution \cite{Freeman1977} of the active set is quite different from those one of the whole graph. All these empirical observations are in agreement with what has been discussed in \cite{samp_comm_struc}. In this work the authors observed that selecting a sub-graph which captures the community structure of the entire graph often does not implies that the sub-graph is also representative of the same degree or betweenness centrality distribution. On the other hand, if we compare the degree distribution associated to the full kernel matrix $\Omega$ and the out-of-sample degree distribution related to the test kernel matrix $\Omega_{test}$, they are quite similar (see figure \ref{out_degree}). This indicates that the active selected subset is meaningful and allows our model to correctly generalize to unseen nodes.

\begin{figure}[htbp]
\centering
\begin{tabular}{c}
\begin{psfrags}
\psfrag{x}[t][t]{{Number of clusters $k$ }}
\psfrag{y}[b][b][1][180]{{BLF}}
\psfrag{t}[b][b]{{Number of detected clusters: $3$ or $4$}}
\includegraphics[width=3in]{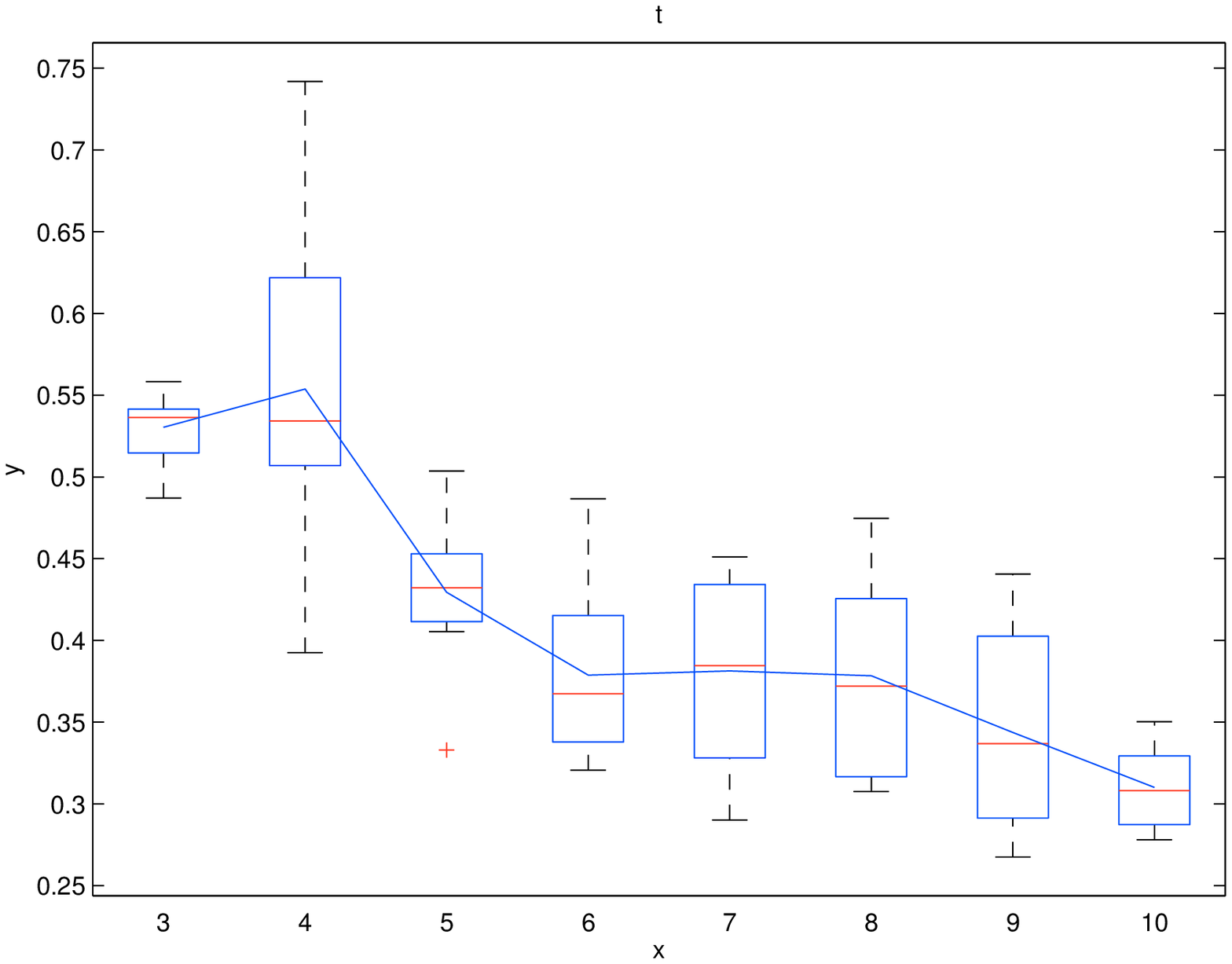}
\end{psfrags}\\\\
\begin{psfrags}
\psfrag{x}[t][t]{{Number of clusters $k$  }}
\psfrag{y}[b][b][1][180]{{Modularity}}
\psfrag{t}[b][b]{{Number of detected clusters: $2$ or $3$}}
\includegraphics[width=3in]{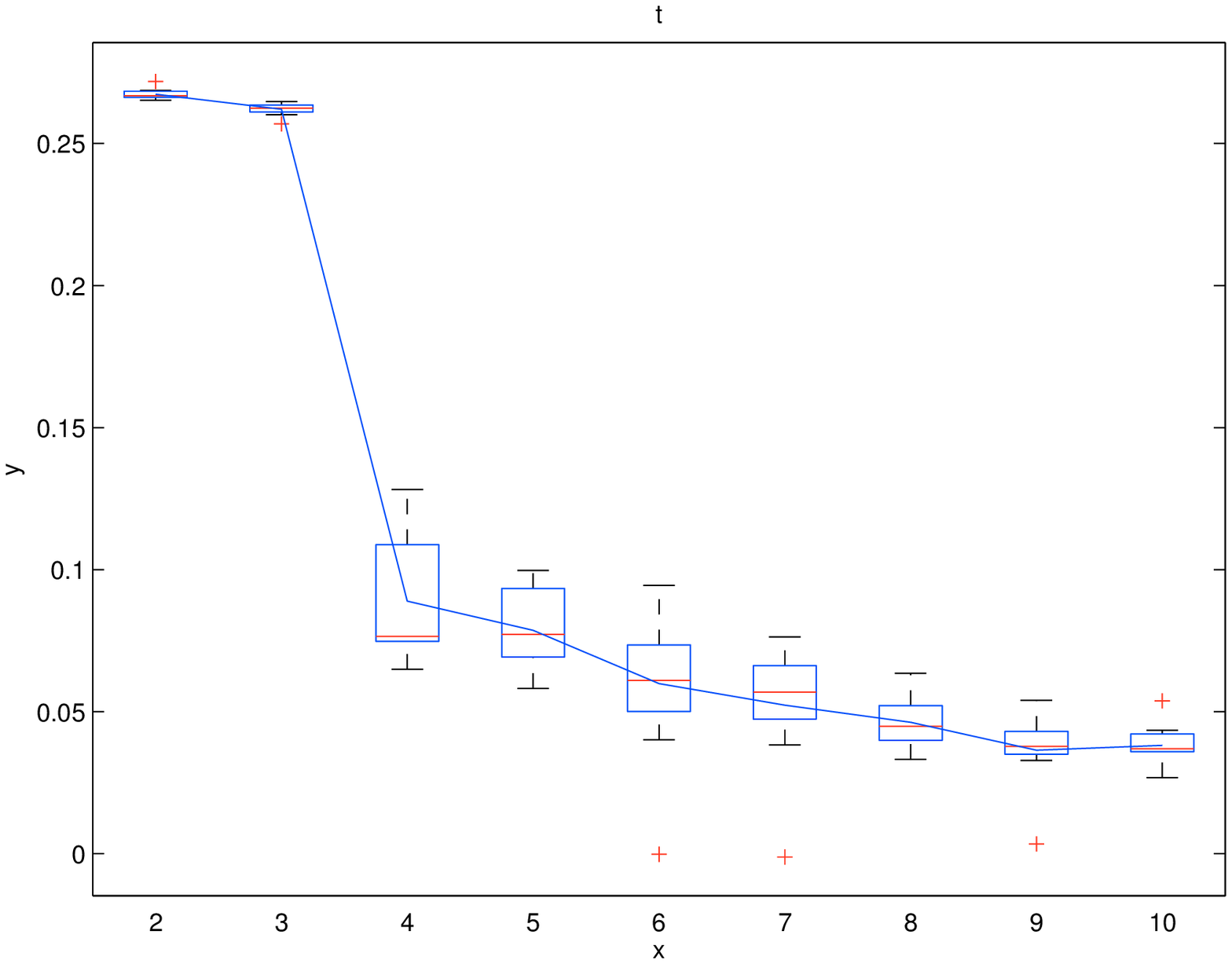}
\end{psfrags}
\end{tabular}
\caption{\textbf{Model selection \textit{Net\_{unw3C1000ov}}}. It is interesting to notice that in this case BLF also suggests to select $4$ clusters. On the other hand Modularity criterion correctly identifies the possible presence of $3$ clusters without falling in this confusion.}\label{fig_3unw}
\end{figure}

\begin{figure}[htbp]
\centering
\subfloat[$3$ clusters]{
\begin{psfrags}
\psfrag{x}[t][t]{{\tiny$e^{(1)}$}}
\psfrag{y}[b][b][1][180]{{\tiny$e^{(2)}$}}
\psfrag{t}[b][b]{{}}
\includegraphics[width=3in]{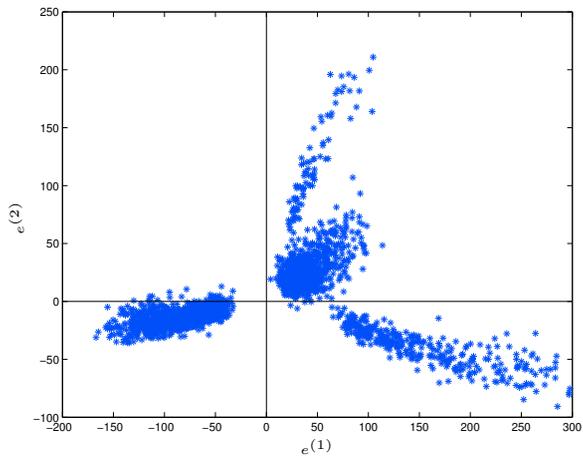}
\end{psfrags}\vspace*{1em}}\\
\subfloat[$4$ clusters]{
\begin{psfrags}
\psfrag{x}[t][t]{{}}
\psfrag{y}[b][b][1][180]{{}}
\psfrag{t}[b][b]{{}}
\includegraphics[width=3in]{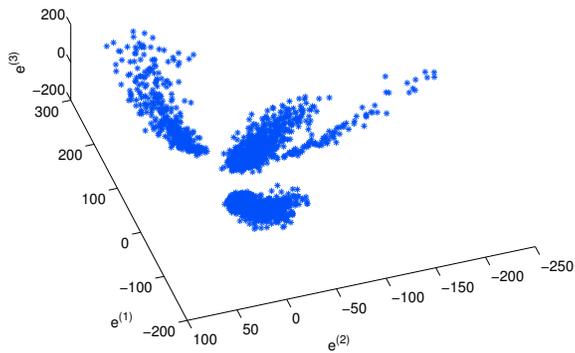}
\end{psfrags}\vspace*{1em}}
\caption{\textbf{Projections space \textit{Net\_{unw3C1000ov}}}. Nodes represented in the space of the latent variables for $3$ and $4$ clusters. We can notice that in the case of $3$ communities there are not $3$ clear lines, while in the case of $4$ communities the line structure is more evident. Probably the BLF criterion is considering the overlapping nodes as belonging to a separate community.}\label{fig_4unw}
\end{figure}

\begin{figure}[htbp]
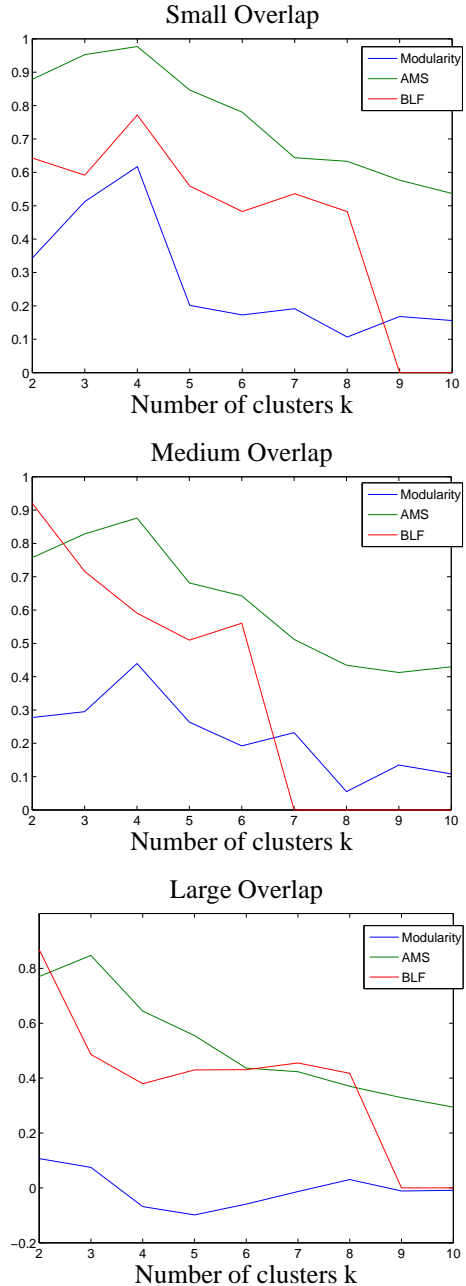

\centering
\begin{tabular}{c}
\begin{psfrags}
\psfrag{x}[t][t]{{Number of clusters k}}
\psfrag{y}[b][b]{{}}
\psfrag{t}[b][b]{{Small Overlap}}
\includegraphics[width=2.5in]{small_overlap_net_MS.eps}
\end{psfrags}\\\\
\begin{psfrags}
\psfrag{x}[t][t]{{Number of clusters k}}
\psfrag{y}[b][b]{{}}
\psfrag{t}[b][b]{{Medium Overlap}}
\includegraphics[width=2.5in]{medium_overlap_net_MS.eps}
\end{psfrags}\\\\
\begin{psfrags}
\psfrag{x}[t][t]{{Number of clusters k}}
\psfrag{y}[b][b]{{}}
\psfrag{t}[b][b]{{Large Overlap}}
\includegraphics[width=2.5in]{big_overlap_net_MS.eps}
\end{psfrags}
\end{tabular}
\caption{\textbf{Model selection for the synthetic network \textit{Net\_{unw4C}}}. BLF (red), Modularity (blue) and AMS (green) are compared. The true number of communities is $k=4$.}\label{fig_netMS}
\end{figure}

\begin{figure}[htbp]
\centering
\begin{tabular}{c}
\begin{psfrags}
\psfrag{x}[t][t]{{Number of clusters $k$ }}
\psfrag{y}[b][b][1][180]{{BLF}}
\psfrag{t}[b][b]{{Number of detected clusters: $5$, $\sigma = 105.07$}}
\includegraphics[width=3in]{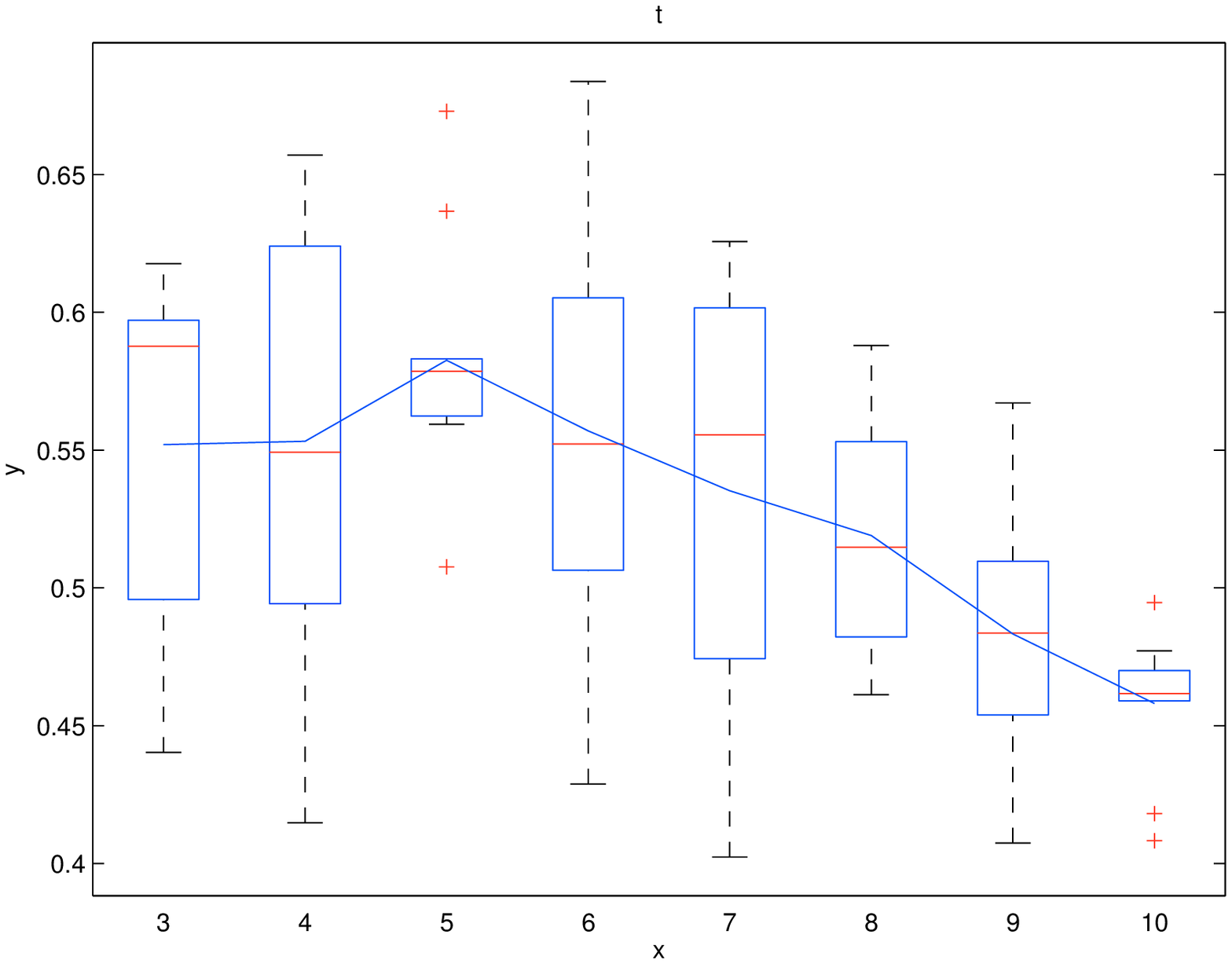}
\end{psfrags}\\\\
\begin{psfrags}
\psfrag{x}[t][t]{{Number of clusters $k$  }}
\psfrag{y}[b][b][1][180]{{Modularity}}
\psfrag{t}[b][b]{{Number of detected clusters: $6$, $\sigma = 127.29$ }}
\includegraphics[width=3in]{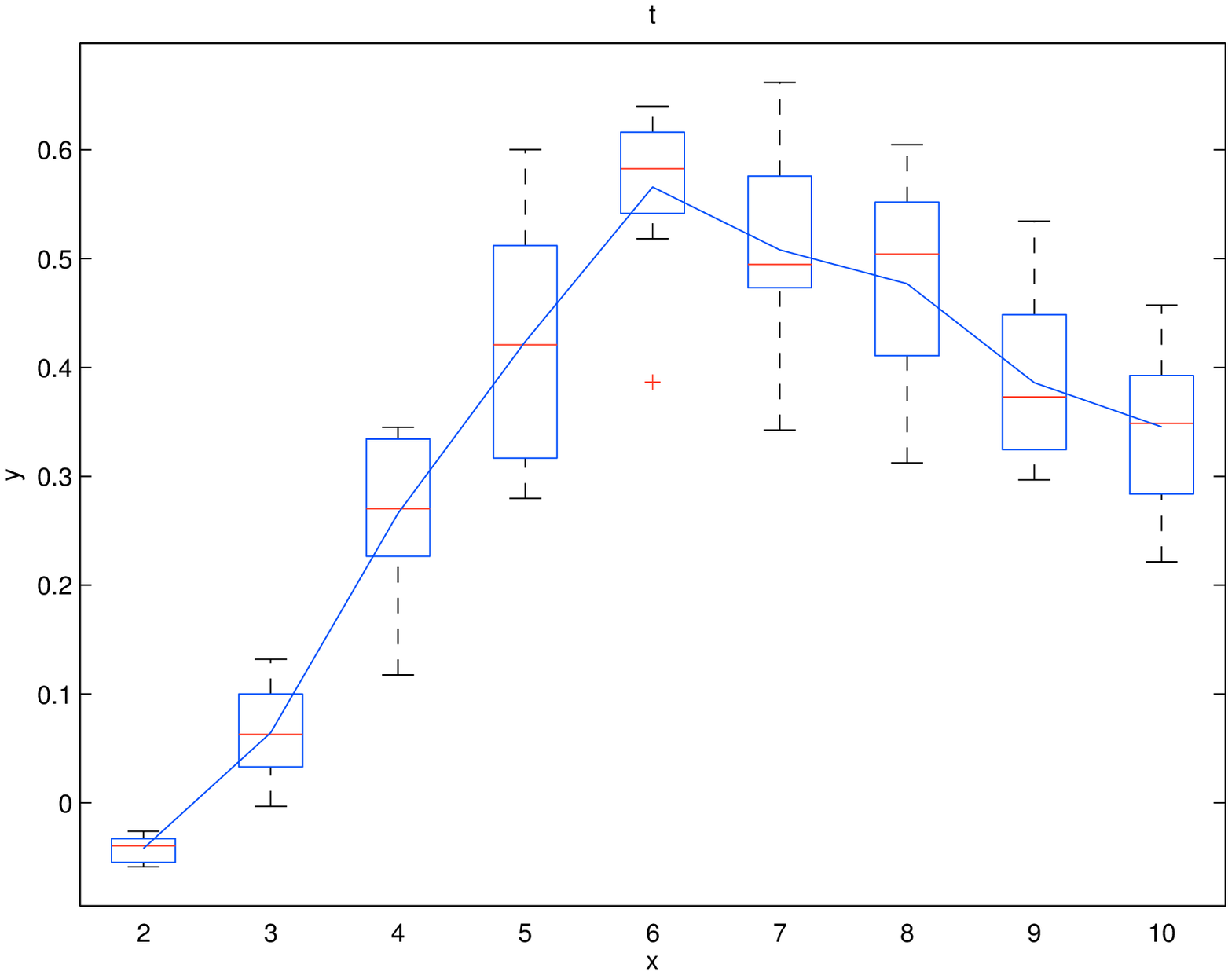}
\end{psfrags}
\end{tabular}
\caption{\textbf{Model selection network \textit{Net\_{w6C}}}. BLF and Modularity are contrasted in relation to their ability of selecting the proper number of communities, which is $6$. Moreover since we deal with a weighted network, we use the RBF kernel to describe the similarity between the nodes and its bandwidth $\sigma$ must also be carefully tuned.}\label{fig_2w}
\end{figure}

\begin{figure}[htbp]
\centering
\begin{tabular}{c}
\begin{psfrags}
\psfrag{x}[t][t]{{Number of clusters $k$ }}
\psfrag{y}[b][b][1][180]{{BLF}}
\psfrag{t}[b][b]{{Number of detected clusters: $3$ , $\sigma = 104.58$}}
\includegraphics[width=3in]{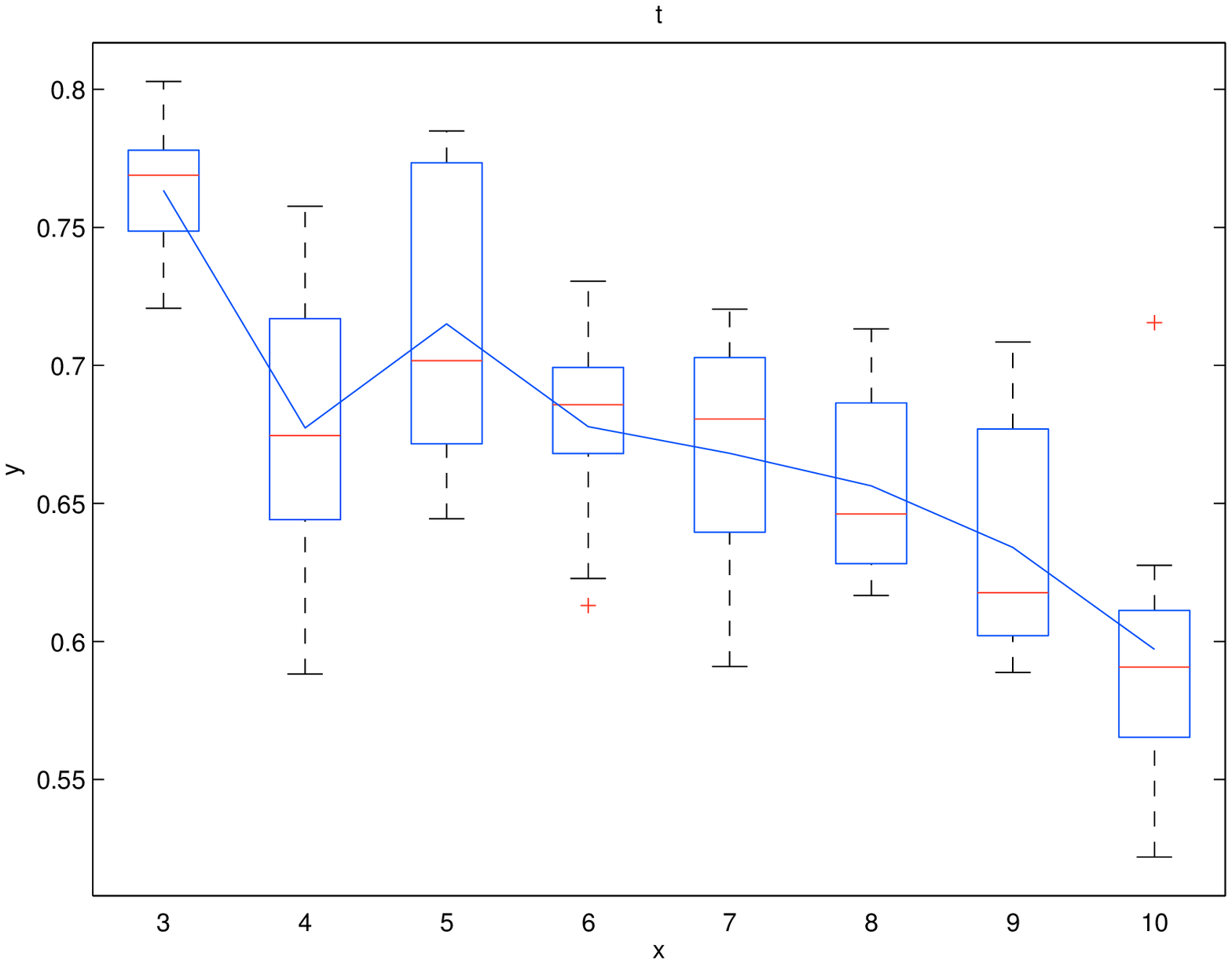}
\end{psfrags}\\\\
\begin{psfrags}
\psfrag{x}[t][t]{{Number of clusters $k$  }}
\psfrag{y}[b][b][1][180]{{Modularity}}
\psfrag{t}[b][b]{{Number of detected clusters: $5$, $\sigma = 117.80$}}
\includegraphics[width=3in]{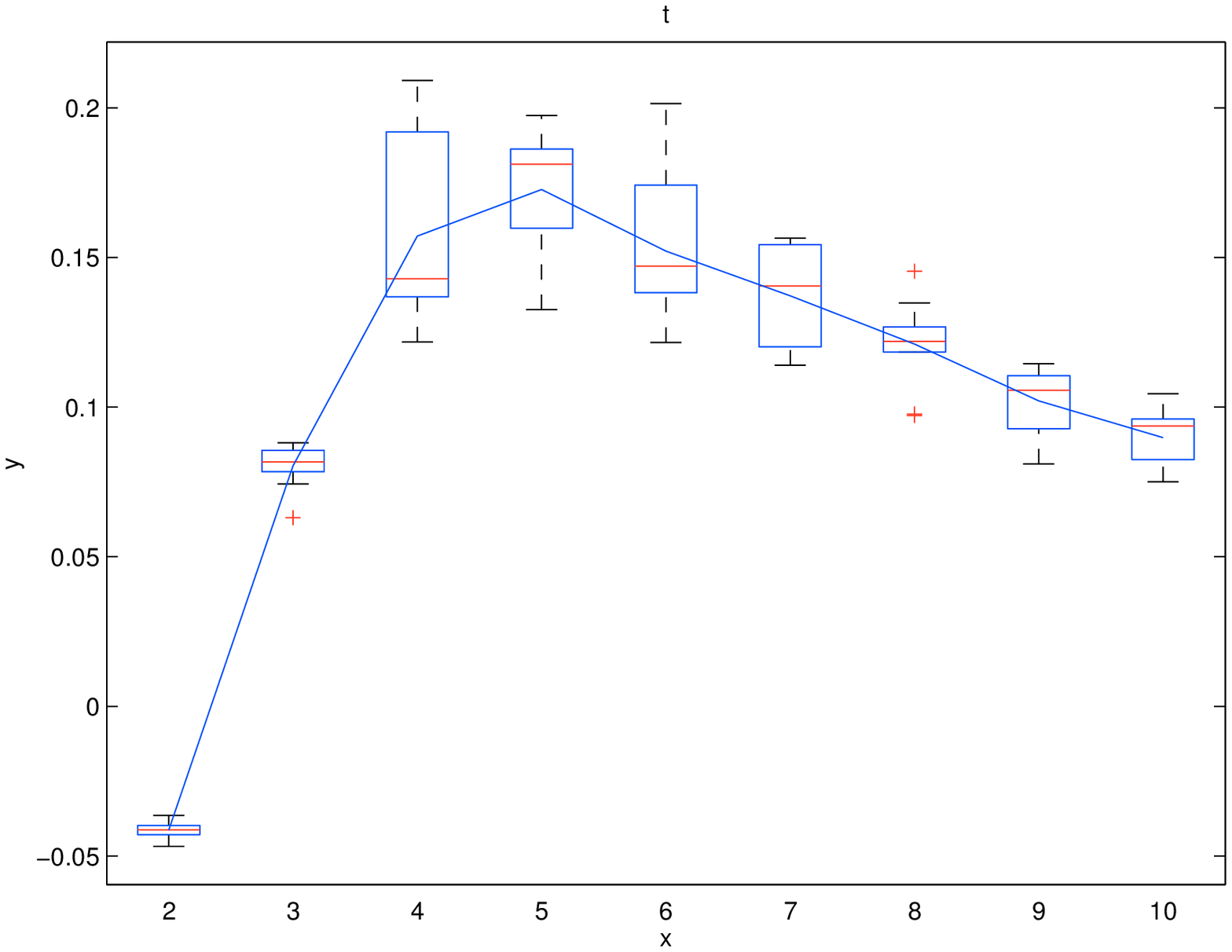}
\end{psfrags}
\end{tabular}
\caption{\textbf{Model selection network \textit{Net\_{w4C1000ov}}}. BLF and Modularity are compared. Both criteria do not seem to be able to detect the $4$ overlapping clusters. This is likely due to the not strong community structure testified by quite low values of Modularity corresponding to the various partitions.}\label{fig_3w}
\end{figure}

Regarding the model selection issue, we compare the BLF, Modularity and AMS criteria for some of the synthetic networks reported in table \ref{table_synthnet}. Roughly speaking, the main observation we can draw is that in the most difficult and more realistic cases where a certain overlap between the communities is present, Modularity and AMS outperform BLF. For instance in the analysis of \textit{Net\_{unw3C1000ov}} depicted in figure \ref{fig_3unw}, from the boxplot it could seem that the BLF is slightly more likely to detect $3$ communities rather than $4$ (less variability). However, by viewing the results from another perspective, we can be convinced that this is not true. In fact in the space of the latent variables, as we already mentioned in section \ref{sc:BLF}, in the ideal case of perfectly separated clusters every line represents a different community. In figure \ref{fig_4unw} we compare the line structure in this space for $3$ and $4$ clusters, showing as in the latter case the line structure is more clear. This is an indication that the BLF criterion is probably detecting $4$ clusters rather than $3$. Also in figure \ref{fig_netMS} we can observe that concerning the networks \textit{Net\_{unw4CmuS}}, \textit{Net\_{unw4CmuM}} and \textit{Net\_{unw4CmuL}} the BLF criterion works only in case of small overlap, while Modularity and AMS criteria correctly identify $4$ communities in the medium overlap case. Moreover, BLF and Modularity fail in case of large overlap, while AMS detects $3$ communities, giving then an useful indication of the community structure to the user. Finally, the model selection results related to the weighted networks are depicted in figures \ref{fig_2w} and \ref{fig_3w}. In both cases AMS suggests a number of clusters closer to the true community structure.

\begin{figure}[htbp]
\centering
\begin{tabular}{c}
\begin{psfrags}
\psfrag{x}[t][t]{{Number of clusters $k$ }}
\psfrag{y}[b][b][1][180]{{BLF}}
\psfrag{t}[b][b]{{Number of detected clusters: $8$ or $9$}}
\includegraphics[width=3in]{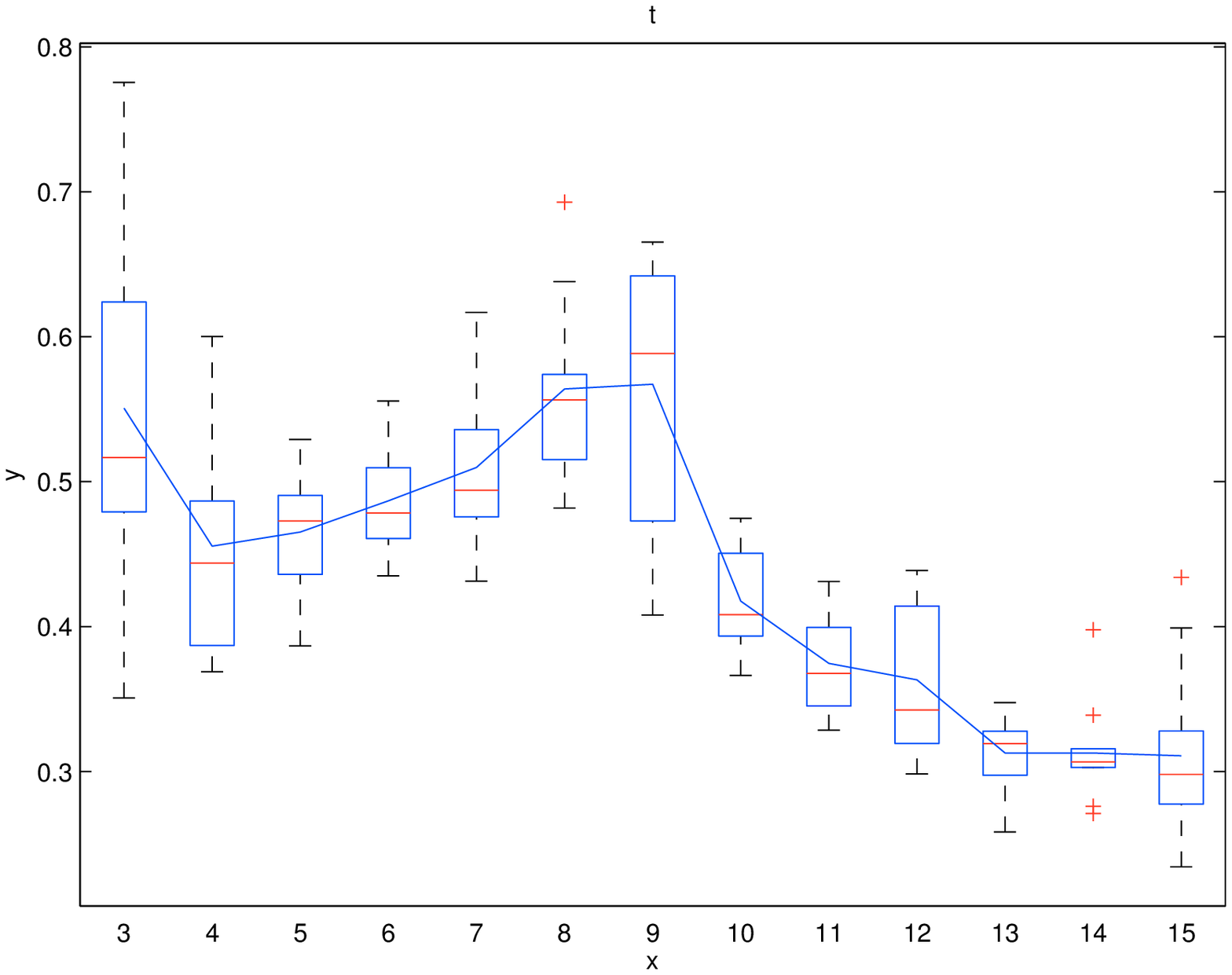}
\end{psfrags}\\\\
\begin{psfrags}
\psfrag{x}[t][t]{{Number of clusters $k$  }}
\psfrag{y}[b][b][1][180]{{Modularity}}
\psfrag{t}[b][b]{{Number of detected clusters: $8$ or $9$}}
\includegraphics[width=3in]{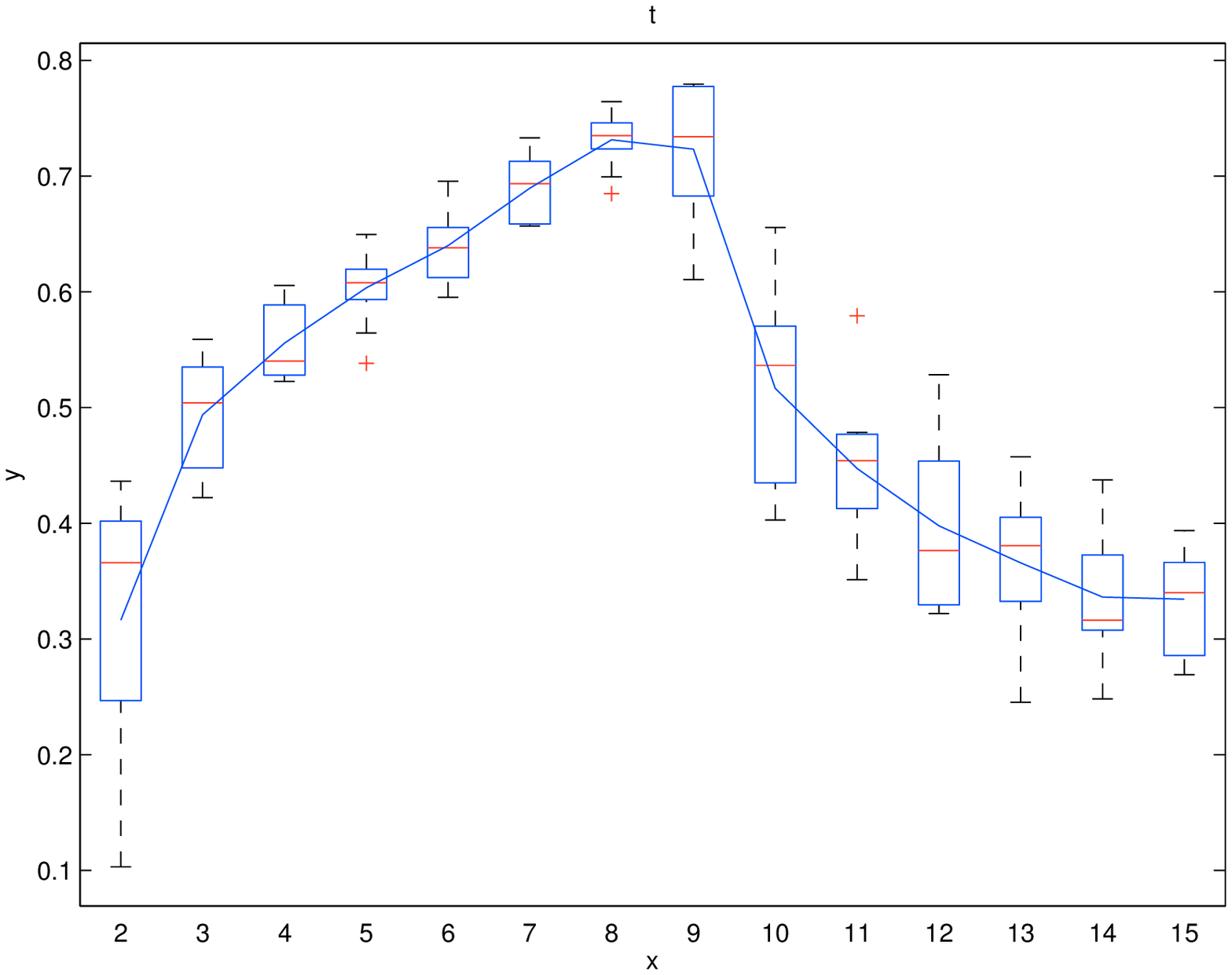}
\end{psfrags}
\end{tabular}
\caption{\textbf{Model selection Net\_{unw9C}}. Modularity and BLF criteria are compared in terms of their ability of detecting the right number of communities, which is $9$ in this case.}\label{fig_2unw}
\end{figure}

In table \ref{table_synthnet_results} we evaluate the ability of KSC and SKSC of correctly partitioning the computer generated networks described in table \ref{table_synthnet}. Average values of Modularity, Conductance and ARI have been used as cluster quality measures. A comparison with the spectral clustering using Nystr\"om approximation for the out-of-sample extension named SC-Nystr\"om is performed. The Nystr\"om method is a technique for finding numerical approximations of eigenfunctions. It has been proposed in \cite{sc_nystrom} to reduce the computational burden in spectral clustering eigenvalue problems. In fact, it allows one to extrapolate the complete grouping solution using only a small subset extracted randomly from the whole dataset to partition. From table \ref{table_synthnet_results} we can notice that the kernel spectral clustering approach in general performs better than SC-Nystr\"om for most of the graphs and is faster for larger networks. 

%\begin{comment}

\begin{table}[t]
\begin{center}
\begin{tabular}{|c|c|c|c|c|c|} 
\hline 
\textbf{Network} & \textbf{Algorithm} & \textbf{ARI}  & \textbf{Mod} & \textbf{Cond} & \textbf{CPU time (s)}\\
\hline
\multirow{2}{*}{\textbf{Net\_{unw9C}}} & {\textbf{KSC}} & $\mathbf{0.92}$ & $\mathbf{0.75}$ & $\mathbf{0.0021}$ & $8.28$ \\
& {\textbf{SC-Nystr\"om}} & $0.85$ & $0.69$ & $0.0025$ & $\mathbf{7.82}$ \\
\hline
\multirow{2}{*} {\textbf{Net\_{unw13C}}} & {\textbf{KSC}} & $\mathbf{0.90}$ & $\mathbf{0.62}$ & $\mathbf{0.0015}$ & $\mathbf{187.03}$  \\
& {\textbf{SC-Nystr\"om}} & $0.87$ & $0.56$ & $0.0019$ & $457.29$  \\
\hline
\multirow{2}{*}{\textbf{Net\_{unw22C}}} & {\textbf{KSC}} & $0.69$ & $0.53$ & $0.0008$ & $\mathbf{9778.57}$ \\
& {\textbf{SC-Nystr\"om}} & $\mathbf{0.79}$ & $\mathbf{0.58}$ & $0.0008$ & $14\,889.68$ \\
\hline
\multirow{2}{*}{\textbf{Net\_{unw3C1000ov}}} & {\textbf{KSC}} & $\mathbf{0.55}$ & $\mathbf{0.27}$ & $\mathbf{0.0061}$ & $8.49$ \\
& {\textbf{SC-Nystr\"om}} & $0.46$ & $0.25$ & $0.0064$ & $\mathbf{7.89}$\\
\hline
\multirow{2}{*}{\textbf{Net\_{unw4CmuS}}} & {\textbf{SKSC}} & $\mathbf{0.95}$ & $\mathbf{0.66}$ & $\mathbf{0.0039}$ & $6.28$ \\
& {\textbf{SC-Nystr\"om}} & $0.89$ & $0.58$ & $0.0040$ & $\mathbf{5.71}$\\
\hline
\multirow{2}{*}{\textbf{Net\_{unw4CmuM}}} & {\textbf{SKSC}} & $\mathbf{0.91}$ & $\mathbf{0.50}$ & $0.0042$ & $6.27$ \\
& {\textbf{SC-Nystr\"om}} & $0.84$ & $0.45$ & $0.0042$ & $\mathbf{5.73}$\\
\hline
\multirow{2}{*}{\textbf{Net\_{unw4CmuL}}} & {\textbf{SKSC}} & $\mathbf{0.62}$ & $\mathbf{0.31}$ & $0.0043$ & $6.29$ \\
& {\textbf{SC-Nystr\"om}} & $0.58$ & $0.26$ & $0.0043$ & $\mathbf{5.74}$\\
\hline
\multirow{2}{*}{\textbf{Net\_{w6C}}} & {\textbf{KSC}} & $\mathbf{0.61}$ & $\mathbf{0.58}$ & $\mathbf{0.0019}$ & $\mathbf{4.33}$ \\
& {\textbf{SC-Nystr\"om}} & $0.56$ & $0.55$ & $0.0020$ & $4.51$\\
\hline
\multirow{2}{*}{\textbf{Net\_{w4C1000ov}}} & {\textbf{KSC}} & $\mathbf{0.42}$ & $\mathbf{0.23}$ & $0.0029$ & $4.41$ \\
& {\textbf{SC-Nystr\"om}} & $0.39$ & $0.21$ & $0.0029$ & $\mathbf{4.34}$\\
\hline
\end{tabular}
\end{center}
\caption{\textbf{Synthetic networks results}. The performance of KSC, SKSC and spectral clustering with the Nystr\"om method used for out-of-sample extension are reported. The results are evaluated according to several measures like ARI, Modularity, Conductance (see Appendix and references therein). Moreover also the time required for execution of the algorithms is illustrated. In the majority of the cases the kernel spectral clustering is more accurate and faster for bigger network data.}
\label{table_synthnet_results}
\end{table}

\section{Real-World Applications}
In order to test KSC and SKSC algorithms on real-life data the following graphs have been used:
\begin{itemize}
\item \textbf{Yeast\_{pro}}: interaction network data for yeast formed by $2\,114$ nodes and $ 4\,480 $ edges. As explained in \cite{barabasi_protein} proteins can have direct or indirect interactions with one another. Indirect interaction refers to being a member of the same functional module but without directly binding to one another. In contrast, direct interaction, refers to two amino acid chains that bind to each other. Obviously, many of these interactions reflect the dynamic state of the cell and are present or absent depending on the particular environment or developmental status of the cell.
\item \textbf{Power grid}: the network of Western USA power grid \cite{wattstro} formed by $4\,941$ nodes and $6\,594$ edges. The vertices represent generators, transformers and substations, and edges represent high voltage transmission lines between them.
\item \textbf{Karate}: the Zachary's karate club network \cite{karate_net} consists of 34 member nodes, and splits in two smaller clubs after a dispute arose during the course of Zachary's study between the administrator and the instructor. 
\item \textbf{Football}: this network \cite{Girvan11062002} describes American college football games and is formed by $115$ nodes (the teams) and $616$ edges (the games). It can be divided into $12$ communities according to athletic conferences.    
\item \textbf{Dolphins}: the dataset, described in \cite{dolphins_net}, depicts the associations between $62$ dolphins with ties between dolphin pairs representing statistically significant frequent associations. As pointed out in \cite{dolphins_net}, this network can be naturally split in two main groups (female-female and male-male associations), even if another cluster of mixed sex associations could also be considered.     
\end{itemize}  
These networks have been widely studied and represent standard real-world benchmarks for testing community detection algorithms, since for most of them the underlying community structure is known.    

\begin{figure}[htbp]
\centering
\begin{tabular}{cc}
\begin{psfrags}
\psfrag{x}[t][t]{{Number of clusters k}}
\psfrag{y}[b][b][1][360]{{Modularity}}
\psfrag{t}[b][b]{{Yeast\_{pro}}}
\includegraphics[width=2.25in]{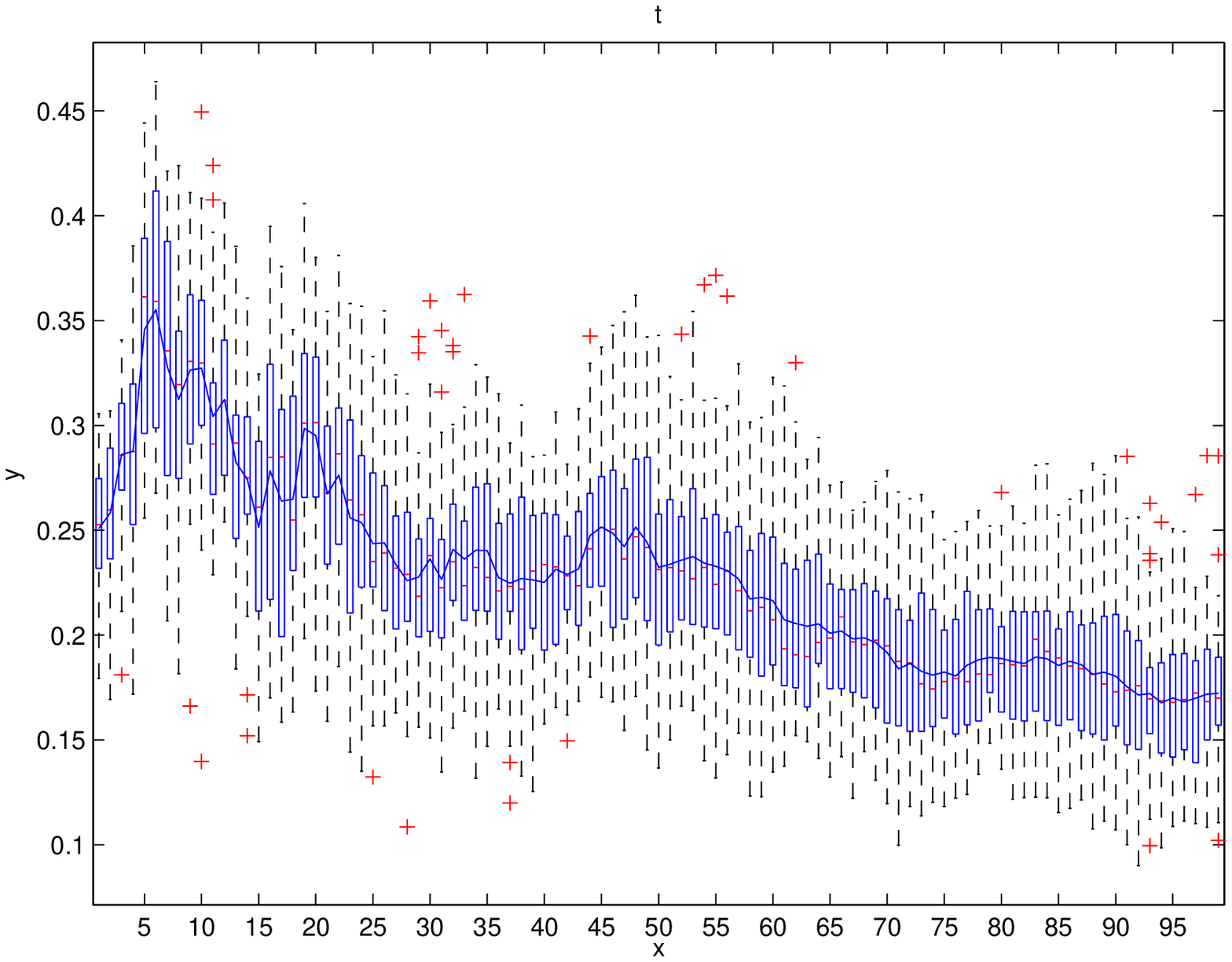}
\end{psfrags}&
\begin{psfrags}
\psfrag{x}[t][t]{{Number of clusters k}}
\psfrag{y}[b][b][1][360]{{Modularity}}
\psfrag{t}[b][b]{{Power grid}}
\includegraphics[width=2.25in]{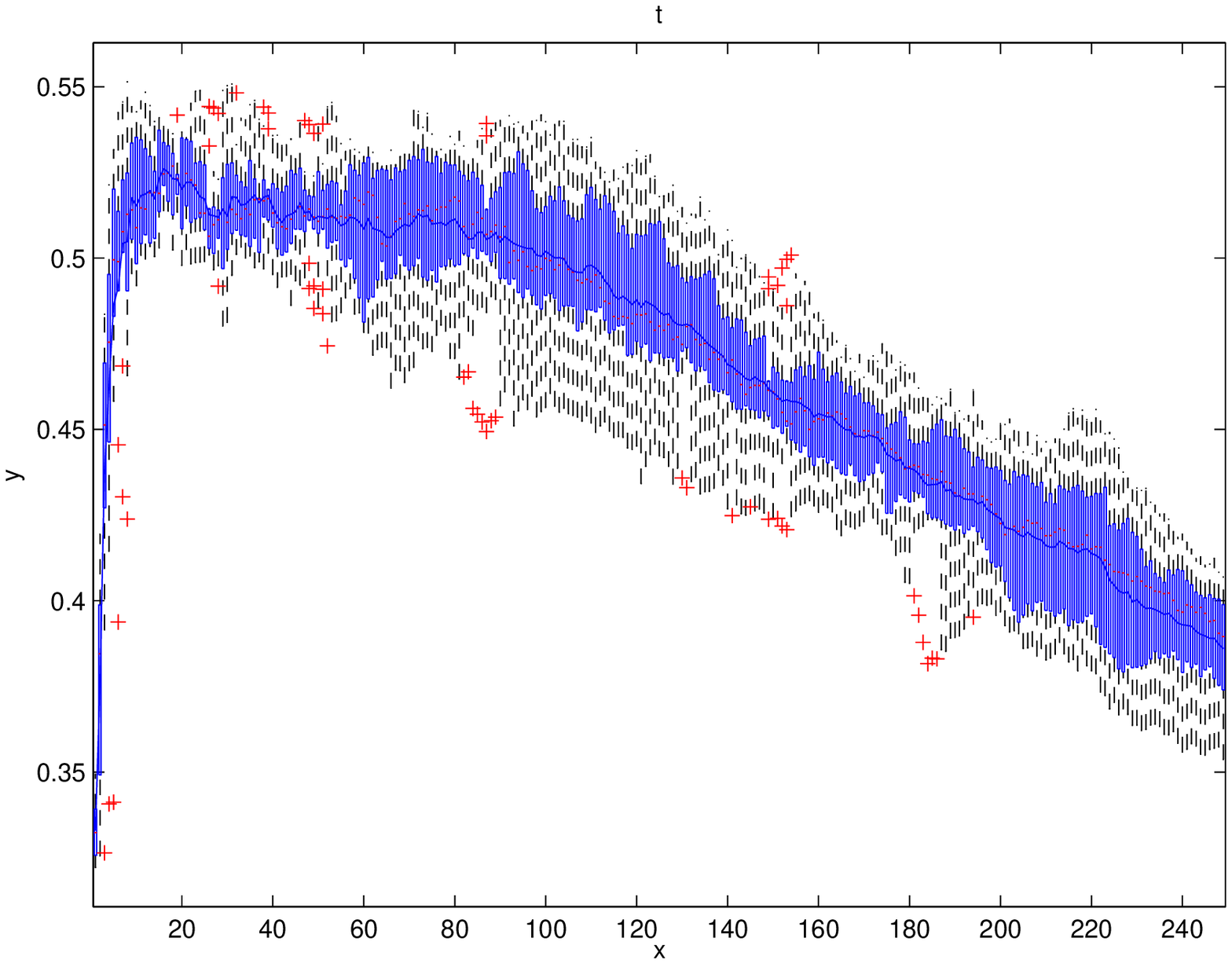}
\end{psfrags}\\\\
\begin{psfrags}
\psfrag{x}[t][t]{{Number of clusters k}}
\psfrag{y}[b][b]{{}}
\psfrag{t}[b][b]{{Karate}}
\includegraphics[width=2.25in]{karate_MS.eps}
\end{psfrags}&
\begin{psfrags}
\psfrag{x}[t][t]{{Number of clusters k}}
\psfrag{y}[b][b]{{}}
\psfrag{t}[b][b]{{Football}}
\includegraphics[width=2.25in]{football_MS.eps}
\end{psfrags}\\\\
\multicolumn{2}{c}{ 
\begin{psfrags}
\psfrag{x}[t][t]{{Number of clusters k}}
\psfrag{y}[b][b]{{}}
\psfrag{t}[b][b]{{Dolphins}}
\includegraphics[width=2.25in]{dolphins_MS.eps}
\end{psfrags} }
\end{tabular}
\caption{\textbf{Validation procedure for the real-world networks}. BLF, Modularity and AMS model selection criteria are compared.}\label{fig_MSreal}
\end{figure}

The model selection outcomes are depicted in figure \ref{fig_MSreal}. For the first two networks Modularity and BLF have been compared. Since BLF did not give any good indication we only show the results related to Modularity. The latter suggests the presence of $7$ and $16$ communities for the \textit{Yeast\_{pro}} and \textit{Power grid} graphs respectively. Regarding \textit{Karate}, \textit{Football} and \textit{Dolphins} networks AMS, BLF and Modularity model selection criteria have been contrasted. In the \textit{Karate} network case all methods detect $2$ communities. For what concerns the \textit{Football} dataset BLF fails, while Modularity and AMS select $10$ communities (AMS suggests also $3$). Finally in the \textit{Dolphins} network AMS selects $3$ communities, Modularity $6$ and BLF $2$. In all these cases, given the limited size of the graphs, the training and validation sub-graphs have been chosen to be $70 \%$ and $30 \%$.
 
\begin{figure}[htbp]
\centering
\begin{tabular}{c}
\includegraphics[width=3.5in]{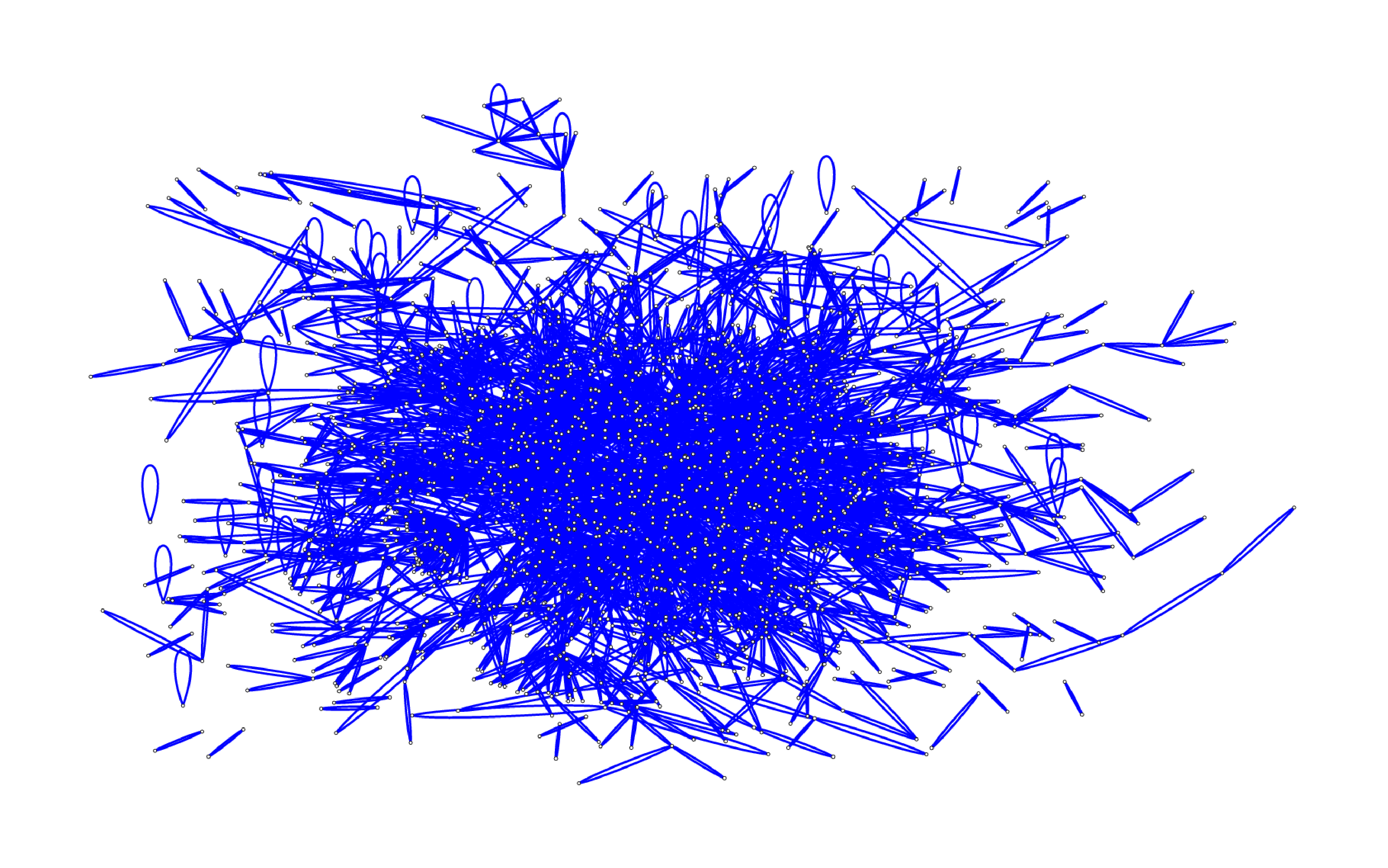}\\
\includegraphics[width=3.5in]{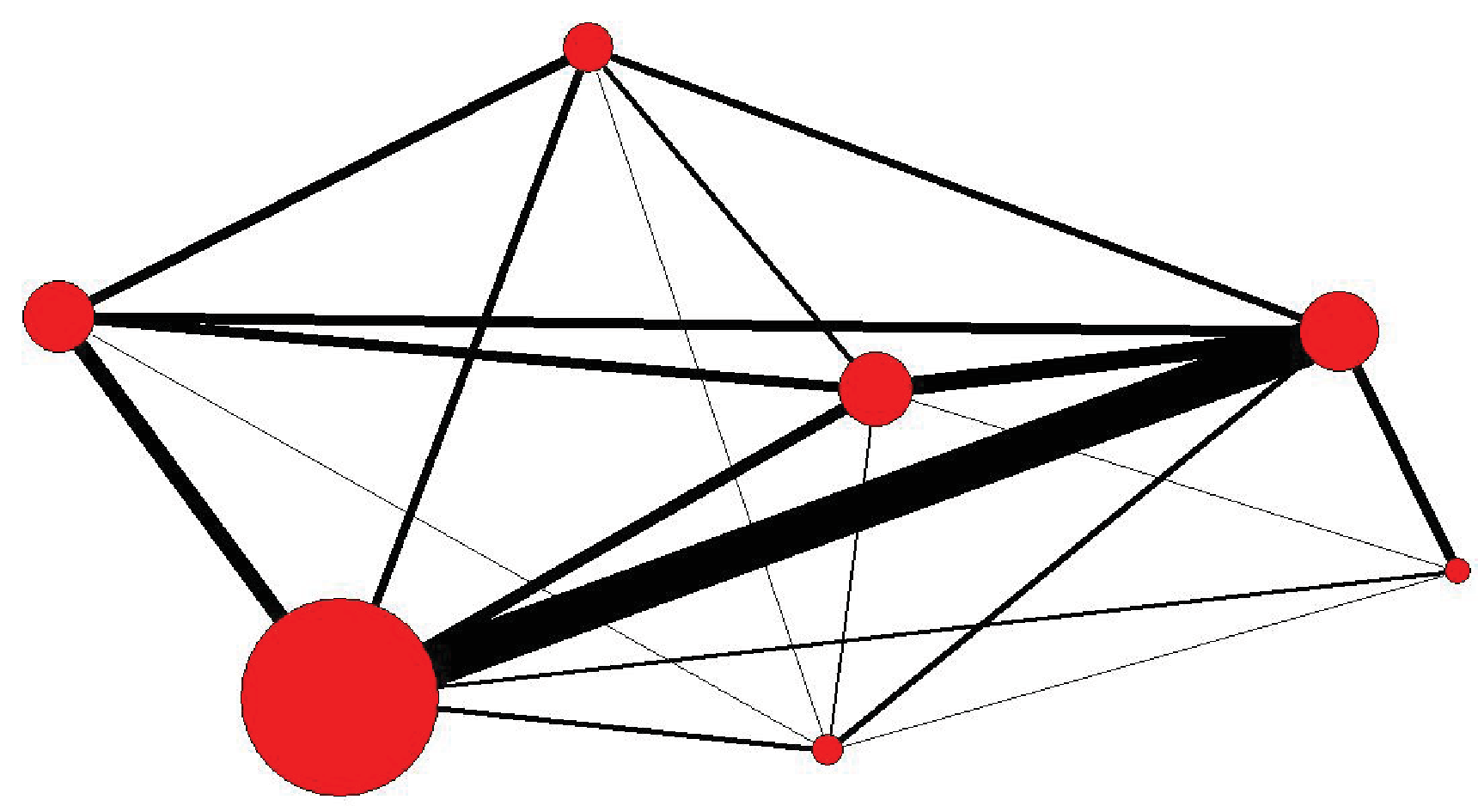}
\end{tabular}
\caption{\textbf{\textit{Yeast\_{pro}} network visualization}. Original network (top) and partition discovered by KSC for the network of interacting proteins of yeast (bottom). Every red circle represents a cluster with size related to the number of nodes belonging to it. The position of the circles is not relevant. The edges are the links between nodes belonging to different communities, with thickness proportional to the number of these inter-community edges. The nodes and edges in each detected community, for simplicity, are not shown. The figure has been made by using the software for large network analysis \texttt{Pajek} (see \textit{http://pajek.imfm.si/doku.php}).} \label{pajek_yeast_protein}
\end{figure}

\begin{figure}[htbp]
\centering
\begin{tabular}{c}
\includegraphics[width=3.5in]{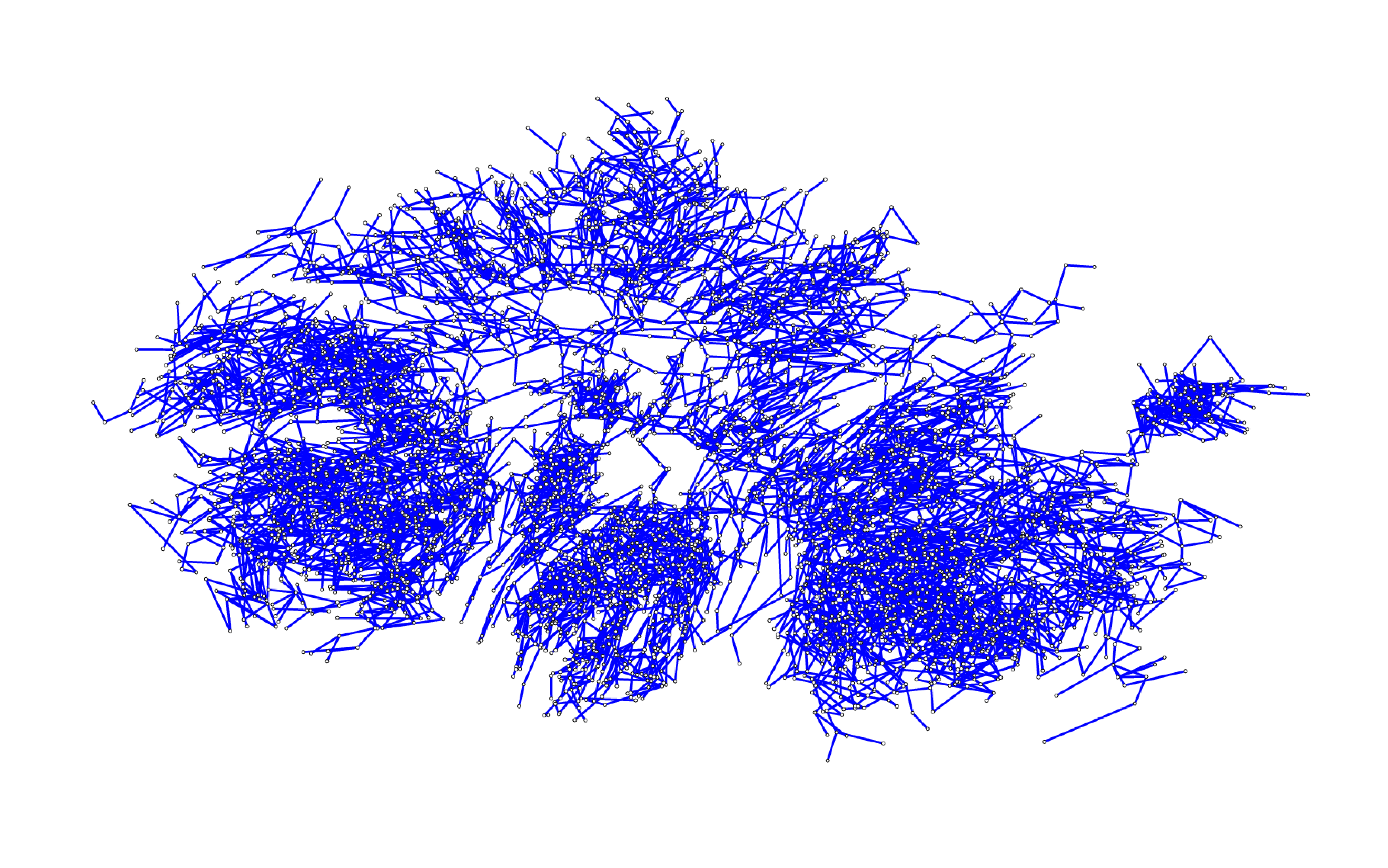}\\
\includegraphics[width=3.5in]{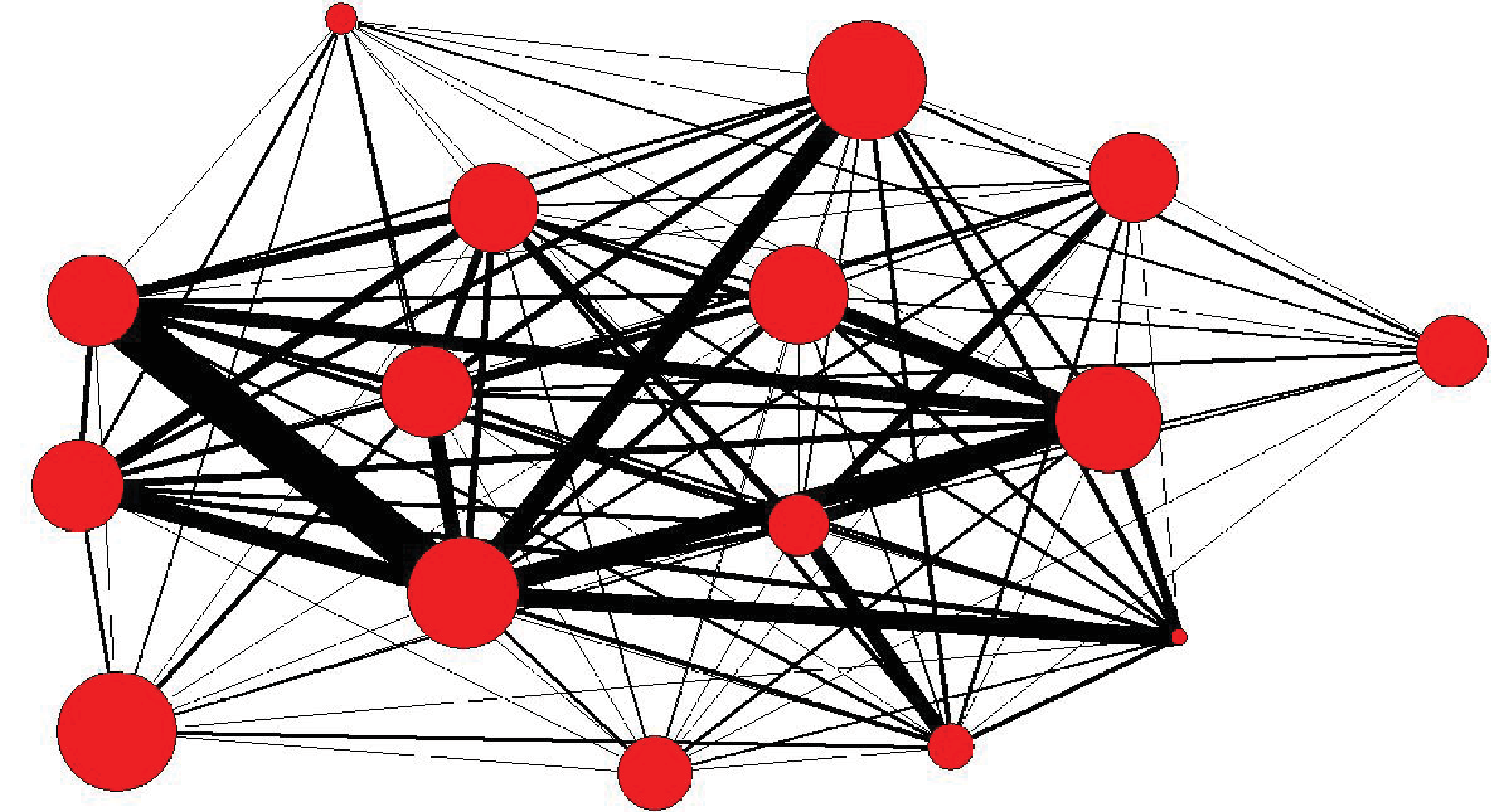}
\end{tabular}
\caption{\textbf{\textit{Power grid} network visualization}. Original network (top) and community structure of the Western USA power grid discovered by the kernel spectral clustering model (bottom). The comments made for Figure \ref{pajek_yeast_protein} are still valid here.} \label{pajek_powergrid}
\end{figure}

In table \ref{table_results_real} the partitions found by KSC or SKSC are evaluated according to Conductance, Modularity and ARI. We can notice how for all the networks the proposed kernel-based approach for community detection is able to discover a meaningful modular structure.
Regarding the \textit{Yeast\_{pro}} graph, unlike the analysis performed in \cite{barabasi_protein} we consider both direct and indirect interactions, for a total of $2\,114$ nodes instead of $1\,870$. We found $7$ clusters, with the largest component containing about the $60\%$ of the linked proteins (and all the proteins with indirect interactions). This outcome is shown in the bottom of figure \ref{pajek_yeast_protein} (only the directly interacting proteins are considered). The community structure found by our algorithm seems quite attractive for its simplicity, but needs further investigation in order to assess a meaningful biological interpretation of the discovered modules. The same considerations can be drawn for the Power grid network, which has been partitioned into $16$ distinct communities as illustrated in the bottom of figure \ref{pajek_powergrid}. Regarding the analysis of the smaller networks, more observations can be done by considering the moderated outputs produced by SKSC:     
\begin{itemize}
\item Karate network: node $31$ has the most certain membership, being assigned with probability $1$ to its community. This is reasonable since it is linked to the highest degree nodes present in the same module  
\item Football network: node $25$ and $59$ are assigned to every cluster with nearly the same probability (i.e. they are overlapping nodes)
\item Dolphins network: the membership of node $39$ is equally shared between the $2$ communities of the network.  
\end{itemize}
However, these insights need to be further investigated in order to better interpret the results.

Finally, an exhaustive comparison between the proposed KSC-based method for community detection explained in this chapter and other state-of-the-art algorithms has been performed in \cite{raghvendra_bigdata}. In the cited work we contrasted  our technique with the Louvain method \cite{louvainmethod}, Infomap \cite{infomap} and CNM \cite{CNM} on several synthetic and real-world graphs. In general, we observed that KSC produces a smaller number of clusters with better (i.e. lower) Conductance, while the Louvain method is the best w.r.t. the Modularity quality metric.
 
\begin{table}[htbp]
\begin{center}
\begin{tabular}{|c|c|c|c|c|} 
\hline 
\textbf{Network} & \textbf{Algorithm}  & \textbf{Mod} &  \textbf{Cond} & \textbf{ARI} \\
\hline
\textbf{Yeast\_{pro}} & \textbf{KSC} & $0.35$ & $0.029$ & $-$ \\
\hline
\textbf{Power grid} & \textbf{KSC} & $0.52$  & $0.033$ & $-$ \\
\hline
\textbf{Karate} & \textbf{SKSC} & $0.37$ & $0.083$ & $1$ \\
\hline
\textbf{Football} & \textbf{SKSC} & $0.58$ & $0.76$ & $0.77$ \\
\hline
\textbf{Dolphins} & \textbf{KSC} & $0.35$ & $0.13$ & $0.91$ \\
\hline
\end{tabular}
\end{center}
\caption{\textbf{Evaluation partitions of the real-world graphs}. The community structure detected by the proposed kernel- based approach (SKSC or KSC) is evaluated in terms of Modularity, Conductance and ARI (when a sort of ground-truth describing the real splitting is available).}
\label{table_results_real}
\end{table}

\section{Conclusions}
This chapter focused on the community detection problem. The latter has been the subject of an extensive research in recent years, due to the large availability of network data coming from many sources. After discussing the literature in the field, we have introduced our approach to deal with community detection, which is based on the KSC and SKSC methods presented in the previous chapter. In particular, we have presented a model selection criterion based on the Modularity quality function, which is more suited for network data. Moreover, a sampling method based on the greedy maximization of the expansion factor is used to extract a small sub-graph representative of the community structure of the entire network. This sub-graph is used to train the KSC model, and in a second phase the out-of-sample extension property allows to predict the memberships of the remaining nodes. We showed how our technique produces a high-quality partitioning both in terms of Modularity and Conductance, and it outperforms spectral clustering using the Nystr\"om method for out-of-sample extension. Finally, we discussed the computational complexity of our approach.

%%%%%%%%%%%%%%%%%%%%%%%%%%%%%%%%%%%%%%%%%%%%%%%%%
% Keep the following \cleardoublepage at the end of this file, 
% otherwise \includeonly includes empty pages.
\cleardoublepage

% vim: tw=70 nocindent expandtab foldmethod=marker foldmarker={{{}{,}{}}}
%

%
  \graphicspath{{\chaptersdir/chapter4/image/}}%
  %\cleardoublepage%
  \chapter{Clustering Evolving Networks}
\label{ch:evclustering}
Evolving graphs describe many natural phenomena changing over time, such as social relationships, trade markets, metabolic networks etc. In this chapter we describe a new model for clustering evolving networks, where the smoothness of the clustering results over time can be considered as a valid prior knowledge. This algorithm, called kernel spectral clustering with memory effect (MKSC), is based on a constrained optimization formulation typical of LS-SVM, where the objective function is designed to explicitly incorporate temporal smoothness. The latter allows the model to cluster the current data well and to be consistent with the recent past. We also introduce new cluster quality measures that can be used for model selection or better assess the performance of a clustering algorithm in a dynamic scenario. Moreover, we consider two frameworks:
\begin{itemize}
\item framework $1$: we assume that at every time step the partitioning has to be smoothed, and we focus on the case where the number of nodes and communities is not changing across time.
\item framework $2$: the smoothness parameter $\nu$ is tuned, and the number of nodes and communities is varying over time.
\end{itemize}
Also a compact visualization of the cluster dynamics in a 3D embedding is proposed. 
Finally, the MKSC model is successfully applied to a number of toy problems and a real-world network. We also compare the novel algorithm with some state-of-the-art methods like Evolutionary Spectral Clustering (ESC \cite{evol_spec_clustering}), the Louvain method applied separately in each snapshot, and Adaptive Evolutionary Clustering (AFFECT \cite{adapt_evol_clu_Xu}). 

\section{Literature Review}
In many practical applications we deal with community detection in dynamic scenarios, which recently has been the subject of a common research endeavour in the science community. The first article that formalized the problem is \cite{evol_clustering_chakrabarti}, where dynamic community detection is named evolutionary clustering. This work is based on the intuition that if the new data does not deviate from the recent history the clustering should be similar to that performed for the previous data. However, if the data changes significantly, the clustering must be modified
to reflect the new structure. This temporal smoothness between clusters in successive time-steps is also the main principle  behind the methods introduced in \cite{evol_spec_clustering}, \cite{Lin:2009:ACE:1514888.1514891} and \cite{adapt_evol_clu_Xu}. In particular, in \cite{evol_spec_clustering} the evolutionary spectral clustering algorithm (ESC) has been proposed, which aims to optimize the cost function $J_{tot} = \eta J_{\textrm{temp}} + (1-\eta)J_{\textrm{snap}}$. $J_{\textrm{snap}}$ describes the classical spectral clustering objective related to each snapshot of an evolving graph. $J_{\textrm{temp}}$ measures the cost of applying the partitioning found at time $t$ to the snapshot at time $t-1$, penalizing then clustering results that disagree with the recent past. In \cite{adapt_evol_clu_Xu} an evolutionary clustering framework that adaptively estimates the optimal smoothing parameter using shrinkage estimation is presented. The method, called AFFECT, allows to extend a number of static clustering algorithms into evolutionary clustering techniques. In \cite{Mucha14052010} the authors developed a methodology that generalizes the determination of community structure to multi-slice networks that are defined by coupling multiple adjacency matrices. By encoding in the inter-slice connections variations across time, it is possible to analyse evolving networks. The work presented in \cite{Palla07quantifyingsocial} discovered an interesting relationship between the size of a community, its lifetime and stationarity. A typical small and stationary community is observed to undergo minor changes, while living for a long time. In contrast, a small community with high turnover of its members tends to have a small lifetime. The opposite is observed for large communities: a big stationary community disintegrates early, while a large non-stationary community whose members change dynamically has a longer lifetime. In \cite{Asur:2009:EFC:1631162.1631164} the authors perform clustering separately for each snapshot of an evolving network and develop a framework for capturing and identifying interesting events from them. They use then these events to characterize complex behavioural patterns of nodes and communities over time. Also \cite{Greene2011} introduces a method for efficiently identifying and tracking dynamic communities, which involves matching communities found at consecutive time steps in the individual snapshot graphs.

In this chapter, we describe a new method for clustering dynamic networks called kernel spectral clustering with memory effect (MKSC). Our technique is developed in the LS-SVM learning framework, where we incorporate the temporal smoothness between clusters in successive time-steps at the primal level. In this way the model is able to track the long-term trend and at the same time it reduces the short-term variation due to noise, similarly to what happens with moving averages in time-series analysis. Moreover, a precise model selection scheme and the out-of-sample extension to new nodes is presented. Up to our knowledge, all these features make MKSC unique in its kind. Finally, we consider two main frameworks:
\begin{itemize}
\item \textbf{framework 1}: we assume that at every time step the partitioning has to be smoothed. Thus, we fix the regularization constant $\nu$ (which acts as a smoothness parameter) to $\nu=1$. Then the amount of smoothness is chosen by the user through the memory $M$. Moreover, we focus on the case where the number of nodes and communities is not changing across time.
\item \textbf{framework 2}: the smoothness parameter $\nu$ is tuned, meaning that the memory activates automatically only when is needed, i.e. when the network undergoes major changes in its community structure. Moreover, we allow MKSC to deal with nodes entering and leaving over time and to recognize a rich variety of events (splitting, merging, dissolving etc.) by introducing a simple tracking mechanism.         
\end{itemize}

\section{The MKSC Model}
A dynamic network is a sequence of networks $\mathcal{S} = \{\mathcal{G}_t = (\mathcal{V}_t,\mathcal{E}_t)\}_{t=1}^T$ over time $T$,  where $t$ indicates the time index. The symbol $\mathcal{V}_t$ indicates the set of nodes in the graph $\mathcal{G}_t$ and $\mathcal{E}_t$ the related set of edges. In what follows we assume that $\vert \mathcal{V}_t \vert$ is constant, that is all the graphs in the sequence have the same number of nodes (the symbol $\vert \cdot \vert$ indicates the cardinality of a set).      
The MKSC model implements a trade-off between the current clustering and the previous partitioning by incorporating the temporal smoothness as prior knowledge. 

Given an evolving network with $N$ nodes, for each snapshot the primal problem of the MKSC model, where $N_{\textrm{Tr}}$ nodes are used for training, can be stated as follows in matrix notation \cite{Langone_physicA}:
\begin{equation}
%\tiny 
\label{primalMKSC}
\begin{aligned}
& \underset{w^{(l)},e^{(l)},b_l}{\text{min}}
& & \frac{1}{2}\sum_{l=1}^{k-1}w^{(l)^T}w^{(l)}-\frac{\gamma}{2{N_{\textrm{Tr}}}}\sum_{l=1}^{k-1} e^{(l)^T}D_{\textrm{Mem}}^{-1}e^{(l)} -\nu \sum_{l=1}^{k-1}w^{(l)^T}w_{\textrm{old}}^{(l)} \\
& \text{subject to}
& & e^{(l)}=\Phi w^{(l)}+b_l1_{N_{\textrm{Tr}}}.
\end{aligned}
\end{equation}
The first term in the objective (\ref{primalMKSC}) indicates the minimization of the model complexity, while the second term casts the clustering problem in a weighted kernel PCA formulation as in \cite{multiway_pami}. The third term, i.e. $\sum_{l=1}^{k-1}w^{(l)^T}w_{\textrm{old}}^{(l)}$, describes the correlation between the actual and the previous models, which we want to maximize. In this way it is possible to introduce temporal smoothness in our formulation, such that the current partition does not deviate too dramatically from the recent past. The subscripts \textit{old} and \textit{Mem} refer to time steps $t-1,\hdots,t-M$, where $M$ indicates the memory, that is the past information we want to consider when performing the clustering at the actual time step $t$. The symbols have the following meaning:
\begin{itemize}
\item $e^{(l)}$ represent the $l$-th binary clustering model for the $N$ points and are referred interchangeably as projections, latent variables or score variables. 
\item the index $l=1,\hdots,k-1$ indicates the score variables needed to encode the $k$ clusters to find via an Error Correcting Output Codes (ECOC) encoding-decoding procedure. In other words, $e_i^{(l)}=w^{(l)^T}\varphi(x_i)+b_l$ are the latent variables of a set of $k-1$ binary clustering indicators given by $\textrm{sign}(e_i^{(l)})$. The binary indicators are combined to form a codebook $\mathcal{CB} = \{c_p\}_{p = 1}^k$, where each codeword is a binary string of length $k-1$ representing a cluster.
\item $w^{(l)} \in \mathbb{R}^{d_h}$ and $b_l$ are the parameters of the model at time $t$, and $w^{(l)}_{\textrm{old}} = \sum_{i=1}^{M}w^{(l)}_{\textrm{prev,i}} = \sum_{i=1}^{M}\Phi^{(l)}_{\textrm{prev,i}}\alpha^{(l)}_{\textrm{prev,i}}$ are the model parameters related to the $M$ previous snapshots $\mathcal{G}_{t-1}, \hdots, \mathcal{G}_{t-M}$. The subscript \textit{prev,i} means time step $t-i$. By considering $M$ time stamps in the past, we say that our model has a memory of $M$ snapshots. 
\item $D^{-1} \in \mathbb{R}^{N_{\textrm{Tr}} \times N_{\textrm{Tr}}}$ is the inverse of the degree matrix $D$ related to the actual kernel matrix $\Omega$, i.e. $D_{ii} = \sum_j\Omega _ {ij}$, while $D_ {\textrm{Mem}}^{-1} \in \mathbb{R}^{{N_{\textrm{Tr}}} \times {N_{\textrm{Tr}}}}$ is the inverse of the degree matrix $D_ {\textrm{Mem}} = D + \sum_{r=1}^{M} D_{\textrm{new-prev,r}}$, which is the sum of the actual degree matrix $D$ and the $M$ previous degree matrices, each with entries $D_{\textrm{new-prev},ii} = \sum_j\Omega _ {\textrm{new-prev},ij}$.
\item $\Phi$ is the $N_{\textrm{Tr}} \times d_h$ feature matrix $\Phi=[\varphi(x_1)^T;\hdots;\varphi(x_{N_{\textrm{Tr}}})^T]$ which expresses the relationship between each pair of data objects in a high dimensional feature space $\varphi:\mathbb{R}^d\rightarrow\mathbb{R}^{d_h}$. 
\item $\gamma \in\mathbb{R}^+$ and $\nu \in\mathbb{R}^+$ are regularization constants. In particular $\nu$ can be thought as a kind of smoothness parameter, since it enforces the current model to resemble the old models, that is the ones developed for the previous $M$ snapshots. 
\end{itemize}
The dual solution to the constrained optimization problem (\ref{primalMKSC}) is formalized in the following Lemma.\\\\
\textbf{Lemma \cite{Langone_physicA}} \textit{Given a positive definite kernel function $K: \mathbb{R}^d \times \mathbb{R}^d \rightarrow \mathbb{R}$, with $K(x_i,x_j) = \varphi(x_i)^T \varphi(x_j)$, non-negative regularization constants $\gamma$ and $\nu$, the inverse of the degree matrix $D_ {\textrm{Mem}}$ which is diagonal with positive entries, the Karush-Kuhn-Tucker (KKT) optimality conditions of the Lagrangian of (\ref{primalMKSC}) are satisfied by the following set of linear systems:}\\    
\begin{equation}
\label{dualMKSC}
\begin{split}
(D_ {\textrm{Mem}}^{-1}M_{D_ {\textrm{Mem}}}\Omega - \frac{I}{\gamma})\alpha ^{(l)}=& - \nu D_ {\textrm{Mem}}^{-1}M_{D_ {\textrm{Mem}}} \Omega _ {\textrm{new-old}} \alpha ^{(l)} _ {\textrm{old}}\\=&-\nu D_ {\textrm{Mem}}^{-1}M_{D_ {\textrm{Mem}}} \sum_{r=1}^{M}\Omega _ {\textrm{new-prev,r}} \alpha ^{(l)} _ {\textrm{prev,r}}
\end{split}
\end{equation} 
where 
\begin{itemize}
\item $\Omega$ indicates the current kernel matrix with $ij$-th entry $\Omega_{ij}=K(x_i,x_j) = \varphi(x_i)^T \varphi(x_j)$. $\Omega _{\textrm{new-old}}$ captures the similarity between the objects of the current snapshot and the ones of the previous $M$ snapshots, and has the $ij$-th entry $\Omega _{\textrm{new-old},ij}=\sum_{r=1}^{M}K(x_i^{\textrm{new}},x_j^{\textrm{prev,r}})$.
\item $M_{D_ {\textrm{Mem}}}$ is the centering matrix equal to $M_{D_ {\textrm{Mem}}} = I_{N_{\textrm{Tr}}} - \frac{1}{1_{N_{\textrm{Tr}}}^TD_ {\textrm{Mem}}^{-1}1_{N_{\textrm{Tr}}}}1_{N_{\textrm{Tr}}}1_{N_{\textrm{Tr}}}^TD_ {\textrm{Mem}}^{-1}$.
\end{itemize}

\textit{Proof:} The Lagrangian of the problem (\ref{primalMKSC}) is:
\begin{equation*}
\label{LagrangianMKSC}
\begin{split}
\mathcal{L}(w^{(l)},e^{(l)},b_l;\alpha ^{(l)})=& \frac{1}{2}\sum_{l=1}^{k-1}w^{(l)^T}w^{(l)}-\frac{\gamma}{2{N_{\textrm{Tr}}}}\sum_{l=1}^{k-1} e^{(l)^T}D_{\textrm{Mem}}^{-1}e^{(l)} \\ &-\nu  \sum_{l=1}^{k-1}w^{(l)^T}w_{\textrm{old}}^{(l)} + \alpha ^{(l)^T} (e^{(l)} - \Phi w^{(l)} - b_l1_{N_{\textrm{Tr}}}).
\end{split}
\end{equation*} 
The KKT optimality conditions are:\\\\
$\frac{\partial \mathcal{L}}{\partial w^{(l)}} = 0 \rightarrow w^{(l)} = \Phi ^T \alpha ^{(l)} + \nu w^{(l)}_ {\textrm{old}}$,\\ $\frac{\partial \mathcal{L}}{\partial e^{(l)}} = 0 \rightarrow \alpha^{(l)} = \gamma D_ {\textrm{Mem}}^{-1}e^{(l)}$,\\ $\frac{\partial \mathcal{L}}{\partial b_l} = 0 \rightarrow 1_{N_{\textrm{Tr}}}^T \alpha^{(l)} = 0$,\\ $\frac{\partial \mathcal{L}}{\partial \alpha^{(l)}} = 0 \rightarrow e = \Phi w^{(l)} + b_l1_{N_{\textrm{Tr}}}$.\\\\
From $w^{(l)}_{\textrm{old}} = \Phi _{\textrm{old}}^T \alpha^{(l)} _{\textrm{old}}$, the bias term becomes 

$b_l = -\frac{1}{1_{N_{\textrm{Tr}}}^TD_{\textrm{Mem}}^{-1}1_{N_{\textrm{Tr}}}}1_{N_{\textrm{Tr}}}^TD_{\textrm{Mem}}^{-1}(\Omega \alpha^{(l)} + \nu \Omega _{\textrm{new-old}} \alpha^{(l)} _{\textrm{old}})$.
Eliminating the primal variables $e^{(l)}$, $w^{(l)}$, $b_l$ leads to the dual problem (\ref{dualMKSC}) \hfill $\Box$\\\\
The cluster indicators for the training data are:
\begin{equation}
\label{eq_proj_MKSC}
\textrm{sign}(e^{(l)}) = \textrm{sign}(\Omega \alpha^{(l)} + \nu \Omega _{\textrm{new-old}}\alpha ^{(l)}_ {\textrm{old}}  + b1_{N_{\textrm{Tr}}}). 
\end{equation}
The score variables for test points are defined as follows: 
\begin{equation}
\label{eq_proj_MKSC_test}
e^{(l)}_{\textrm{test}} = \Omega ^{\textrm{test}} \alpha ^{(l)} + \nu \Omega _{\textrm{new-old}}^{\textrm{test}}\alpha^{(l)}_ {\textrm{old}}  + b_l1_{N_{\textrm{test}}}. 
\end{equation}
Thus, once we have properly trained our model, the cluster memberships of new points can be predicted by projecting the test data into the solution vectors $\alpha ^{(l)}$ and $\alpha^{(l)}_ {\textrm{old}}$ via eq. (\ref{eq_proj_MKSC_test}).

\subsection{Cluster Quality Measures in a Dynamic Scenario}
\label{smoothed_measures}
In order to assess the quality of a partition related to an evolving network, new measures are introduced. The new measures are the weighted sum of the snapshot quality and the temporal quality. The former only measures the quality of the current clustering with respect to the current data, while the latter measures the temporal smoothness in terms of the ability of the actual model to cluster the historic data. A partition related to a particular snapshot will then receive a score that is higher as it is more consistent with the past, for a particular value of the parameter $\eta$. For a given cluster quality criterion $\textrm{CQ}$, we can define its smoothed version related to the actual snapshot $\mathcal{G}_t$ in this way:        
\begin{equation}
\label{dyn_crit}
\textrm{CQ}_{\textrm{Mem}}(X_{\alpha},\mathcal{G}_t) = \eta \textrm{CQ}(X_{\alpha} ,\mathcal{G}_t) + (1 - \eta)\textrm{CQ}(X_{\alpha} ,\mathcal{G}_{t-1}),
\end{equation}
where $X_{\alpha}$ indicates the cluster indicator matrix calculated by using the current solution vectors $\alpha^{(l)}$.
With $\eta$ we denote a user-defined parameter which takes values in the range $[0,1]$ and reflects the emphasis given to the snapshot quality and the temporal smoothness, respectively. 

The smoothed counterparts of the model selection criteria introduced for KSC and SKSC, that is $\textrm{BLF}_{\textrm{Mem}}$, $\textrm{Mod}_{\textrm{Mem}}$ and $\textrm{AMS}_{\textrm{Mem}}$ can be used also to select optimal parameters for the MKSC model. Nevertheless, in the experiments described later, we did not notice big differences in using the static or smoothed measures when performing the model selection for MKSC. 

\subsubsection{Model selection}
The model selection scheme for MKSC is summarized in algorithm \ref{alg:alg1}.         
\begin{algorithm}[t]
\small
\LinesNumbered
\KwData{ Training sets and validation sets (actual and previous snapshots), kernel function $K: \mathbb{R}^d \times \mathbb{R}^d \rightarrow \mathbb{R}$ positive definite and localized.}
\KwResult{ Selected number of clusters $k$, kernel parameters (if any), $\gamma$, $\nu$.}
define a grid of values for the parameters to select\\
select training and validation sets\\
train the related MKSC model\\
compute the cluster indicator matrix corresponding to the validation data\\
for every partition calculate the related score by using the chosen criterion\\
select the model with the highest score.
\caption{Model selection algorithm for MKSC \cite{Langone_physicA}}
\label{alg:alg1}
\end{algorithm}
This procedure is standard but often we do not need to tune all the parameters. For instance in case of network data we can use the cosine kernel and the community kernel which are parameter-free. Moreover we have observed by extensive simulations that we can fix one of the two regularization constants and tune only the other one. In particular, in \textit{framework 1} we tune $\gamma$ and we fix $\nu$ and vice-versa in \textit{framework 2}.

\subsection{Computational Complexity}
\begin{figure}[htbp]
\centering
\begin{tabular}{c}
\begin{psfrags}
\psfrag{x}[t][t]{{Number of nodes ($N$)}}
\psfrag{y}[b][b]{{Runtime (s)}}
\psfrag{t}[b][b]{{}}
\includegraphics[width=3.5in]{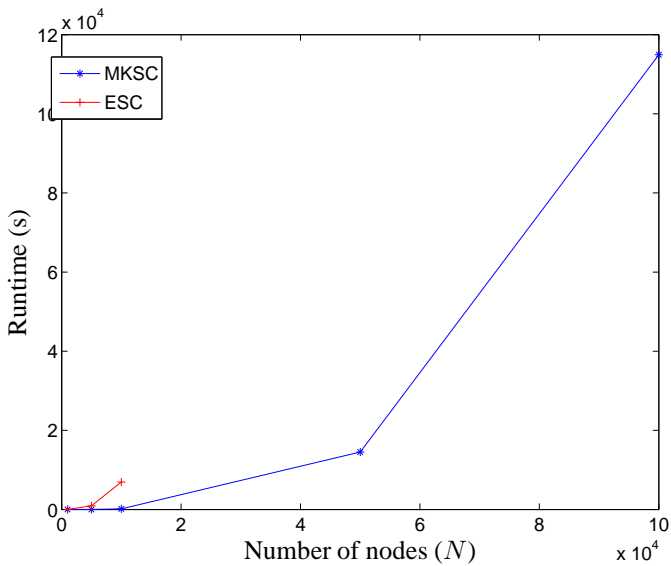}
\end{psfrags}
\end{tabular}
\caption{\textbf{MKSC computational complexity.} Evolution of the speed with the size of a benchmark network for ESC (until $N = 10^3$ because of memory problems) and MKSC. Although MKSC is faster than ESC, the runtime of the former seems to scale as $O(N^3)$. The is due to the usage of a non-sparse representation of the variables and an inefficient implementation of the community kernel function.}
\label{cpu_complexity}
\end{figure}

The time required to solve the set of linear systems (\ref{dualMKSC}) scales as $O(N_{\textrm{Tr}}^3)$. However, the runtime can be reduced by exploiting the matrix inversion Lemma \cite{Higham:1996:ASN:525601}. We can solve the dual problem of MKSC for a smaller training set and then calculate iteratively the $N_{\textrm{Tr}} \times 1 $ solution vector $\alpha^{(l)}$, $\forall l$, by using the Woodbury formula. Considering the whole network as test set, the time needed to compute the membership for all the $N$ nodes in each snapshot scales as $O(N_{\textrm{Tr}}N)$ if we use a sparse representation. Moreover, in case the matrices  $\Omega ^{\textrm{test}}$ and $\Omega _{\textrm{new-old}}^{\textrm{test}}$ cannot fit the memory we can divide the test set into blocks and perform the testing operations iteratively on a single computer or in parallel in a distributed environment, as shown in \cite{raghvendra_bigdata,raghvendra_sclara} for kernel spectral clustering (KSC). However, in the worst case scenario (non-sparse representation of the variables and inefficient implementation of the kernel function), the complexity can rise to $O(N^2)$ or $O(N^3)$. Finally, in figure \ref{cpu_complexity} the runtime of MKSC and ESC needed for clustering some artificial evolving networks of different size are compared. 

\section{Framework 1}
In this section we describe the experimental results when considering networks where the number of nodes and communities is not varying across time. Moreover we require that the clustering results are smoothed at each time stamp by fixing $\nu=1$. The method that we use can be summarized in algorithm \ref{alg:alg2}.  
\begin{algorithm}[t]
\small
\LinesNumbered
\KwData{Training sets $\mathcal{D}=\{x_i\}_{i=1}^{N_{\textrm{Tr}}}$ and $\mathcal{D_\textrm{old}}=\{x_i^{\textrm{old}}\}_{i=1}^{N_{\textrm{Tr}}}$, test sets $\mathcal{D^{\textrm{test}}}=\{x_m^{\textrm{test}}\}_{m=1}^{N_{\textrm{test}}}$ and $\mathcal{D^{\textrm{test}}_\textrm{{old}}}=\{x_m^{\textrm{test,old}}\}_{m=1}^{N_{\textrm{test}}}$, $\alpha ^{(l)} _ {\textrm{old}}$ (the $\alpha ^{(l)}$ calculated for the previous $M$ snapshots), positive definite kernel function $K: \mathbb{R}^d \times \mathbb{R}^d \rightarrow \mathbb{R}$ such that $K(x_i,x_j) \rightarrow 0$ if $x_i$ and $x_j$ belong to different clusters, kernel parameters (if any), number of clusters $k$, regularization constants $\gamma$ and $\nu$ found using algorithm \ref{alg:alg1}.}
\KwResult{Clusters $\{\mathcal{C}^t_1,\hdots,\mathcal{C}^t_p\}$, cluster codeset $\mathcal{CB} = \{c_p\}_{p = 1}^k$, $c_p \in \{-1,1\}^{k-1}$.}
\eIf {t==1}{
Initialization by using kernel spectral clustering (KSC \cite{multiway_pami}).
}{
%\textbf{For every snapshot from $t-1$ to $t-M$ rearrange the data matrices and the solution vectors  $\alpha^{(l)}$ as explained in Section \ref{changing_objects}.}\\
Compute the solution vectors $\alpha^{(l)}$, $l = 1,\hdots,k-1$, related to the linear systems described by eq. (\ref{dualMKSC}): $$(D_ {\textrm{Mem}}^{-1}M_{D_ {\textrm{Mem}}}\Omega - \frac{I}{\gamma})\alpha ^{(l)}= - \nu D_ {\textrm{Mem}}^{-1}M_{D_ {\textrm{Mem}}} \Omega _ {\textrm{new-old}} \alpha ^{(l)} _ {\textrm{old}}$$.\\ 
Binarize the solution vectors: $\textrm{sign}(\alpha_i^{(l)})$, $i = 1,\hdots,{N_{\textrm{Tr}}}$, $l = 1,\hdots,k-1$, and let $\textrm{sign}(\alpha_i) \in \{-1,1\}^{k-1}$ be the encoding vector for the training data point $x_i$.\\     
Count the occurrences of the different encodings and find the $k$ encodings with most occurrences. Let the codeset be formed by these $k$ encodings: $\mathcal{CB} = \{c_p\}_{p = 1}^k$, with $c_p \in \{-1,1\}^{k-1}$.\\ 
$\forall i$, assign $x_i$ to $C_{p^*}$ where $p^* = \textrm{argmin}_pd_H(\textrm{sign}(\alpha_i),c_p)$ and $d_H(\cdot, \cdot)$ is the Hamming distance.\\
Binarize the test data projections $\textrm{sign}(e_m^{(l)})$, $m = 1,\hdots,N_{\textrm{test}}$, $l = 1,\hdots,k-1$ and let  $\textrm{sign}(e_m) \in \{-1,1\}^{k-1}$ be the encoding vector of $x_m^{\textrm{test}}$, $m = 1,\hdots,N_{\textrm{test}}$.\\
$\forall m$, assign $x_m^{\textrm{test}}$ to $C^t_{p^*}$ using an ECOC decoding scheme, i.e. $p^* = \textrm{argmin}_pd_H(\textrm{sign}(e_m),c_p)$.\\
%\textbf{Match the actual clusters $\{\mathcal{C}^t_1,\hdots,\mathcal{C}^t_p\}$ with the previous partition $\{\mathcal{C}^{t-1}_1,\hdots,\mathcal{C}^{t-1}_q\}$ using the tracking scheme described in Section \ref{clu_matching} and summarized in algorithm \ref{algo:algo1}.}  
}  
\caption{Clustering evolving networks: framework 1 \cite{Langone_physicA}}
\label{alg:alg2}
\end{algorithm}

\subsection{Artificial Examples}
\label{artificial_net_f1}
\begin{figure}[htbp]
\centering
\includegraphics[width=3in]{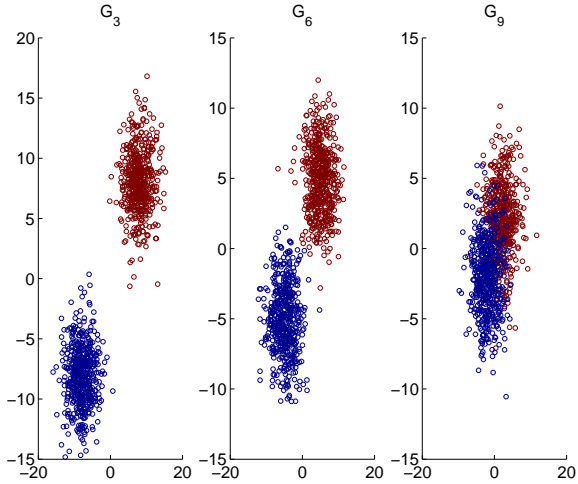}
\caption{\textbf{MG2 dataset}. The snapshots number $3$, $6$ and $9$ of the two moving Gaussians synthetic example are shown. The axes represent the two dimensions of the $N$ input vectors $x_i \in \mathbb{R}^2, i=1,\hdots,N$. }\label{fig_2Gaussian}
\end{figure}

\begin{figure}[htbp]
\centering
\includegraphics[width=3in]{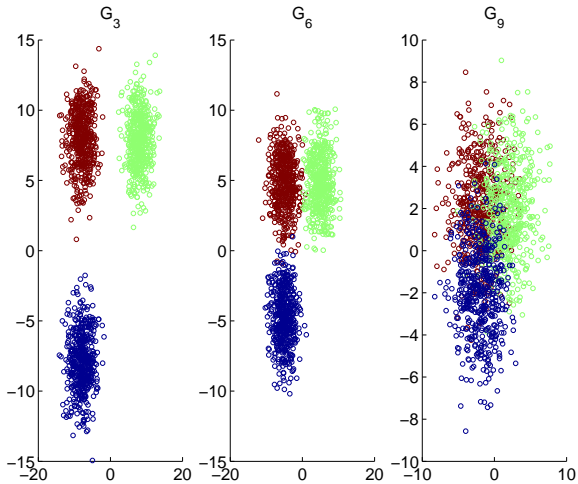}
\caption{\textbf{MG3 dataset}. The snapshots number $3$, $6$ and $9$ of the three moving Gaussians artificial example are depicted. The axes represent the two dimensions of the $N$ input vectors $x_i \in \mathbb{R}^2, i=1,\hdots,N$.}\label{fig_3Gaussian}
\end{figure}

The synthetic datasets used in the simulations can be described as follows:  
\begin{itemize}
\item \textbf{MG2}: the setup for this experiment is shown in figure \ref{fig_2Gaussian}. We generate $1000$ samples from a mixture of two 2D Gaussian clouds, with $500$ samples drawn from each component of the mixture. From time-steps $1$ to $10$ we move the means of the two Gaussian clouds towards each other until they overlap almost completely. This phenomenon can be also described by considering the points of the two Gaussians as nodes of a weighted network, where the weights of the edges change across time.
\item \textbf{MG3}: three merging Gaussian clouds, shown in figure \ref{fig_3Gaussian}. 
\item \textbf{SwitchingNet}: we build up $9$ snapshots of a network of $1000$ nodes formed by $2$ communities. At each time-step some nodes switch their membership between the two clusters. We used the software related to \cite{Greene2011} to generate this benchmark.
\item \textbf{ECNet}: a network with $5$ communities experiences over time $24$ expansion events and $16$ contractions of its communities. 
\end{itemize}
In the next sections we describe the model selection issue and we assess the quality of the partitions produced by MKSC. We compare MKSC with KSC and ESC. In general, as expected, we observe that the dynamic clustering algorithms MKSC and ESC produce better clustering results over time than KSC.  

\subsubsection{Tuning the hyper-parameters}
\begin{figure}[htbp]
\centering
\begin{tabular}{c}
\begin{psfrags}
\psfrag{x}[t][t]{{$\sigma ^2$}}
\psfrag{y}[b][b]{{Number of clusters $k$}}
\psfrag{t}[b][b]{{BLF, Optimal values: $k=2$, $\sigma ^2 = 8.5$}}
\includegraphics[width=3in]{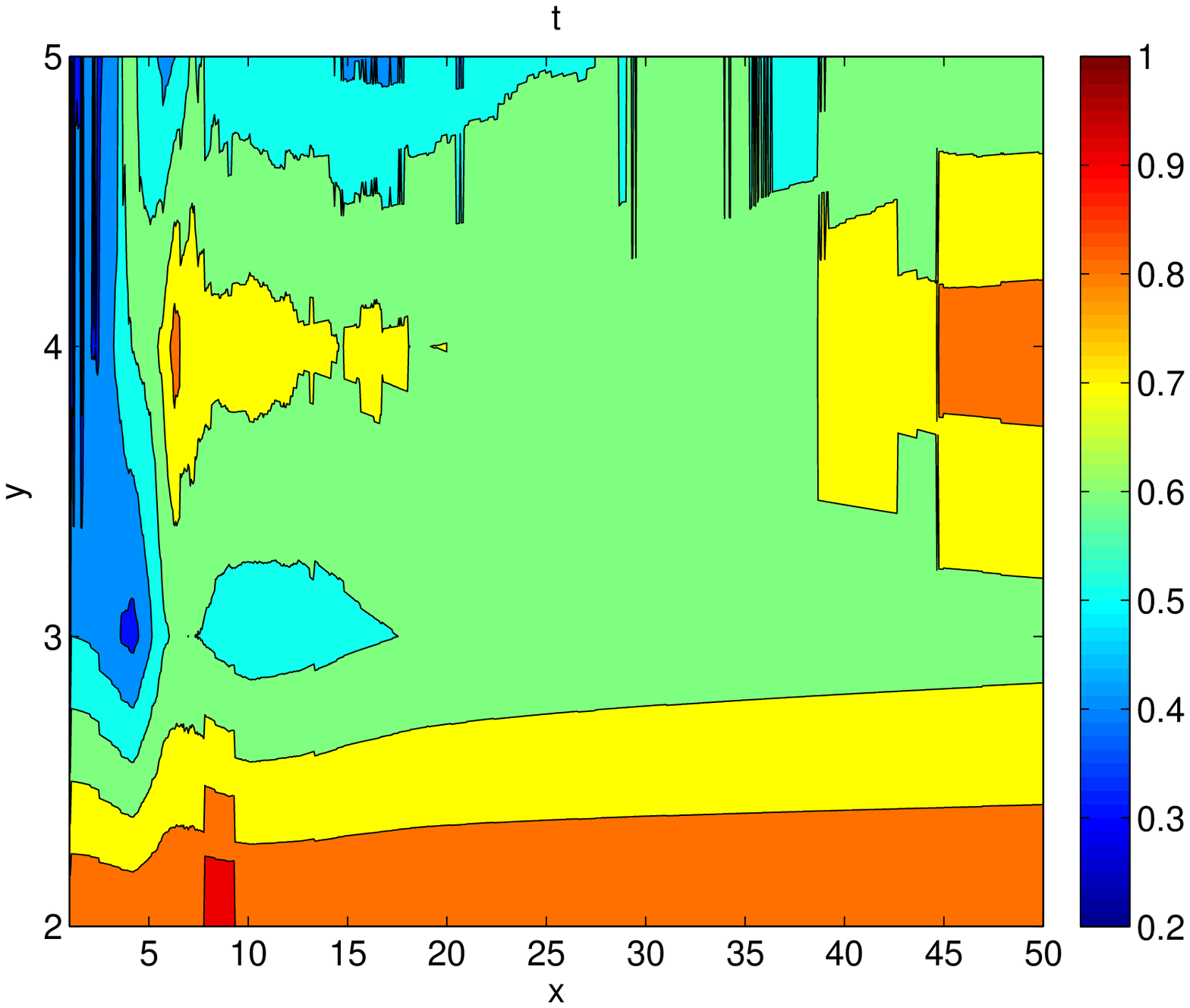}
\end{psfrags}\\
\begin{psfrags}
\psfrag{x}[t][t]{{Snapshot}}
\psfrag{y}[b][b]{{$\gamma$}}
\psfrag{t}[b][b]{{}}
\includegraphics[width=3in]{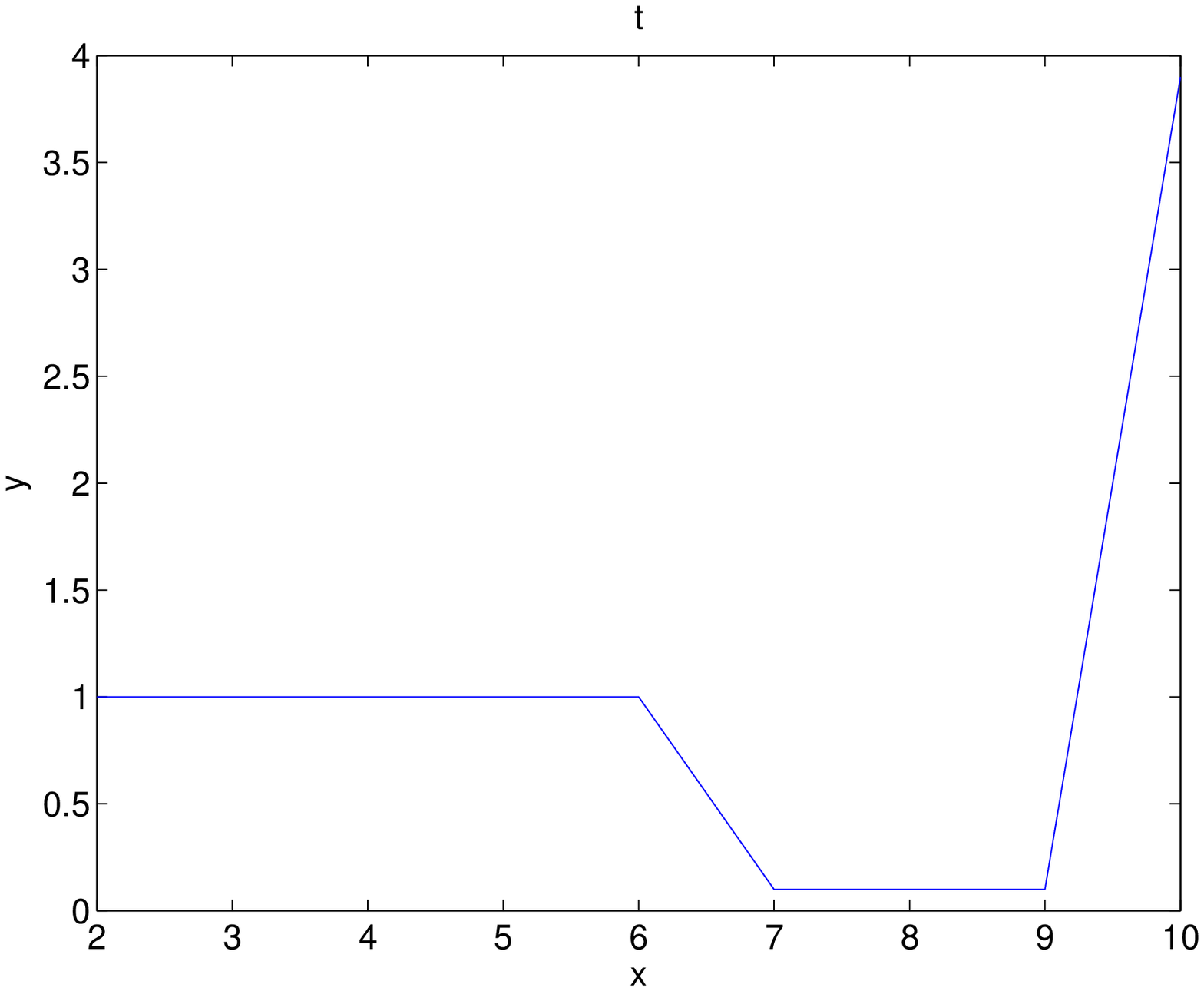}
\end{psfrags}
\end{tabular}
\caption{\textbf{Model selection MG2}. Top: tuning of the number of clusters $k$ and the RBF kernel parameter $\sigma ^2$ related to the first snapshot of the two moving Gaussians experiment, for KSC. The optimal $\sigma ^2$ does not change over time (the model selection procedure gives similar results also for the other snapshots). Moreover, this value is also used for the MKSC model. Bottom: optimal value of $\gamma$ over time for MKSC, tuned using the $BLF_{\textrm{Mem}}$ method.}\label{MS_gaussian}
\end{figure}

\begin{figure}[htbp]
\centering
\begin{tabular}{c}
\begin{psfrags}
\psfrag{x}[t][t]{{$\sigma ^2$}}
\psfrag{y}[b][b]{{Number of clusters $k$}}
\psfrag{t}[b][b]{{BLF, Optimal values: $k=3$, $\sigma ^2 = 3$}}
\includegraphics[width=3in]{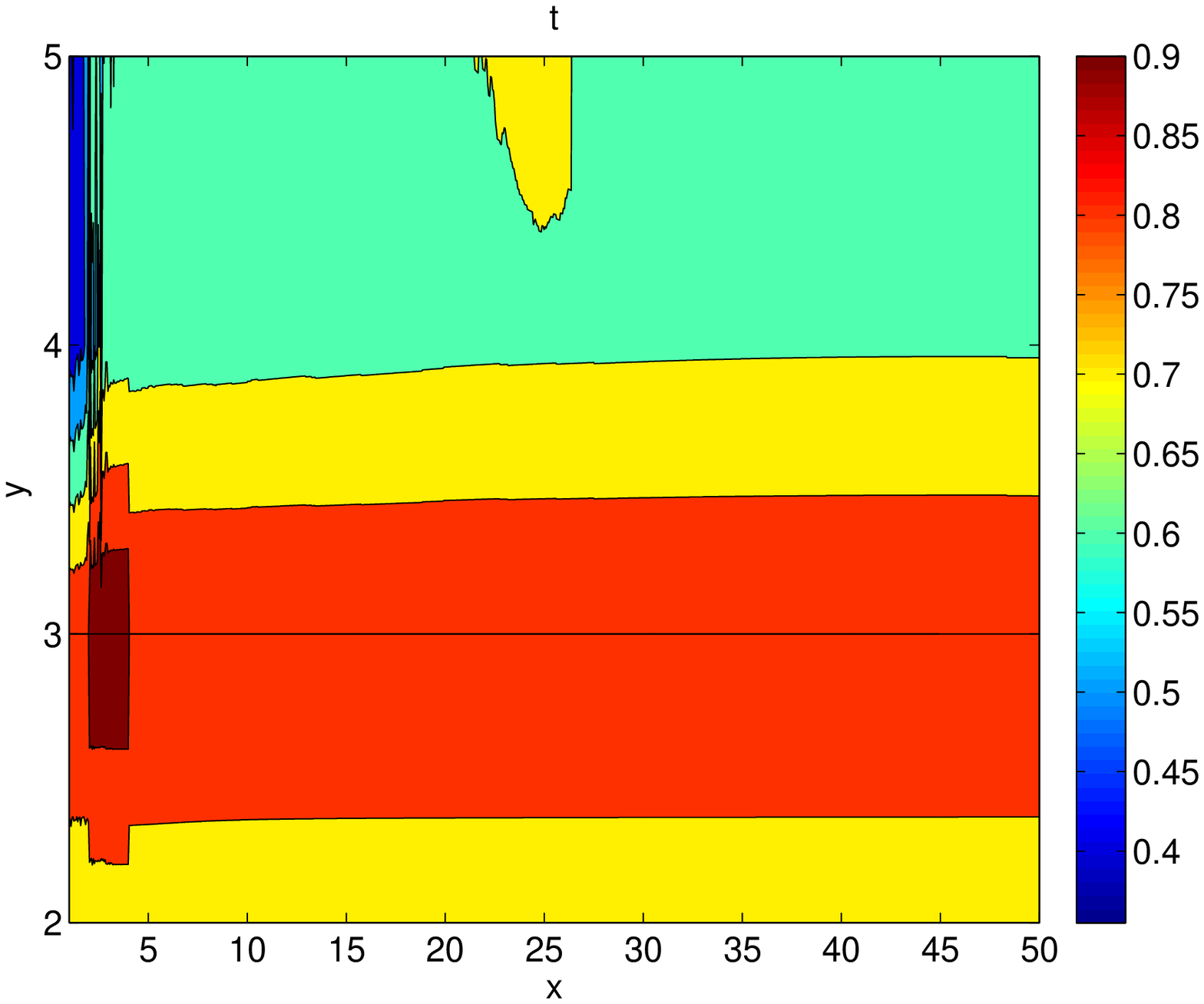}
\end{psfrags}\\
\begin{psfrags}
\psfrag{x}[t][t]{{Snapshot}}
\psfrag{y}[b][b]{{$\gamma$}}
\psfrag{t}[b][b]{{}}
\includegraphics[width=3in]{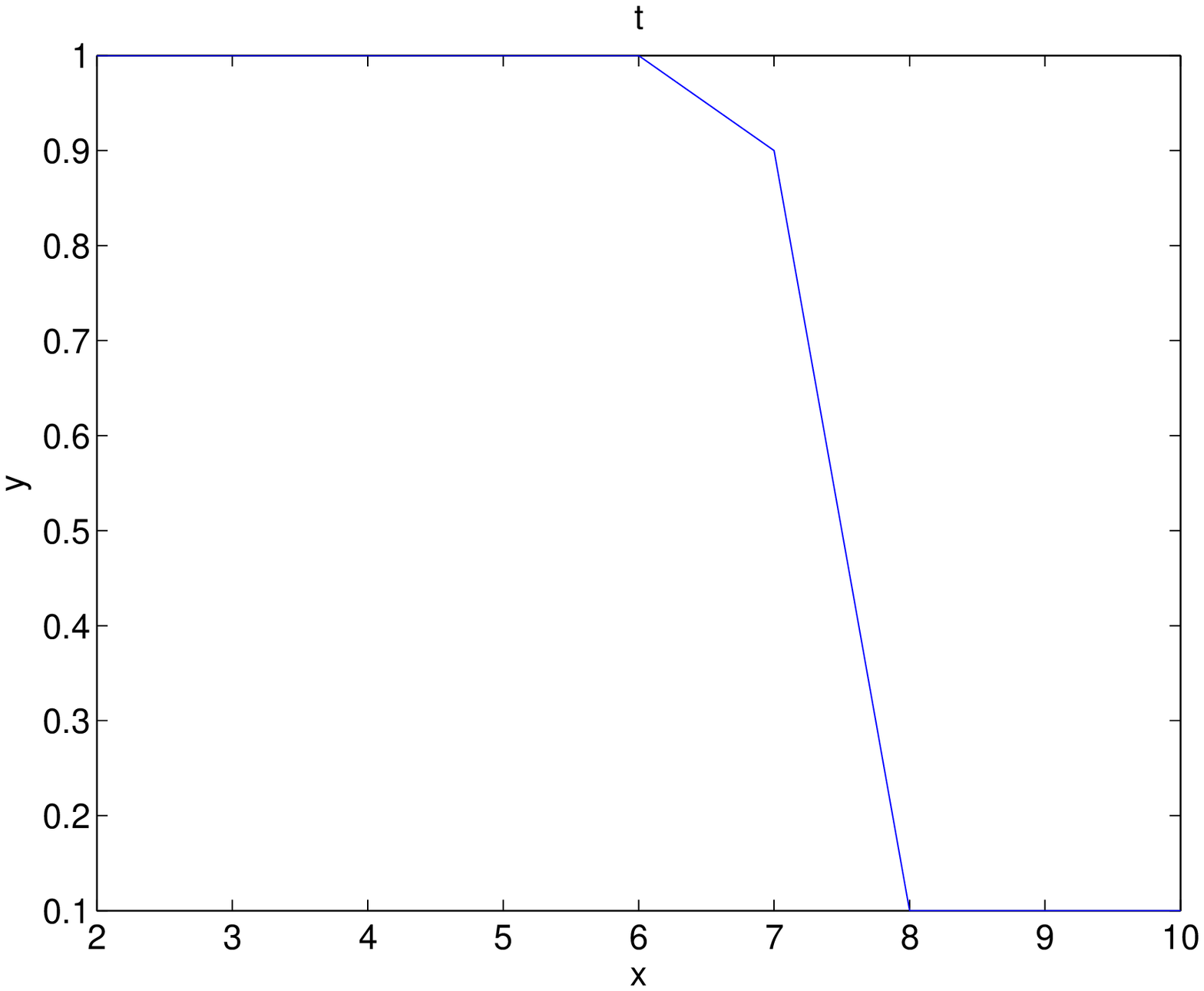}
\end{psfrags}
\end{tabular}
\caption{\textbf{Model selection MG3}. Top: tuning of the RBF kernel parameter $\sigma ^2$ and the number of clusters $k$ related to the first snapshot of the three moving Gaussians experiment, for KSC. Bottom: tuning of $\gamma$ for MKSC. The same comments made for Fig. \ref{MS_gaussian} are still valid here.}\label{MS_3gaussian}
\end{figure}

\begin{figure}[htbp]
\centering
\begin{tabular}{c}
\begin{psfrags}
\psfrag{x}[t][t]{{Number of communities $k$}}
\psfrag{y}[b][b]{{Modularity}}
\psfrag{t}[b][b]{{Optimal $k = 2$}}
\includegraphics[width=3in]{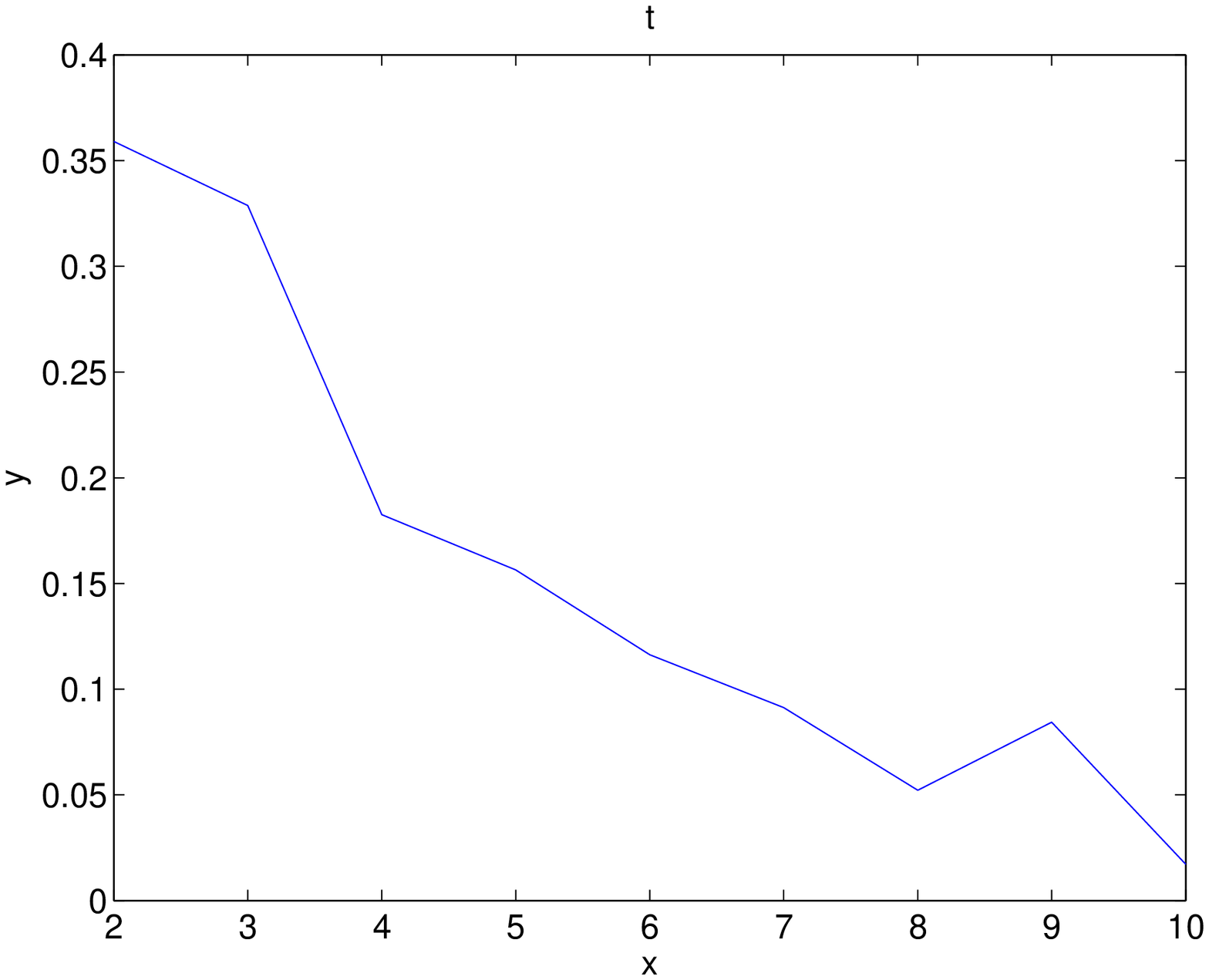}
\end{psfrags}\\
\begin{psfrags}
\psfrag{x}[t][t]{{Snapshot}}
\psfrag{y}[b][b]{{$\gamma$}}
\psfrag{t}[b][b]{{}}
\includegraphics[width=3in]{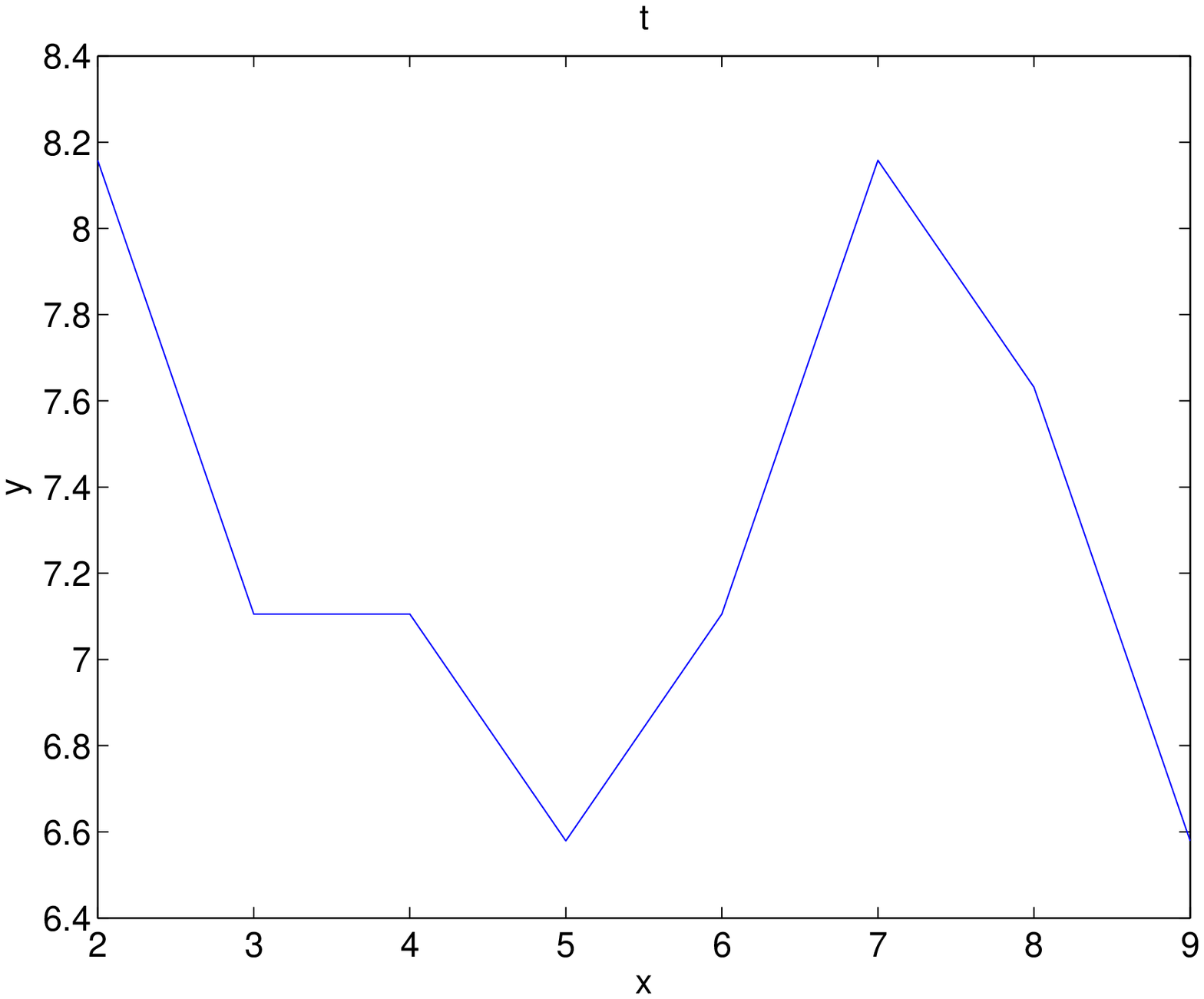}
\end{psfrags}
\end{tabular}
\caption{\textbf{Model selection SwitchingNet}. Top: tuning of the number of clusters $k$ for the switching network, related to the first snapshot and the KSC model. The results are similar for the other snapshots and this $k$ is also used as input to the MKSC algorithm. Bottom: optimal value of $\gamma$ over time for the MKSC model, selected by using the $Mod_{\textrm{Mem}}$ criterion.}\label{MS_switch_fig}
\end{figure}

\begin{figure}[htbp]
\centering
\begin{tabular}{c}
\begin{psfrags}
\psfrag{x}[t][t]{{Number of communities $k$}}
\psfrag{y}[b][b]{{Modularity}}
\psfrag{t}[b][b]{{Optimal $k = 5$}}
\includegraphics[width=3in]{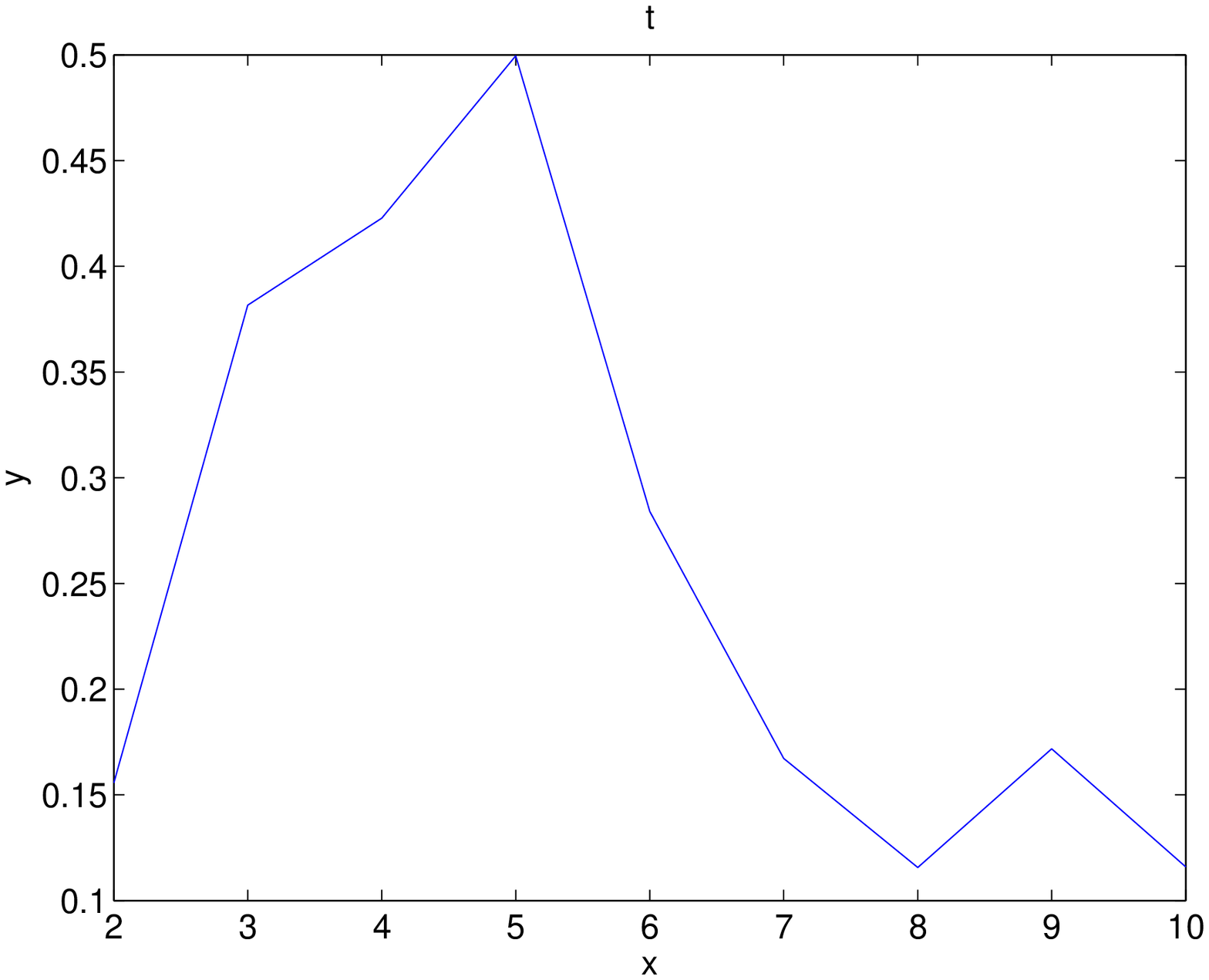}
\end{psfrags}\\
\begin{psfrags}
\psfrag{x}[t][t]{{Snapshot}}
\psfrag{y}[b][b]{{$\gamma$}}
\psfrag{t}[b][b]{{}}
\includegraphics[width=3in]{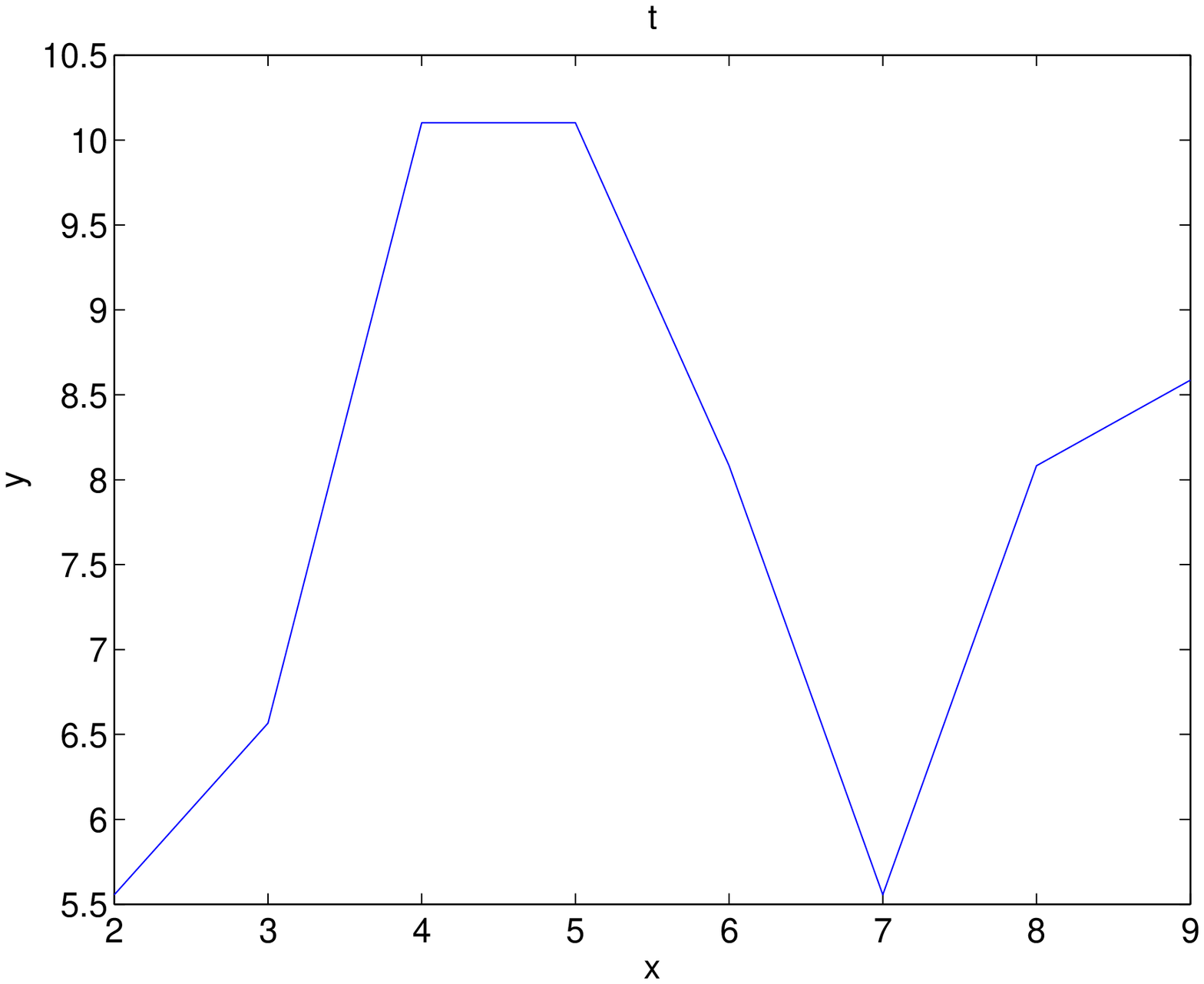}
\end{psfrags}
\end{tabular}
\caption{\textbf{Model selection ECNet}. Top: tuning of the number of clusters $k$ for the expanding/contracting network, related to the first snapshot and the KSC algorithm. Bottom: tuning of $\gamma$ for MKSC. The comments made for Fig. \ref{MS_switch_fig} hold also in this case.}\label{MS_expand_fig}
\end{figure}

Regarding KSC we use the BLF criterion described in section \ref{sc:BLF} to tune the number of clusters $k$ and the hyper-parameter $\sigma$ of the RBF kernel in the two and three moving Gaussians experiments. The results are shown at the top of figures \ref{MS_gaussian} and \ref{MS_3gaussian} and refer to the first snapshot (for the other snapshots the plots are similar). For the SwitchingNet graph in the first row of figure \ref{MS_switch_fig} we illustrate how the Modularity-based model selection scheme introduced in section \ref{sc:MS_community_detection} correctly identifies the possible presence of two communities. In this case $k$ is the only parameter to tune since the community kernel used in this case is parameter-free as explained in section \ref{kernel_description}. Also in the case of the ECNet artificial network the $5$ communities are correctly detected, as illustrated at the top of figure \ref{MS_expand_fig}.    

Concerning the MKSC model, for all the datasets we use the same values of $\sigma$ and $k$ found for KSC and we need to tune $\nu$ and $\gamma$. For simplicity we fix the value of $\nu$ to $1$ and we tune only $\gamma$. The optimal $\gamma$ over time for the two and three moving Gaussian experiments are shown respectively in the second row of figures \ref{MS_gaussian} and \ref{MS_3gaussian}, where the $\textrm{BLF}_{\textrm{Mem}}$ criterion is used for tuning. At the bottom of figure \ref{MS_switch_fig} the optimal value of $\gamma$ across time for SwitchingNet suggested by the Smoothed Modularity-based model selection criterion sketched in algorithm \ref{alg:alg1} is shown. The bottom side of figure \ref{MS_expand_fig} depicts the optimal $\gamma$ for the ECNet synthetic network. 

\subsubsection{Evaluation of the results}
\begin{figure}[htbp]
\centering
%\footnotesize
\begin{tabular}{m{0.8in}m{1in}}
Smoothed ARI &
\begin{psfrags}
\psfrag{x}[t][t]{{time}}
\psfrag{y}[b][b]{{$ARI_{\textrm{Mem}}$}}
\psfrag{t}[b][b]{{Two moving Gaussians}}
\includegraphics[width=1.5in]{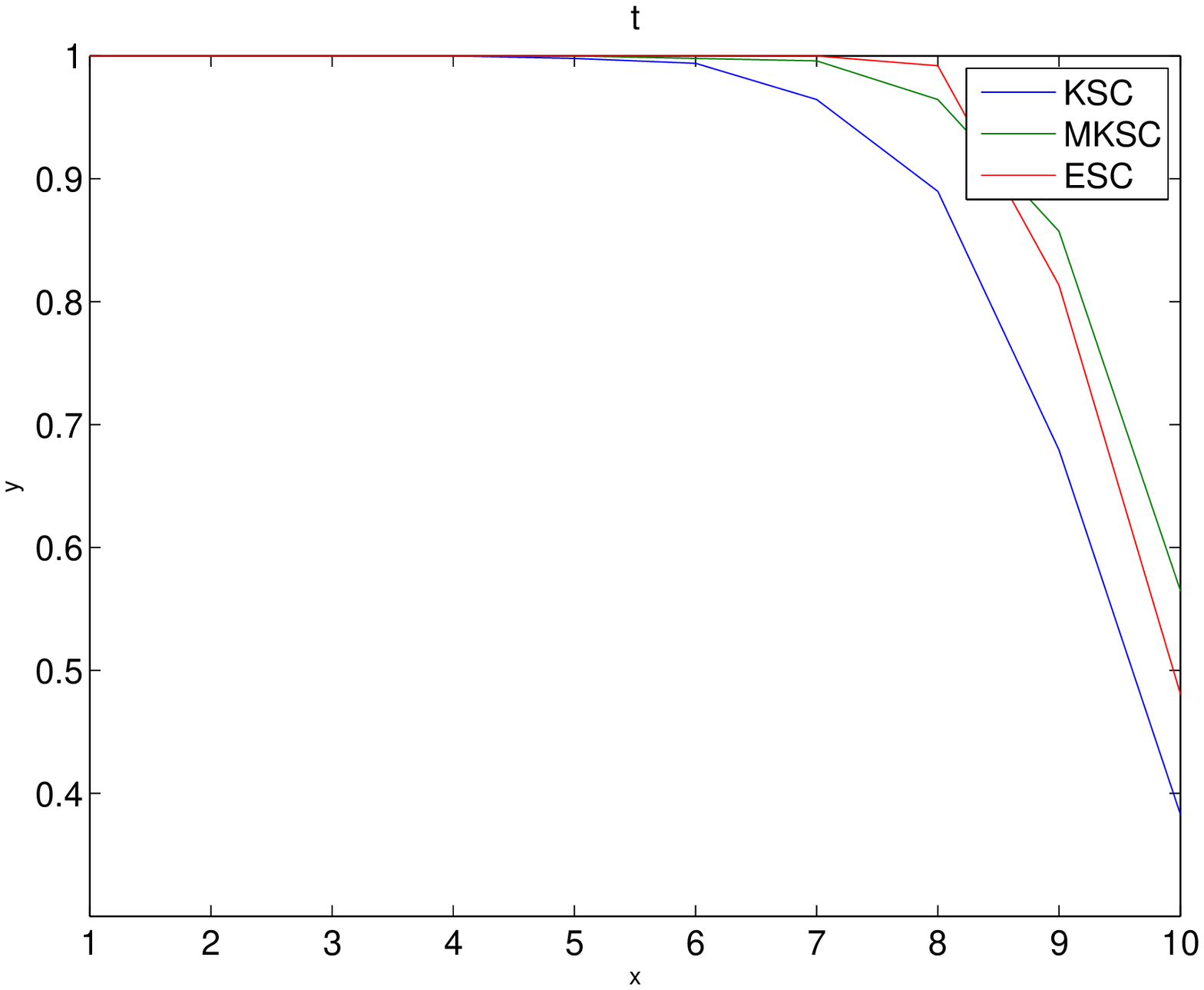}
\end{psfrags}\\\\
NMI between $2$ consecutive partitions &
\begin{psfrags}
\psfrag{x}[t][t]{{time}}
\psfrag{y}[b][b]{{NMI}}
\psfrag{t}[b][b]{{}}
\includegraphics[width=1.5in]{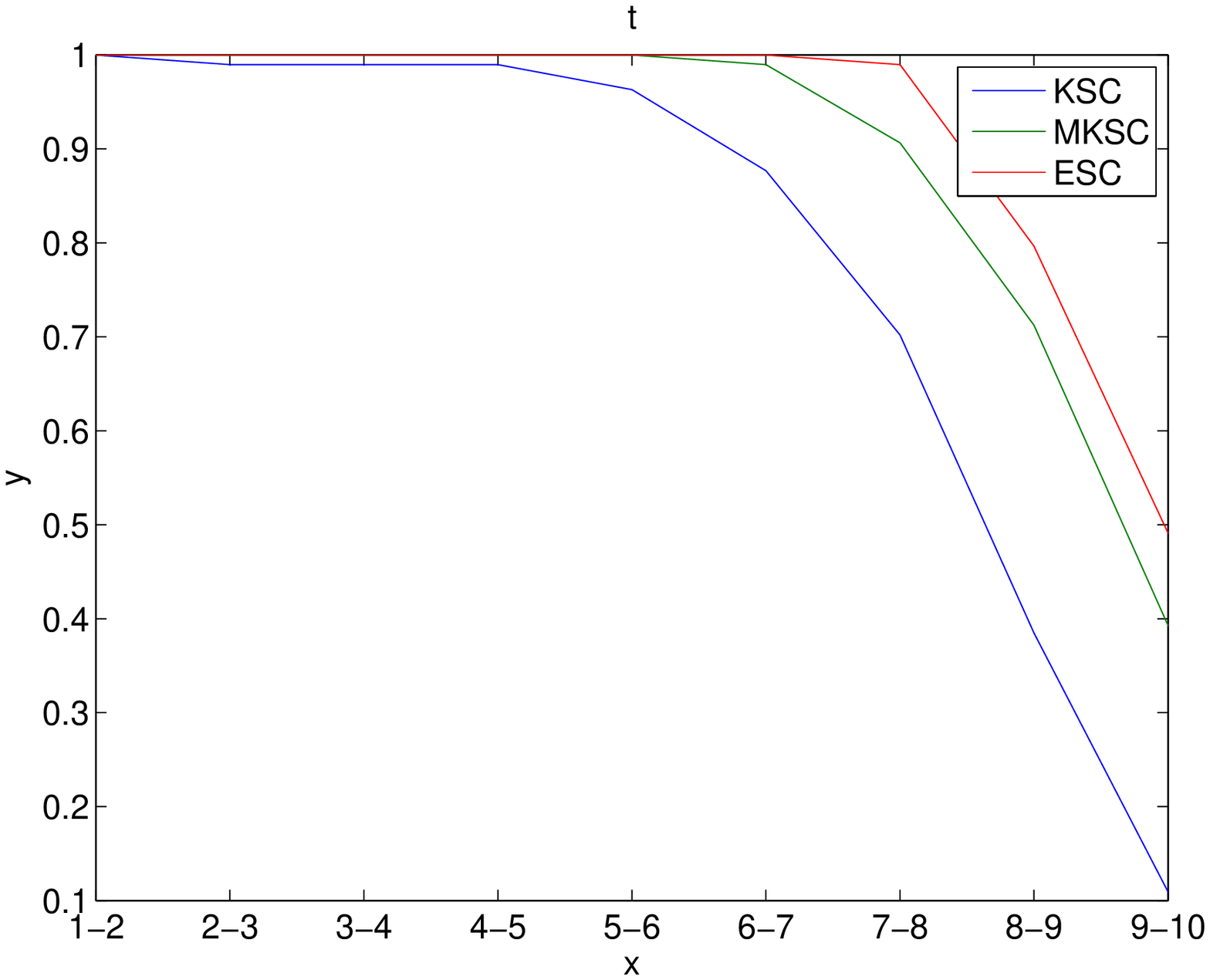}
\end{psfrags}\\
MKSC &
\begin{psfrags}
\psfrag{x}[t][t]{{}}
\psfrag{y}[b][b]{{}}
\psfrag{t}[b][b]{{}}
\includegraphics[width=1.5in]{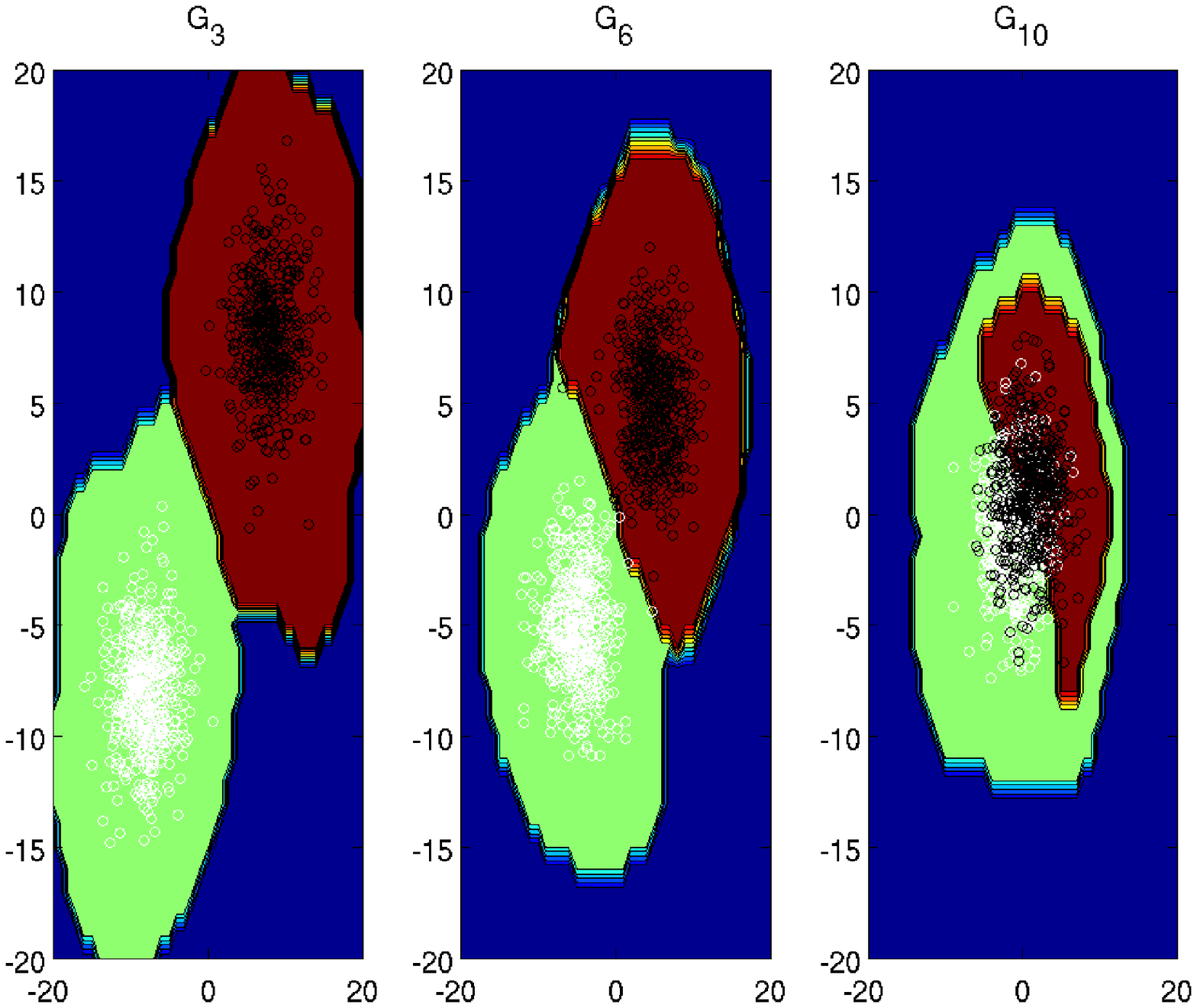}
\end{psfrags}\\
KSC &
\begin{psfrags}
\psfrag{x}[t][t]{{}}
\psfrag{y}[b][b]{{}}
\psfrag{t}[b][b]{{}}
\includegraphics[width=1.5in]{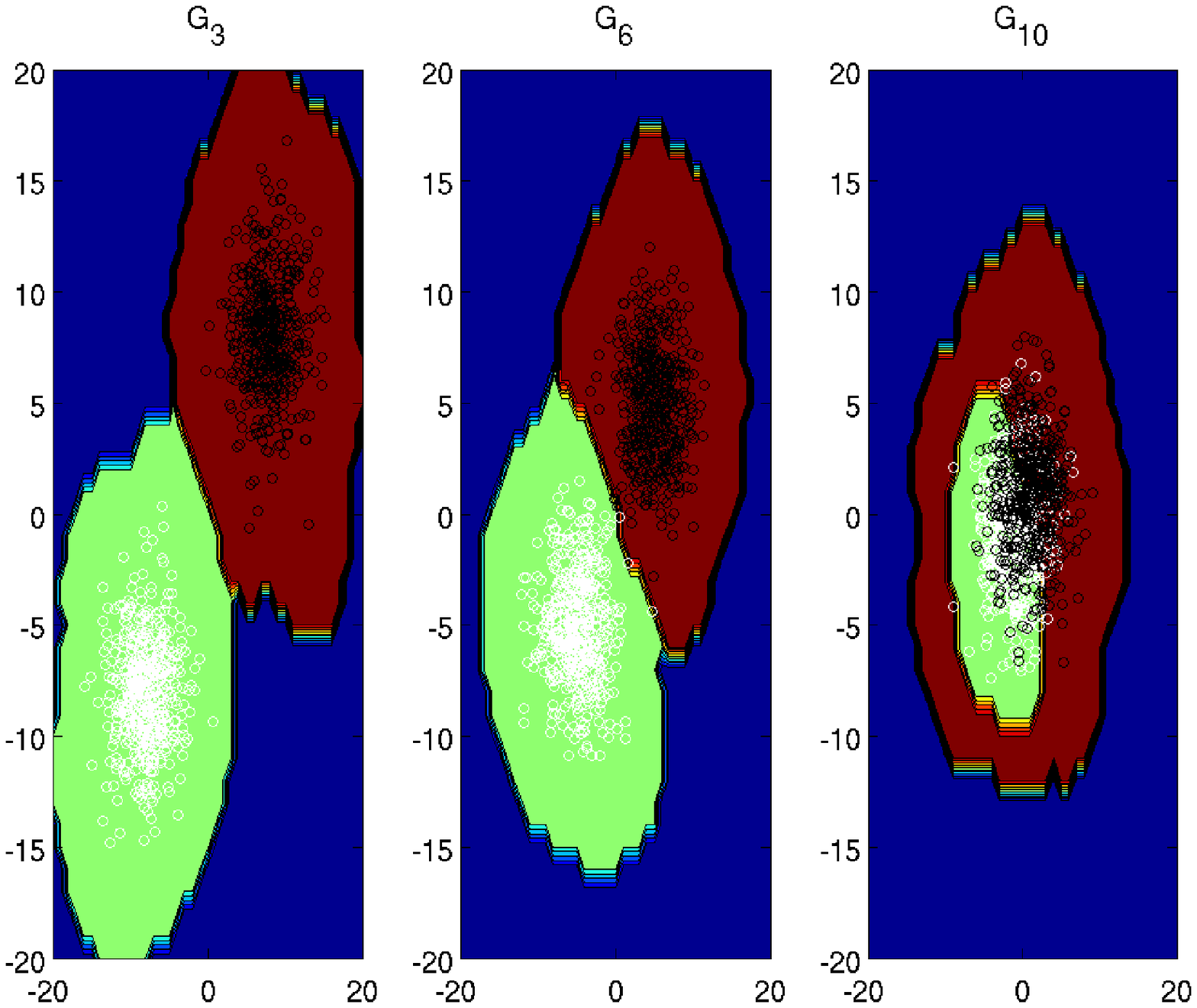}
\end{psfrags}\\
Ground truth &
\begin{psfrags}
\psfrag{x}[t][t]{{}}
\psfrag{y}[b][b]{{}}
\psfrag{t}[b][b]{{}}
\includegraphics[width=1.5in]{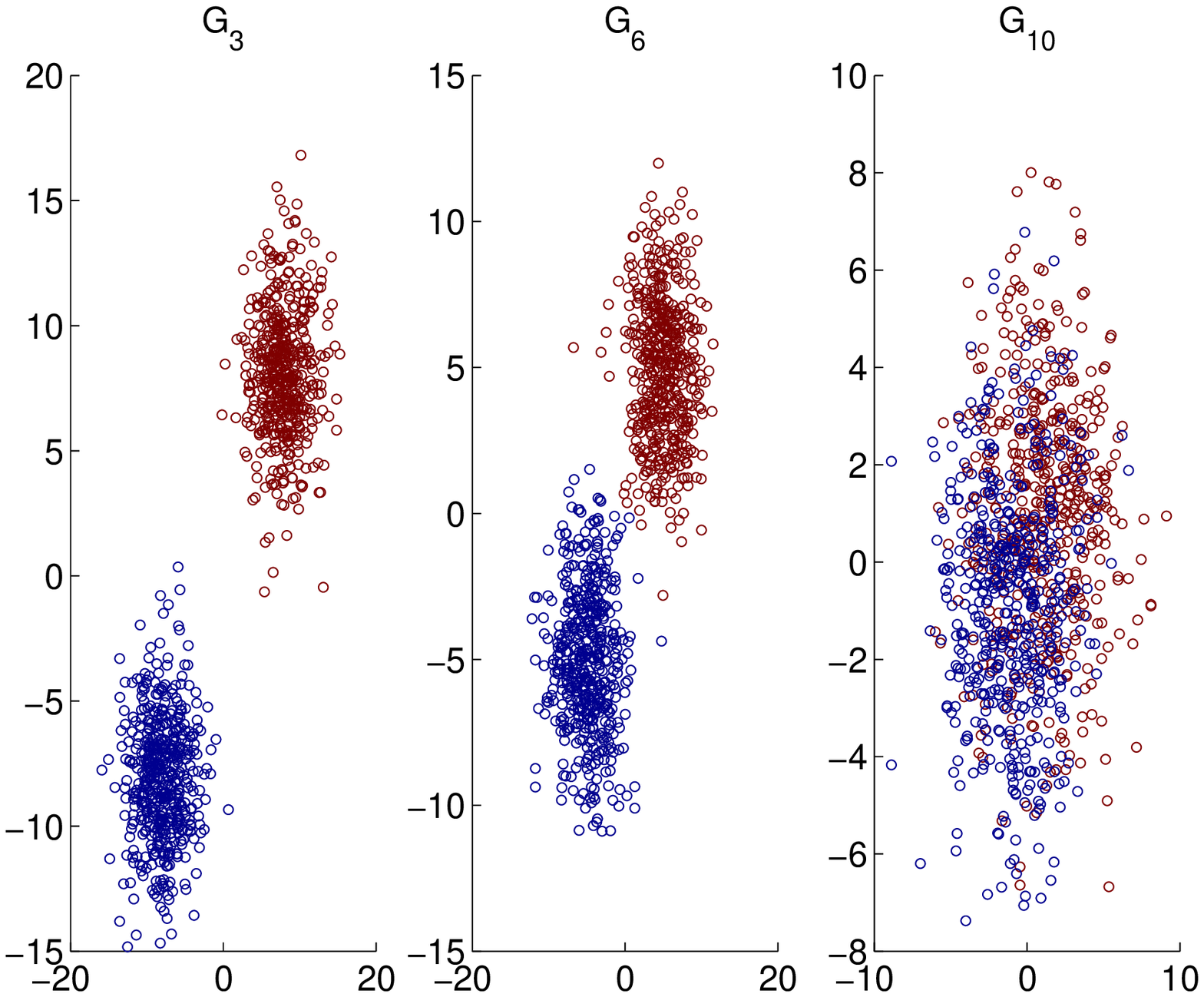}
\end{psfrags}
\end{tabular}
\caption{\scriptsize{\textbf{Clustering results MG2}: performance of MKSC, KSC and ESC in the two moving Gaussians experiment (first row) and out-of-sample plot for MKSC and KSC (only the results related to snapshots $3$, $6$ and $10$ are shown). The true partitioning is depicted in the fifth row. The smoothed ARI plot and the NMI trend tell us that, as expected, the models with temporal smoothness are more able than KSC to produce clustering results which are more similar to the ground truth and also more consistent and smooth over time. However, in the out-of-sample plot we cannot visually appreciate the better performance of MKSC with respect to KSC.}}\label{res_gauss}
\end{figure}

\begin{figure}[htbp]
\centering
\begin{tabular}{m{0.8in}m{1in}}
Smoothed ARI &
\begin{psfrags}
\psfrag{x}[t][t]{{time}}
\psfrag{y}[b][b]{{$\textrm{ARI}_{\textrm{Mem}}$}}
\psfrag{t}[b][b]{{Three moving Gaussians}}
\includegraphics[width=1.5in]{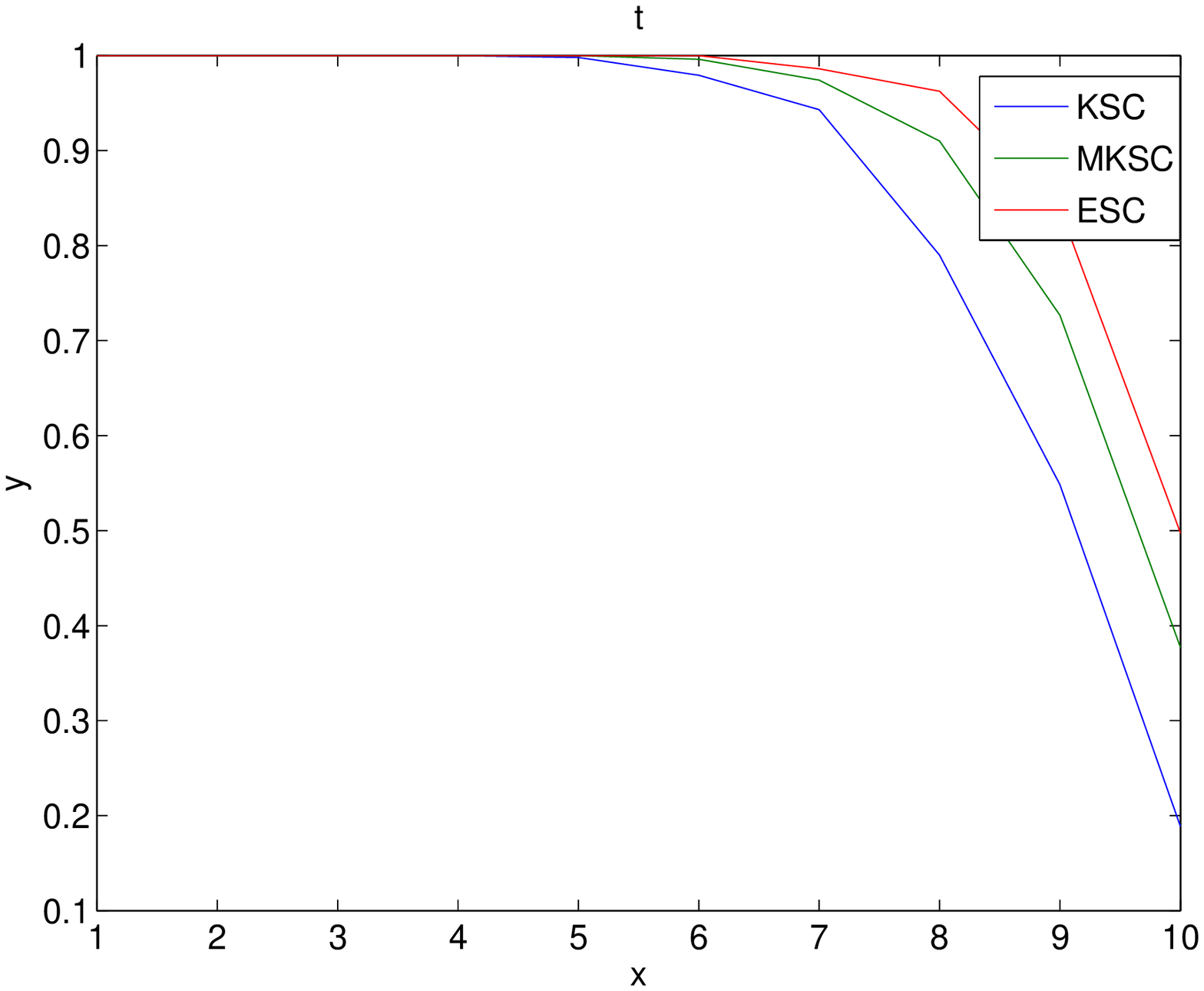}
\end{psfrags}\\\\
NMI between $2$ consecutive partitions &
\begin{psfrags}
\psfrag{x}[t][t]{{time}}
\psfrag{y}[b][b]{{NMI}}
\psfrag{t}[b][b]{{}}
\includegraphics[width=1.5in]{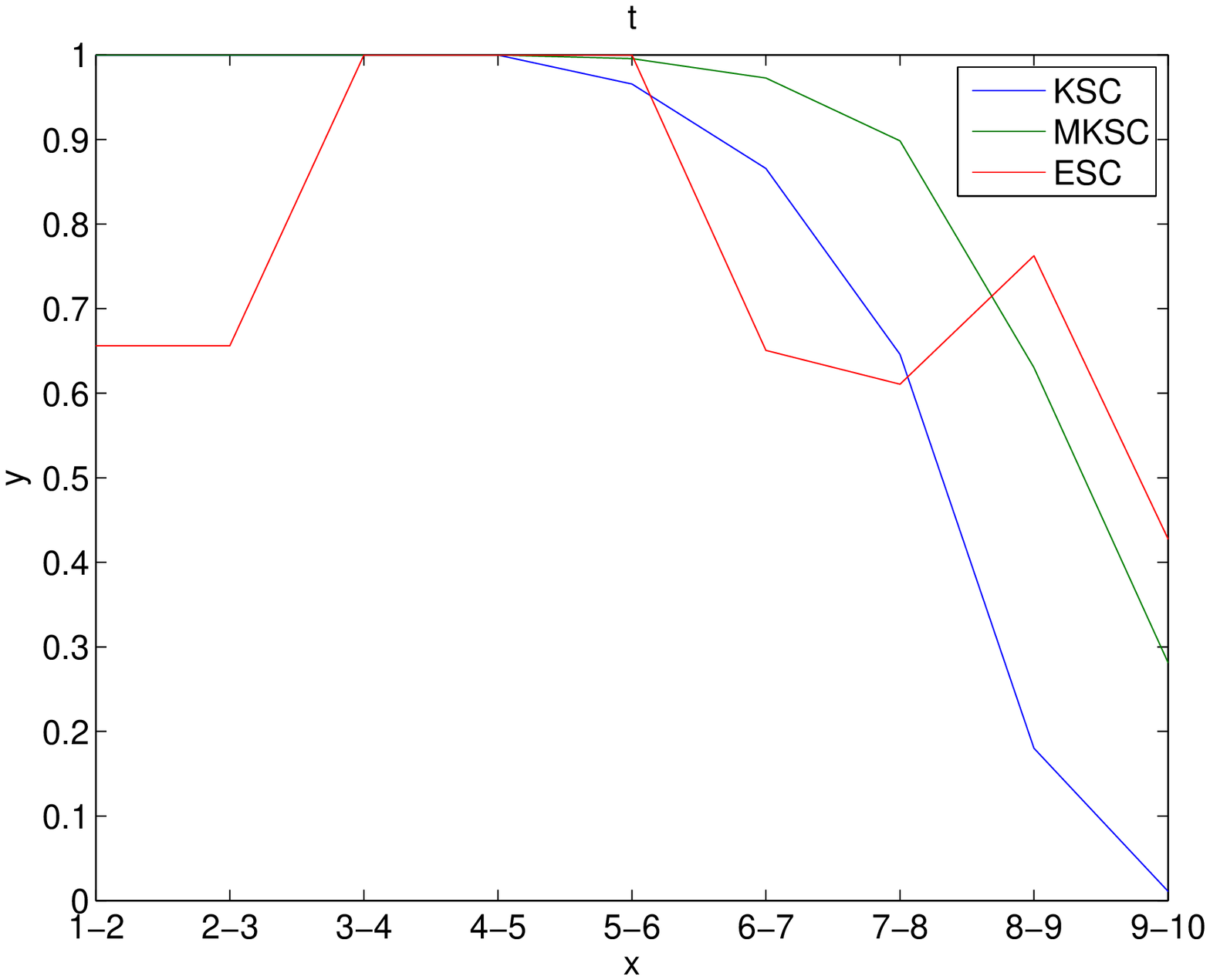}
\end{psfrags}\\
MKSC &
\begin{psfrags}
\psfrag{x}[t][t]{{}}
\psfrag{y}[b][b]{{}}
\psfrag{t}[b][b]{{}}
\includegraphics[width=1.5in]{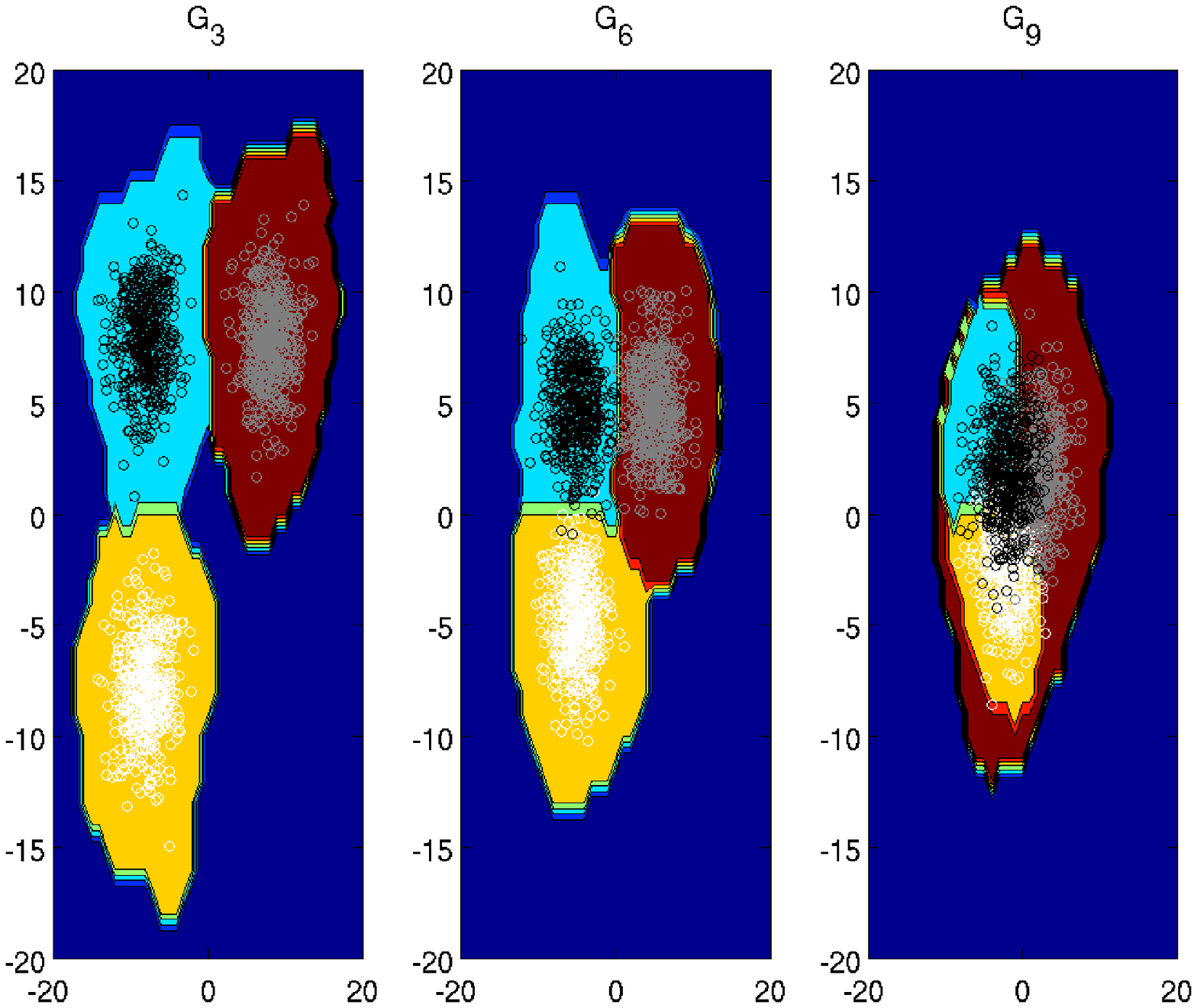}
\end{psfrags}\\
KSC &
\begin{psfrags}
\psfrag{x}[t][t]{{}}
\psfrag{y}[b][b]{{}}
\psfrag{t}[b][b]{{}}
\includegraphics[width=1.5in]{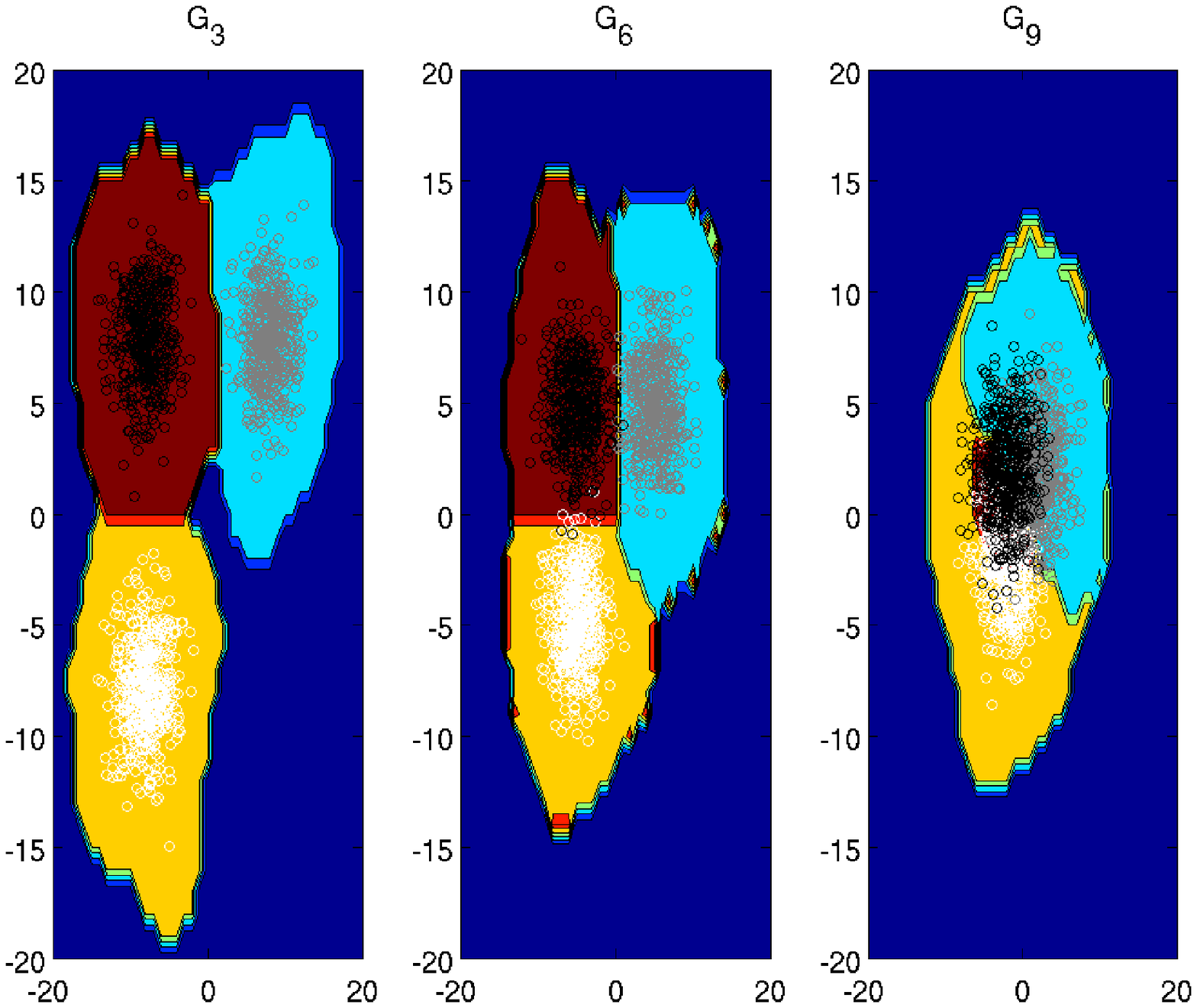}
\end{psfrags}\\
Ground truth &
\begin{psfrags}
\psfrag{x}[t][t]{{}}
\psfrag{y}[b][b]{{}}
\psfrag{t}[b][b]{{}}
\includegraphics[width=1.5in]{3gaussians_merging_ok.eps}
\end{psfrags}
\end{tabular}
\caption{\scriptsize{\textbf{Clustering results MG3}: Performance of MKSC, KSC and ESC in the three moving Gaussians experiment in terms of smoothed ARI and NMI between two consecutive partitions, and out-of-sample plot for MKSC and KSC (only the results related to snapshots $3$, $6$ and $9$ are shown). The true partitioning is depicted in the last row. The same observations made for Fig \ref{res_gauss} are still valid here. In this case, however, in the out-of-sample plots we can better recognize that MKSC, thanks to the memory effect introduced in the formulation of the primal problem, is more able than KSC to remember the old clustering boundaries and produces then smoother results over time (consider in particular the $9$-th snapshot). Finally, from the NMI plot we can notice that sometimes the ESC algorithm produces unstable results.}}\label{res_3gauss}
\end{figure}

\begin{figure}[htbp]
\centering
\begin{tabular}{c}
\begin{psfrags}
\psfrag{x}[t][t]{{Snapshot}}
\psfrag{y}[b][b]{{$\textrm{ARI}_{\textrm{Mem}}$}}
\psfrag{t}[b][b]{{Switching network}}
\includegraphics[width=2.5in]{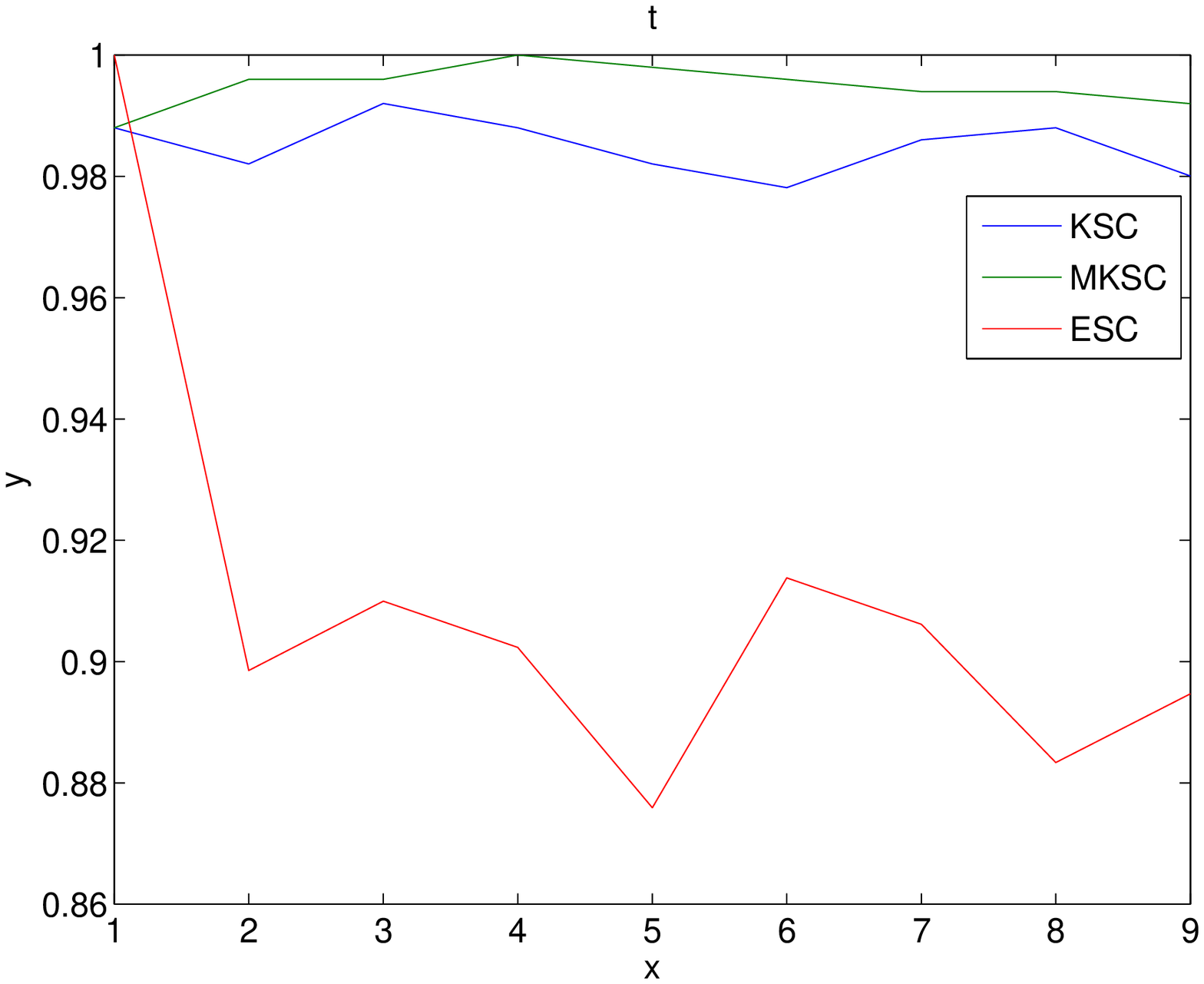}
\end{psfrags}\\\\
\begin{psfrags}
\psfrag{x}[t][t]{{Snapshot}}
\psfrag{y}[b][b]{{$\textrm{Mod}_{\textrm{Mem}}$}}
\psfrag{t}[b][b]{{}}
\includegraphics[width=2.5in]{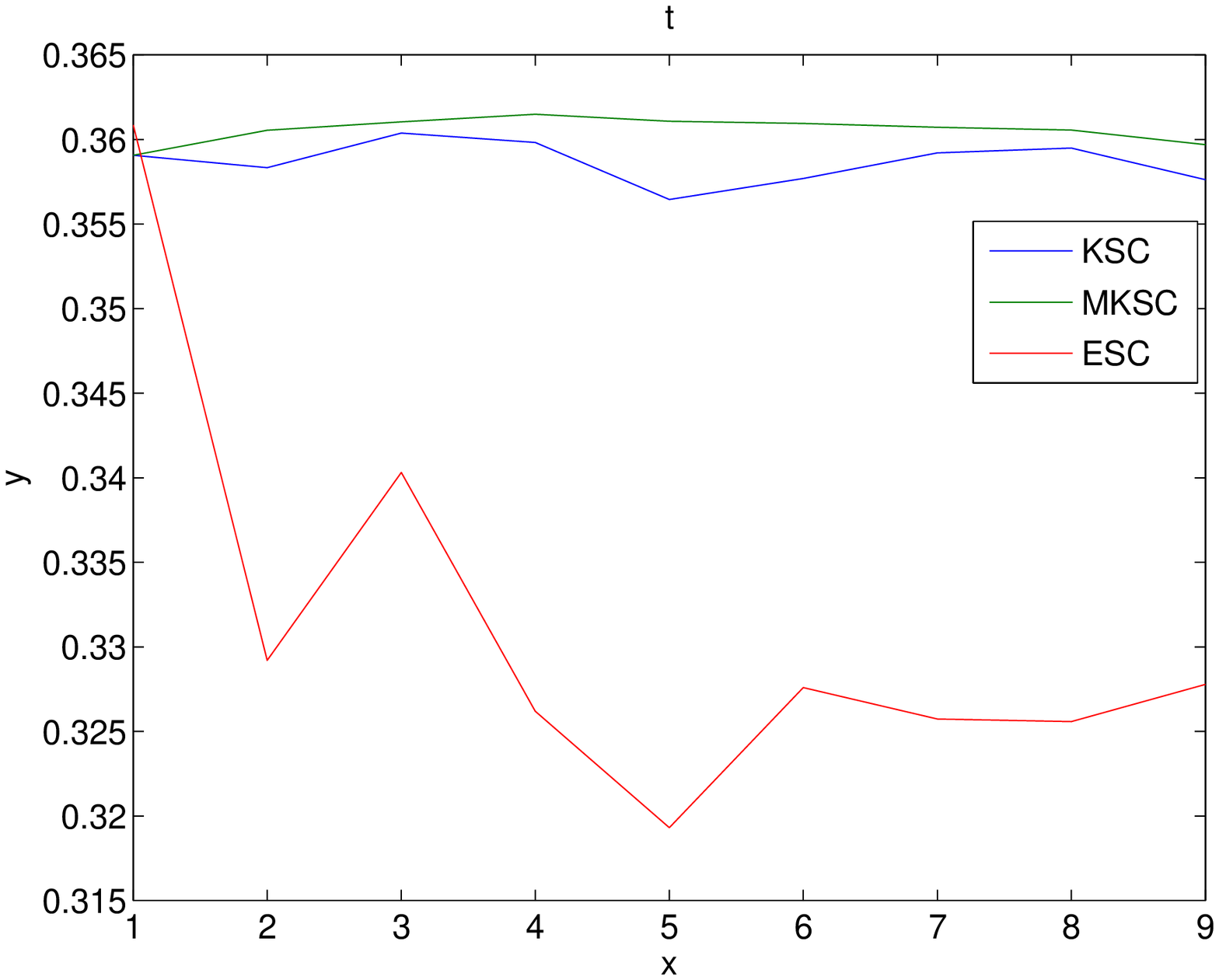}
\end{psfrags}\\\\
\begin{psfrags}
\psfrag{x}[t][t]{{Snapshot}}
\psfrag{y}[b][b]{{NMI}}
\psfrag{t}[b][b]{{}}
\includegraphics[width=2.5in]{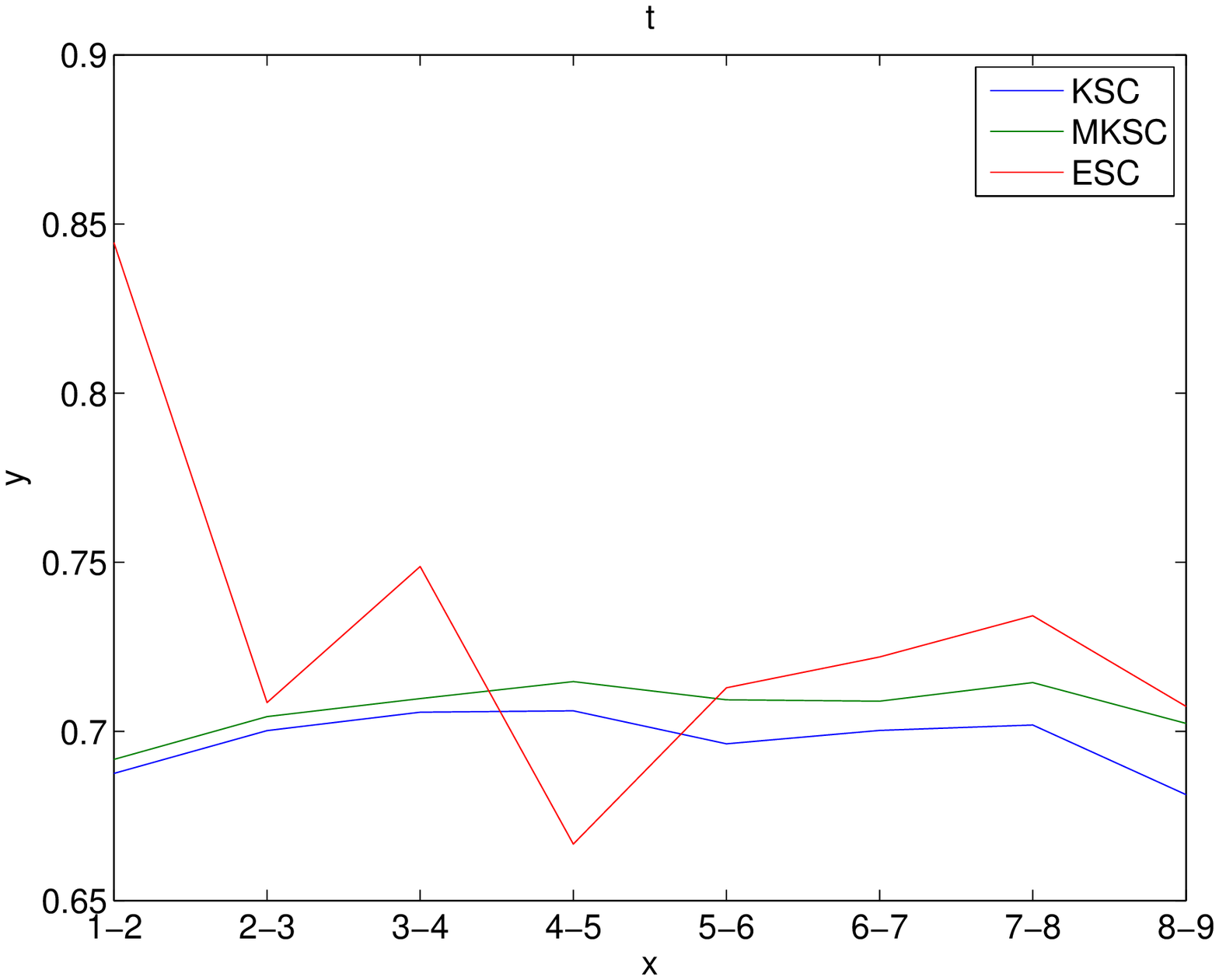}
\end{psfrags}
\end{tabular}
\caption{\textbf{Clustering results SwitchingNet}: performance of MKSC, KSC and ESC on the artificial evolving network with $2$ communities in terms of the new smoothed cluster measures explained in Section \ref{smoothed_measures}. Here, surprisingly, KSC produces better results than the ESC model according to the smoothed ARI and Modularity. However MKSC performs better than KSC.}\label{res_switch}
\end{figure}

\begin{figure}[htbp]
\centering
\begin{tabular}{c}
\begin{psfrags}
\psfrag{x}[t][t]{{Snapshot}}
\psfrag{y}[b][b]{{$\textrm{ARI}_{\textrm{Mem}}$}}
\psfrag{t}[b][b]{{Expanding/contracting network}}
\includegraphics[width=2.5in]{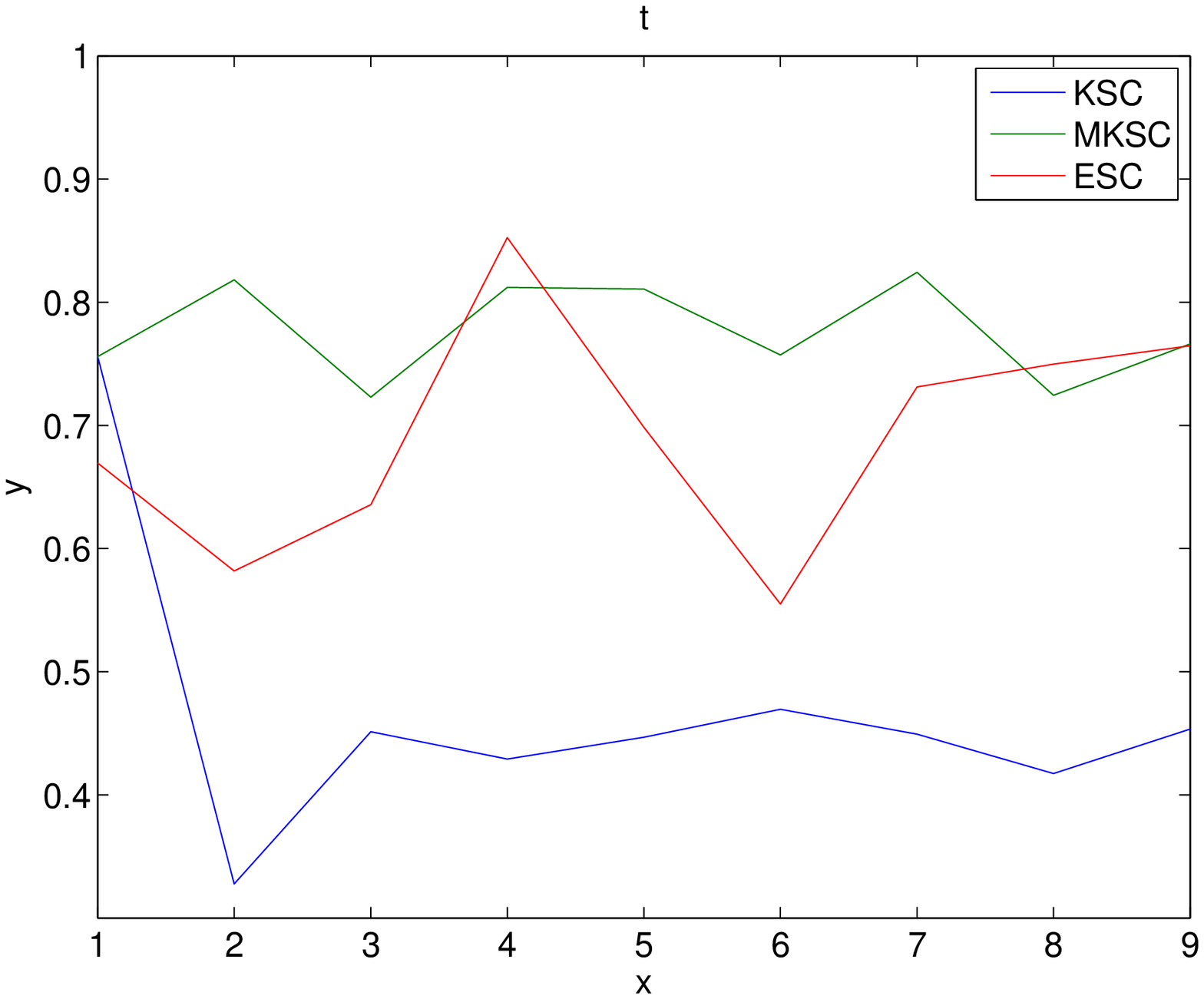}
\end{psfrags}\\\\
\begin{psfrags}
\psfrag{x}[t][t]{{Snapshot}}
\psfrag{y}[b][b]{{$\textrm{Mod}_{\textrm{Mem}}$}}
\psfrag{t}[b][b]{{}}
\includegraphics[width=2.5in]{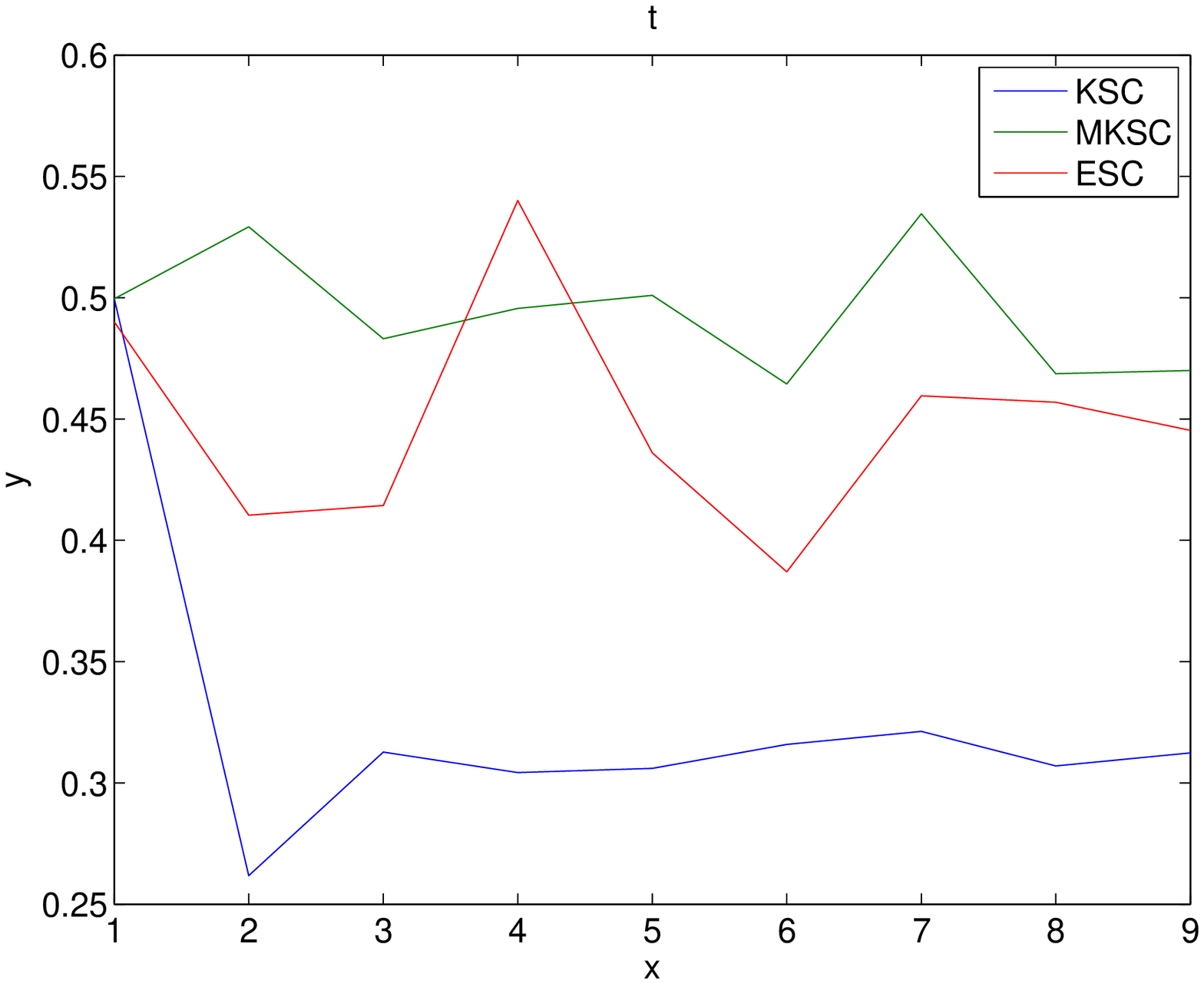}
\end{psfrags}\\\\
\begin{psfrags}
\psfrag{x}[t][t]{{Snapshot}}
\psfrag{y}[b][b]{{NMI}}
\psfrag{t}[b][b]{{}}
\includegraphics[width=2.5in]{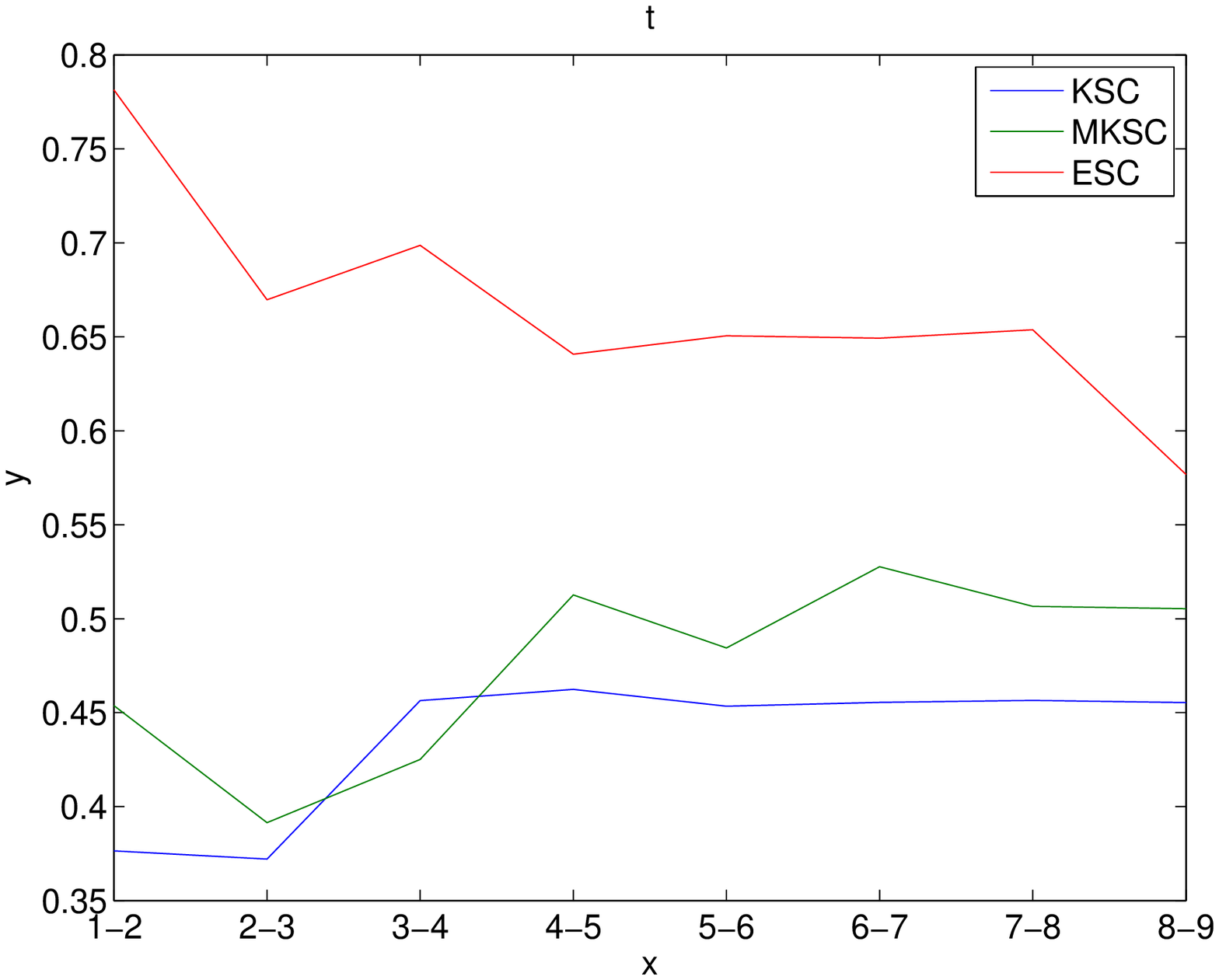}
\end{psfrags}
\end{tabular}
\caption{\scriptsize{\textbf{Clustering results ECNet}: performance of MKSC, KSC and ESC on the artificial evolving network with $5$ communities. The models with temporal smoothness produce partitions of higher quality than KSC (according to the smoothed measures introduced in Section \ref{smoothed_measures}), encouraging more consistent clustering over time. If we consider the NMI plot, ESC is the best method while MKSC outperforms all the others in terms of the smoothed measures.}}\label{res_expand}
\end{figure}

Here we present the simulation results. For what concerns the models with temporal smoothness MKSC and ESC the first partition is found by applying the corresponding static model KSC and SC to the first snapshot since we do not have any information from the past. Then we move along the next snapshots, and we apply the ESC method and MKSC algorithm with memory of $M_1 = 1$, $M_2 = 2$ and $M_3 = 3$ snapshots.     
In figures \ref{res_gauss}, \ref{res_3gauss}, \ref{res_switch} and \ref{res_expand} we present the performance of KSC, MKSC with memory $M = M_1$ and ESC in analysing the four artificial datasets under study, in terms of the smoothed cluster quality measures introduced before. Moreover, for what concerns the two and three Gaussians experiments, we also show the out-of-sample clustering results evaluated on grid points surrounding the Gaussian clouds. By looking at the figures, we can draw the following observations:
\begin{itemize}
\item \textit{MG2} experiment: the models with temporal smoothness (ESC and MKSC) can better distinguish between the two Gaussians while they are very overlapping with respect to the static model KSC, and obtain comparable results.
\item \textit{MG3} dataset: the same consideration valid for the two moving Gaussians can be drawn (here the ESC algorithm obtains the best results). In this case, however, we can notice also from the out-of-sample plot that MKSC, thanks to the memory effect introduced in the formulation of the primal problem, remembers the old clustering boundaries compared to KSC (see in particular the results related to the $9$-th snapshot).  
\item \textit{SwitchingNet}: MKSC performs slightly better than KSC and much better than ESC. The bad results obtained by ESC are quite unexpected and need further investigation. Probably they can be explained by considering that the community structure is quite different from snapshot to snapshot and while MKSC is flexible in adapting to this situation, ESC is not.
\item \textit{ECNet} graph: as expected the models with temporal smoothness (MKSC and ESC) obtain better results than the static KSC model. MKSC produces the best performances. 
\end{itemize}
In table \ref{table_budapest} the results regarding MKSC with memory $M_1$, $M_2$ and $M_3$ are summarized, together with the memory requirement and the computation time. For the two and three moving Gaussians datasets we can notice how increasing the memory of the MKSC model can give smoother and better clustering results over time, measured in terms of Normalized Mutual Information (NMI) between consecutive partitioning and the smoothed ARI. For what concerns \textit{SwitchingNet} and \textit{ECNet} networks, the more memory we add the more similar are the clustering results over time. Moreover it seems that one snapshot of memory is enough to have good performances. Finally it has to be mentioned that ESC provides unstable results, since sometimes the performances can decrease in quality (see for example the NMI plot in figure \ref{res_3gauss}). This is possibly due to the use of $K$-means to produce the final clustering. Indeed it is well known that the $K$-means algorithm depends on a random initialization which can lead sometimes to suboptimal results. On the other hand, MKSC does not suffer from this drawback.   

%\end{comment}

\begin{table}[htbp]
\begin{center}
\begin{tabular}{|c|c|c|} 
\hline
\multicolumn{3}{|c|}{\hspace{2 mm}\textbf{MG2}\hspace{2 mm}}\\
\hline 
\textbf{MEASURE}  & \textbf{MKSC} &  \textbf{ESC}\\
\hline
$\mathbf{ARI_{\textrm{Mem}}}$ & $\mathbf{0.88}$ ($M_3$) & $0.87$\\
\hline
\textbf{NMI} & $\mathbf{0.86}$ ($M_3$) & $0.85$ \\
\hline
\textbf{CPU time (s)} & $\mathbf{0.53}$ ($M_1$)& $5.36$\\
\hline
\textbf{\hspace{0.6 mm}Memory requirement (s)\hspace{0.6 mm}} & $\mathbf{0.29}$ ($M_1$) & $2.92$\\
\hline
\end{tabular}\\
\begin{tabular}{|c|c|c|} 
\hline
\multicolumn{3}{|c|}{\textbf{MG3}}\\
\hline 
\textbf{MEASURE}  & \textbf{MKSC} &  \textbf{ESC}\\
\hline
$\mathbf{ARI_{\textrm{Mem}}}$ & $\mathbf{0.92}$ ($M_3$) & $0.81$ \\
\hline
\textbf{NMI} &  $\mathbf{0.91}$ ($M_3$) & $0.75$\\
\hline
\textbf{CPU time (s)} & $\mathbf{1.30}$ ($M_1$) & $18.95$ \\
\hline
\textbf{Memory requirement (s)} & $\mathbf{0.73}$ ($M_1$) & $10.77$\\
\hline
\end{tabular}\\
\begin{tabular}{|c|c|c|} 
\hline
\multicolumn{3}{|c|}{\textbf{SwitchingNet}}\\
\hline 
\textbf{MEASURE}  & \textbf{MKSC} &  \textbf{ESC}\\
\hline
$\mathbf{ARI_{\textrm{Mem}}}$ & $\mathbf{0.99}$ ($M_1$) & $0.91$\\
\hline
$\mathbf{Mod_{\textrm{Mem}}}$ & $\mathbf{0.36}$ ($M_1$)& $0.31$\\
\hline
\textbf{NMI} & $0.73$ ($M_3$) & $0.73$\\
\hline
\textbf{CPU time (s)} & $52.23$ ($M_1$) & $\mathbf{5.56}$\\
\hline
\textbf{Memory requirement (s)} & $4.31$ ($M_1$) & $\mathbf{3.06}$\\
\hline
\end{tabular}\\
\begin{tabular}{|c|c|c|} 
\hline
\multicolumn{3}{|c|}{\textbf{ECNet}}\\
\hline 
\textbf{MEASURE}  & \textbf{MKSC} &  \textbf{ESC}\\
\hline
$\mathbf{ARI_{\textrm{Mem}}}$ & $0.76$ ($M_1$) & $\mathbf{0.78}$ \\
\hline
$\mathbf{Mod_{\textrm{Mem}}}$ & $\mathbf{0.51}$ ($M_1$) & $0.50$ \\
\hline
\textbf{NMI} & $0.45$ ($M_1$)  & $\mathbf{0.61}$\\
\hline
\textbf{CPU time (s)} & $51.83$ ($M_1$) & $\mathbf{5.08}$ \\
\hline
\textbf{Memory requirement (s)} & $\mathbf{2.57}$ ($M_1$) & $2.79$\\
\hline
\end{tabular}
\end{center}
\caption{\textbf{Clustering results synthetic datasets}. For each evaluation measure the values represent an average over time (i.e. the mean value per snapshot). The NMI is calculated between consecutive partitions, while the other quality functions are evaluated in each snapshot. Regarding the MKSC model, between parenthesis we indicate with which amount of memory we obtained the best results ($1$, $2$ or $3$ snapshots of memory, that is $M_1$, $M_2$, or $M_3$). Concerning the datasets MG2 and MG3, MKSC is the best performer in terms of all the evaluation measures. Since an efficient implementation of the RBF kernel and a small training set are used, the runtime required by MKSC is rather small. On the other hand, we can notice how the CPU time required by MKSC to analyse the synthetic network data is much higher than the time needed by ESC. This can be explained by considering that in the ESC method it is not necessary to construct a kernel matrix, which in this case represents a computational burden for MKSC, where the community kernel has been used.}
\label{table_budapest}
\end{table}

\subsection{Real-Life Application}
\label{cellphone_net_f1}
The \textbf{CellphoneNet} has been investigated. This dataset records the cellphone activity for students and staff from two different labs in MIT \cite{MIT_realitymining}. It is constructed on users whose cellphones periodically scan for nearby phones over Bluetooth at five minute intervals. The similarity between two users is related to the number of intervals where they were in physical proximity. Each graph snapshot is a weighted network corresponding to $1$ week activity. In particular we consider $42$ nodes, representing students always present during the fall term of the academic year $2004$-$2005$, for a total of $12$ snapshots.

\begin{figure}[htbp]
\centering
\begin{psfrags}
\psfrag{x}[t][t]{{Week}}
\psfrag{y}[b][b]{{$\sigma ^2$}}
\psfrag{t}[b][b]{{}}
\includegraphics[width=3in]{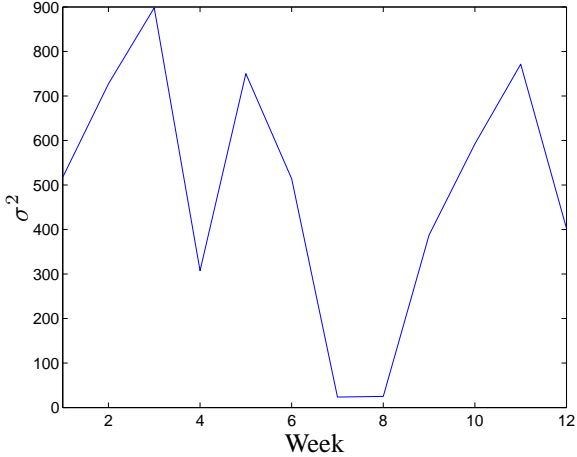}
\end{psfrags}
\caption{\textbf{Model selection CellphoneNet}. Optimal $\sigma^2$ over time for the cellphone network, related to MKSC. The number of clusters is $k = 2$, $\gamma = 1$ is an optimal value for all the snapshots.}\label{MS_cellphone}
\end{figure}

Regarding the model selection, we have a partial ground truth, namely the affiliations of each participant. In particular, as observed in \cite{MIT_realitymining} and in \cite{graphscope}, $2$ dominant clusters could be identified from the Bluetooth proximity data, corresponding to new students at the Sloan business school and co-workers who work in the same building. Then for this experiment we perform clustering with number of clusters $k = 2$, while the optimal $\sigma$ over time is estimated by using the Modularity-based model selection algorithm on each snapshot (see figure \ref{MS_cellphone}). So $k = 2$ and $\gamma = 1$ are optimal hyper-parameters for each of the $12$ weeks, while the values of $\sigma ^2$ over time are reported in figure \ref{MS_cellphone}.     

\begin{figure}[htbp]
\centering
\begin{tabular}{c}
\begin{psfrags}
\psfrag{x}[t][t]{{Week}}
\psfrag{y}[b][b]{{$\textrm{Mod}_{\textrm{Mem}}$}}
\psfrag{t}[b][b]{{Cellphone network}}
\includegraphics[width=3in]{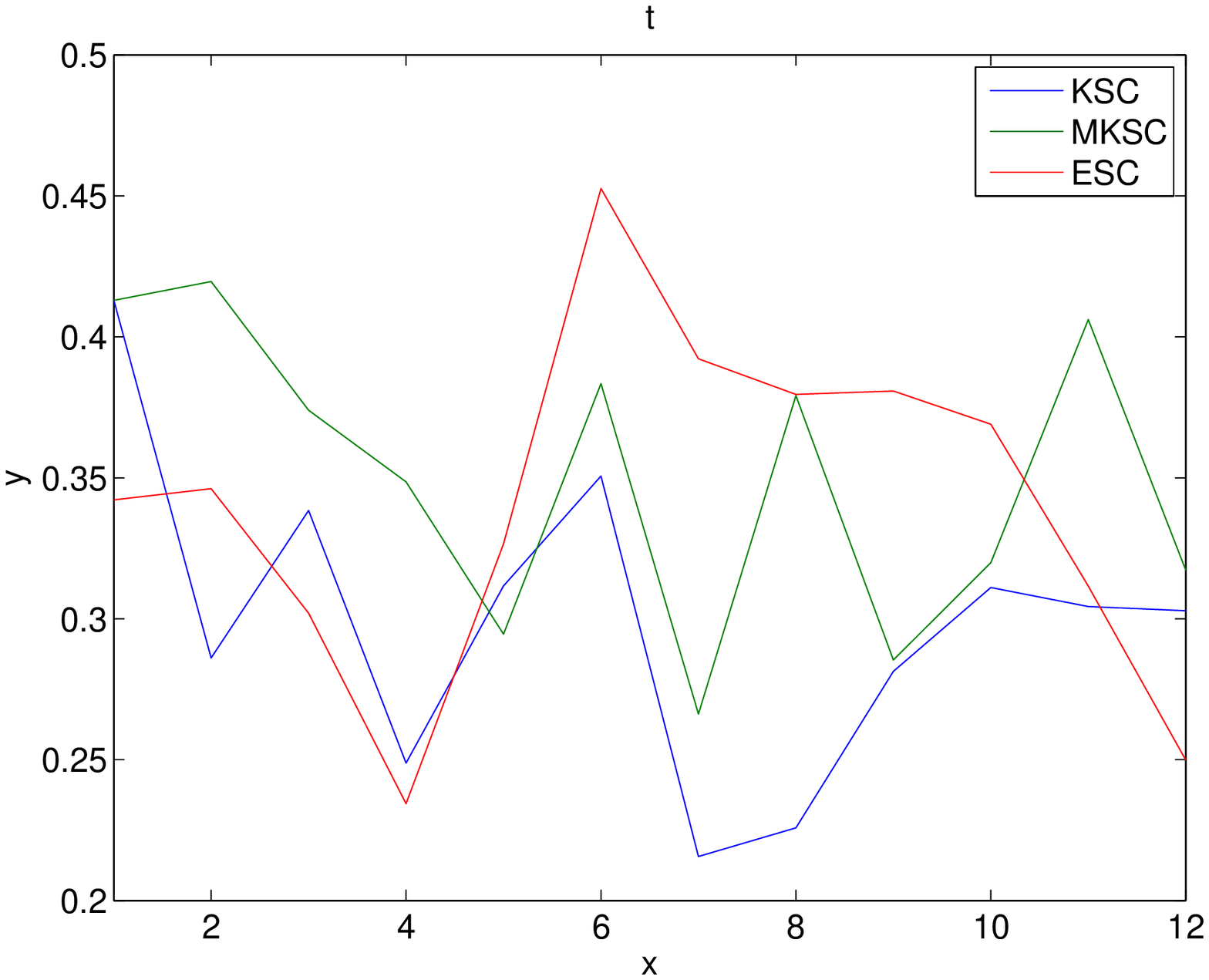}
\end{psfrags}\\\\
\begin{psfrags}
\psfrag{x}[t][t]{{Week}}
\psfrag{y}[b][b]{{NMI}}
\psfrag{t}[b][b]{{}}
\includegraphics[width=3.4in]{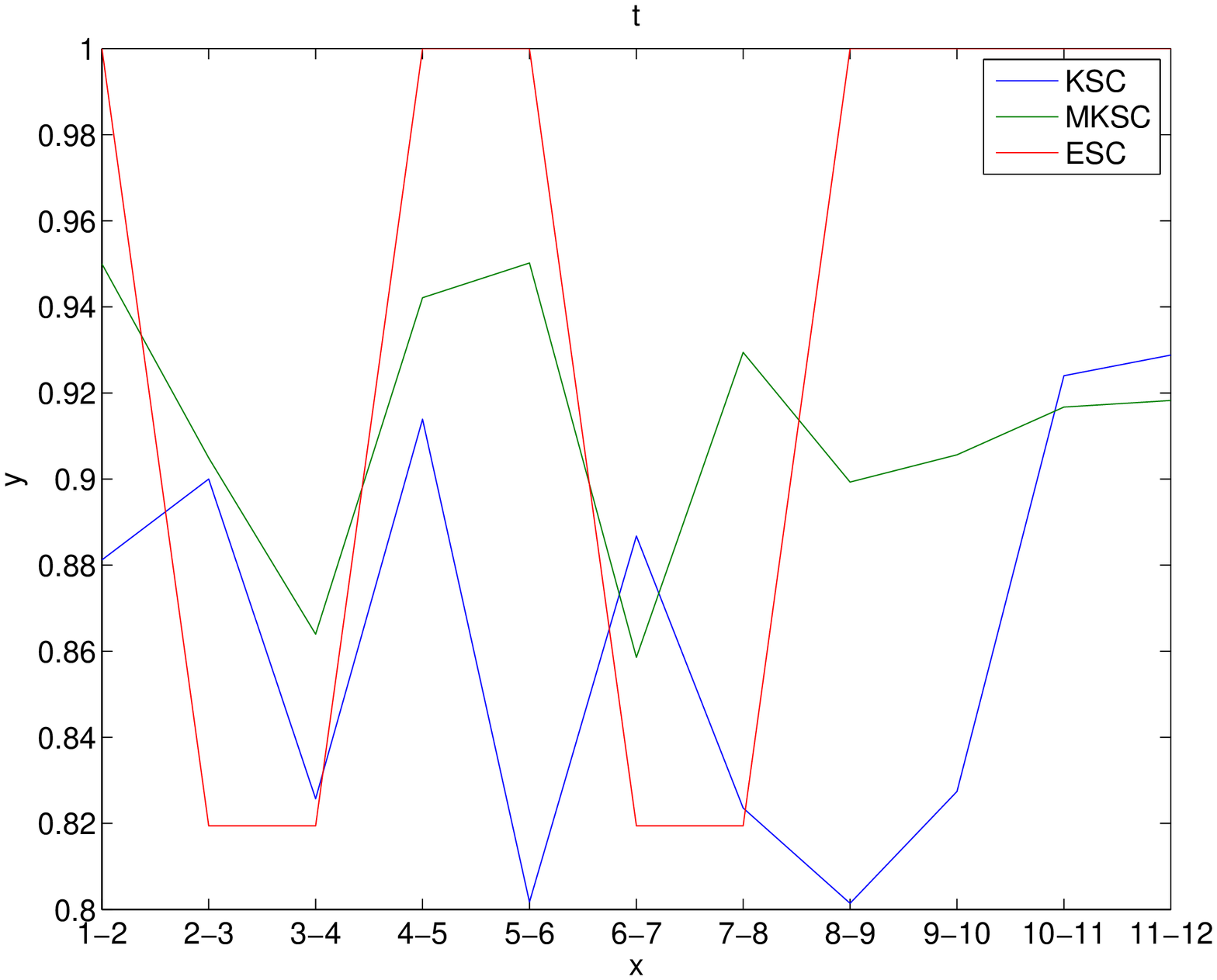}
\end{psfrags}
\end{tabular}
\caption{\textbf{Clustering results CellphoneNet}: performance of MKSC, KSC and ESC on the cellphone network in terms of the smoothed Modularity and the NMI between consecutive clustering results. Also in this case the models with temporal smoothness MKSC and ESC, in most of the time period, perform better than the static KSC.}\label{res_cellphone}
\end{figure}

For what concerns the final partitioning the results are illustrated in figure \ref{res_cellphone}. MKSC performs better than ESC in some periods and worse in other ones. Both obtain better performances than KSC in terms of the smoothed Modularity and the NMI between consecutive partitions.    

\section{Framework 2}
In this section we describe the experimental results related to networks where nodes are entering and leaving over time and the number of communities is changing. Moreover we do not assume that the clustering results must be smoothed at each time stamp, but the smoothness is activated by tuning the regularization constant $\nu$. Thus, in contrast to \textit{framework 1} we now select $\nu$ and we fix $\gamma=1$. The method that we use can be summarized in algorithm \ref{alg:alg3}.
\begin{algorithm}[t]
\small
\LinesNumbered
\KwData{Training sets $\mathcal{D}=\{x_i\}_{i=1}^{N_{\textrm{Tr}}}$ and $\mathcal{D_\textrm{old}}=\{x_i^{\textrm{old}}\}_{i=1}^{N_{\textrm{Tr}}}$, test sets $\mathcal{D^{\textrm{test}}}=\{x_m^{\textrm{test}}\}_{m=1}^{N_{\textrm{test}}}$ and $\mathcal{D^{\textrm{test}}_\textrm{{old}}}=\{x_m^{\textrm{test,old}}\}_{m=1}^{N_{\textrm{test}}}$, $\alpha ^{(l)} _ {\textrm{old}}$ (the $\alpha ^{(l)}$ calculated for the previous $M$ snapshots), positive definite kernel function $K: \mathbb{R}^d \times \mathbb{R}^d \rightarrow \mathbb{R}$ such that $K(x_i,x_j) \rightarrow 0$ if $x_i$ and $x_j$ belong to different clusters, kernel parameters (if any), number of clusters $k$, regularization constants $\gamma$ and $\nu$ found using algorithm \ref{alg:alg1}.}
\KwResult{Clusters $\{\mathcal{C}^t_1,\hdots,\mathcal{C}^t_p\}$, cluster codeset $\mathcal{CB} = \{c_p\}_{p = 1}^k$, $c_p \in \{-1,1\}^{k-1}$.}
\eIf {t==1}{
Initialization by using kernel spectral clustering (KSC \cite{multiway_pami}).
}{
For every snapshot from $t-1$ to $t-M$ rearrange the data matrices and the solution vectors  $\alpha^{(l)}$ as explained in Section \ref{changing_objects}.\\
Use steps $4-10$ of algorithm \ref{alg:alg2}.\\
Match the actual clusters $\{\mathcal{C}^t_1,\hdots,\mathcal{C}^t_p\}$ with the previous partitioning $\{\mathcal{C}^{t-1}_1,\hdots,\mathcal{C}^{t-1}_q\}$ using the tracking scheme described in Section \ref{clu_matching} and summarized in algorithm \ref{algo:algo1}.  
}  
\caption{Clustering evolving networks: framework 2 \cite{paper_KDD2014}}
\label{alg:alg3}
\end{algorithm}
In the experimental results reported later we utilize the Average Membership Strength (AMS) criterion \cite{rocco_IJCNN2013} and the Modularity quality function \cite{newmanmod2,rocco_IJCNN2011} to perform model selection. Moreover, the Fast and Unique Representative Subset Selection (FURS \cite{FURS_paper}) is used to select the training and validation sets, in the proportion of $15 \% $ and $30 \% $ respectively. Thus, the results are related to a single run of the MKSC algorithm since the subset selection performed by the FURS method is deterministic.  

\subsection{Object appearing and leaving over time}
\label{changing_objects}
When performing the clustering for the actual data snapshot present at time $t$, two possible situations can arise: new data points are introduced or some existing objects may be disappeared. To cope with the first scenario, the rows of the old data matrices corresponding to the new points can be set to zero, as well as the related components of the solution vectors $\alpha^{(l)}_ {\textrm{old}}$. In this way, when solving problem (\ref{dualMKSC}), the components of $\alpha^{(l)}$ related to the new objects have no influence from the past. On the other hand, data points that were present in the previous snapshots but not in the actual one, can simply be removed in order for the the previous $M$ snapshots\footnote{We remind that the memory $M$ indicates the amount of past information to carry along when clustering the current data matrix.} $\mathcal{G}_{t-1},\hdots,\mathcal{G}_{t-M}$ to have the same dimensions as the data matrix $\mathcal{G}_t$.                   

\subsection{Tracking the clusters} 
\label{clu_matching}
Several events that happen during the evolution of clusters are continuing, shrinking, growing, splitting, merging, dissolving and forming of clusters. In order to recognize these circumstances a tracking algorithm that matches the partitions found at each time step is needed, like the ones proposed in \cite{Greene2011} and \cite{GED_brodka}. In this realm, we introduce a tracking method that is based on a maximum weight matching mechanism \cite{Kuhn_55}, as depicted in algorithm \ref{algo:algo1}.
\begin{algorithm}[ht]
\SetAlgoLined
    \KwData{At a given time stamp $t$, take the clustering information of time stamp $t-1$ and $t$ i.e. $C^{t-1}$ and $C^{t}$.}
    \KwResult{A weighted directed network $W_{N}^{t}$ tracking the relationship between the clusters at time stamps $t-1$ and $t$.}
    \ForEach { $C^{t-1}_{j} \subset C^{t-1}$ } {
        \ForEach { $C^{t}_{k} \subset C^{t}$ }{
            \If{ Nodes with labels $c^{t-1}_{j}$ at $t-1$ have the label $c^{t}_{k}$ at $t$}    {
            Create temporary edge $v^{t}(j,k)$ between $C^{t-1}_{j}$ and $C^{t}_{k}$.\\
            $n(j,k)$ = Number of nodes with labels $c^{t-1}_{j}$ at $t-1$ which have the label $c^{t}_{k}$ at $t$.\\
            Weight of the edge $v^{t}(j,k) = \frac{n(j,k)}{|C^{t}_{j}|}$ \\
            }
        }
    }
    Keep the edge which has maximum weight (in case of multiple edges keep all) w.r.t. $C^{t-1}_{j}$.\\
    Add this weighted edge to the graph $W_{N}^{t}$.\\
    \ForEach { $C^{t}_{k} \subset C^{t}$ } {
        \If {$C^{t}_{k}$ is isolated} {
            Select the edge $v^{t}(j,k)$ which had maximum incoming weight w.r.t. $C^{t}_{k}$.
            Add this weighted edge to the graph $W_{N}^{t}$.\\
            \tcc{This is done in order to prevent isolated nodes in $W_{N}^{t}$.}
        }
    }    
\caption{Cluster Tracking Algorithm \cite{paper_KDD2014}}
\label{algo:algo1}
\end{algorithm} 

\begin{figure}[htbp]
\centering
\includegraphics[width=4in]{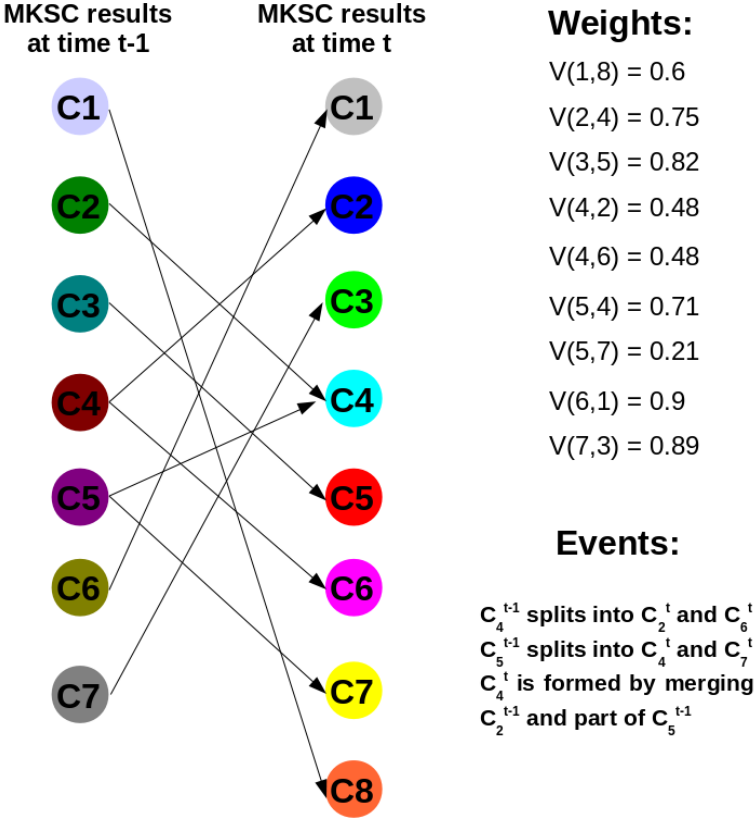}
\caption{\textbf{Illustrative example of the cluster matching procedure}. Since the labelling at each time step is arbitrary, the clusters found by MKSC at successive time stamps have to be matched to keep track of their evolution. In this specific case, for instance, it is clear that cluster $3$ at time $t$ should be labelled as cluster $7$, cluster $5$ as cluster $3$ and so on.}\label{fig_tracking}
\end{figure}

We generate a directed weighted network $W_N^{t}$ from the clusters at two consecutive timestamps $t$ and $t+1$. Thus, if we have $T$ timestamps we generate a set ${\cal W_N} = \{W_N^{1},\ldots,W_N^{T-1} \}$ of directed weighted networks. Each directed weighted network $W_N^{t}$ creates a map between the clusters at time stamp $t$ i. e. $C^{t}$ and time stamp $t+1$ i. e. $C^{t+1}$, which form the edges of the network. The weight $v^t(j,k)$ of an edge between two clusters is equivalent to the fraction of nodes in cluster $C_{j}$ at time stamp $t$ which are assigned to cluster $C_{k}$ at time stamp $t+1$. An edge exists between two clusters $C_{j}$ and $C_{k}$ only if $v^t(j,k)>0$. Thus, if the number of edges going out of a node of $W_N^{i}$ is greater than $1$ it indicates a split whereas if the number of edges entering a node is greater than $1$ then it indicates a merge. In case $v^t(j,k)=1.0$ then the cluster remains unchanged between the $2$ time intervals $t$ and $t+1$. In order to tackle the birth and death of clusters, we add a $C_{0}$ cluster for each time stamp $t$. For the network $W_N^{t}$, if $C^{t}_{0}$ is isolated then no new clusters were generated at time $t$ and if $C^{t+1}_{0}$ is isolated then none of the clusters present at time stamp $t$ dissolved in the next snapshot. However, if we have outgoing edges from $C^{t}_{0}$ then there was birth of new clusters. Similarly, if we have incoming edges to $C^{t+1}_{0}$ in $W_N^{t}$ then some clusters dissolved at time $t$. In Figure \ref{fig_tracking} an example of the matching mechanism is given. 

\subsection{Description of the data sets}
\label{data_sets_f2}
The artificial benchmarks consist of evolving networks generated by the software related to \cite{Greene2011}:
\begin{itemize}
\item \textbf{MergesplitNet}: an initial network of $1000$ nodes formed by $7$ communities evolves over $5$ time steps. At each time step there are $2$ splitting events and $1$ merging event. The number of nodes remains unchanged.  
\item \textbf{BirthdeathNet}: a starting network with $13$ communities experiences at each time one cluster death and one cluster generation, while the number of nodes decreases from $1000$ to $866$ as time increases from $1$ to $5$. 
\item \textbf{HideNet}: at each time step a community of an initial network with $1000$ nodes and $7$ communities dissolves, and the number of nodes also varies over time. 
\end{itemize}   
To analyse these data we use the cosine or normalized linear kernel defined as $\Omega_{ij} = \frac{x_i^Tx_j}{||x_i||||x_j||} $. So at each time step, when performing model selection, we only have to detect the optimal number of clusters $k$ and tune the smoothness parameter $\nu$.   

The real-world dataset is the \textbf{CellphoneNet} network described in Section \ref{cellphone_net_f1}. However, unlike for \textit{framework 1} now we consider the complete dataset of $46$ weeks since we can handle nodes entering and leaving across time. In total there are $94$ nodes, but not all the nodes are present in every snapshot. In particular, the smallest network comprises $21$ people and the largest has $88$ nodes. 

\subsection{Experiments}
\label{experimental_results} 
In this Section an exhaustive study about the ability of the MKSC algorithm to perform dynamic clustering is performed. First we discuss the model selection issue: different criteria are contrasted and the outcomes are analysed. Then the clustering results are evaluated according to a number of cluster quality measures and the MKSC method is compared with the AFFECT algorithm \cite{adapt_evol_clu_Xu} and the ESC \cite{evol_spec_clustering} technique. Finally, a simple 3D visualization of the clusters evolution over time is presented.   

\subsubsection{Model selection}
\label{model_sel}
Here the AMS criterion \cite{rocco_IJCNN2013} and the Modularity criterion \cite{rocco_IJCNN2011} are compared for tuning the number of clusters $k$ and the smoothness parameter $\nu$.

\begin{figure*}[htbp]
\centering
\begin{tabular}{cc}
\includegraphics[width=2.25in]{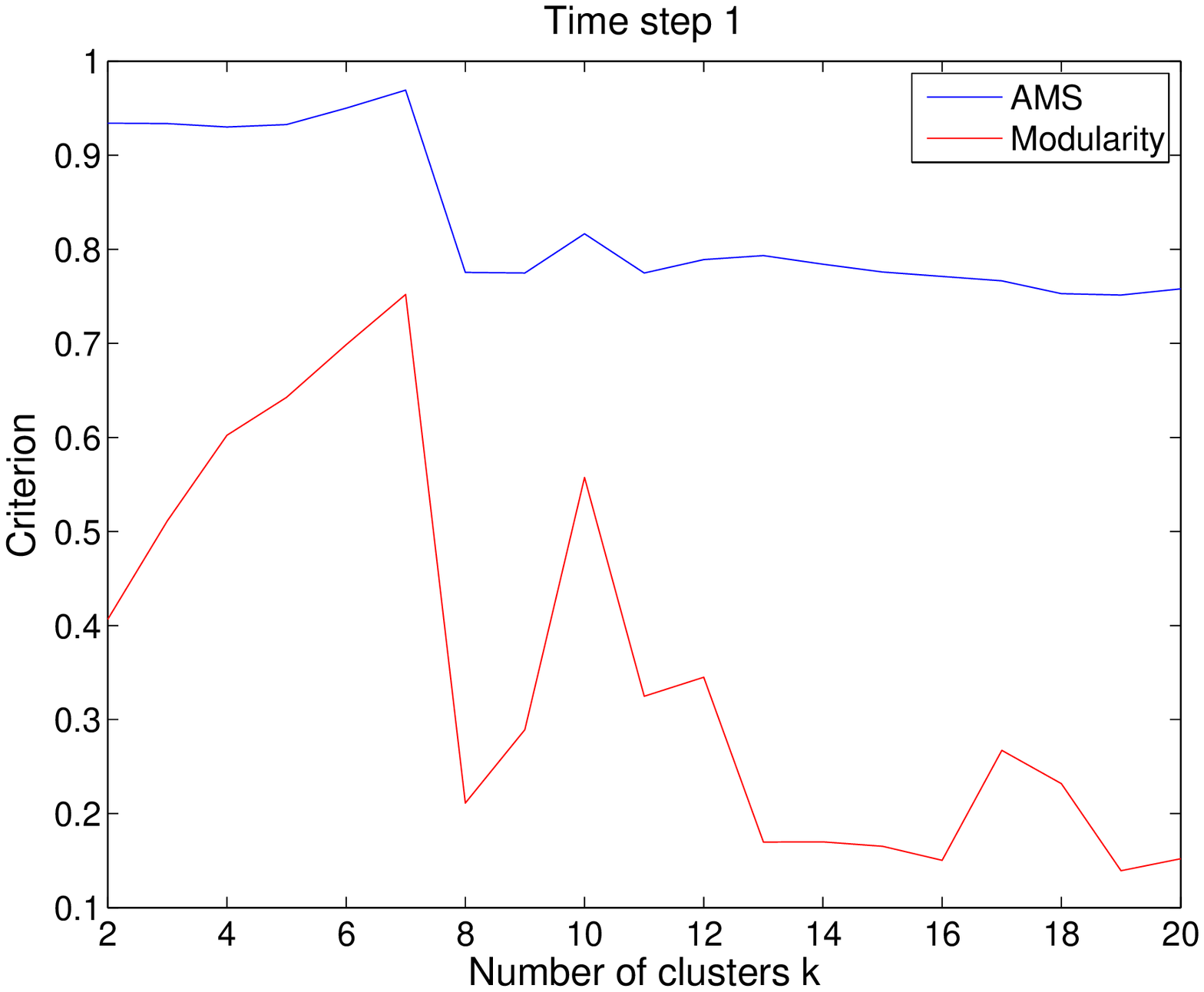}&
\includegraphics[width=2.25in]{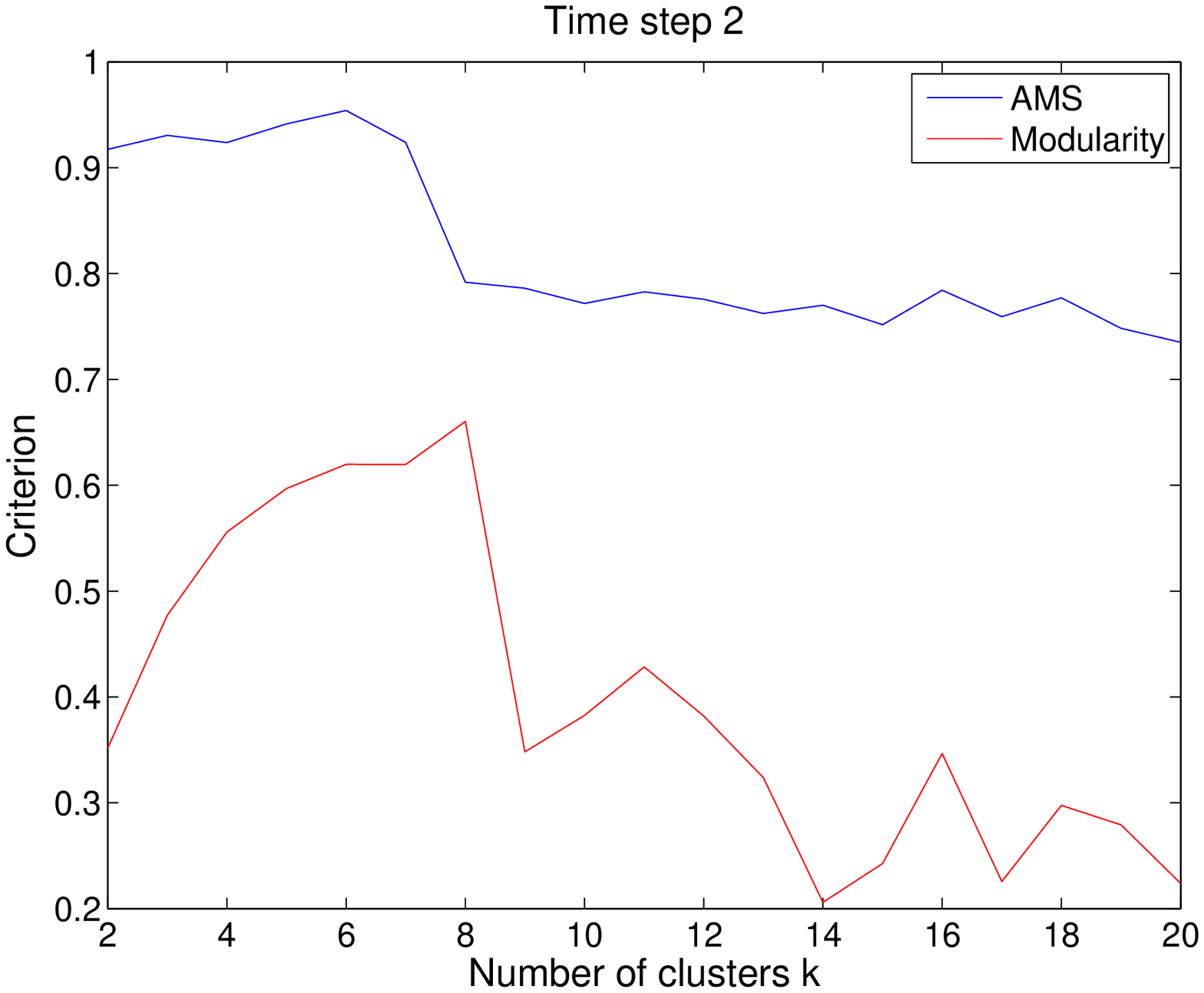}\\
\includegraphics[width=2.25in]{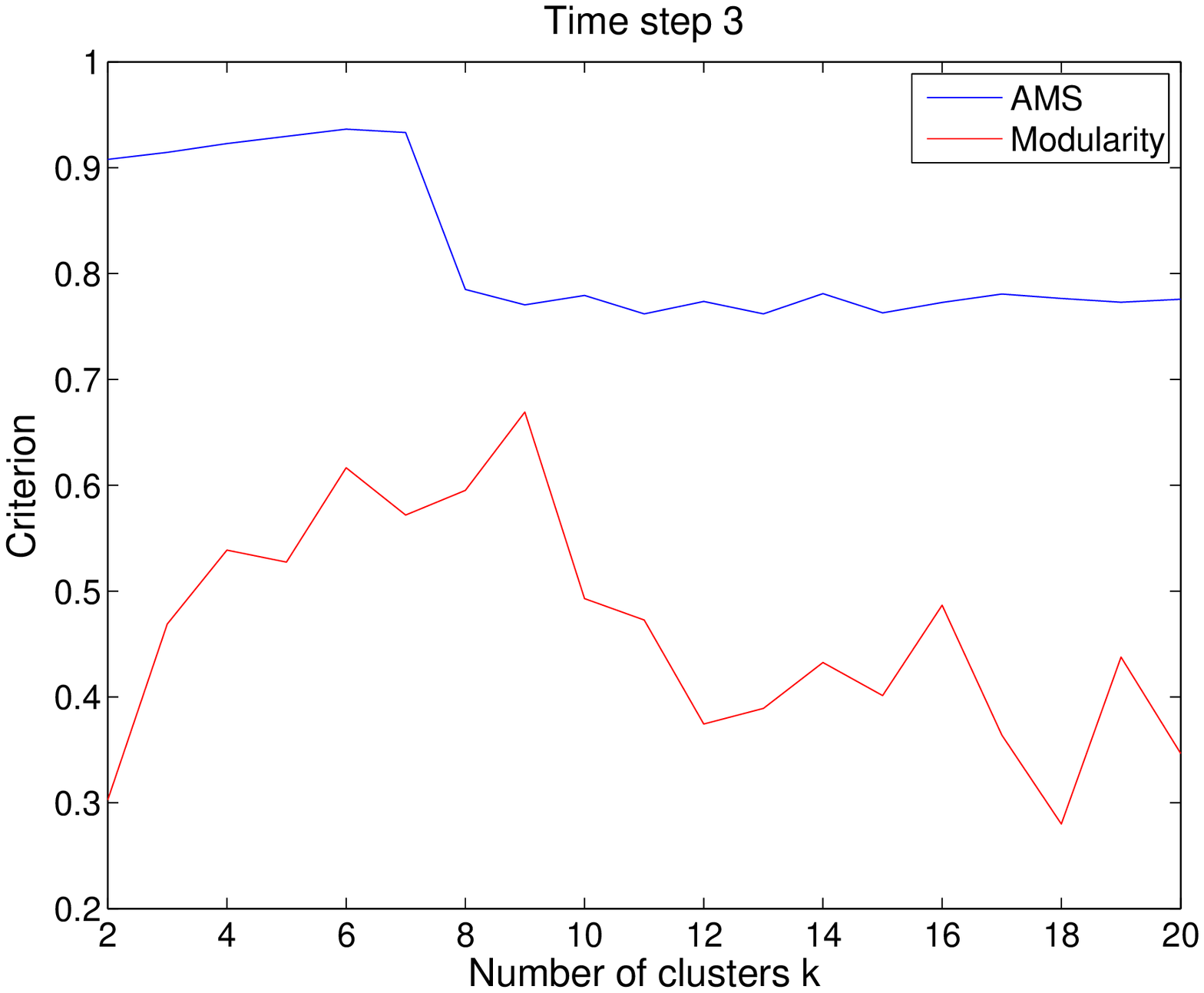}&
\includegraphics[width=2.25in]{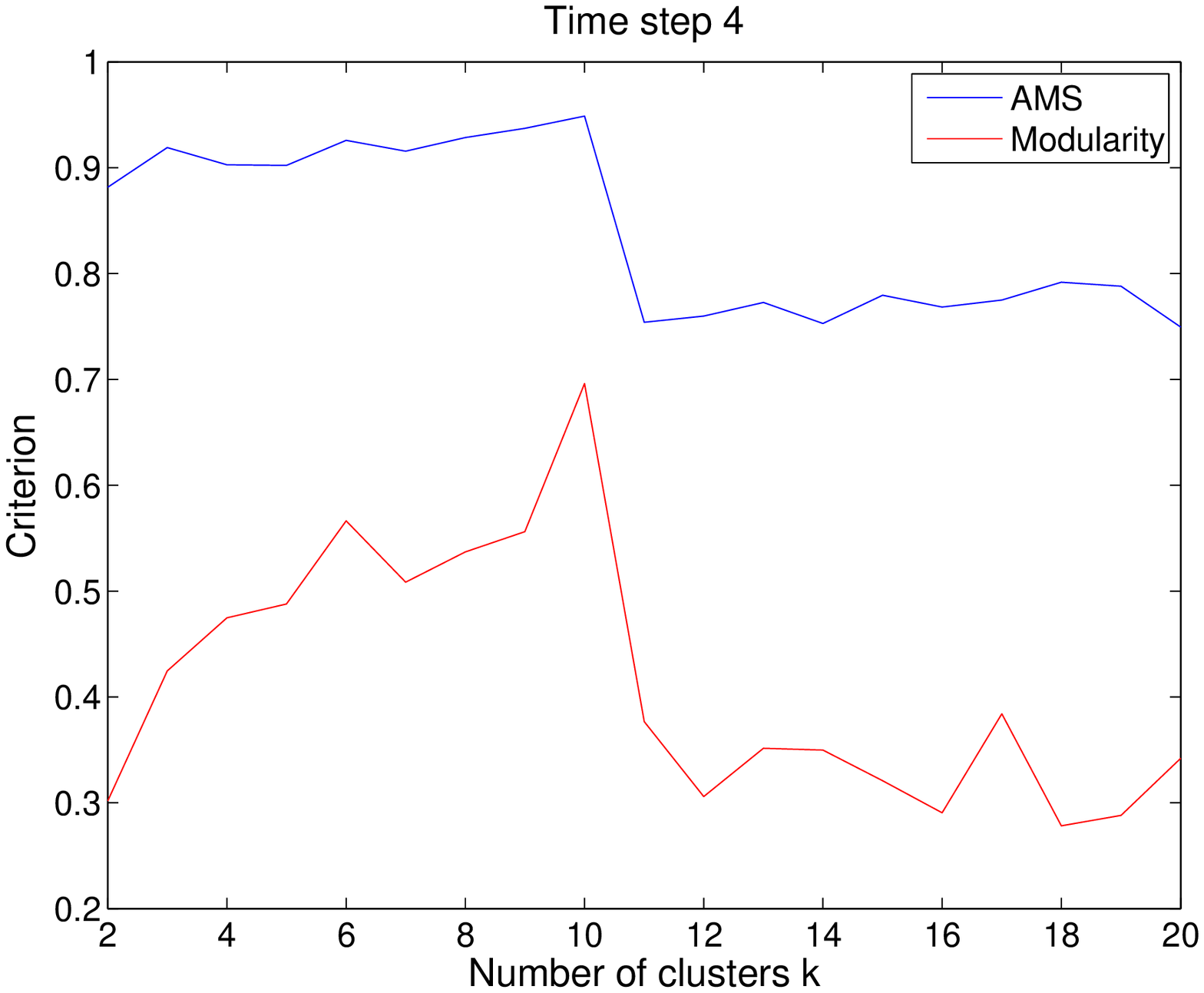}\\
\multicolumn{2}{c}{\includegraphics[width=2.25in]{MS_mergesplit_t4.eps}} 
%\multicolumn{3}{c}{\includegraphics[width=6in]{Mergesplit_Network_Evolution.eps}}
\end{tabular}
\caption{\textbf{MergesplitNet: selection of the number of clusters}. The true number of clusters over time is $7$, $8$, $9$, $10$, $11$ at times $T_1,\hdots,T_5$. Modularity (red) perfectly detects the right number of clusters in all the time steps, while AMS (blue) selects  $7$, $6$, $6$, $10$, $9$. In this case $\nu$ has been tuned using Modularity and is zero for all time steps except $t=T_4$, where $\nu = 0.1$. This indicates that at each time from $t=T_1$ to $t=T_3$ the current network is not enough different from the previous one to require smoothing. This can be noticed also by looking at the 3D visualization shown in Figure \ref{fig:subfig3}.}\label{MS_mergesplit}
\end{figure*}

\begin{figure*}[htbp]
\centering
\begin{tabular}{cc}
\includegraphics[width=2.25in]{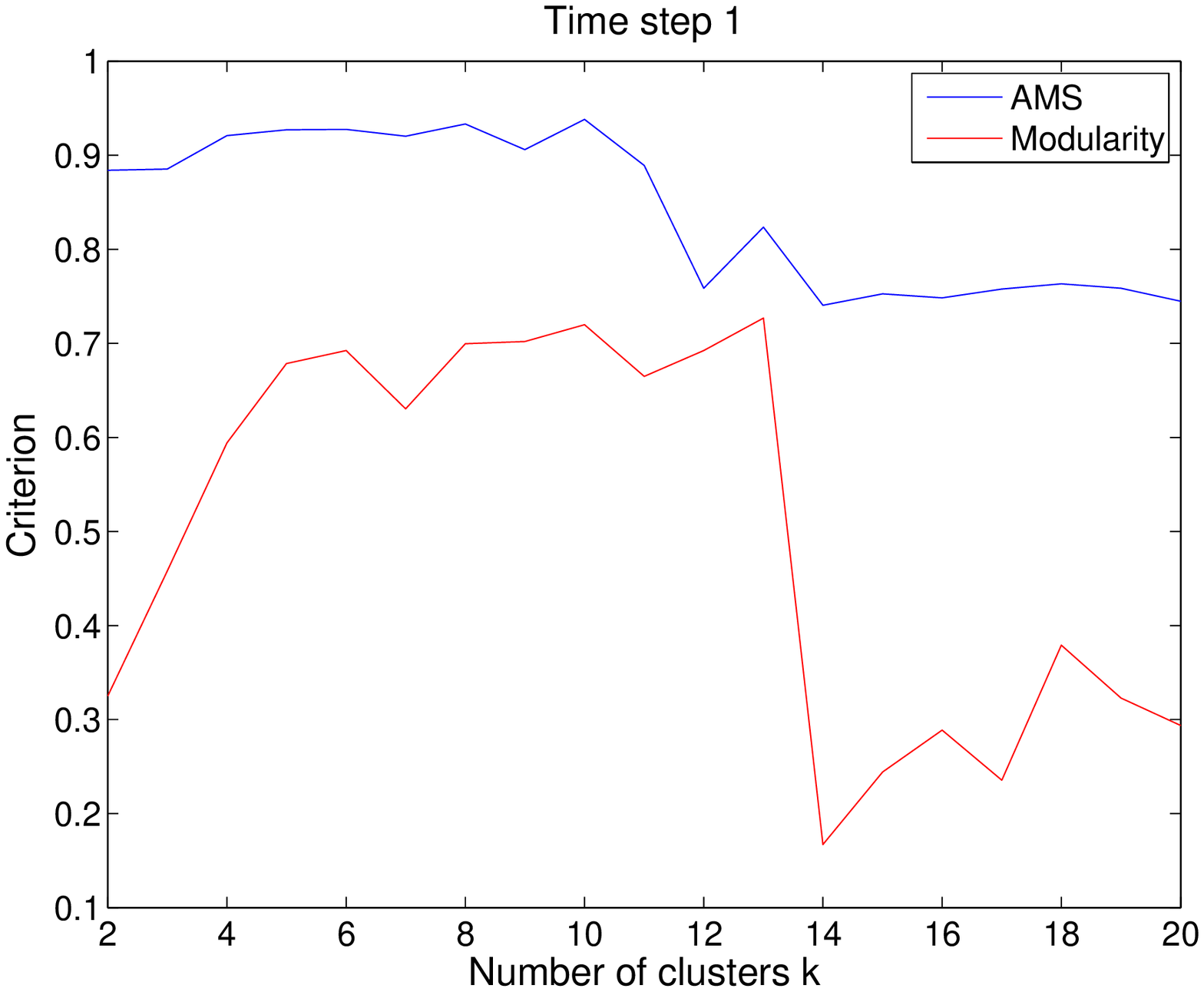}&
\includegraphics[width=2.25in]{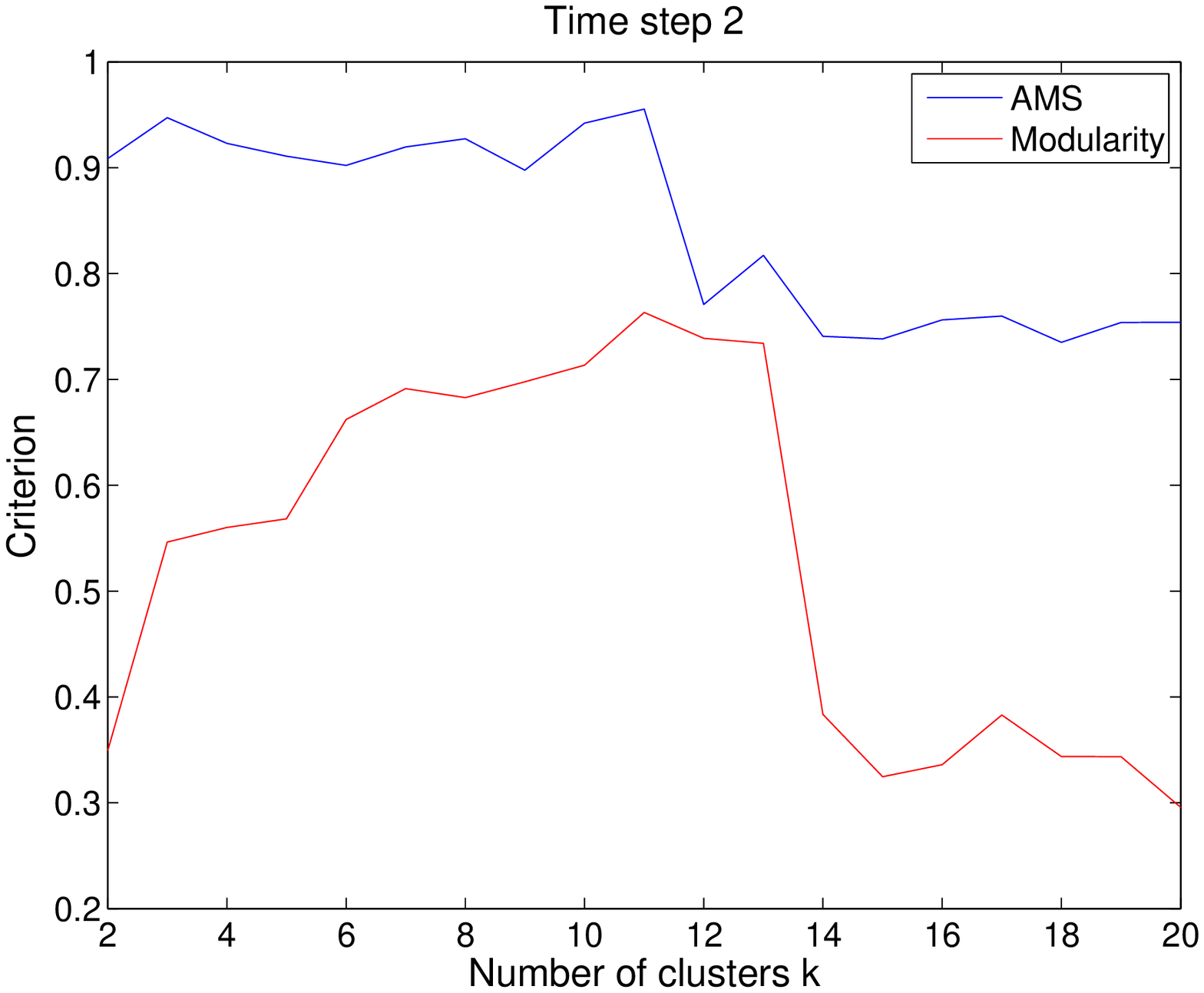}\\
\includegraphics[width=2.25in]{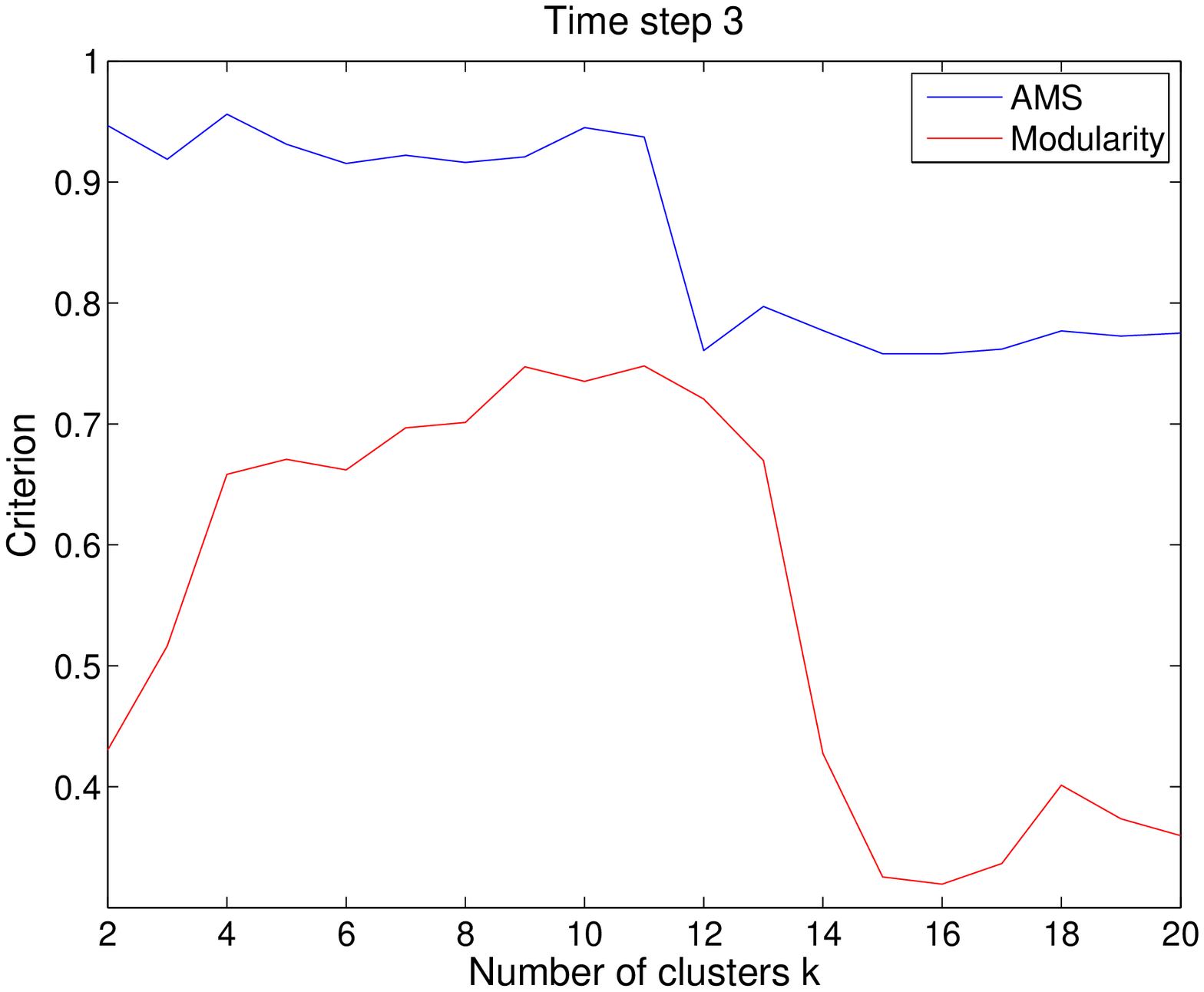}&
\includegraphics[width=2.25in]{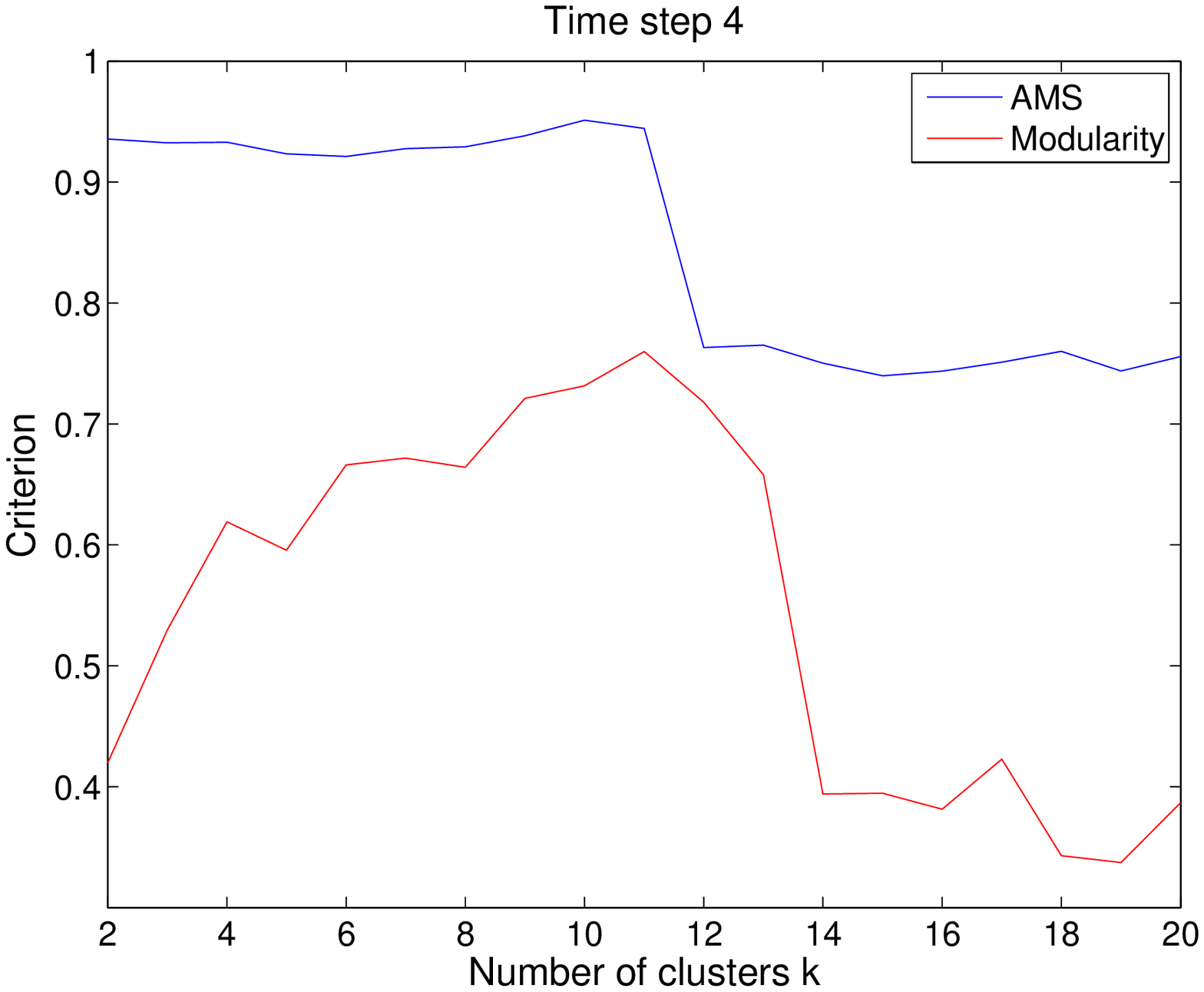}\\
\multicolumn{2}{c}{\includegraphics[width=2.25in]{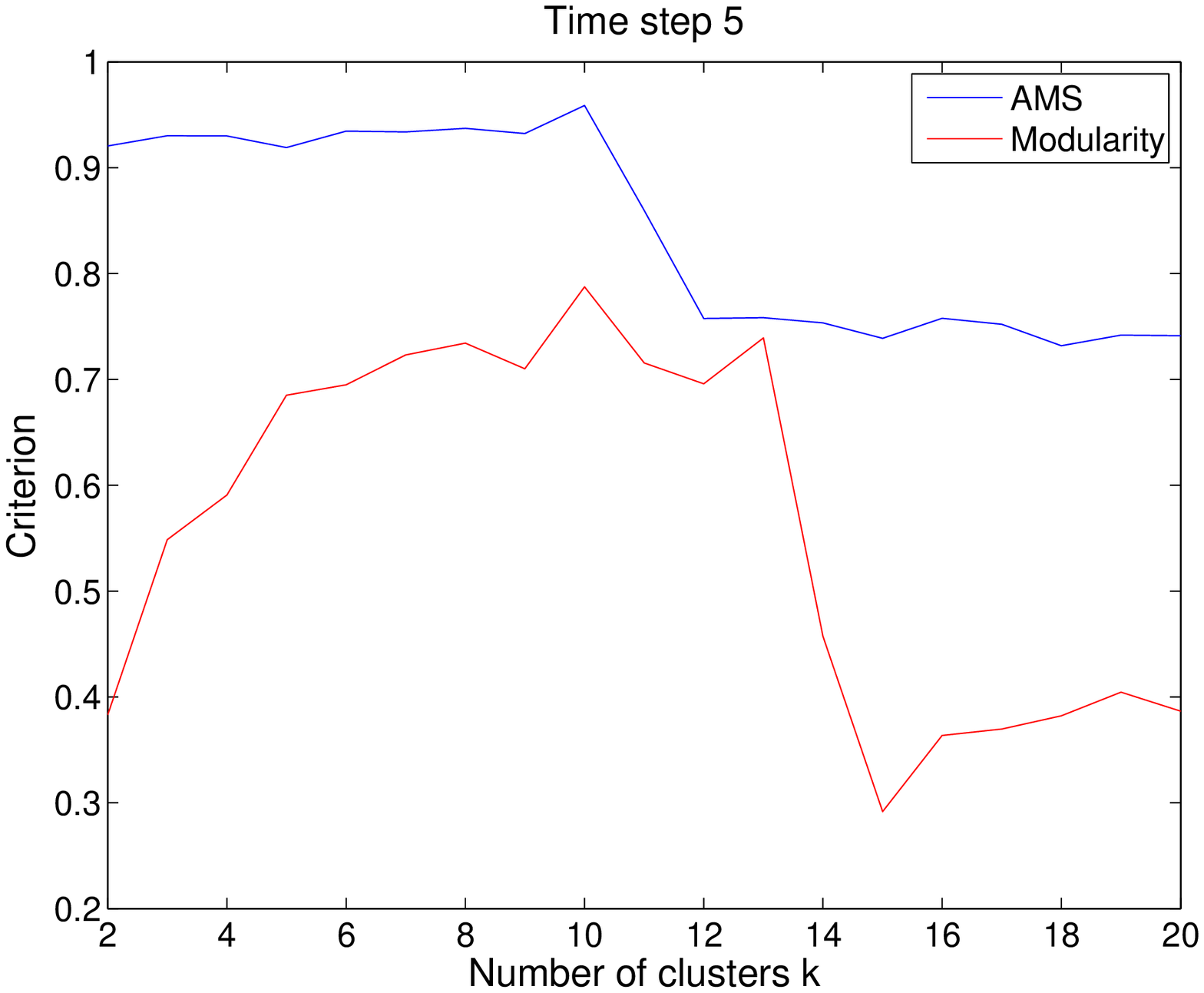}}
\end{tabular}
\caption{\textbf{BirthdeathNet: selection of the number of clusters}. The true number of clusters over time is $k=13$ in every snapshot. Modularity (red) detects $13$, $11$, $11$, $11$, $10$. AMS (blue) selects $10$, $11$, $3$, $10$, $10$: however, if we look carefully at the figure, we can notice some peaks also at $k=13$.}\label{MS_birthdeath}
\end{figure*}

\begin{figure*}[htbp]
\centering
\begin{tabular}{cc}
\includegraphics[width=2.25in]{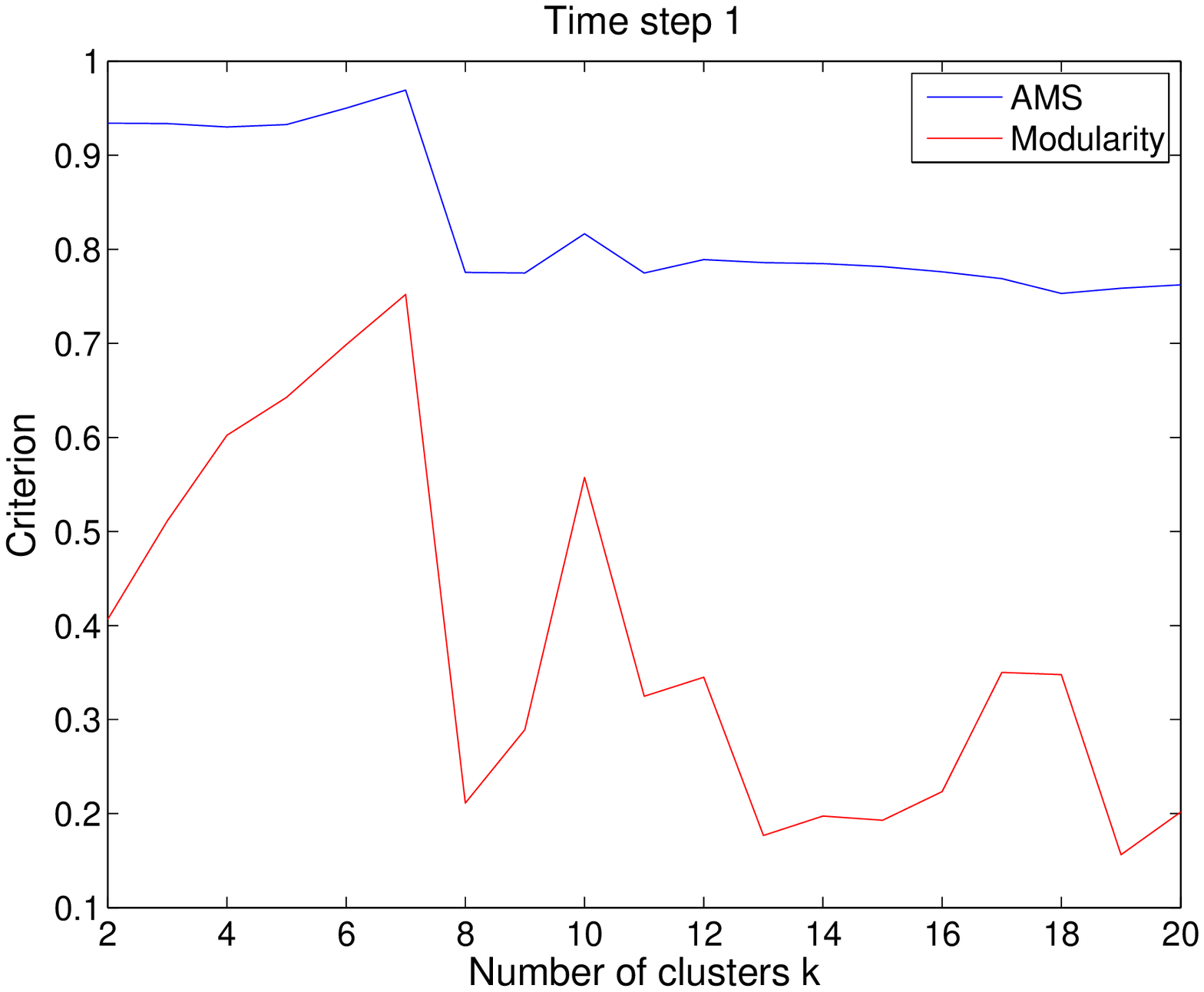}&
\includegraphics[width=2.25in]{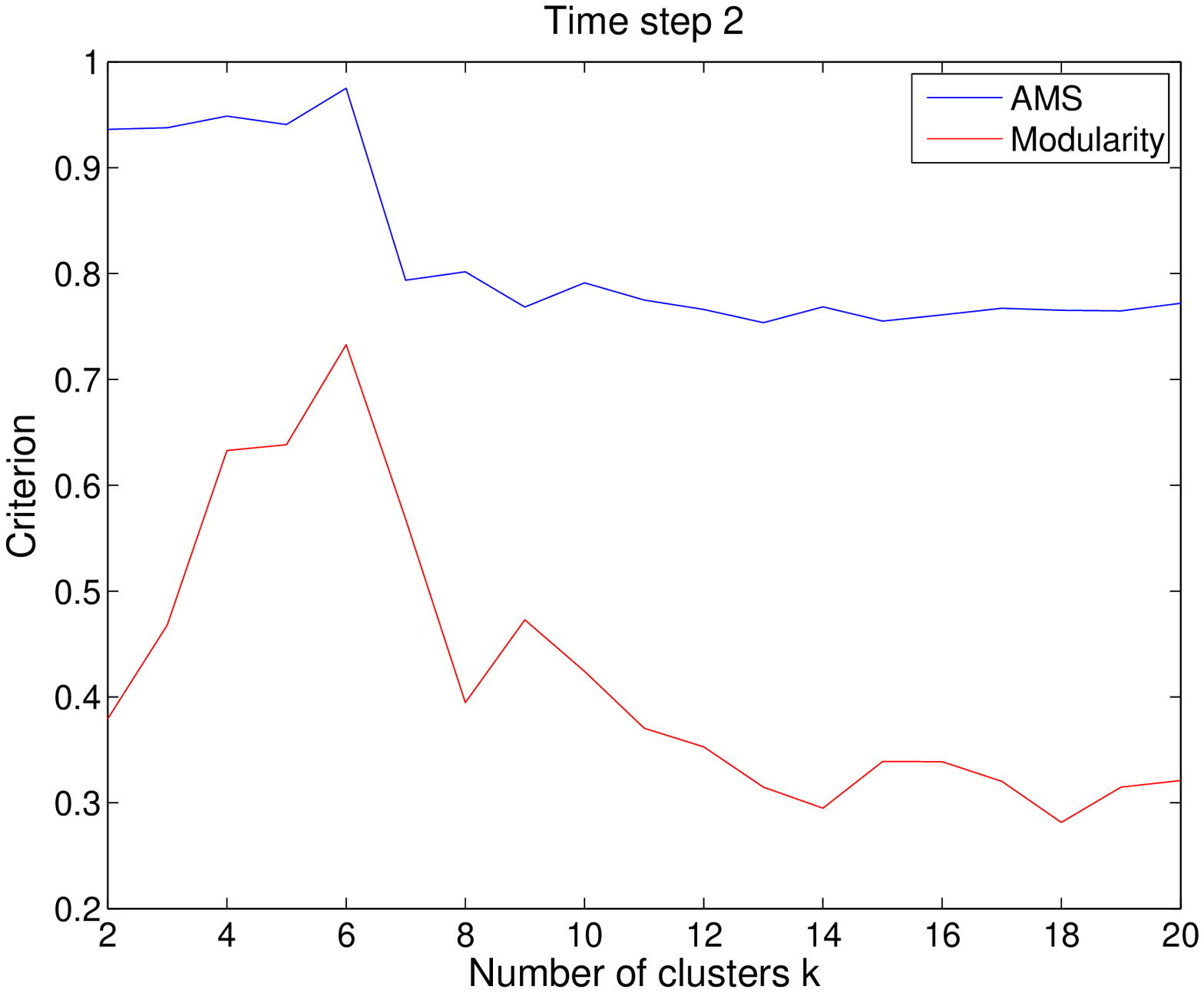}\\
\includegraphics[width=2.25in]{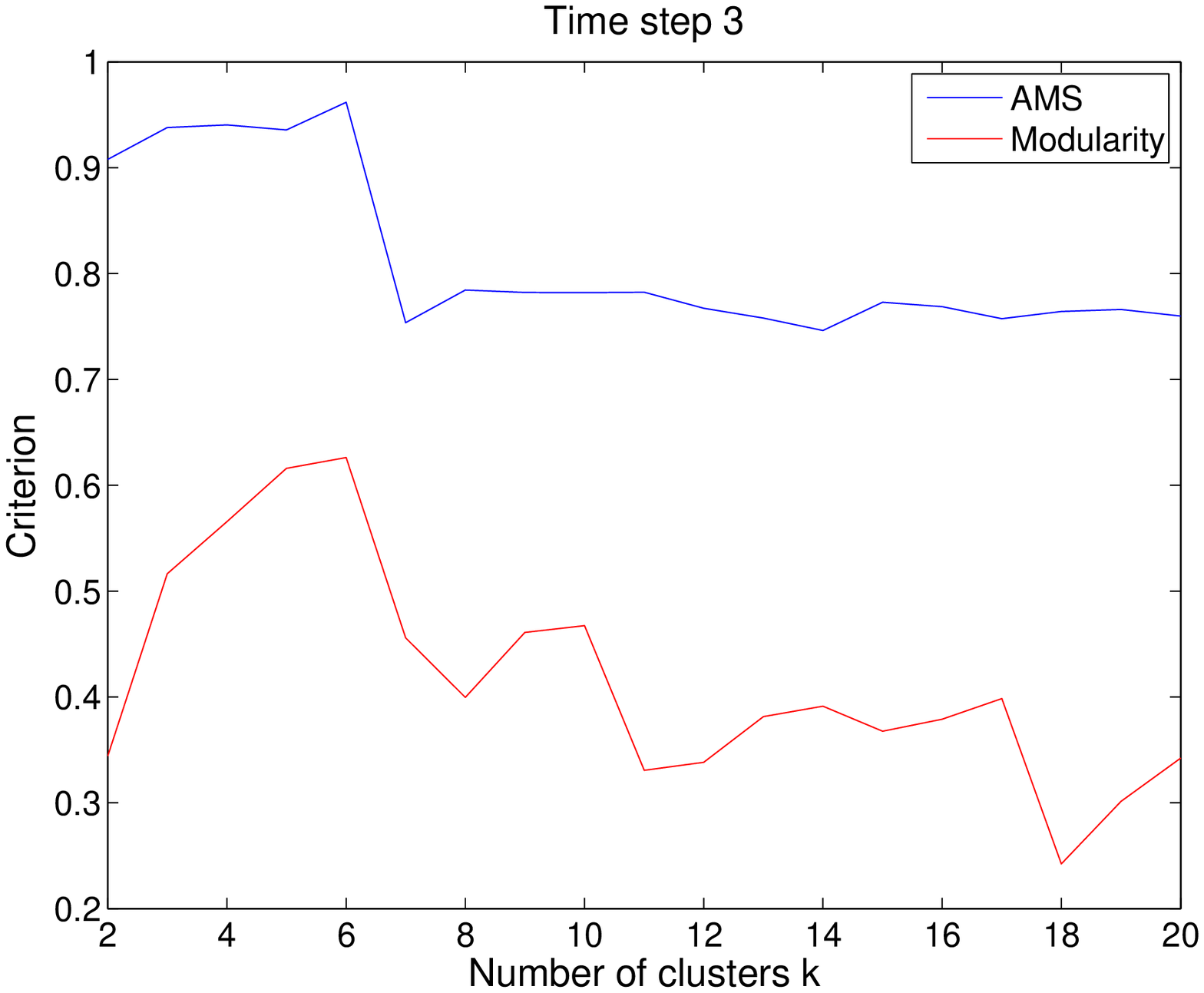}&
\includegraphics[width=2.25in]{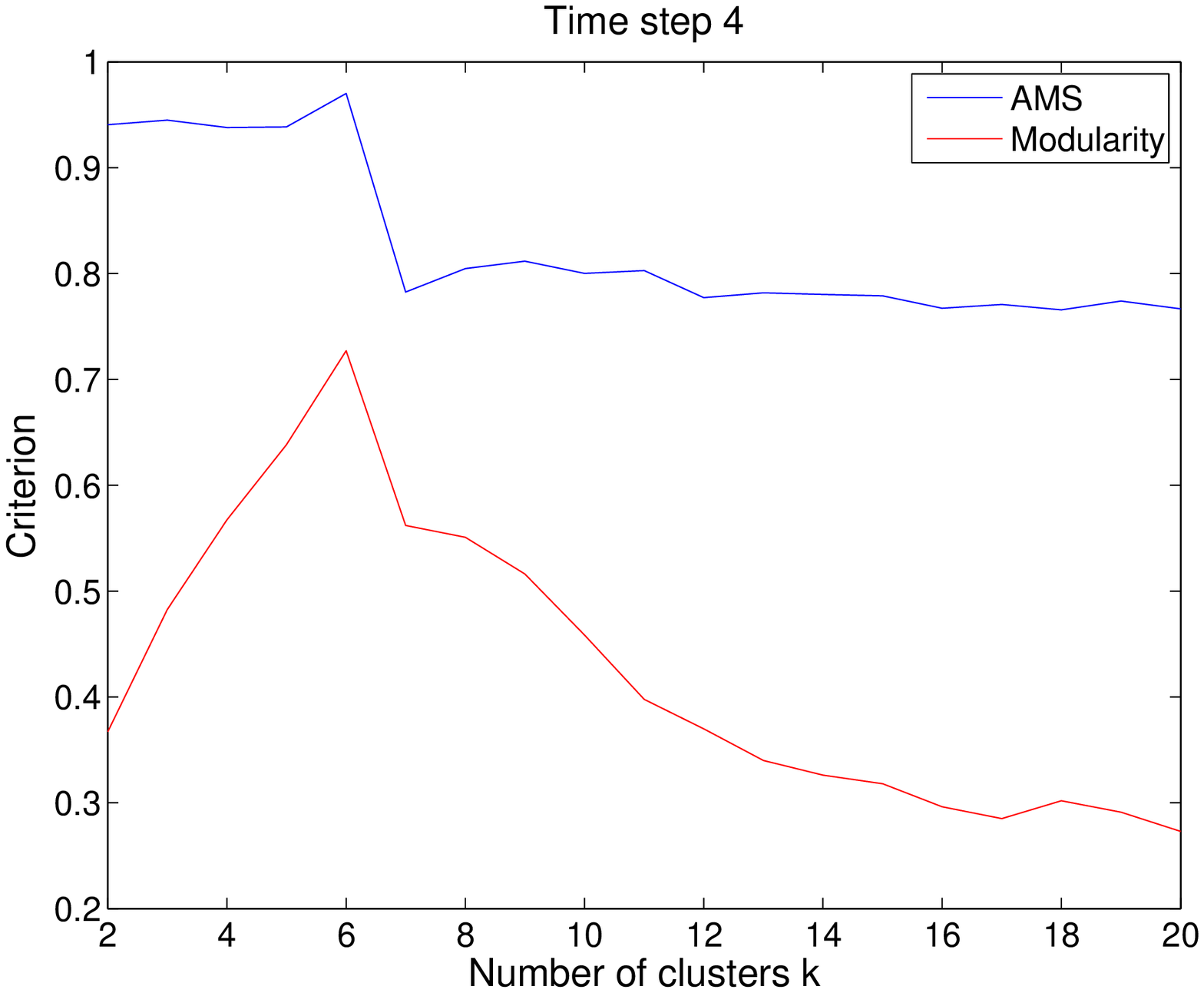}\\
\multicolumn{2}{c}{\includegraphics[width=2.25in]{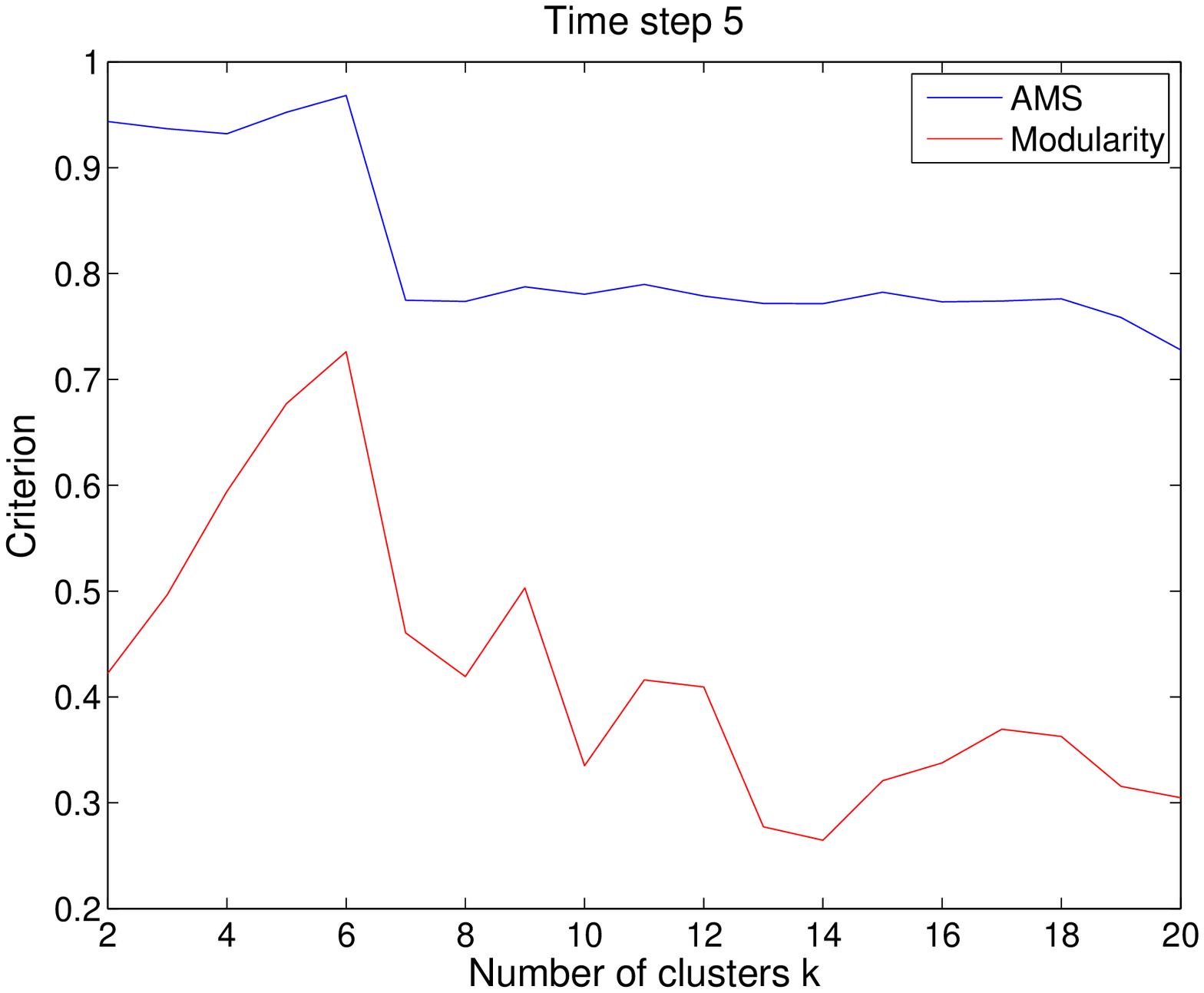}}
\end{tabular}
\caption{\textbf{HideNet: selection of the number of clusters}. The true number of clusters over time is $7$, $6$, $6$, $6$, $6$. Both Modularity and AMS are able to recognize the right sequence.}\label{MS_hide}
\end{figure*}

\begin{figure*}[htbp]
\centering
\begin{tabular}{c}
\includegraphics[width=3in]{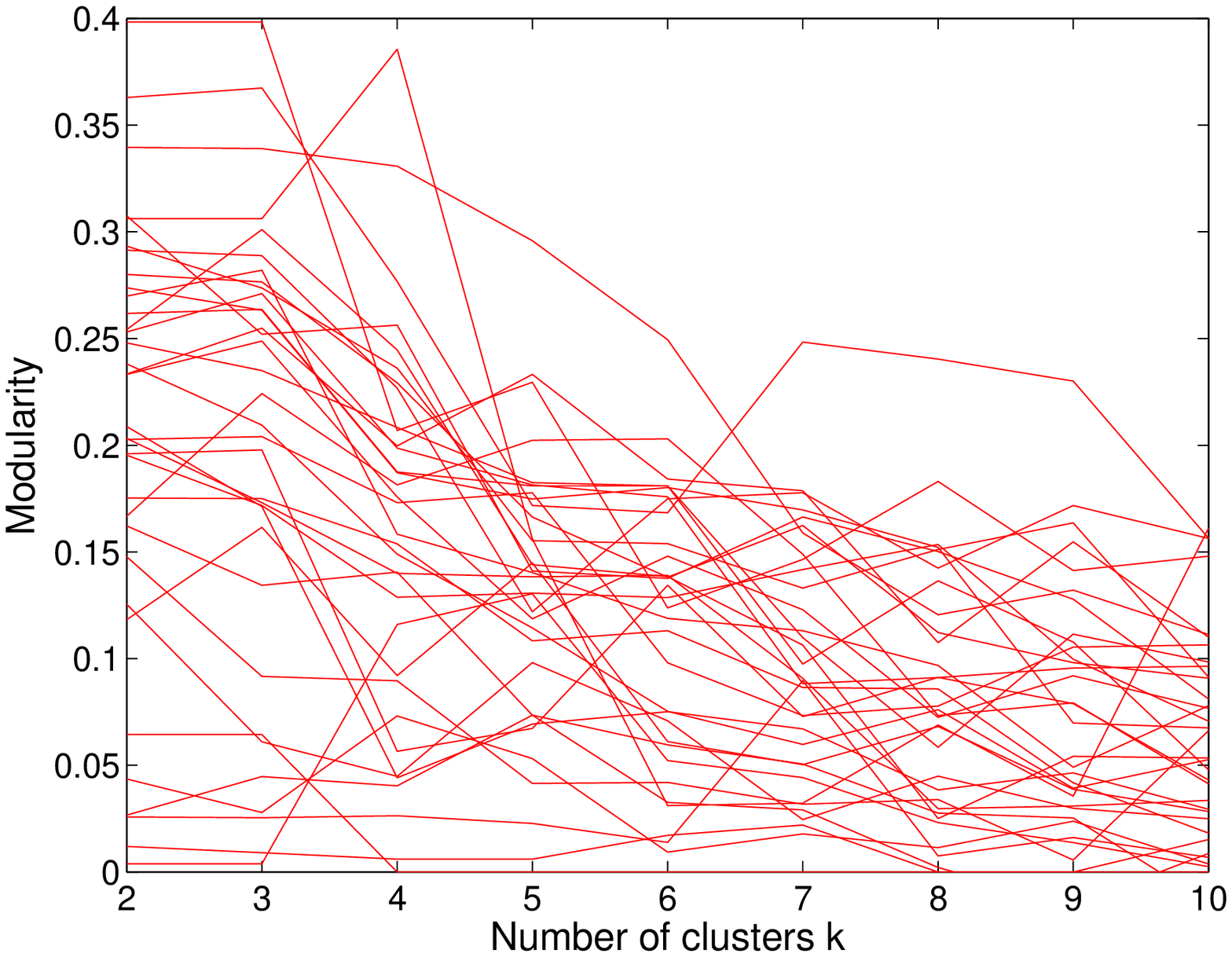}\\
\includegraphics[width=3in]{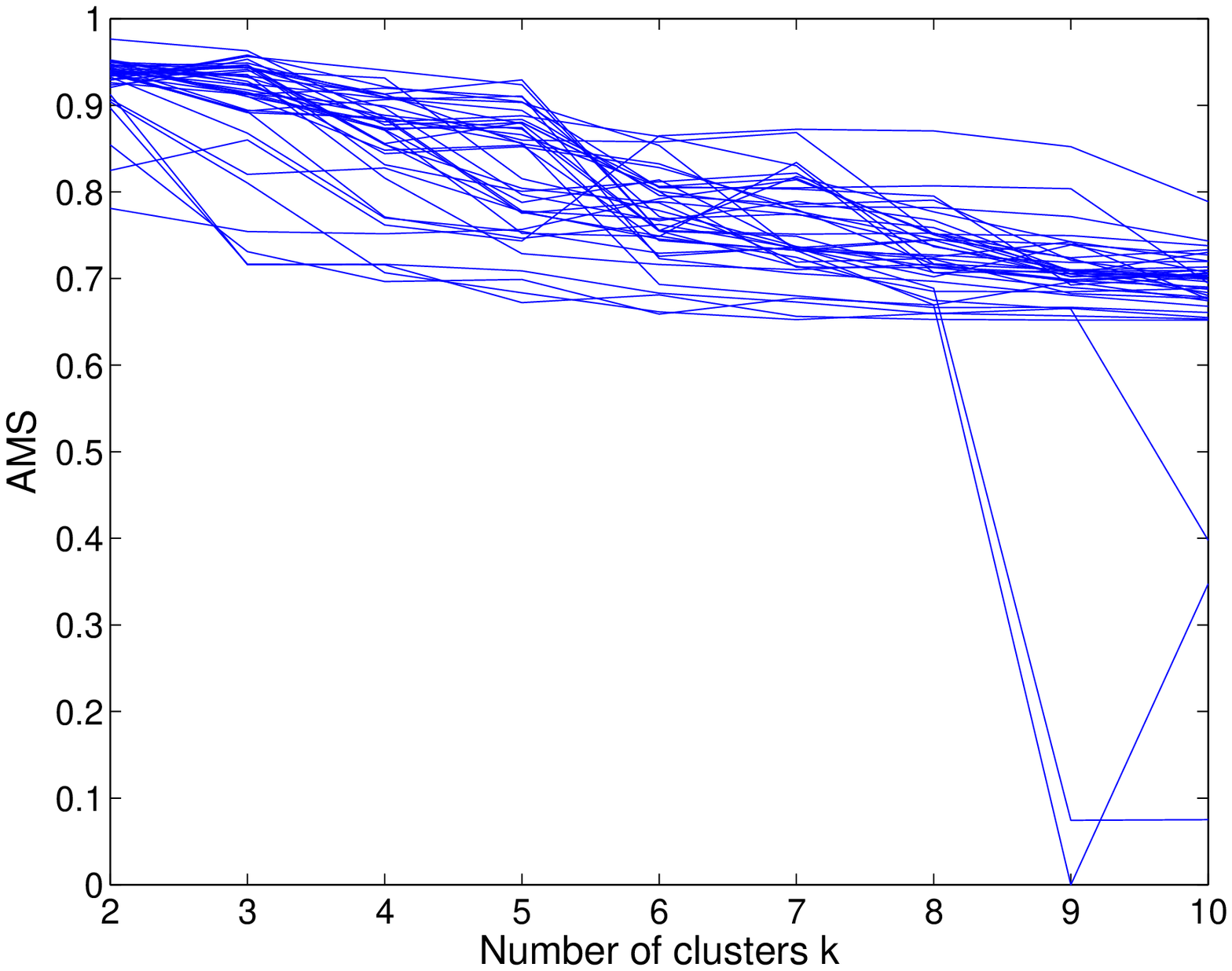}
\end{tabular}
\caption{\textbf{CellphoneNet: selection of the number of clusters}. For this network it has been suggested to consider $k=2$ as the ground-truth for the entire period of $46$ weeks \cite{MIT_realitymining}. This is due to the fact that the people representing the nodes of the network belong to $2$ different departments at MIT. However we have noticed that the network does not have a clear community structure in every snapshot (for instance in some snapshots the maximum value of the Modularity quality function for $k=2$ approaches zero). Moreover, both Modularity (top) and AMS (bottom) do not always select $k=2$: the former mainly detects $k=4,5,6$, and the latter $k=2$ but also $k=3$.}\label{MS_realdata}
\end{figure*}

The results related to the selection of $k$ for the synthetic networks are depicted in Figures \ref{MS_mergesplit}-\ref{MS_hide}. In general Modularity, AMS or both are able to suggest the right number of communities or a partitioning close to the ground truth. The results concerning the \textit{CellphoneNet} are depicted in Figure \ref{MS_realdata}. Only AMS mostly selects $k=2$ along all the time period, in agreement with the ground truth suggested in \cite{MIT_realitymining} and in \cite{graphscope}. As mentioned before, in the cited works it has been proposed to group students and staff at MIT according to their belonging to the Sloan business school or being co-workers in the same building. 

\begin{figure*}[htbp]
\centering
\begin{tabular}{c}
\includegraphics[width=4in]{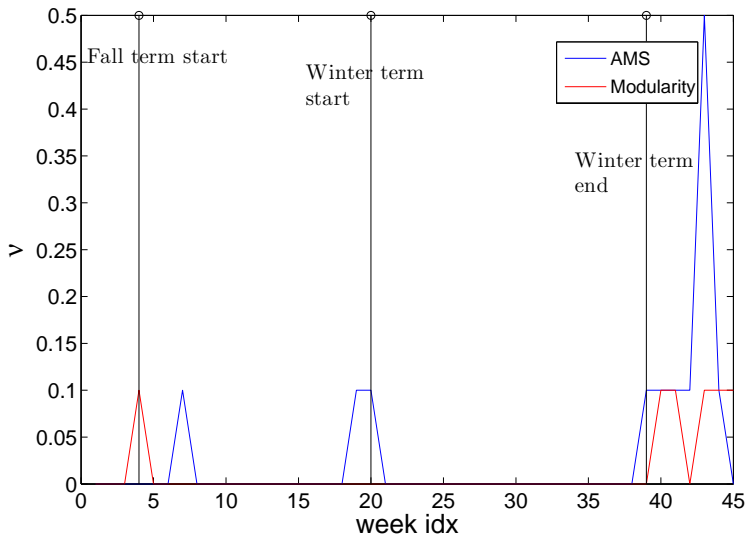}
\end{tabular}
\caption{\textbf{CellphoneNet: selection of the smoothness parameter $\nu$}. The regularization constant $\nu$ selected by AMS (blue) and Modularity (red). In both cases some peaks are present around important dates which are labelled in the plot. }\label{fig_nu_realdata}
\end{figure*}

Regarding the smoothness parameter $\nu$, the results related to the \textit{CellphoneNet} are plotted in Figure \ref{fig_nu_realdata}. We can notice how the regularization constant has some small peaks around important dates like beginning of fall and winter term and end of winter term. Thus, it seems that $\nu$ is behaving as a kind of change indicator measure.  

\subsubsection{Evaluation of the results}
\label{results}
In this Section the clustering results are evaluated according to the Adjusted Rand Index (ARI \cite{arindex}) when the true memberships are available and the smoothed Conductance (Cond\_Mem).

\begin{table*}[htbp]
%\scriptsize{
\begin{center}
\begin{tabular}{|c|c|c|} 
\hline
\multicolumn{3}{|c|}{\hspace{2 mm}\textbf{MergeplitNet}\hspace{2 mm}}\\
\hline 
\textbf{MEASURE}  & \textbf{MKSC} &  \textbf{AFFECT \cite{adapt_evol_clu_Xu}}\\
\hline
\textbf{ARI} & $\mathbf{0.90\pm 0.03}$ (MOD) & $0.73 \pm 0.01$ (MOD) \\
\hline
\textbf{Cond\_Mem} & $0.0112 \pm 0.0001$ (AMS) & $\mathbf{0.0038 \pm 0.0005}$ (SIL) \\
\hline
\multicolumn{3}{|c|}{\hspace{2 mm}\textbf{BirthdeathNet}\hspace{2 mm}}\\
\hline 
\textbf{MEASURE}  & \textbf{MKSC} &  \textbf{AFFECT \cite{adapt_evol_clu_Xu}}\\
\hline
\textbf{ARI} & $\mathbf{0.80\pm 0.02}$ (MOD) & $0.76 \pm 0.03$ (MOD) \\
\hline
\textbf{Cond\_Mem} & $\mathbf{0.036 \pm 0.003}$ (AMS) & $0.052 \pm 0.002$ (MOD) \\
\hline
\multicolumn{3}{|c|}{\hspace{2 mm}\textbf{HideNet}\hspace{2 mm}}\\
\hline 
\textbf{MEASURE}  & \textbf{MKSC} &  \textbf{AFFECT \cite{adapt_evol_clu_Xu}}\\
\hline
\textbf{ARI} & $\mathbf{0.97\pm 0.01}$ (AMS, MOD) & $0.85 \pm 0.03$ (MOD) \\
\hline
\textbf{Cond\_Mem} & $0.011 \pm 0.001$ (AMS, MOD) & $\mathbf{0.005 \pm 0.001}$ (SIL) \\
\hline
\end{tabular}
\end{center}
\caption{\textbf{Clustering results synthetic datasets}. Both ARI and Cond\_Mem values represent an average over time, i.e. $5$ snapshots. Moreover, for each snapshot the smoothed Conductance is obtained by taking the mean over $5$ possible values of the parameter $\eta$ in the range $[0,1]$. Between parenthesis we indicate with which model selection criterion the related result has been obtained. Although the model selection problem is out of the scope of \cite{adapt_evol_clu_Xu}, the authors provide two possible way of tuning the number of clusters, using Silhouette (SIL) or Modularity (MOD). All the results are statistically significant according to a Student t-test for the comparison of two means, with p-value $\l 0.05$.}
\label{results_synthetic}
%}
\end{table*}

\begin{table*}[htbp]
%\scriptsize{
\begin{center}
\begin{tabular}{|c|c|c|c|} 
\hline
\multicolumn{4}{|c|}{\hspace{2 mm}\textbf{CellphoneNet}\hspace{2 mm}}\\
\hline 
\textbf{MEASURE}  & \textbf{MKSC} &  \textbf{AFFECT \cite{adapt_evol_clu_Xu}} & \textbf{ESC \cite{evol_spec_clustering}}\\
\hline
\textbf{ARI} & $\mathbf{0.861 \pm 0.051}$ (AMS) & $0.763 \pm 0.001$ & $-$\\
\hline
\textbf{RI} & $\mathbf{0.943 \pm 0.040}$ (AMS) & $0.893$ & $0.861$\\
\hline
\textbf{Cond\_Mem} & $\mathbf{0.0035\pm 0.0001}$ (AMS) & $0.0048\pm 0.0001$ & $-$ \\
\hline
\end{tabular}
\end{center}
\caption{\textit{Clustering results CellphoneNet}. We compare MKSC with AFFECT and ESC by reporting the results shown in Tables 5 of \cite{adapt_evol_clu_Xu}. Regarding ESC, the mean between the best results related to the PCQ and PCM frameworks are considered.  Moreover, for AFFECT and ESC the number of clusters has been fixed to $k=2$, while it is fine tuned in case of MKSC. The best performer is MKSC, according to a Student t-test for the comparison of two means with significance level of $0.05$ (p-value $\l 0.05$).}
\label{results_real}
%}
\end{table*}

The MKSC algorithm is compared with Adaptive Evolutionary Clustering (AFFECT \cite{adapt_evol_clu_Xu}) in case of synthetic data and AFFECT and ESC (Evolutionary Spectral Clustering \cite{evol_spec_clustering}) for \textit{CellphoneNet}. In all the datasets  MKSC produces clustering results closer to the ground truth memberships (higher ARI), as it is shown in Tables \ref{results_synthetic} and \ref{results_real}. Moreover, the differences between the results of the compared methods are all statistically significant, according to a Student t-test for the comparison of two means with significance level of $0.05$. In Figure \ref{MS_mergesplit} an example of the partition obtained by MKSC for the \textit{mergesplitNet} network is shown.

\subsection{Visualizing the clusters evolution}
\label{visualization_3D}
In this Section we show the results obtained by our tracking mechanism summarized in algorithm \ref{algo:algo1} on the networks described in Section \ref{data_sets_f2}. We use the first $3$ dual solution vectors of problem (\ref{dualMKSC}), i.e. $\alpha^{(1)}$, $\alpha^{(2)}$, $\alpha^{(3)}$, to visualize the clusters evolution in $3$D. In order to explicitly show the growth and shrinkage events we plot the clusters as spheres, centered around the mean of all the points in that cluster. The radius is equivalent to the fraction of points belonging to that cluster at that time stamp. Each sphere is given a unique colour at time stamp $t=1$. As the clusters grow or shrink the size of the sphere changes. In case of a split the colour and label of that cluster is transferred to all the clusters obtained as a result of the split. In case of a merge we assign the average colour of the clusters which merge together at time $t$ to the new cluster at time $t+1$. In case of birth of a new cluster we allocate it a new colour and all the nodes which have disappeared at time interval $t$ are depicted as a blue-coloured sphere centered at the origin (dump).

\begin{figure*}[!h]
	%\begin{subfigure}{\textwidth}
    \begin{minipage}{\textwidth}		
	{
		\centering
		\includegraphics[width=\textwidth]{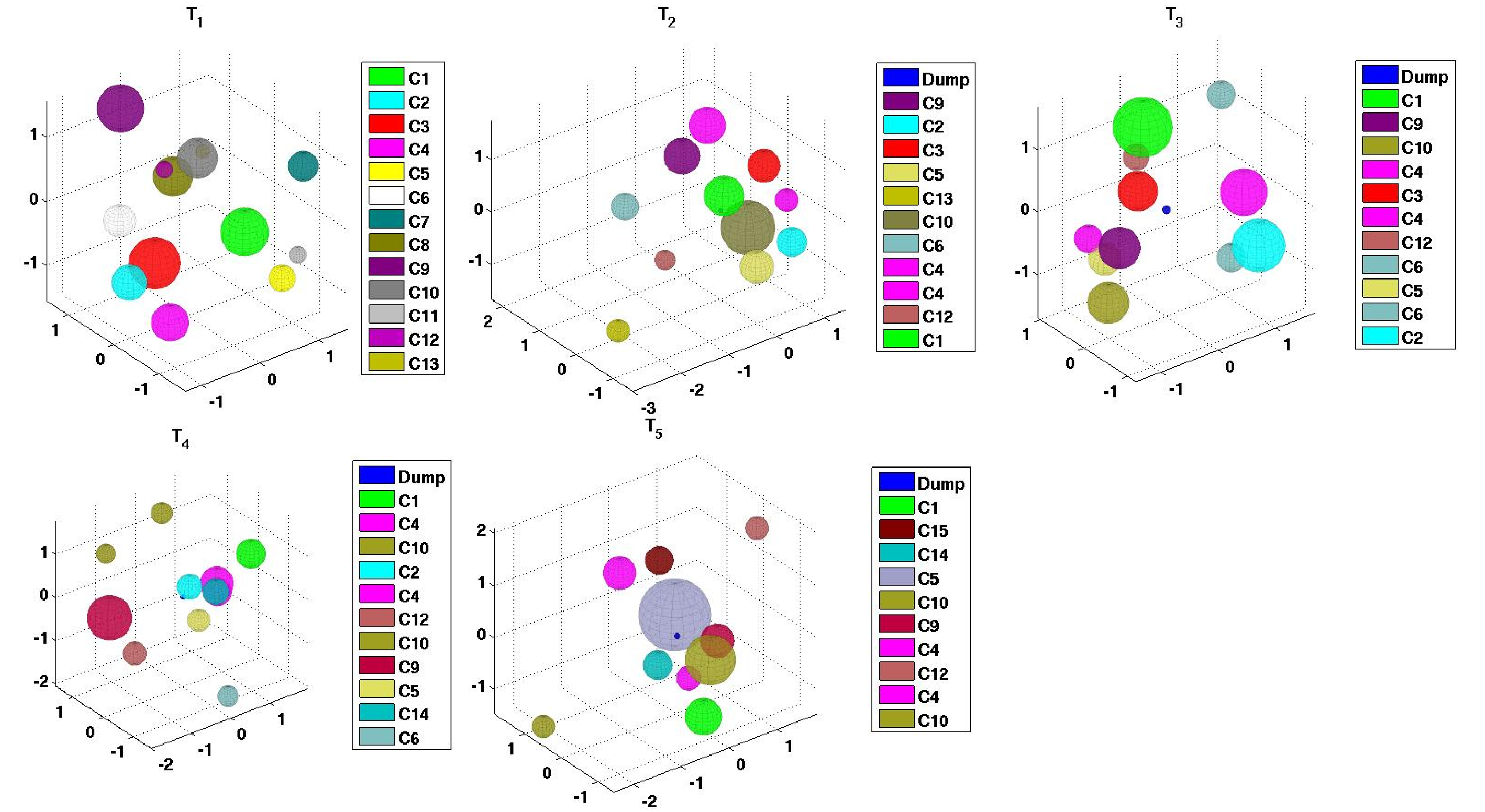}
		\caption{\scriptsize{The \textit{BirthdeathNet} dataset has $13$ clusters at time stamp $T_{1}$. At time stamp $T_{2}$ cluster $C4$ splits into $2$ clusters, clusters $C6$ and $C7$ merge to cluster $C7$, clusters $C8$ and $C10$ merge to cluster $C10$ and $C12$ and $C13$ merge to cluster $C12$. At time stamp $T_{3}$ cluster $C10$ and $C13$ merge to cluster $C10$, cluster $C6$ splits into $2$ clusters. At time stamp $T_{4}$, a new cluster $C14$ appears, cluster $C10$ splits into $2$ clusters, cluster $C3$ and $C9$ merge to cluster $C9$ and one of the $2$ splits of cluster $C6$ dissolves. At the final time interval cluster $C2$ dissolves, cluster $C15$ appears, cluster $C2$ and $C5$ merge to cluster $C5$. At each time step except $T_{1}$ the fraction of nodes which disappeared is represented by blue-coloured sphere. The size of the sphere show the growth of shrinkage of clusters over time.}}
		\label{fig:subfig1}
	}
	%\end{subfigure}
	\end{minipage}
	%\begin{subfigure}{\textwidth}
	\begin{minipage}{\textwidth}
	{
		\centering	
		\includegraphics[width=\textwidth]{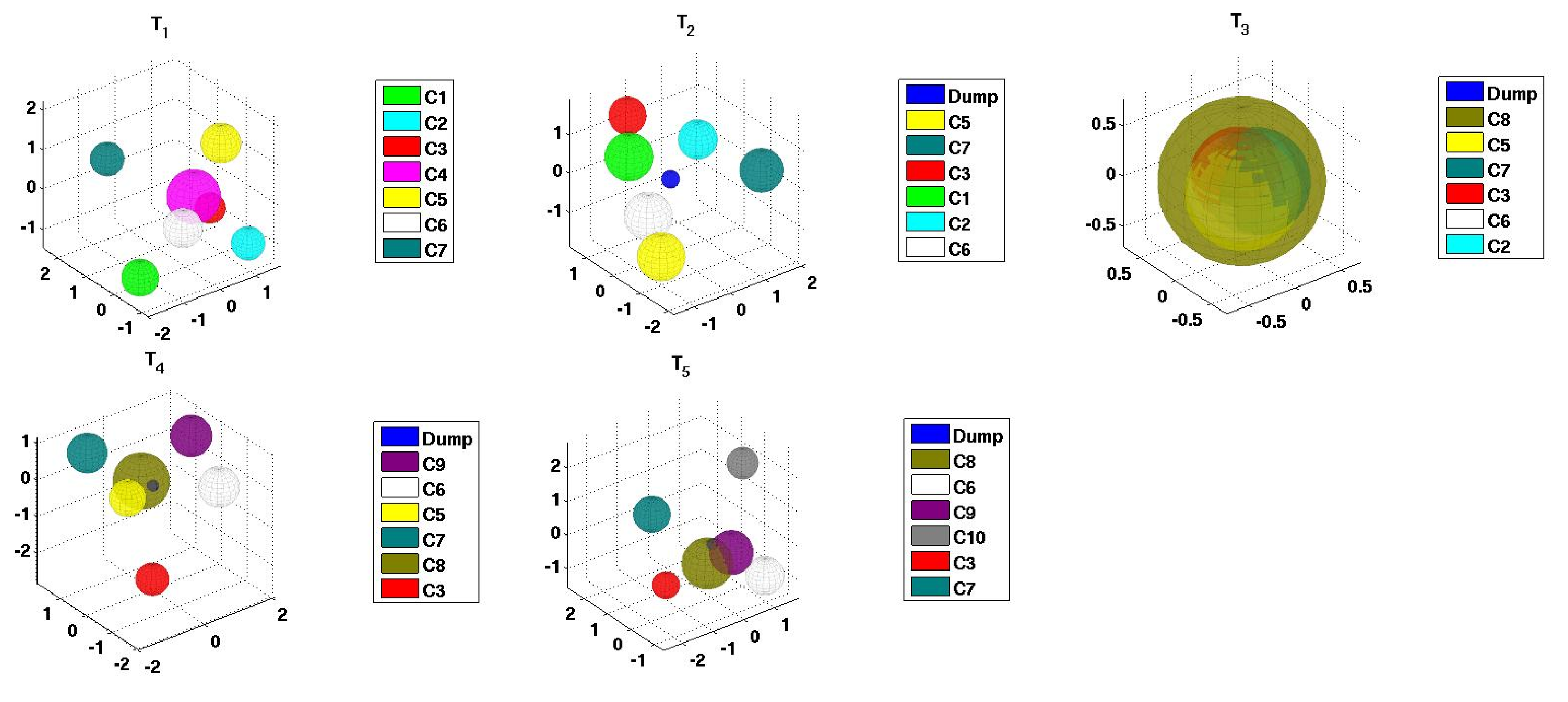}
		\caption{\scriptsize{The \textit{HideNet} dataset has $7$ clusters at time stamp $T_{1}$. Cluster $C4$ dissolves at time stamp $T_{2}$. Cluster $C8$ is born at time stamp $T_{3}$ and cluster $C1$ fades. Cluster $C2$ dies and a new cluster $C9$ is created at time stamp $T_{4}$. In the final phase, a new cluster $C10$ is generated and cluster $C5$ is dumped. The size and position of the spheres change at each time interval showing the growth and shrinkage of the clusters.}}
		\label{fig:subfig2}
	}
	%\end{subfigure}
	\end{minipage}
	\caption{\footnotesize{\textbf{Evolution of clusters found by the MKSC algorithm for \textit{BirthdeathNet} and \textit{HideNet} datasets.}}}
\end{figure*}

\begin{figure*}
	\begin{minipage}{\textwidth}
	{
		\centering
		\includegraphics[width=\textwidth]{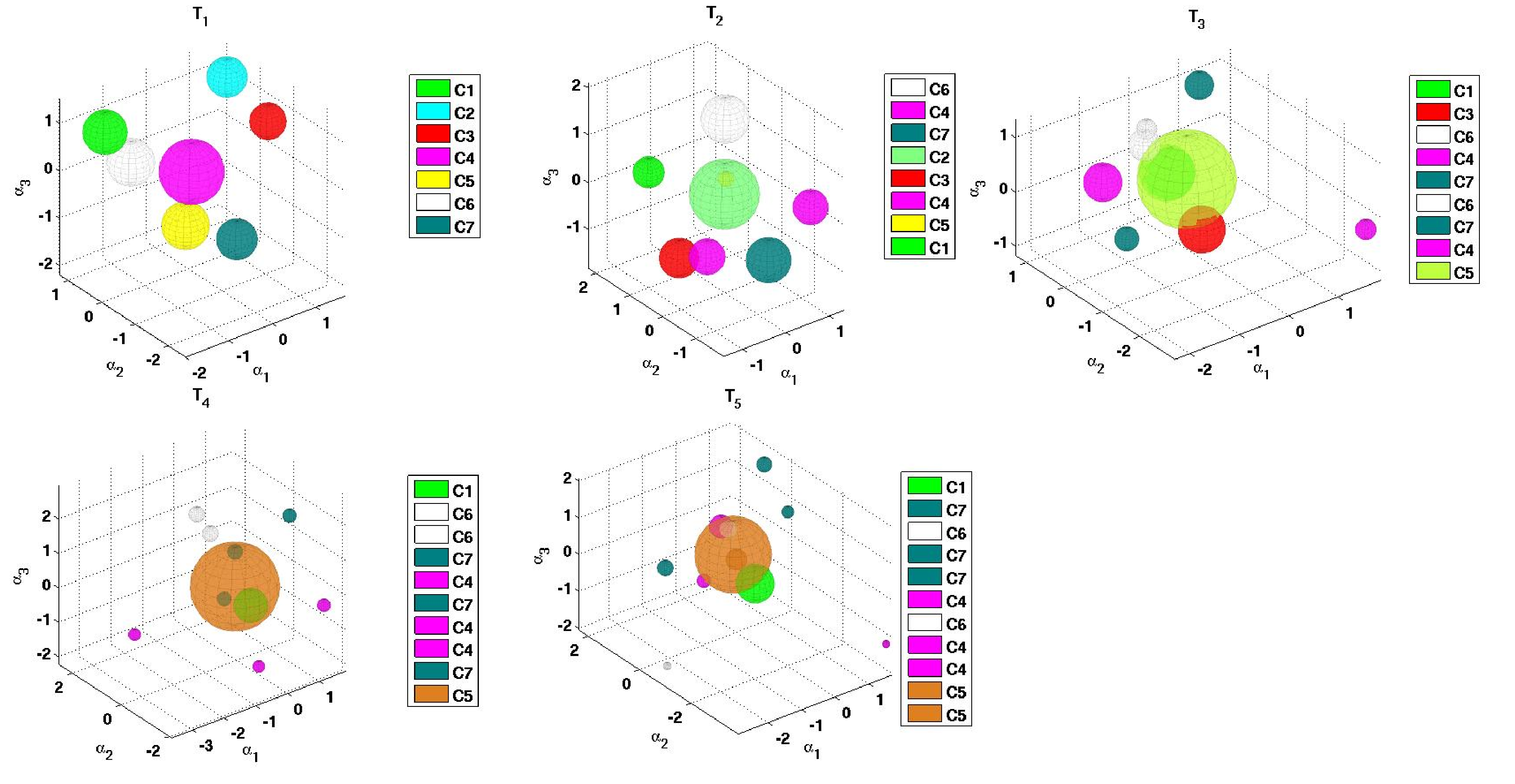}
		\caption{The \textit{MergesplitNet} dataset has $7$ clusters at time stamp $T_{1}$. At time stamp $T_{2}$ cluster $C4$ splits into $2$ clusters, major part of cluster $C5$ merges with cluster $C2$. At time stamp $T_{3}$ cluster $C6$, cluster $C7$ splits into $2$ clusters and clusters $C2$ and $C5$ merge to cluster $C5$. At time stamp $T_{4}$ clusters $C4$ and $C7$ further split to have $3$ clusters each. Cluster $C3$ combines with cluster $C5$. In the final time interval cluster $C5$ splits into $2$ clusters.}
		\label{fig:subfig3}
	}
	\end{minipage}
	\begin{minipage}{\textwidth}
	{
		\centering
		\includegraphics[width=\textwidth]{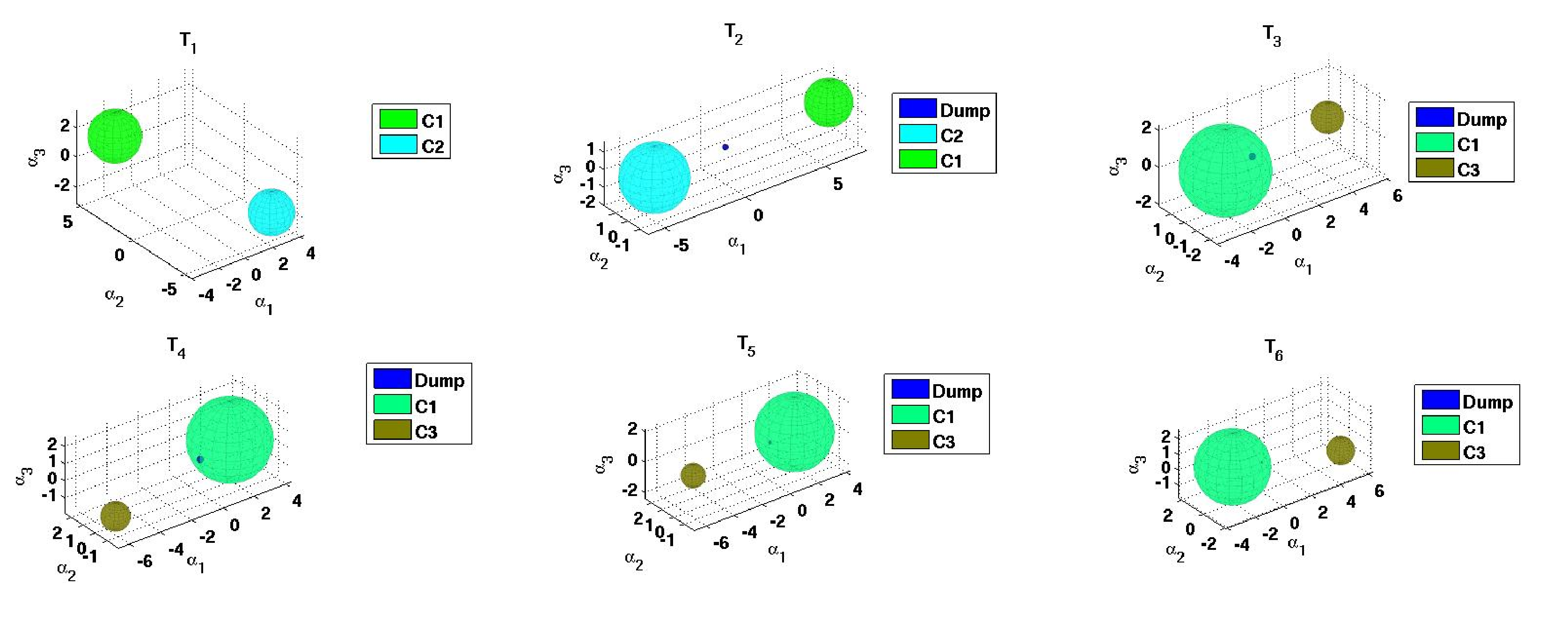}
		\caption{The \textit{CellphoneNet} dataset has $2$ clusters at time stamp $T_{1}$. Some nodes leave the network at time stamp $T_{2}$. At time stamp $T_{3}$ one part of cluster $C2$ is dumped, another part merges with a new cluster $C3$. After this the clusters remain more of less constant till interval $T_{6}$. For the complete evolution of the communities along all the time steps see the video present in the supplementary material.}
		\label{fig:subfig4}
	}
	\end{minipage}
	\caption{\textbf{Visualization of the communities found over time by the MKSC algorithm for the \textit{MergesplitNet} and \textit{CellphoneNet} evolving networks.}}
\end{figure*}

\begin{figure*}[htbp]
\centering
\begin{tabular}{c}
\includegraphics[width=4.75in]{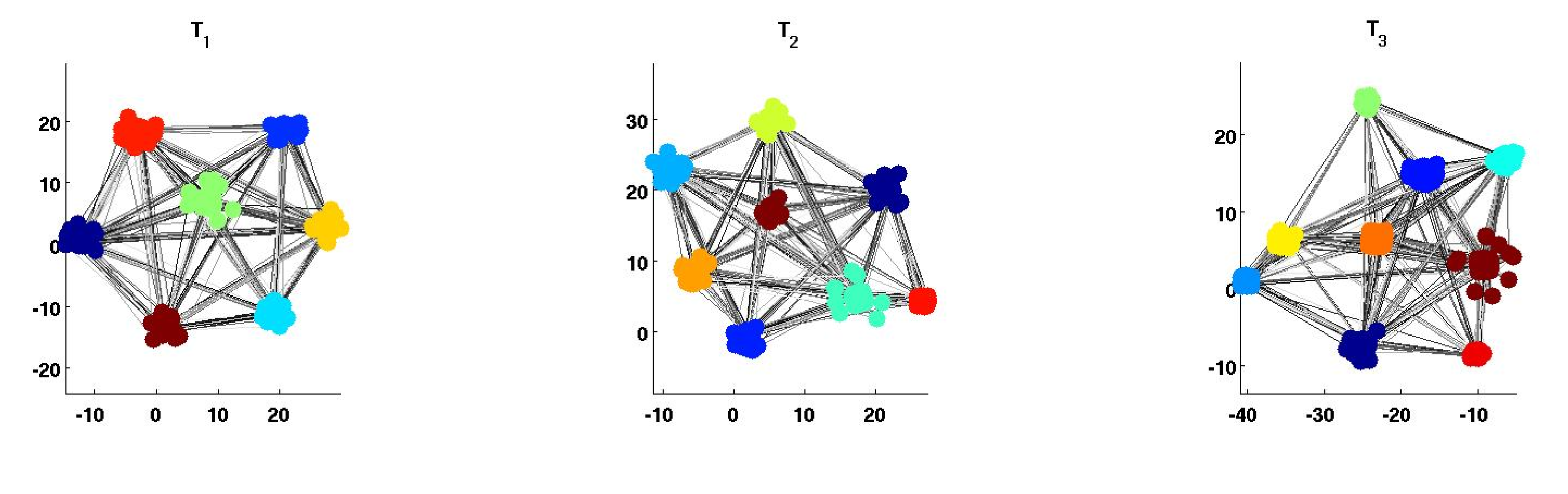}\\
\includegraphics[width=4.75in]{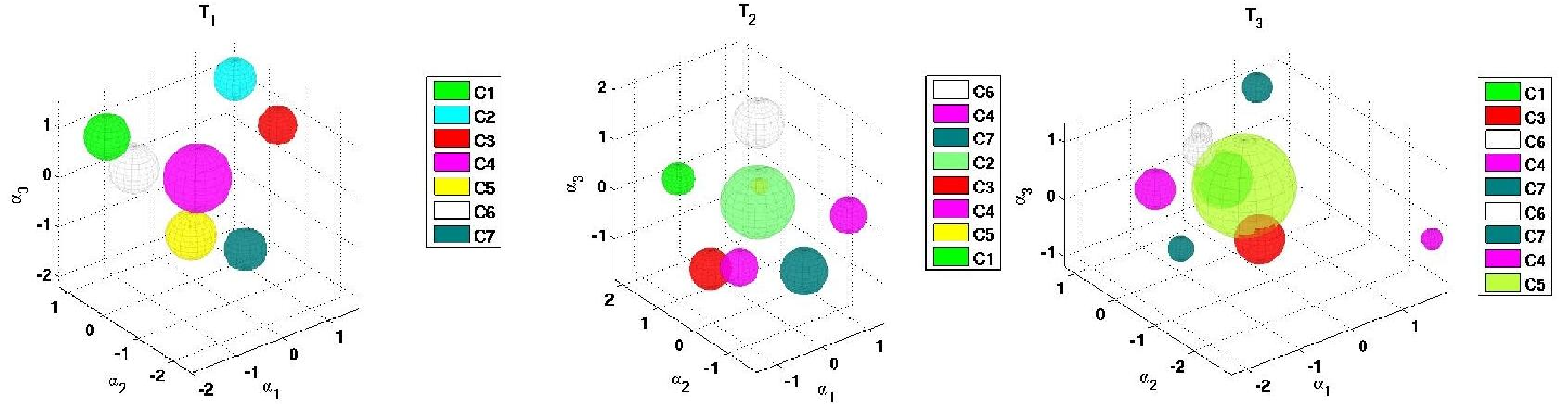}\\
\includegraphics[width=4.75in]{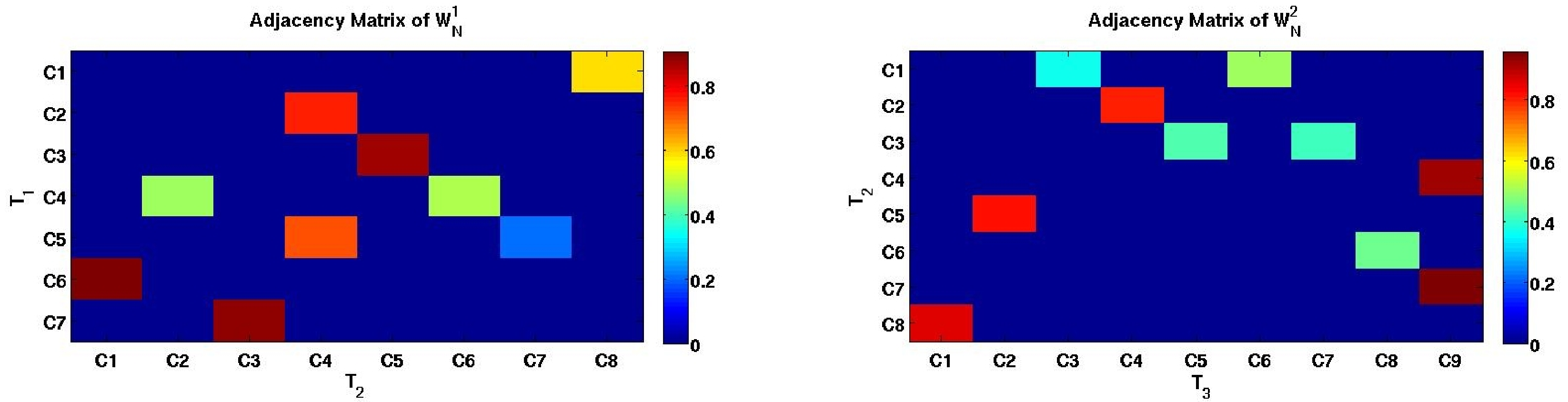}
%\multicolumn{2}{c}{\includegraphics[width=2in]{MS_hide_t5.eps}}
\end{tabular}
\caption{\textbf{MergesplitNet: three possible ways of illustrating the network evolution}. \textbf{(Top)} standard plot of nodes and edges \textbf{(Center)} 3D plot in the $\alpha^{(l)}$ space \textbf{(Bottom)} Affinity matrix of the network $W_N^{t}$ constructed by our tracking mechanism. For simplicity, only the first three time steps are considered.}\label{vis_mergesplit}
\end{figure*}

The visualization of the clusters is shown in Figures \ref{fig:subfig1}-\ref{fig:subfig4}. For all the synthetic networks (although not perfectly) we are able to grasp the main events occurring at each time step. For instance, in case of the \textit{MergesplitNet}, we can appreciate a certain change in the data at time $t=T_4$. In fact, at this time stamp, the tuning algorithm \ref{alg:alg1} selected $\nu = 0.1$, while in the other time steps  $\nu = 0$. Thus, the memory effect got activated to smooth the clustering results. Regarding the \textit{CellphoneNet}, thanks to the proposed visualization tool, it is possible for the user to have an idea of the clusters evolution discovered by the MKSC model. Finally, concerning the \textit{MergesplitNet} dataset, in figure \ref{vis_mergesplit} together with the classical representation of the community structure of a network (top) and the proposed 3D illustration in the $\alpha^{(l)}$ space (center), we show another possible kind of visualization (bottom). The latter depicts the affinity matrix of the network $W_N^{t}$ used by our tracking scheme across time.   

\section{Flexibility of MKSC}
For the analysis of the networks considered in the previous sections we used Modularity or its smoothed version in order to select the optimal parameters for KSC and MKSC. As explained in the appendix, Modularity suffers from some drawbacks like the resolution limit and the high number of local optima \cite{lambiotte_multiscale}. These limitations, however, do not represent an issue in our framework for several reasons:
\begin{itemize}
\item the MKSC model described in equation (\ref{primalMKSC}) is quite general (Modularity is not explicitly optimized as in other algorithms like for example \cite{louvainmethod}, \cite{CNM})
\item Modularity (and its smoothed version) has been used only at the model selection level. However, our framework is quite flexible and allows to plug-in during the validation phase any other quality measure      
\end{itemize}
In order to better understand these issues we present a further analysis based on the Conductance (see appendix). Moreover, in the same way as for BLF, Modularity and AMS we can define the smoothed Conductance $\textrm{Cond}_{\textrm{Mem}}$ of the partition related to the actual snapshot $\mathcal{G}_t$ of an evolving network as:
\begin{equation}
\label{dyn_cond}
\textrm{Cond}_{\textrm{Mem}}(X_{\alpha},\mathcal{G}_t) = \eta \textrm{Cond}(X_{\alpha} ,\mathcal{G}_t) + (1 - \eta)\textrm{Cond}(X_{\alpha} ,\mathcal{G}_{t-1}).
\end{equation}
In table \ref{table_cond} we show the mean smoothed Conductance over-time for the $3$ networks under investigation related to the partitions found by MKSC, ESC, KSC and Louvain method (LOUV) \cite{louvainmethod}. The Louvain method is based on a greedy optimization of Modularity and has a runtime that increases linearly with the number of nodes. From the table we can draw the following considerations:
\begin{itemize}
\item the methods with temporal smoothness (MKSC and ESC) achieve an equal or better score than the static methods (KSC and LOUV)        
\item the Louvain method gives the worst results in terms of Conductance. This is not surprising since it is biased toward partitions maximizing the Modularity, which may not be good in terms of Conductance. On the other hand, as already pointed out, MKSC does not suffer of this drawback since Modularity is used only at the validation level. Moreover for model selection every quality function could in principle be used. It is a user-dependent choice. 
\end{itemize}

\begin{table}[htpb]
%\begin{scriptsize}
\begin{center}
\begin{tabular}{|c|c|} 
\hline 
\textbf{SwitchingNet}  & $\mathbf{Cond_{\textrm{Mem}}}$\\
\hline
\textbf{MKSC} & $\mathbf{0.0020}$\\
\hline
\textbf{ESC} & $0.0022$\\
\hline
\textbf{KSC} & $0.0022$\\
\hline
\textbf{LOUV} & $0.0024$\\
\hline
\end{tabular}\\
\begin{tabular}{|c|c|} 
\hline 
\textbf{\hspace{5 mm}ECNet\hspace{5 mm}}  & $\mathbf{Cond_{\textrm{Mem}}}$\\
\hline
\textbf{MKSC} & $\mathbf{0.0050}$\\
\hline
\textbf{ESC} & $0.0051$\\
\hline
\textbf{KSC} & $0.0051$\\
\hline
\textbf{LOUV} & $0.0184$\\
\hline
\end{tabular}\\
\begin{tabular}{|c|c|} 
\hline 
\textbf{\hspace{-0.5 mm}CellphoneNet}  & $\mathbf{Cond_{\textrm{Mem}}}$\\
\hline
\textbf{MKSC} & $\mathbf{0.0019}$\\
\hline
\textbf{ESC} & $0.0042$\\
\hline
\textbf{KSC} & $0.0056$\\
\hline
\textbf{LOUV} & $0.0153$\\
\hline
\end{tabular}
\end{center}
%\end{scriptsize}
\caption{\textbf{Results network data, Conductance}. Average smoothed Conductance over time for some synthetic graphs and the real network considered in this chapter. ESC, KSC, MKSC and the Louvain Method (LOUV) applied separately in each snapshot are compared.}
\label{table_cond}
\end{table}

\section{Conclusions}
In this chapter the problem of community detection in evolving networks has been addressed. We have introduced a novel method named kernel spectral clustering with memory (MKSC), which is designed to incorporate the temporal smoothness of the clustering results over time. Two main frameworks have been devised: in the first one the smoothness is imposed by fixing the regularization constant $\nu = 1$ and the amount of smoothness is tuned by changing the memory $M$. In the second framework the smoothness get activated automatically only when it is required, that is when an important change in the data structure occurs. As a consequence, $\nu$ can also be used as a change detection measure. We have also proposed a tracking mechanism to detect important events characterizing the evolution of the communities over time. Moreover, two visualization tools based on a 3D embedding in the space of the dual solution vectors and on the network constructed by the tracking algorithm, respectively, are devised. On a number of synthetic and real-life networks we have shown how MKSC is able to produce high-quality partitioning in terms of classical measures such as Modularity and Conductance, and their smoothed version. Finally, comparisons with state-of-the-art methods like evolutionary spectral clustering (ESC), adaptive evolutionary clustering (AFFECT) and the Louvain method (LOUV) have shown that MKSC has a competitive performance.                        

%%%%%%%%%%%%%%%%%%%%%%%%%%%%%%%%%%%%%%%%%%%%%%%%%%
% Keep the following \cleardoublepage at the end of this file, 
% otherwise \includeonly includes empty pages.
\cleardoublepage

% vim: tw=70 nocindent expandtab foldmethod=marker foldmarker={{{}{,}{}}}
%

%
  \graphicspath{{\chaptersdir/chapter5/image/}}%
  %\cleardoublepage%
  \chapter{Predicting Maintenance of Industrial Machines}
\label{ch:POMclustering}

This chapter describes an application of KSC to predictive maintenance. In general, an accurate prediction of forthcoming faults in modern industrial machines plays a key role in reducing downtime, increasing the safety of plant operations, and minimizing manufacturing costs. We use clustering on the sensor data collected from a packing machine to recognize in advance when the machine is entering a faulty regime. In this way an early warning can be raised and optimal maintenance actions can be performed. 

In this framework we assume stationarity, in the sense that a clustering model is trained off-line in order to distinguish between normal operating condition and abnormal situations. Then we use the model in an on-line fashion via the out-of-sample extension property to recognize these two regimes. Moreover, moderated outputs that mimic the degradation process affecting the machine are also provided. Overall, this improves the interpretability of the results and allows us to have further insights in the problem at hand.                    

\section{Problem Description}
\label{introduction_POM}
In industrial processes fault detection, isolation and diagnosis ensure product quality and operational safety. Traditionally, four ways to deal with sensory faults have been used \cite{venka2003a},\cite{venka2003b},\cite{venka2003c}: corrective maintenance, preventive maintenance, manual predictive maintenance, and condition-based maintenance. The first type is performed only when the machine fails, it is expensive and safety and environment issues arise. Preventive maintenance is based on periodic replacement of components. The rough estimation of parts lifetime causes a non-optimal use of parts, and possible unexpected failures can still occur (with downtime, safety and environmental consequences). In predictive maintenance machines are manually checked with expensive monitoring hardware (termography, motor health, bearing health). In this case the components are replaced according to their real status, but the operations are labour intensive and prone to human errors. Condition-based maintenance is receiving increasing attention due to its many advantages. Machines status is automatically collected and centrally analysed, and maintenance is planned based on the results of the analysis. The continuous monitoring of machine parts leads to reliable and accurate lifetime predictions, and maintenance operations can be fully automated and implemented in a cost efficient way.
With the development of information and sensor technology many process variables in a power plant can be sampled, like temperature, pressure, flow rate etc. These measurements give information on the current status of a machine and can be used to predict the faults and plan an optimal maintenance strategy. When a component starts degrading, the related sensor reading shows a deviation from its normal behaviour and this can indicate an incoming failure of the component. So far process models based on the sensor data have been constructed by using exponentially weighted moving average, cumulative sum, principal component analysis (PCA), just to name the most widely used methods \cite{Kourti19953}, \cite{choi2003}. Moreover the problem of discovering the incoming faults can be seen as a special case of outlier detection, since an outlier is an observation which deviates so much from the other observations as to arouse suspicions that it was generated by a different mechanism. In this field supervised, semi-supervised and unsupervised methods are employed \cite{siam_tutorial}.     
In the study that we will discuss in the next sections, we are given sensor data collected from the sealing jaws of a packing machine. Since the data are highly unbalanced for supervised learning, we use clustering for identifying in advance when the machine enters critical conditions.   

\section{Materials and Methods}
\label{data_POM}
\begin{figure}[htbp]
\centering
\includegraphics[width=4.5in]{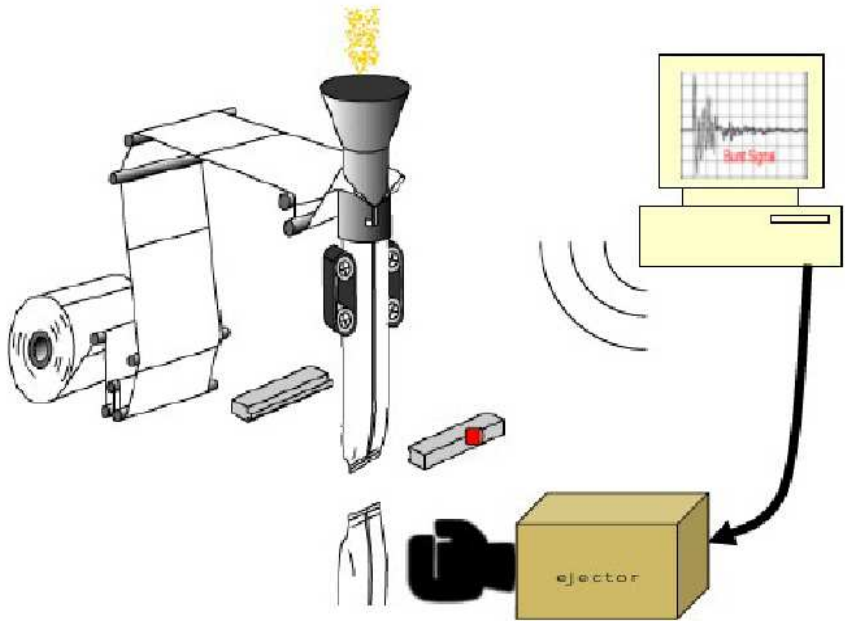}
\caption{\textbf{Vertical Form Fill and Seal (VFFS) machine.} Seal quality monitoring in a packing machine.}
\label{fig:Mebios_machine}
\end{figure}

\begin{figure}[htbp]
\centering
\begin{tabular}{c}
\begin{psfrags}
\psfrag{x}[t][t]{{time}}
\psfrag{y}[b][b]{{Amplitude}}
\psfrag{t}[b][b]{{}}
\includegraphics[width=2.5in]{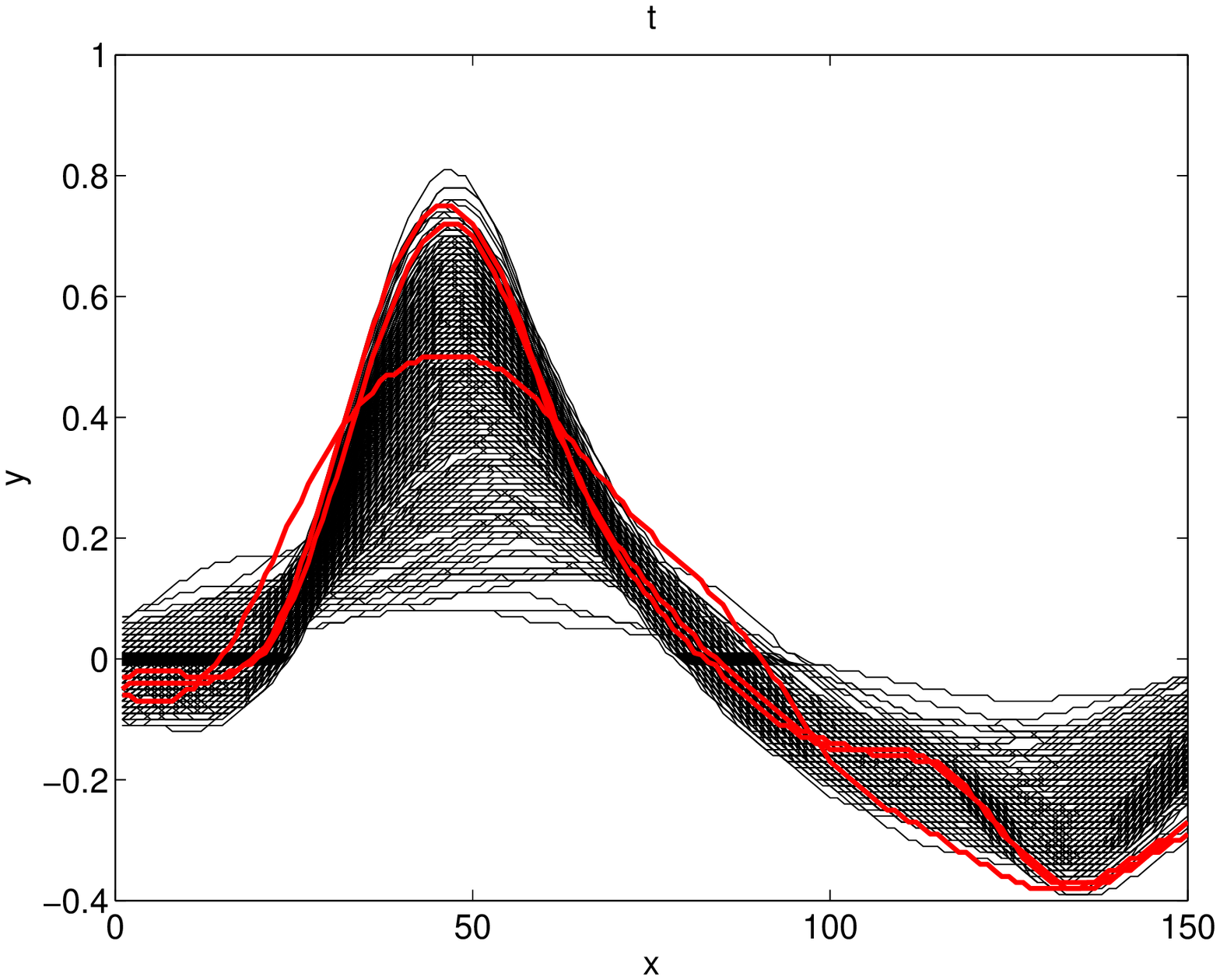}
\end{psfrags}\\
\begin{psfrags}
\psfrag{x}[t][t]{{time}}
\psfrag{y}[b][b]{{Amplitude}}
\psfrag{t}[b][b]{{}}
\includegraphics[width=2.5in]{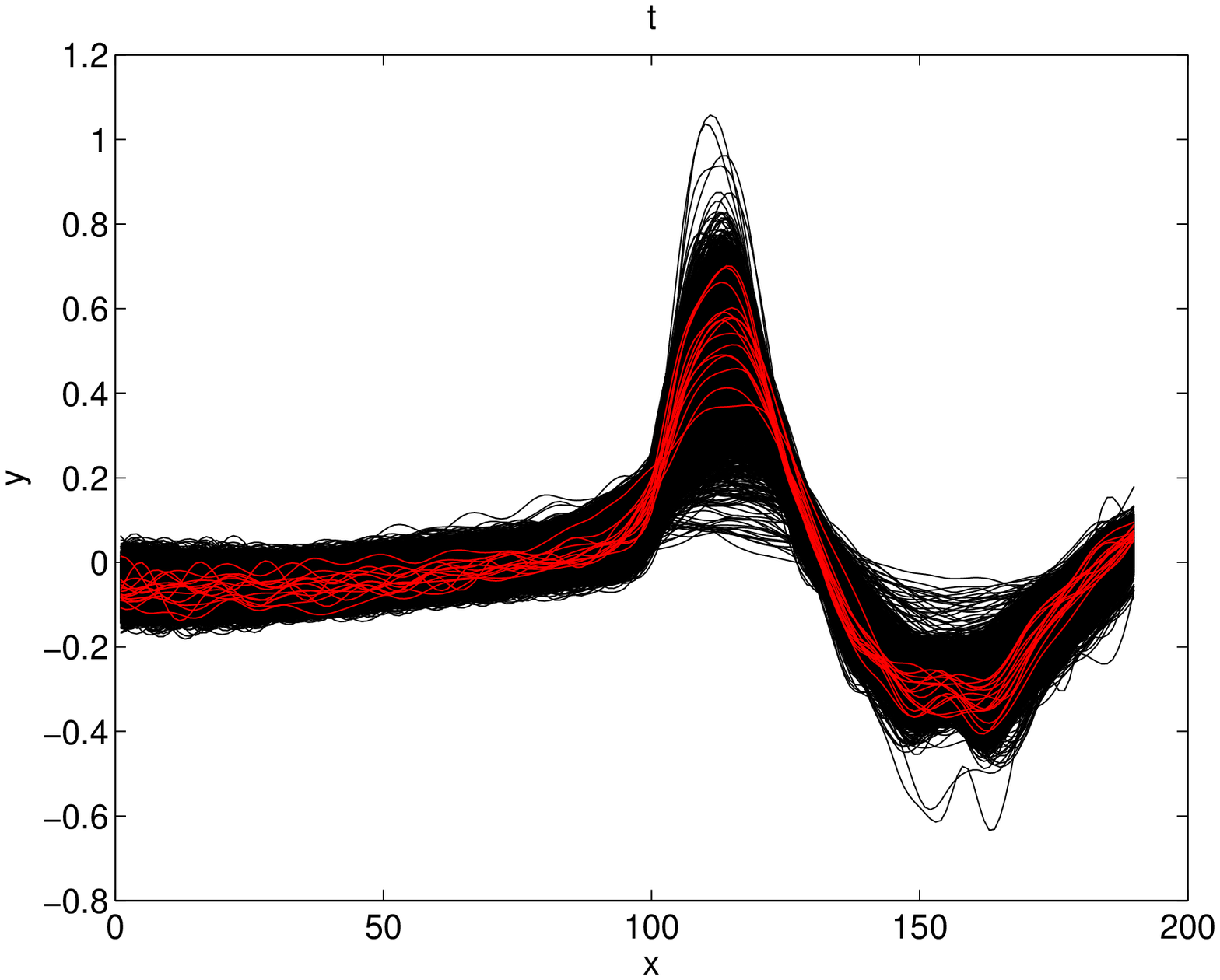}
\end{psfrags}\\
\begin{psfrags}
\psfrag{x}[t][t]{{time}}
\psfrag{y}[b][b]{{Amplitude}}
\psfrag{t}[b][b]{{}}
\includegraphics[width=2.5in]{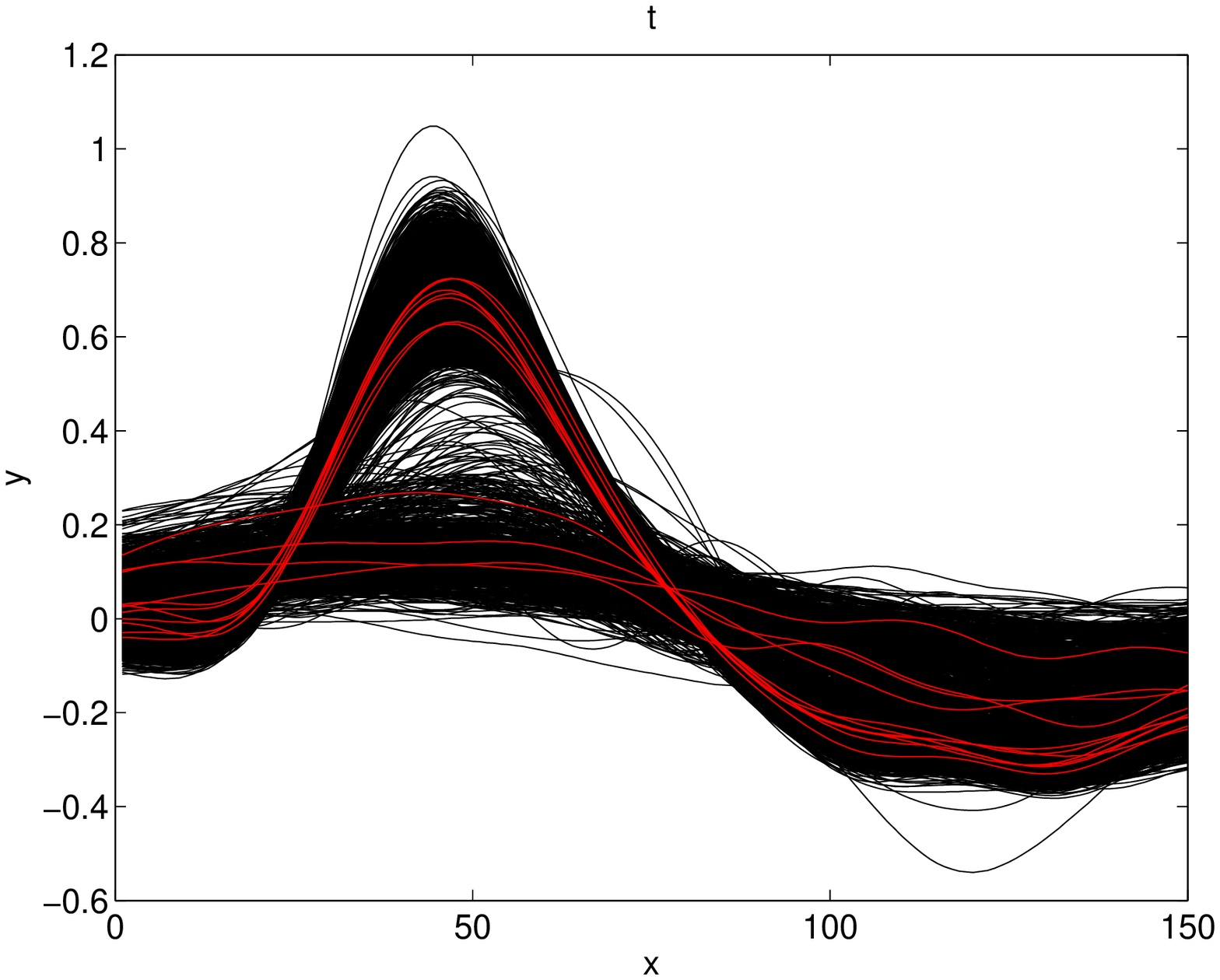}
\end{psfrags}
\end{tabular}
\caption{\textbf{Vibration signals}. \textbf{Top:} Vibration signals for the data-set $\textrm{DS\_I}$. \textbf{Center:} Vibration signals for the data-set $\textrm{DS\_II}$. \textbf{Bottom:} Vibration signals for the data-set $\textrm{DS\_III}$. The signals corresponding to maintenance actions are depicted in red.}
\label{accsignals}
\end{figure}

\begin{figure}[htbp]
\centering
\begin{tabular}{c}
\begin{psfrags}
\psfrag{x}[t][t]{{}}
\psfrag{y}[b][b]{{}}
\psfrag{t}[b][b]{{}}
\includegraphics[width=4in]{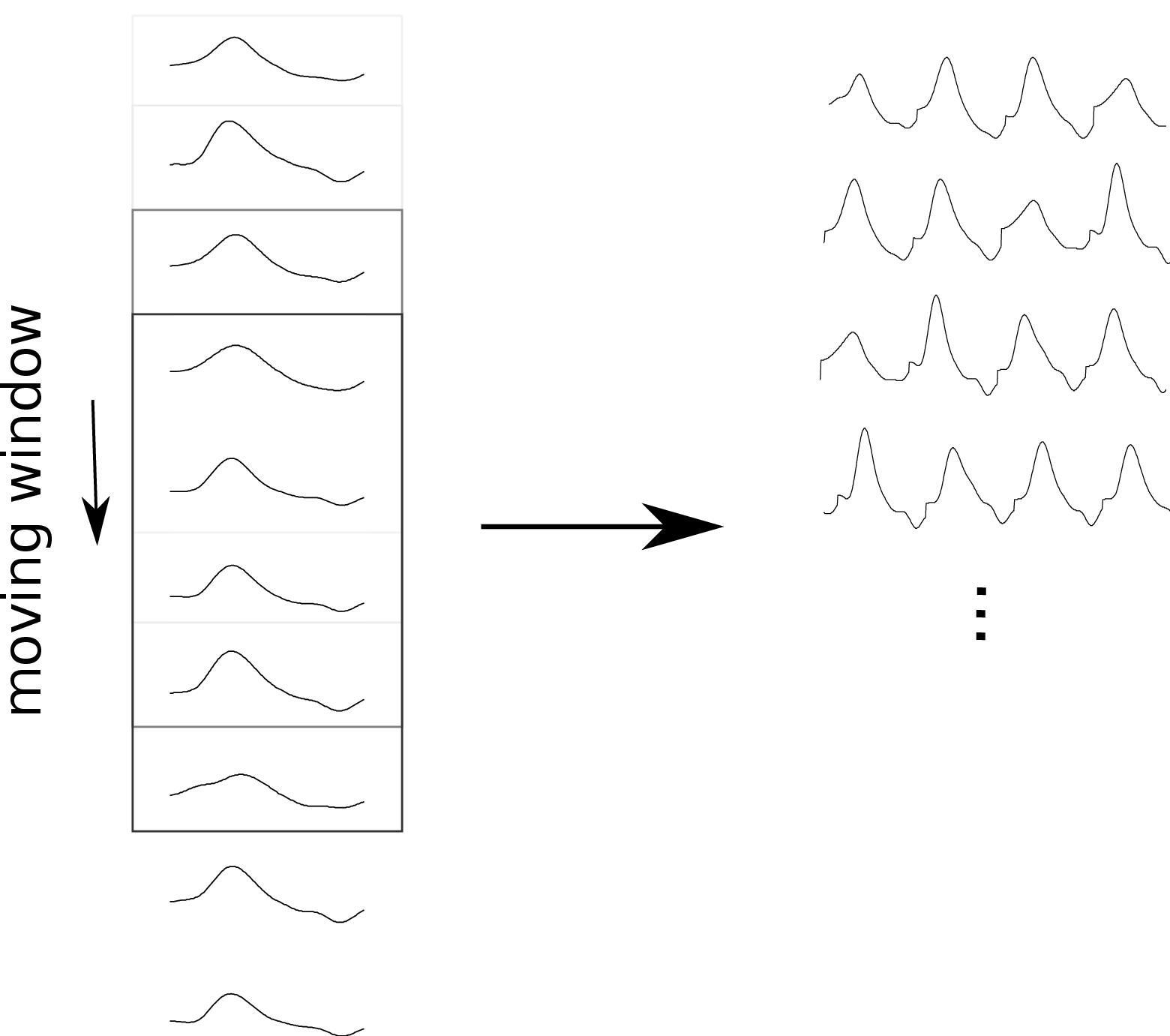} 
\end{psfrags}
\end{tabular}
\caption{\textbf{Concatenation of vibration signals}. After the windowing operation, each data-point is now a time-series of dimension $d=40\times150$ for datasets $\textrm{DS\_I}$ and $\textrm{DS\_III}$ and $d=40\times190$ for the dataset $\textrm{DS\_II}$.}
\label{concat_accsignals}	
\end{figure}

The data are collected from a Vertical Form Fill and Seal (VFFS) machine used to fill and seal packages mainly in the food industry. An illustration of such a machine is given in figure \ref{fig:Mebios_machine}. The VFFS machine supplies film from a roll which is formed into a bag over the vertical cylinder. Sealing jaws close the bag at the bottom before it is filled. At the end of the cycle, the bag is sealed and cut off with a knife. From previous experimental studies the dirt accumulation on the sealing jaws was observed to strongly affect the process quality. For this reason, in the experiments described here the jaws were monitored to predict in advance the maintenance actions. Maintenance consists of stopping the machine and cleaning the sealing jaws. A total of three experiments have been performed, resulting in three datasets:
\begin{itemize}
\item  $\textrm{DS\_I}$: this dataset consists of $771$ events and $3$ external maintenance actions. An event is related to a particular processed bag and takes place every two seconds (i.e. the sampling frequency is $0.5 Hz$). Each event is associated with a $150$-dimensional accelerometer signal, that is each signal is a vector of length $150$ (see top of figure \ref{accsignals}).    
\item  $\textrm{DS\_II}$: it contains a total of $11\,632$ processed bags and $15$ maintenance actions. Here the vibration signals used to monitor the dirt accumulation in the jaws are $190$-dimensional time-series (as shown in the center of figure \ref{accsignals}). This is due to a different setting of the data acquisition system for this experiment.
\item  $\textrm{DS\_III}$: there are $3\,519$ processed bags and $11$ maintenance actions, as depicted at the bottom side of figure \ref{accsignals}.   
\end{itemize}
Moreover, in order to catch the ongoing deterioration process of the jaws we need to use historical values of sealing quality in our analysis. For this purpose we apply a windowing operation on the data, as illustrated in figure \ref{concat_accsignals}. Then we develop off-line a KSC model (see chapter \ref{ch:spectralclustering}) fed with the concatenation of a certain number of accelerometer signals. Once properly trained the model will be used on-line to detect the different behavioural regimes experienced by the running machine.  

\section{Results}
In this section it is shown how KSC can be used to perform just-in-time maintenance, not too early to take full advantage of component lifetime but also not too late to avoid catastrophic failures and unplanned downtime. 

\begin{figure}[htbp]
\centering
\begin{tabular}{c}
\begin{psfrags}
\psfrag{x}[t][t]{{$\sigma$}}
\psfrag{y}[b][b]{{Window size}}
\psfrag{t}[b][b]{{$k=2$}}
\includegraphics[width=3in]{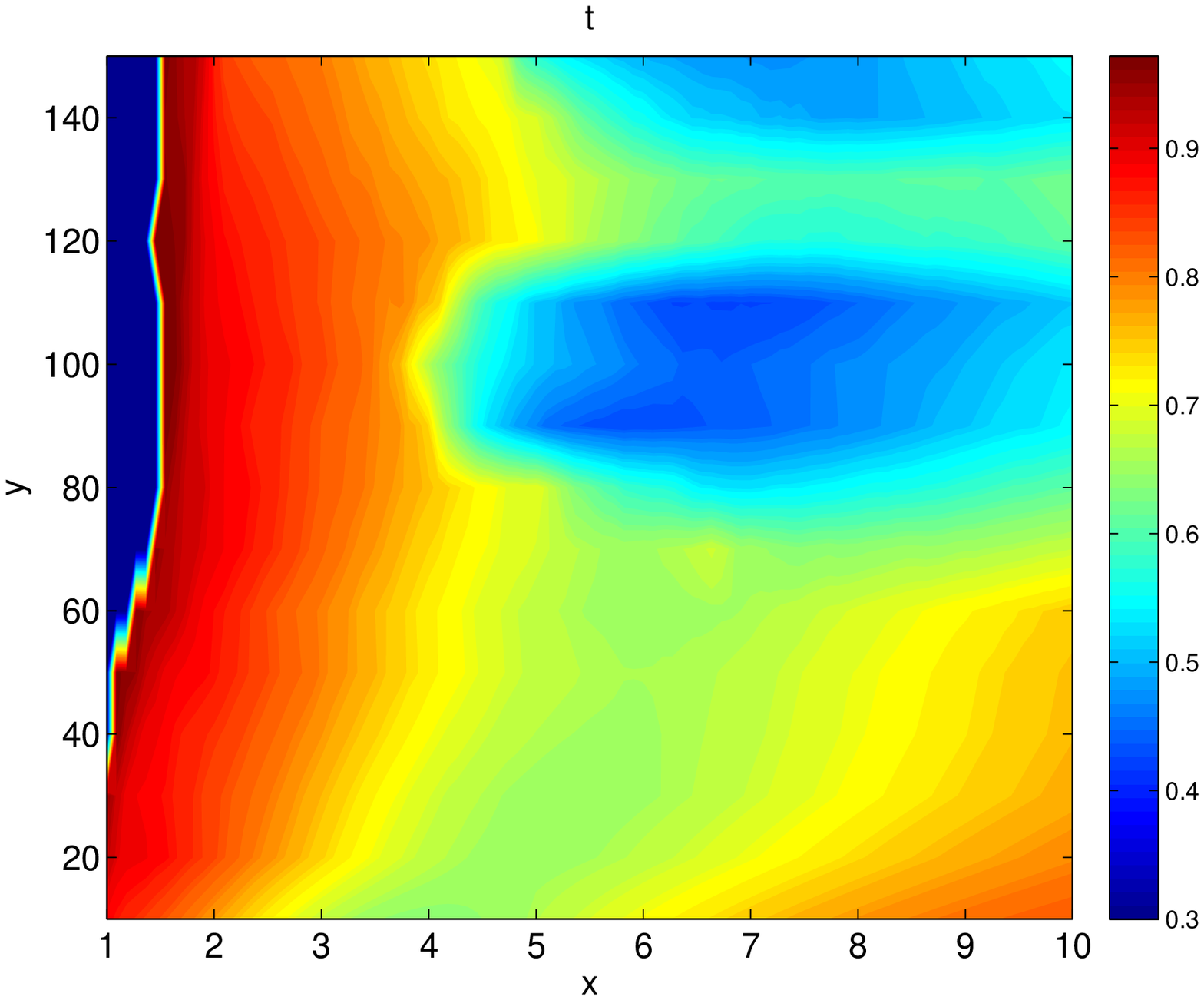}
\end{psfrags}\\
\begin{psfrags}
\psfrag{x}[t][t]{{$\sigma$}}
\psfrag{y}[b][b]{{Window size}}
\psfrag{t}[b][b]{{$k=3$}}
\includegraphics[width=3in]{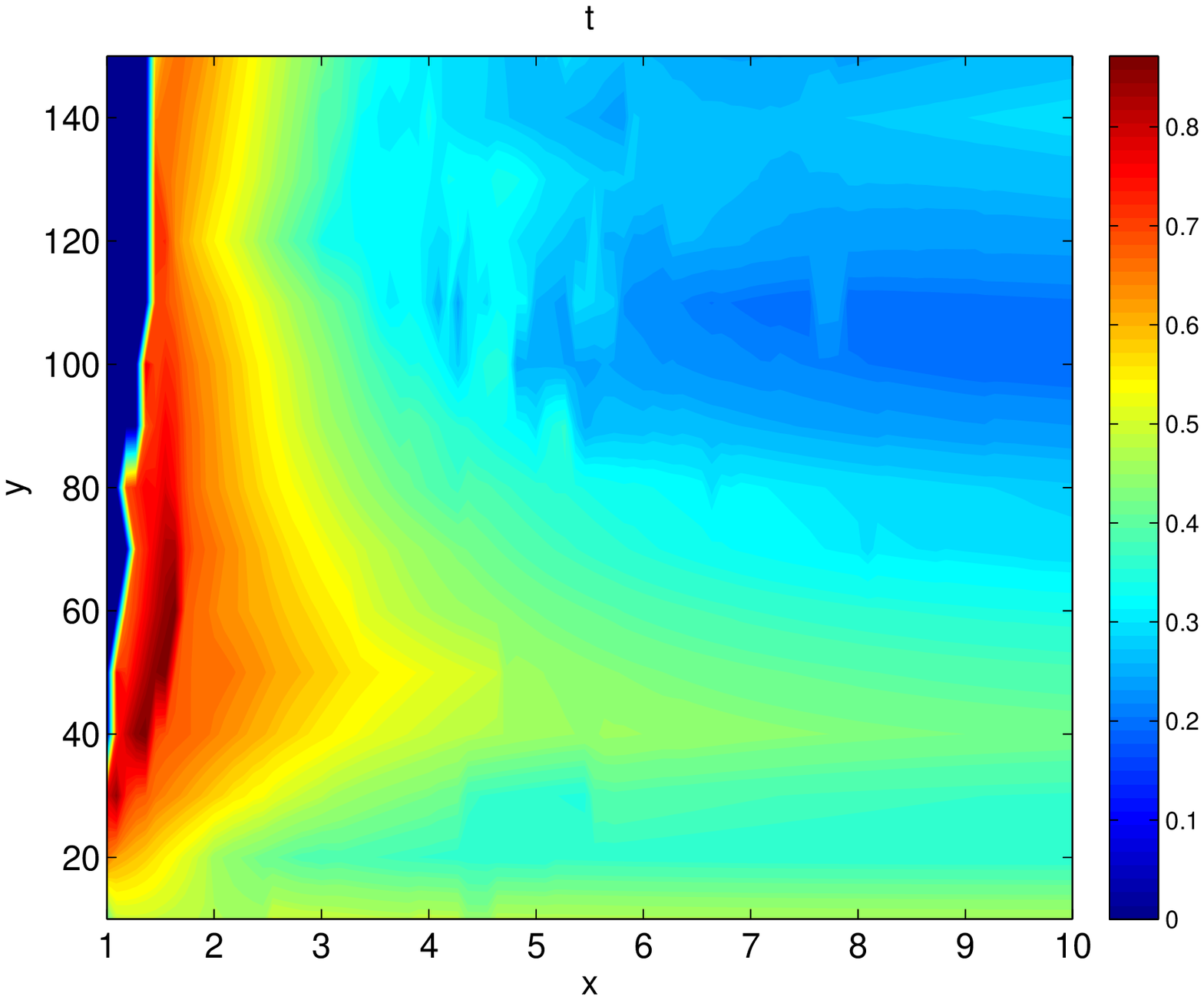}
\end{psfrags}
\end{tabular}
\caption{\textbf{Model selection.} Tuning surfaces for the first dataset, only the results for $k = 2$ and $k =3$ are shown. If we consider more clusters ($k > 3$) the maximum value of the $\textrm{BLF}$ decreases. The outcome is similar for the datasets $\textrm{DS\_II}$ and $\textrm{DS\_III}$.}
\label{tuning_surface}
\end{figure}

Above all, model selection has been performed. In total we have $3$ parameters to determine: the window size (i.e. the number of accelerometer signals to concatenate), the number of clusters $k$ and the RBF kernel parameter $\sigma$. According to the $\textrm{BLF}$ criterion the optimal window size is $40$ and the optimal number of clusters is $k=2$ for the three data-sets, while $\sigma$ is dataset dependent. In figure \ref{tuning_surface} an illustration of the tuning procedure related to the first dataset is given.

After building an optimal KSC model $2$ regimes have been identified, where one of them can be interpreted as normal behaviour and the other as critical conditions (need of maintenance). Moreover, a probabilistic interpretation of the results is also provided, which better describes the degradation process experienced by the sealing jaws of the packing machine. Finally, we compare KSC with $K$-means, which is the most popular method used in industrial applications \cite{p.2006survey}. 

\subsection{Hard Clustering}
In figure \ref{mebios1_results} the KSC prediction for the dataset $\textrm{DS\_I}$ is shown. We can interpret one of the clusters as normal conditions and the other as maintenance cluster. Notice that the KSC model is able to predict some minutes in advance the maintenance actions before they are actually performed by the operator. Concerning the dataset $\textrm{DS\_II}$ we can draw the same comments: KSC is very accurate in predicting the worsening of the packing process around the actual maintenance events (see top of figure \ref{mebios4_results}). Finally, figure \ref{mebios5_results} illustrates the results on dataset $\textrm{DS\_III}$. In this case KSC predicts the need of maintenance also in zones where maintenance has not been really performed. As we will see later, surprisingly it would have been more logical to perform maintenance as suggested by KSC and not as actually done by the operator.

\subsection{Probabilistic Output}
\begin{figure}[htbp]
\centering
\begin{tabular}{cc}
\begin{psfrags}
\psfrag{x}[t][t]{{Event index i}}
\psfrag{y}[b][b]{{Cluster membership}}
\psfrag{t}[b][b]{{}}
\includegraphics[width=3.5in]{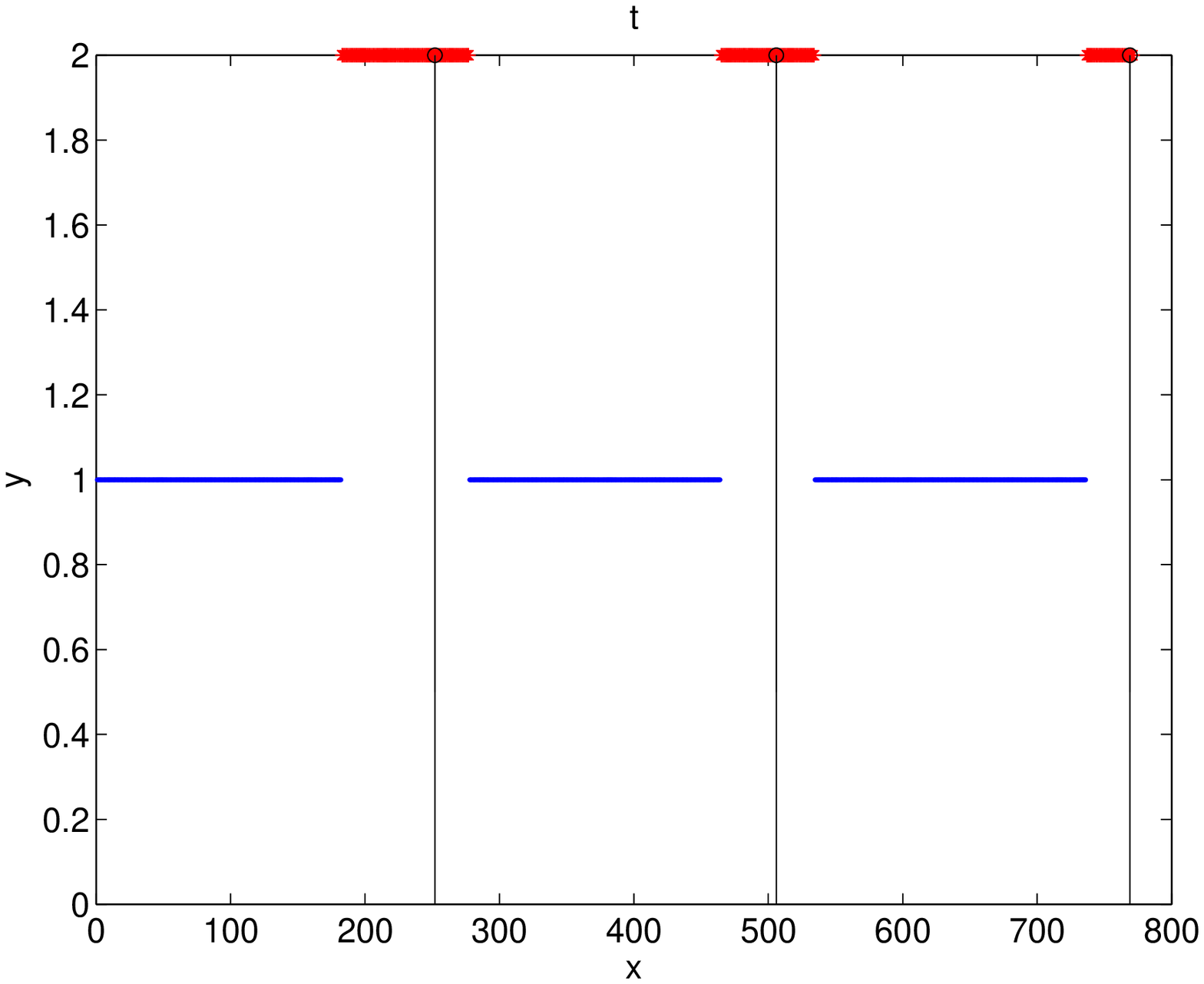}
\end{psfrags}\\
\begin{psfrags}
\psfrag{x}[t][t]{{Event index i}}
\psfrag{y}[b][b]{{Probability to maintenance}}
\psfrag{t}[b][b]{{}}
\includegraphics[width=3.5in]{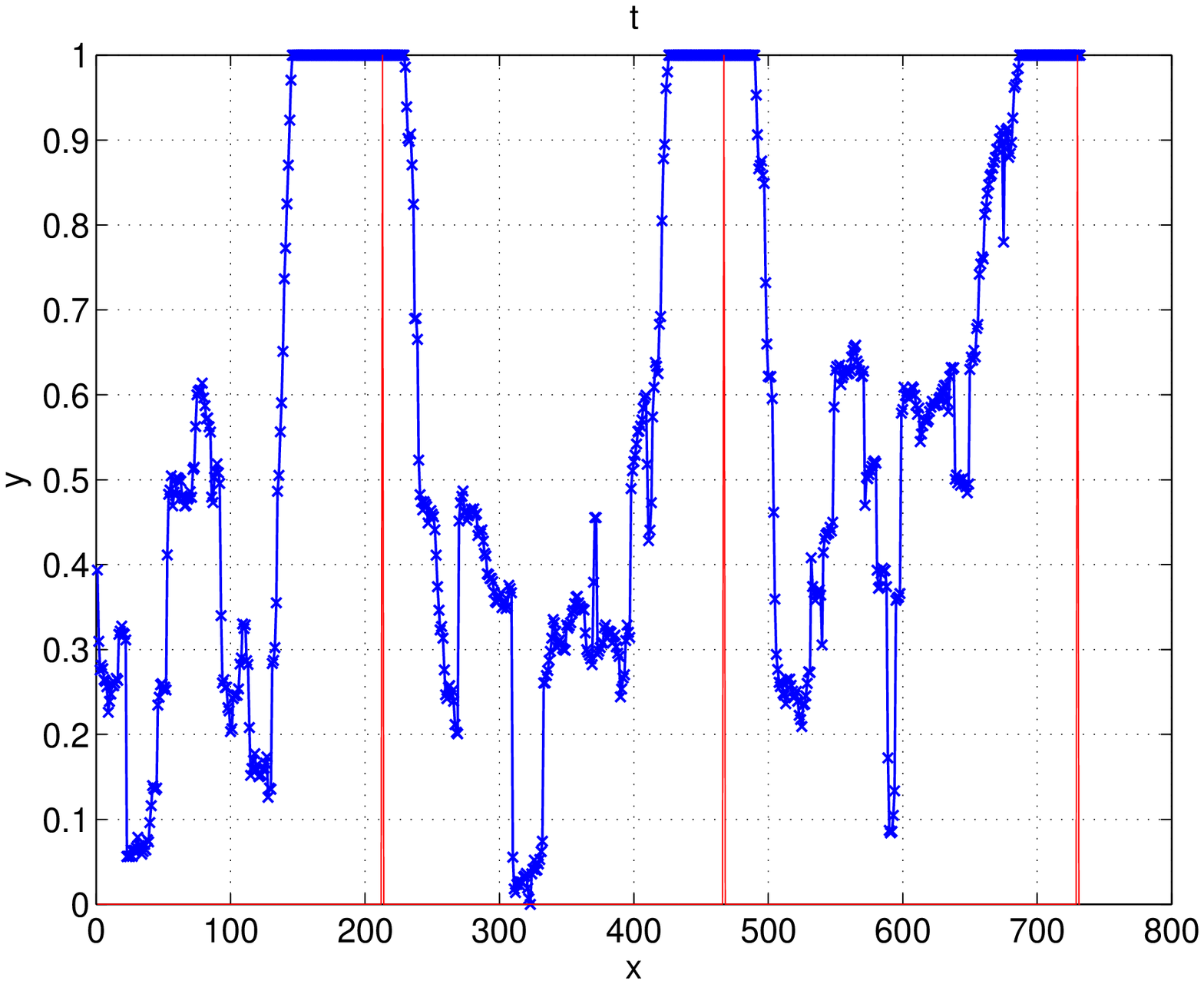}
\end{psfrags}
\end{tabular}
\caption{\textbf{KSC results dataset $\textrm{DS\_I}$.} \textbf{Top:}  Hard clustering results for the whole data-set. Cluster $2$ represents predicted maintenance events. The vertical black lines show the true maintenance. \textbf{Bottom:} Soft clustering results in terms of probability to maintenance.}
\label{mebios1_results}	
\end{figure}

\begin{figure}[htbp]
\centering
\begin{tabular}{cc}
\begin{psfrags}
\psfrag{x}[t][t]{{Event index i}}
\psfrag{y}[b][b]{{Cluster membership}}
\psfrag{t}[b][b]{{}}
\includegraphics[height=3in,width=3.5in]{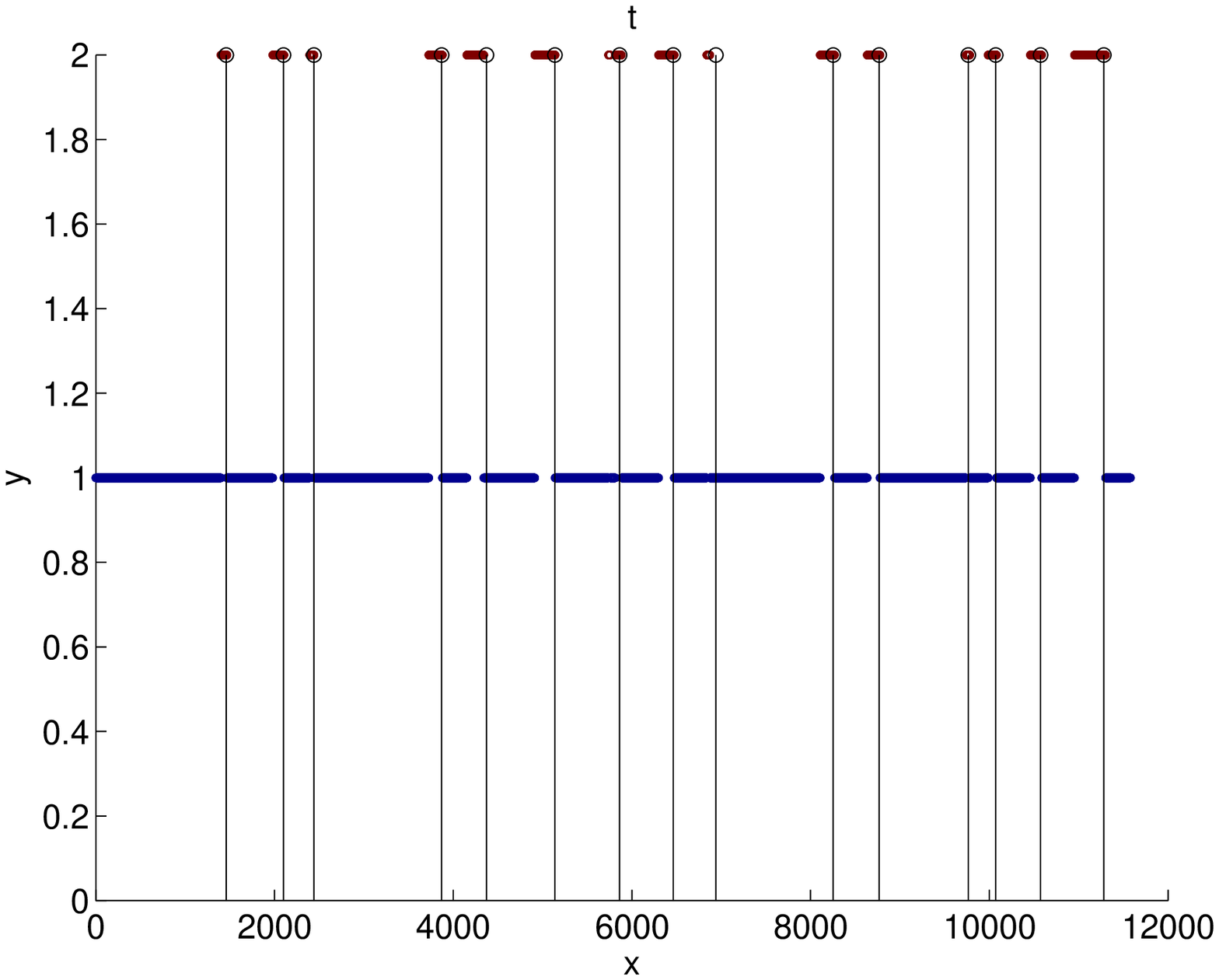} 
\end{psfrags}\\\\
\begin{psfrags}
\psfrag{x}[t][t]{{Event index i}}
\psfrag{y}[b][b]{{\small{Prob. to maintenance}}}
\psfrag{t}[b][b]{{}}
\includegraphics[height=1.5in,width=3.25in]{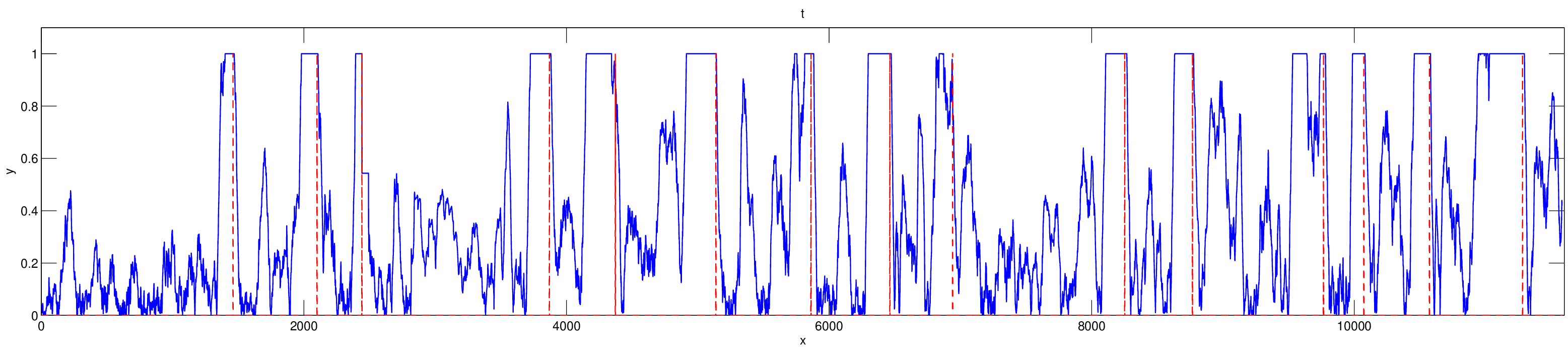}
\end{psfrags}
\end{tabular}
\caption{\textbf{KSC results dataset $\textrm{DS\_II}$.} \textbf{Top:} Hard clustering, cluster $2$ represents predicted maintenance events. \textbf{Bottom:} Soft clustering in terms of probability to maintenance.}
\label{mebios4_results}	
\end{figure}

\begin{figure}[htbp]
\centering
\begin{tabular}{c}
\begin{psfrags}
\psfrag{x}[t][t]{{Event index i}}
\psfrag{y}[b][b]{{Cluster membership}}
\psfrag{t}[b][b]{{}}
\includegraphics[width=3.5in,height=3in]{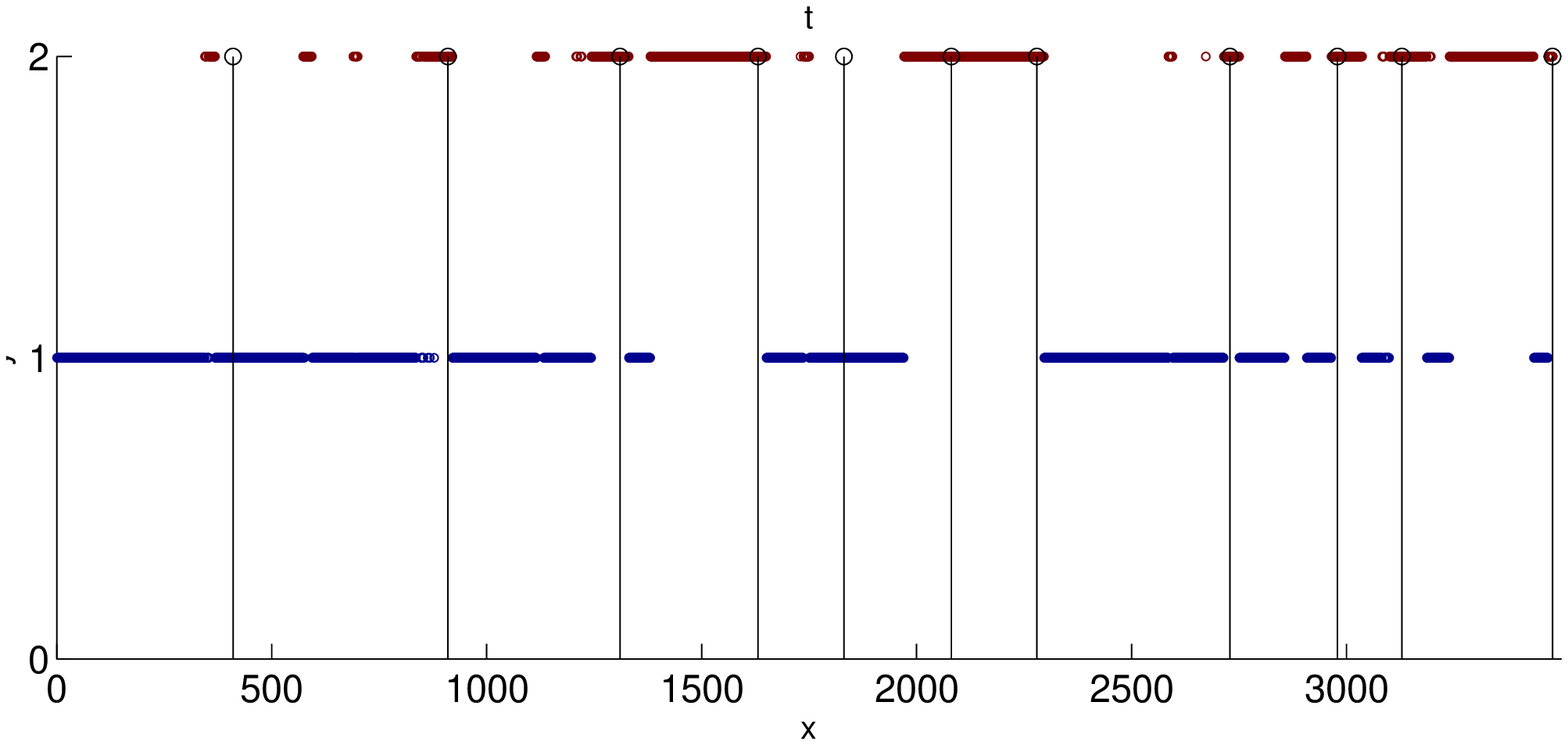}
\end{psfrags}\\
\begin{psfrags}
\psfrag{x}[t][t]{{Event index i}}
\psfrag{y}[b][b]{{Probability to maintenance}}
\psfrag{t}[b][b]{{}}
\includegraphics[width=3.7in,height=3in]{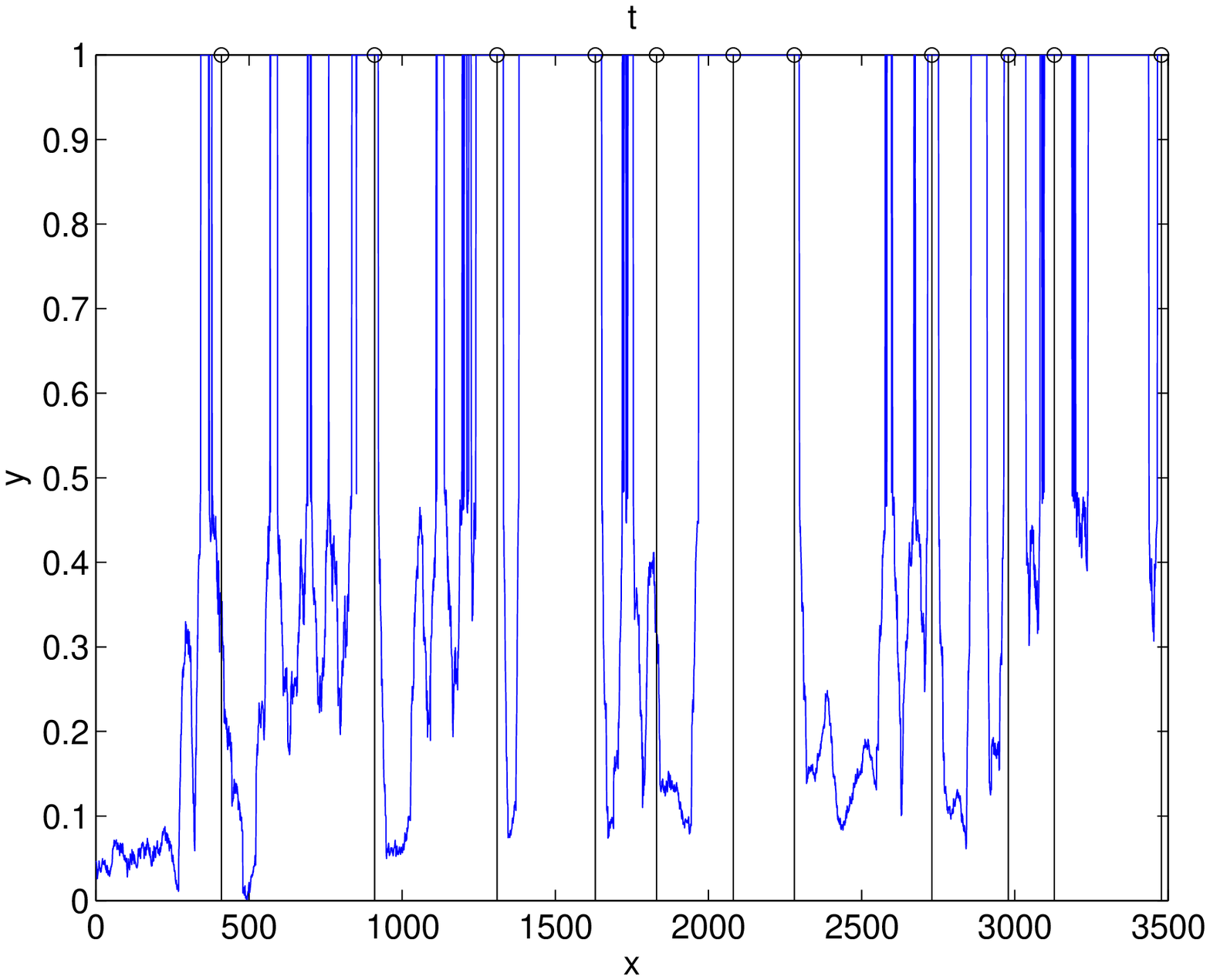}
\end{psfrags}
\end{tabular}
\caption{\textbf{KSC results dataset $\textrm{DS\_III}$.} \textbf{Top:} Hard clustering, cluster $2$ represents predicted maintenance events. \textbf{Bottom:} Soft clustering in terms of probability to maintenance.}
\label{mebios5_results}	
\end{figure}

In the previous section we demonstrated the effectiveness of KSC in predicting in advance the maintenance events. Nevertheless the predicted output is binary (it goes suddenly from normal operation to maintenance). An output of this form does not provide a continuous indicator of the incoming maintenance actions. To solve this issue we can use the latent variable $e(x)$ instead of the binarized clustering output $\textrm{sign}(e(x))$ (see section \ref{KSC_intro}). The latent variable provides a more informative output which can be analysed in order to produce a better prognostic output. Since the range from which it takes a value depends on many factors (e.g., the kernel and its parameters, the number of training data points), the interpretability might be difficult. To improve it, values are normalized between $0$ and $1$. This transformation is based on the structure of the latent variable space. As mentioned in section \ref{sc:BLF}, in this space every cluster is ideally represented as a line. The tips of the lines can be considered prototypes of their cluster, since they have more certainty to belong to it because they are further from the decision boundaries\footnote{However, as we discussed in chapter \ref{ch:spectralclustering}, this procedure is valid only when the amount of overlap between the clusters is negligible. When it is not, it is advisable to use the methodology incorporated in SKSC.} \cite{highlysparse}. Thus, the Euclidean distance from every point to the cluster prototype can be seen as a confidence measure of the cluster membership. The transformed latent variable is depicted at the bottom of figure \ref{mebios1_results} (dataset $\textrm{DS\_I}$), figure \ref{mebios4_results} (dataset $\textrm{DS\_II}$) and figure \ref{mebios5_results} (dataset $\textrm{DS\_III}$). The value can be considered as a soft membership or probability to maintenance \cite{prob_clust_israel}. The latter increases as the number of faulty bags in the window increases. The value can decrease since the window can move onto zones with good seals after a period of bad seals. This is probably due to a self-cleaning mechanism. Maintenance is predicted when the probability reaches the value $1$. 

For datasets $\textrm{DS\_I}$ and $\textrm{DS\_II}$ it can be noticed how KSC is able to discover from the vibration signals registered by the accelerometers the dirt accumulation in the jaws that leads to the maintenance actions. This is very surprising because clustering is an unsupervised technique, and thus does not make use of any information on the location of the maintenance actions (like it occurs for classification). 

\begin{figure}[htbp]
\centering
\begin{tabular}{c}
\begin{psfrags}
\psfrag{x}[t][t]{{Event index i}}
\psfrag{y}[b][b]{{Number of hot area pixels}}
\psfrag{t}[b][b]{{}}
\includegraphics[width=4in]{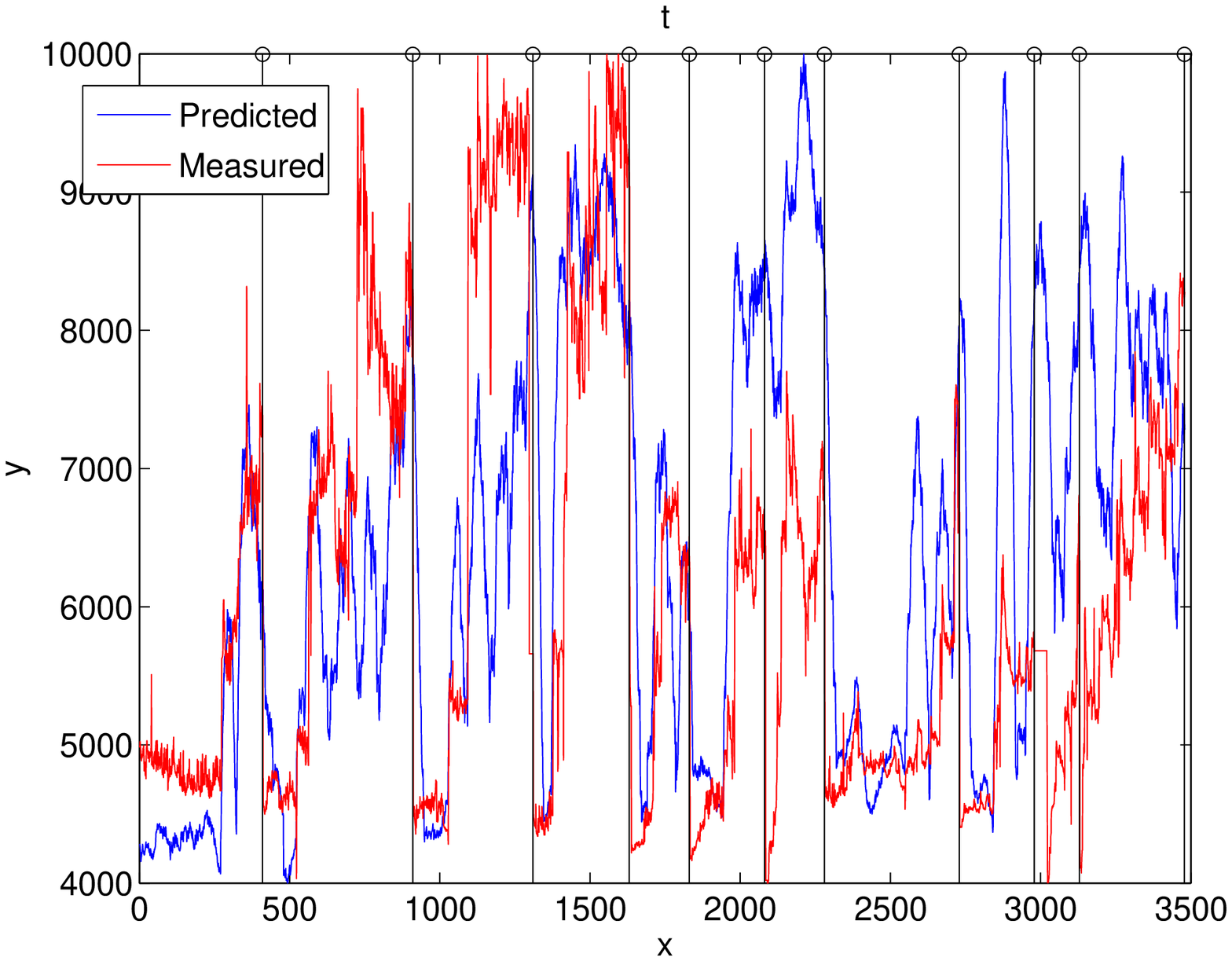}
\end{psfrags}
\end{tabular}
\caption{\textbf{Degradation dataset DS\_{III}}. Degradation inferred by KSC (blue) \textit{Versus} measured degradation (red). We can notice how, although the short-term variability is not detected, KSC can follow the general trend of the measured degradation. This is quite surprising since the clustering model, due to the fact that is unsupervised, did not have any information about the true degradation.}
\label{trueVSKSC_degradation}	
\end{figure}

For what concerns dataset $\textrm{DS\_III}$, also regions where no maintenance actions have been really performed appear associated with high probability to maintenance. Before continuing the discussion, we should mention that in the third experiment the machine is also equipped with a thermal camera, which directly measures the dirt accumulation in terms of number of hot area pixels in the acquired images (with hot we mean that the temperature of the sealing jaws is above a user-defined threshold). Then, by rescaling the probability to maintenance such that it varies in the same range as the measured degradation, we can make a comparison between the two. This is illustrated in figure \ref{trueVSKSC_degradation}. We can recognize similar patterns, meaning that KSC was able to catch the real degradation process also in this last experiment, although in a genuinely unsupervised manner. To conclude, probably the operator should have performed the maintenance operations in a different way, being coherent with his behaviour in the first two experiments.

\section{Comparison with $K$-means}
\begin{figure}[htbp]
\centering
\begin{tabular}{cc}
\begin{psfrags}
\psfrag{x}[t][t]{{Event index i}}
\psfrag{y}[b][b]{{Cluster membership}}
\psfrag{t}[b][b]{{}}
\includegraphics[width=2in]{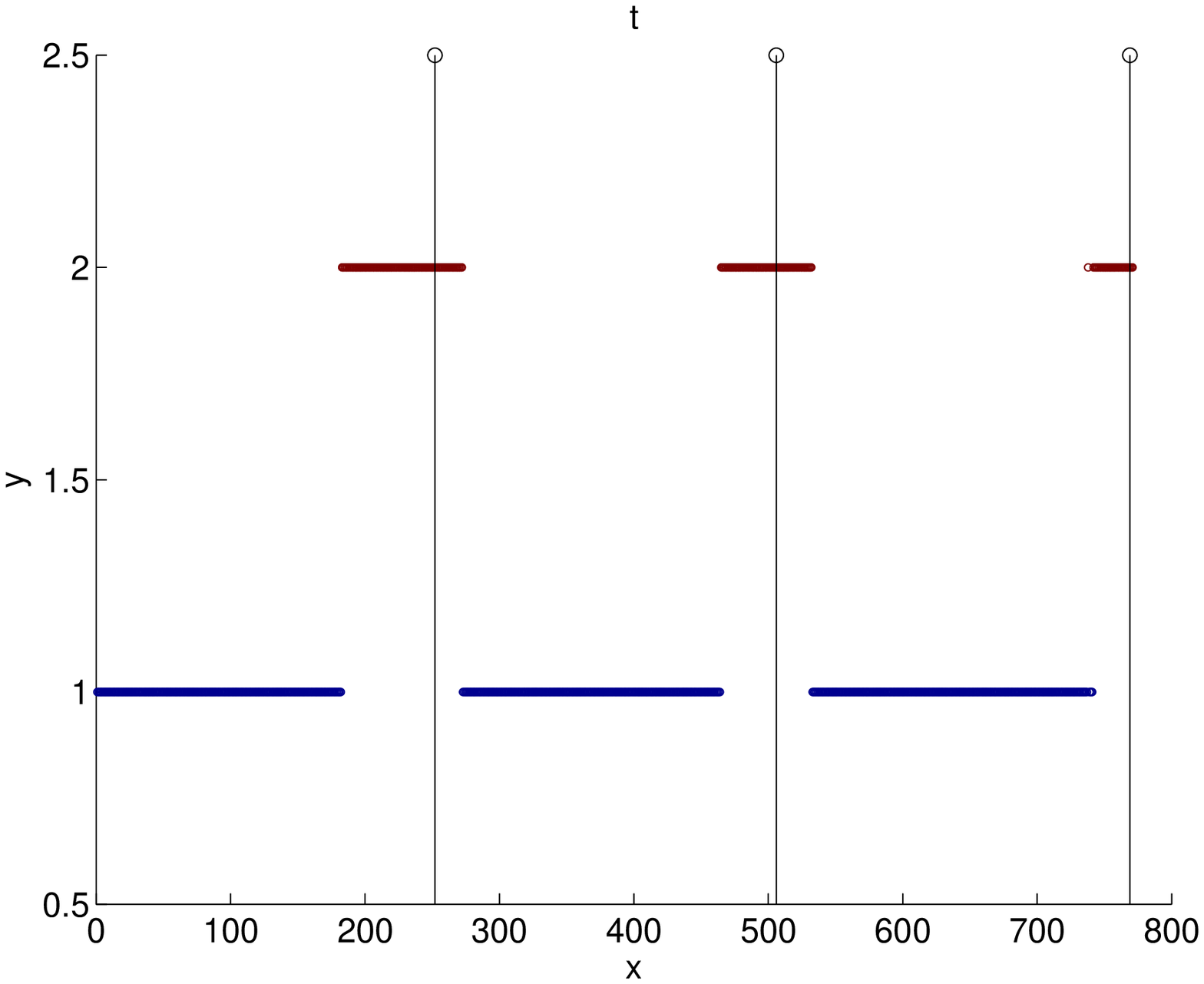}
\end{psfrags}&
\begin{psfrags}
\psfrag{x}[t][t]{{Event index i}}
\psfrag{y}[b][b]{{Prob. to maintenance}}
\psfrag{t}[b][b]{{}}
\includegraphics[width=2.1in]{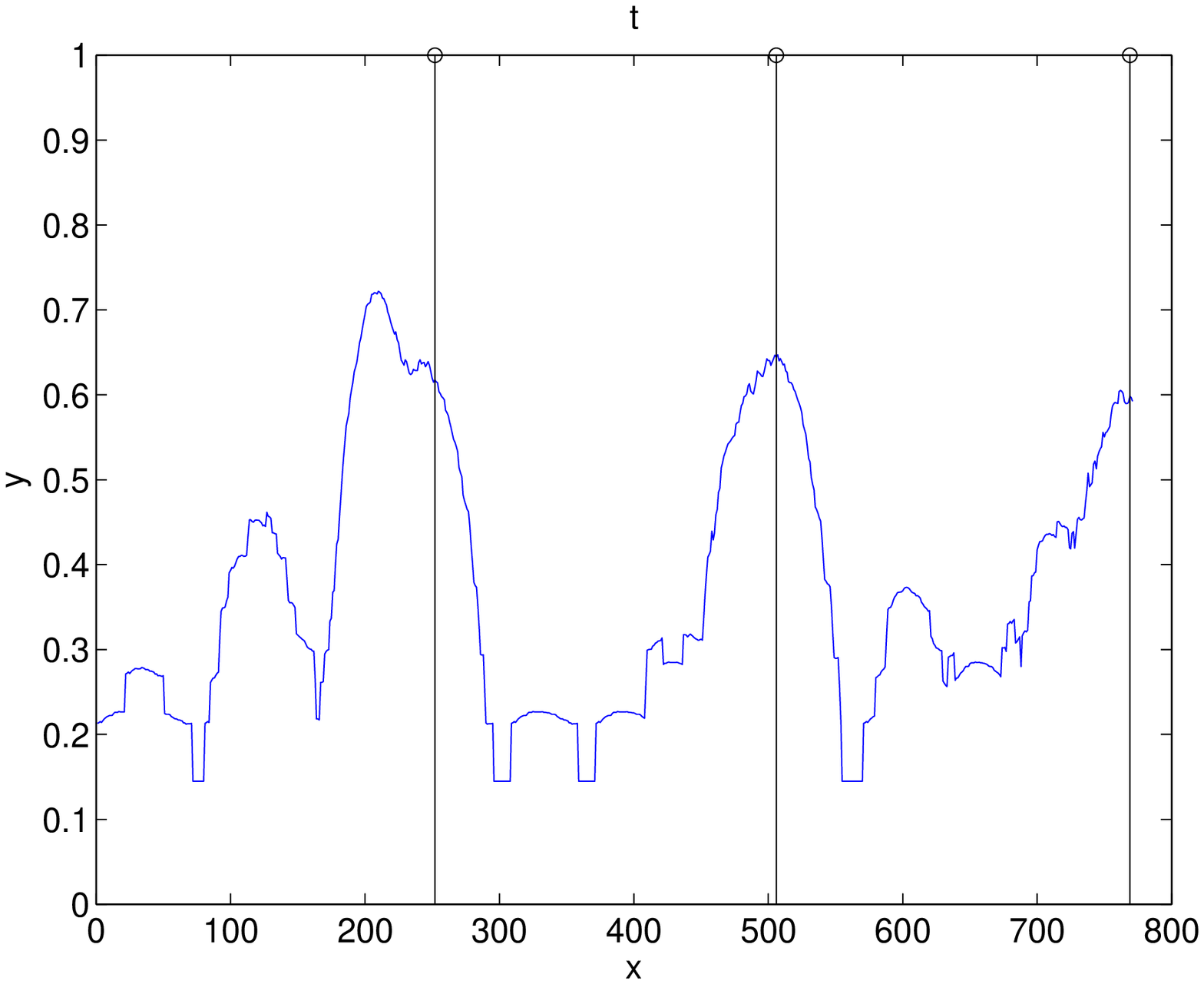}
\end{psfrags}\\
\begin{psfrags}
\psfrag{x}[t][t]{{Event index i}}
\psfrag{y}[b][b]{{Cluster membership}}
\psfrag{t}[b][b]{{}}
\includegraphics[width=2in]{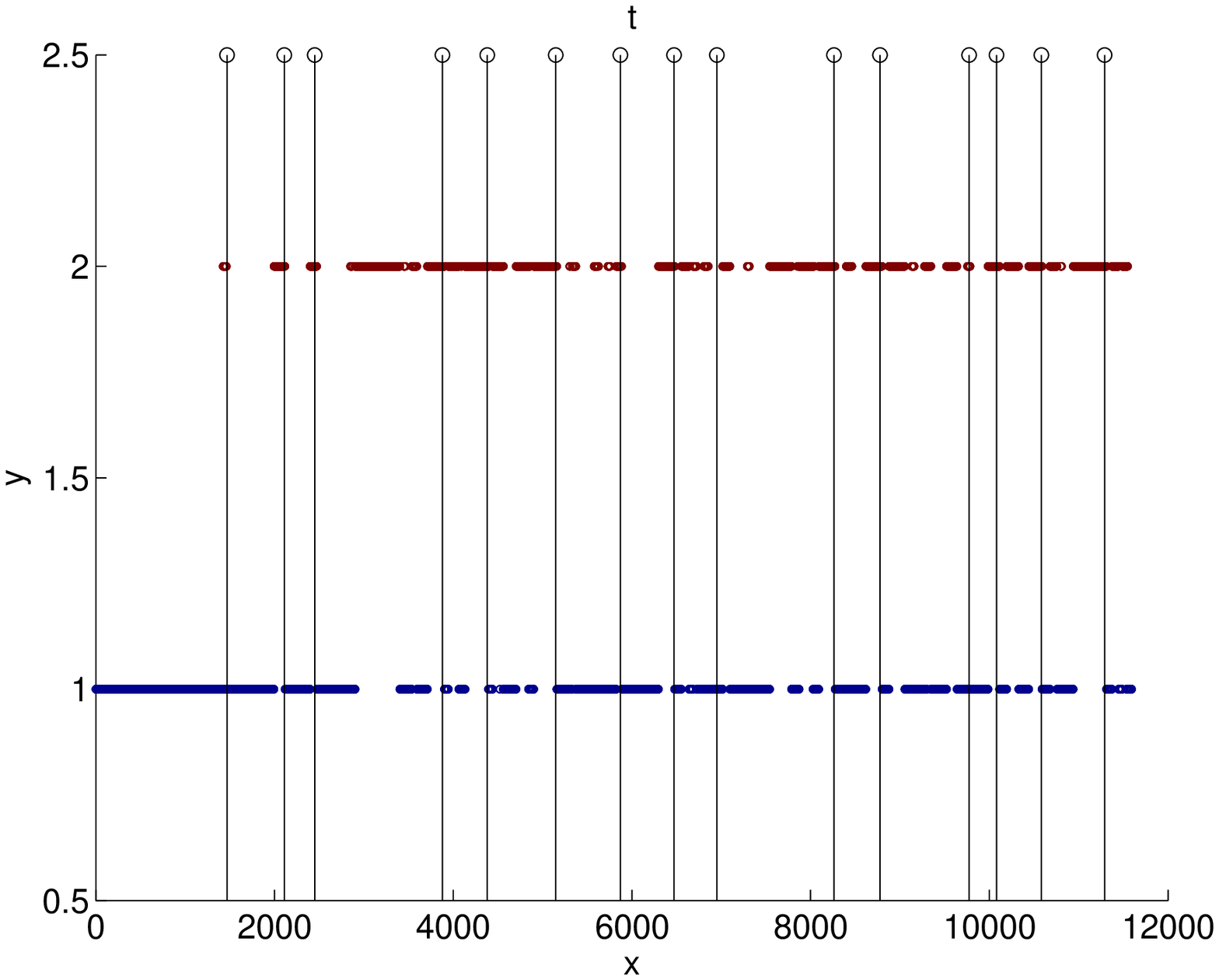}
\end{psfrags}&
\begin{psfrags}
\psfrag{x}[t][t]{{Event index i}}
\psfrag{y}[b][b]{{Prob. to maintenance}}
\psfrag{t}[b][b]{{}}
\includegraphics[width=2.15in]{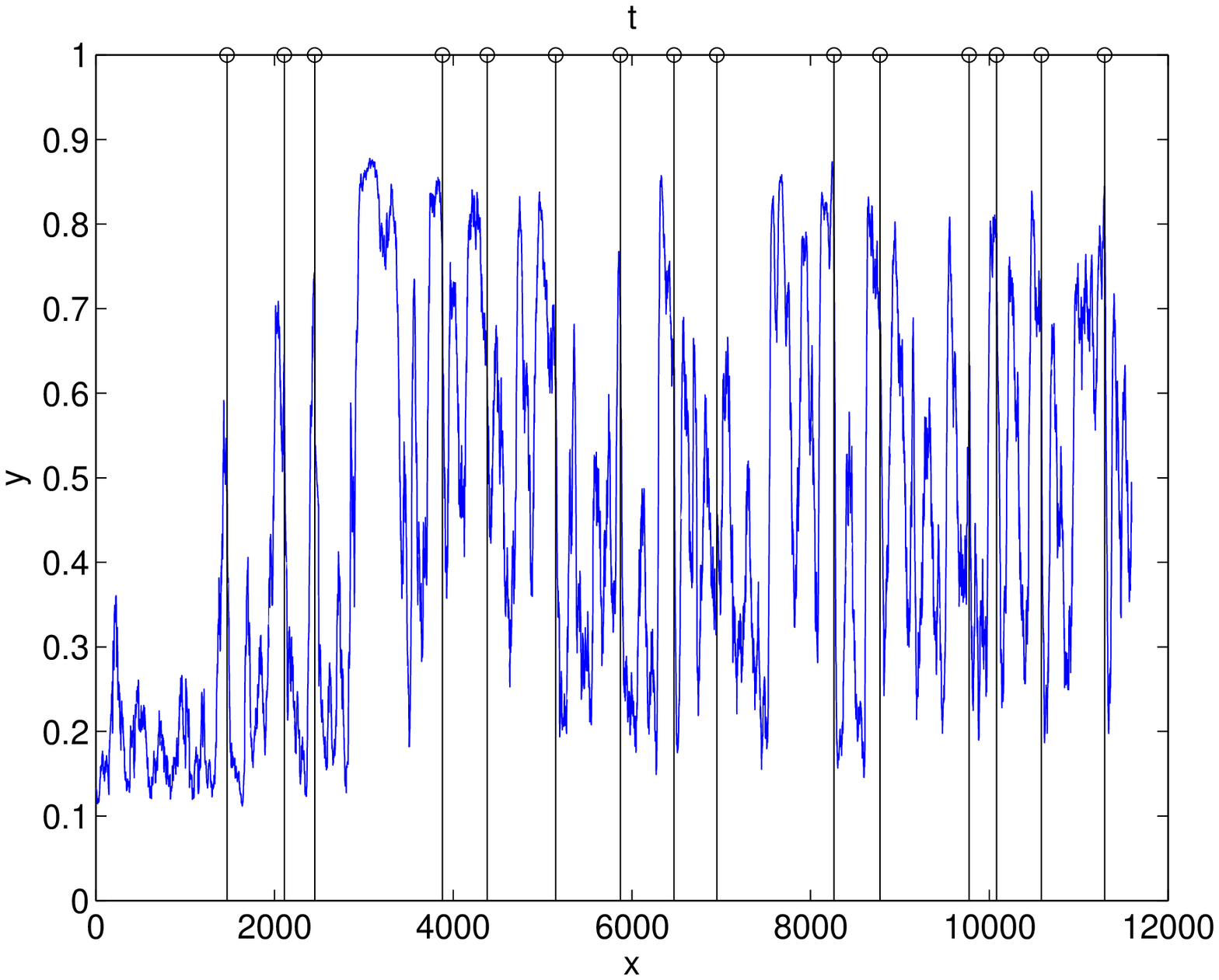}
\end{psfrags}\\\\
\begin{psfrags}
\psfrag{x}[t][t]{{Event index i}}
\psfrag{y}[b][b]{{Cluster membership}}
\psfrag{t}[b][b]{{}}
\includegraphics[width=2in,height=2in]{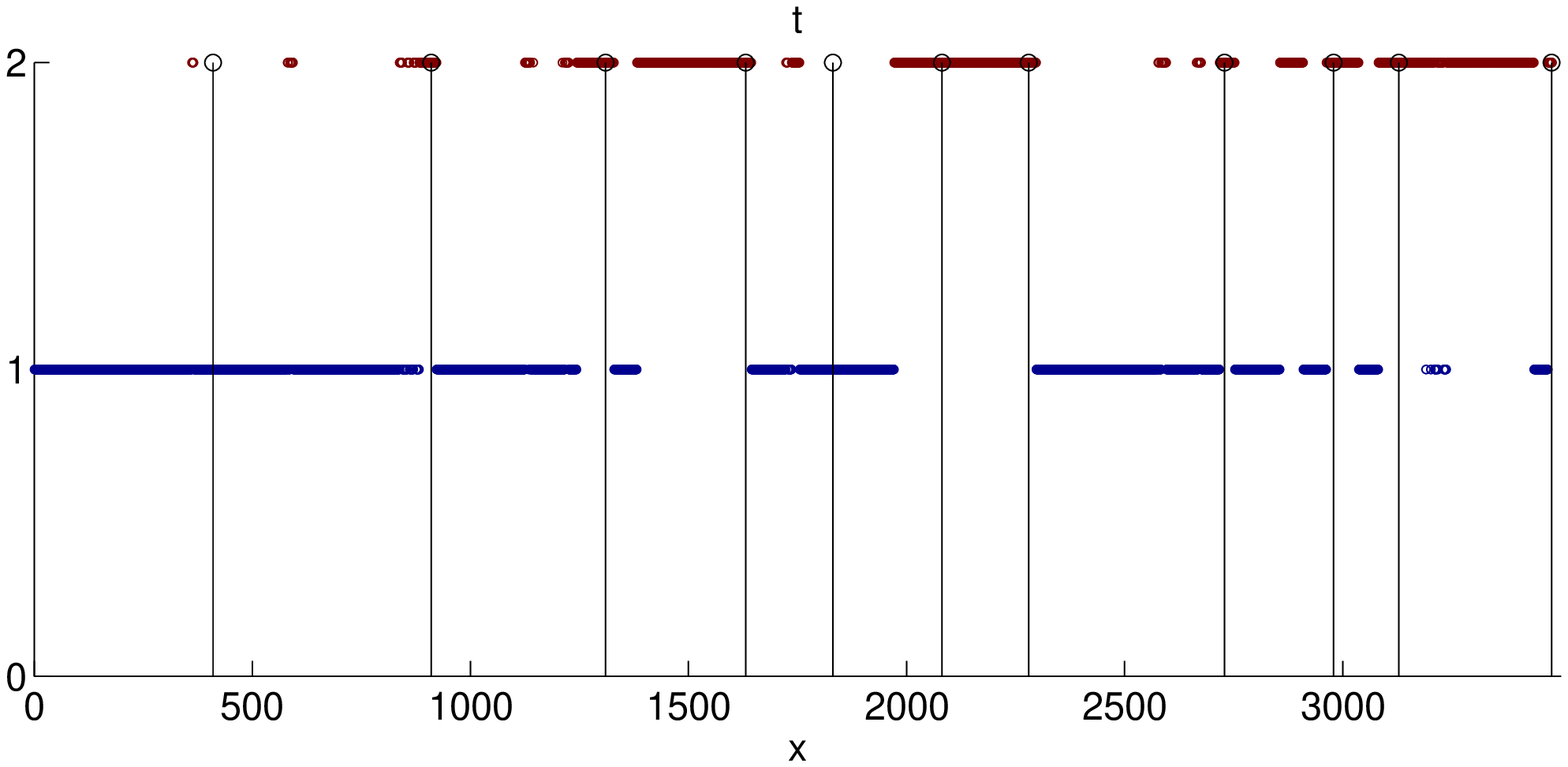}
\end{psfrags}&
\begin{psfrags}
\psfrag{x}[t][t]{{Event index i}}
\psfrag{y}[b][b]{{Prob. to maintenance}}
\psfrag{t}[b][b]{{}}
\includegraphics[width=2.1in,height=2in]{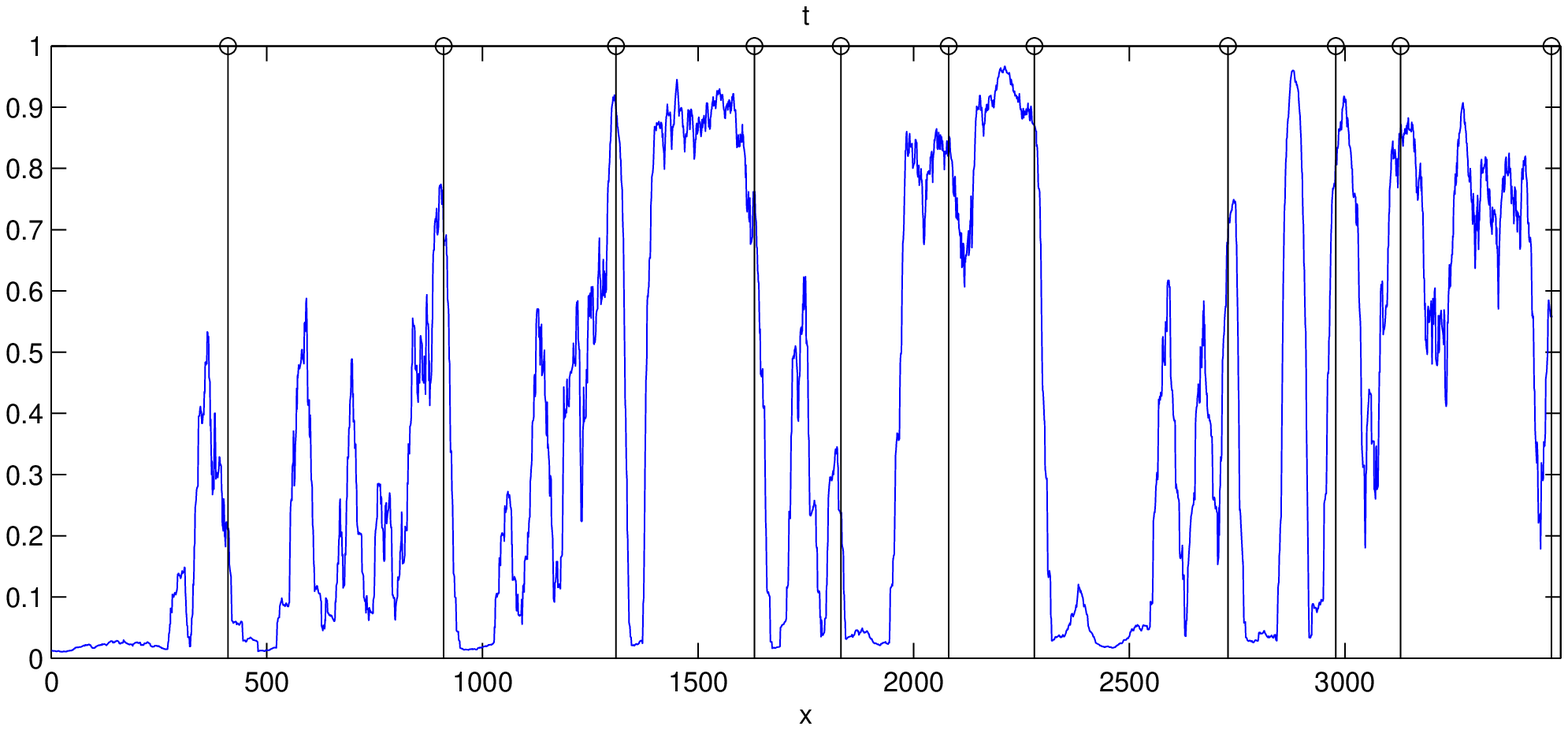}
\end{psfrags}
\end{tabular}
\caption{\footnotesize{\textbf{$K$-means results.} Cluster $1$ symbolizes the normal operating condition, cluster $2$ represents predicted maintenance events. The vertical black lines show the true maintenance. \textbf{Top}: dataset $\textrm{DS\_I}$. The results are similar to KSC outcomes, even if slightly worse since in the end there is a kind of false alarm (a single prediction of maintenance followed by normal behaviour, before the final maintenance cluster). \textbf{Center}: dataset $\textrm{DS\_II}$. In this case $K$-means performs much worse than KSC, suggesting maintenance in regions not corresponding to actual maintenance events. \textbf{Bottom}: dataset $\textrm{DS\_III}$. The results of KSC and $K$-means are very similar. Even if both techniques predict maintenance when it has not been really performed, it is quite likely that in this case the operator should have performed the maintenance actions in a different way he did.}}
\label{mebios_kmeans}	
\end{figure}

$K$-means clustering is a standard method for finding clusters in a set of data-points. After choosing the desired number of cluster centers, the $K$-means procedure iteratively moves the centers to minimize the total within cluster variance. $K$-means has several drawbacks:  
\begin{itemize}
\item the results are strongly influenced by the initialization
\item the number of clusters should be provided by the user (this is done heuristically using a trial and error approach)
\item it can discover only spherical boundaries. 
\end{itemize}
Despite these disadvantages, it is still widely used since it works effectively in many scientific and industrial applications.   
Since it is the most popular method used in the industrial sector, here we present the results of $K$-means applied on the three datasets under analysis. Before discussing the results, it is worth to mention that, thanks to the model selection scheme of KSC described in the previous section, we give optimal parameters to $k$-means (number of clusters = $2$ and window size of concatenated accelerometer signals = $40$). Figure \ref{mebios_kmeans} visualizes the outcomes of the hard and soft clustering. Similarly to KSC, the soft clustering results are based on the distance between the final centroids and the datapoints in the input space (see \cite{prob_clust_israel}). Concerning datasets $\textrm{DS\_I}$ and $\textrm{DS\_III}$ the results of KSC and $K$-means are very similar, while in the dataset $\textrm{DS\_II}$ analysis $K$-means performs worse than KSC, indicating the need of maintenance where it was not really performed. In this case $K$-means would suggest too many maintenance actions, which is not cost-effective. Finally, in table \ref{table_KSC} the performance of KSC and $K$-means are evaluated according to a standard internal cluster quality measures, that is the mean silhouette value (MSV), (see appendix).                 

\section{Conclusions}
Predictive maintenance of industrial plants is receiving increasing attention in the last years due to advantages like cost  efficiency and automation. In this chapter we discussed the usage of the KSC method for maintenance strategy optimization based on real-time condition monitoring of an industrial machine. We used the data collected by accelerometers placed on the jaws of a Vertical Form Fill and Seal (VFFS) machine. After applying a windowing operation on the data in order to catch the deterioration process affecting the sealing jaws, we showed how KSC is able to recognize the presence of at least two working regimes of the VFFS machine, identifiable respectively as normal and critical operating condition. Moreover, we have also proposed a soft clustering output that can be interpreted as "probability" to maintenance. In this way KSC could help to optimize the timing of maintenance actions for the machine under study in order to maximize the production capacity.

\begin{table}[htpb]
\begin{center}
\begin{tabular}{|c|c|c|c|} 
\hline 
\textbf{Algorithm} & \textbf{$\textrm{DS\_I}$}  & \textbf{$\textrm{DS\_II}$} & \textbf{$\textrm{DS\_III}$}\\
\hline
\textbf{KSC} & $\mathbf{0.29}$ & $\mathbf{0.25}$ & $0.27$\\
\hline
\textbf{$K$-means} & $0.28$ & $0.12$ & $0.27$\\
\hline
\end{tabular}
\end{center}
\caption{\textbf{Cluster quality evaluation.} Mean Silhouette value (MSV), with best performance in bold. In the case of dataset $\textrm{DS\_II}$ the low value of MSV indicates that $K$-means does not succeed in correctly separating the normal behaviour and the maintenance cluster, as KSC on the other hand does.}
\label{table_KSC}
\end{table}

%%%%%%%%%%%%%%%%%%%%%%%%%%%%%%%%%%%%%%%%%%%%%%%%%%
% Keep the following \cleardoublepage at the end of this file, 
% otherwise \includeonly includes empty pages.
\cleardoublepage

% vim: tw=70 nocindent expandtab foldmethod=marker foldmarker={{{}{,}{}}}
%

%
  \graphicspath{{\chaptersdir/chapter6/image/}}%
  %\cleardoublepage%
  \chapter{Clustering Non-Stationary Data}
\label{ch:IKSC}
In this chapter we face the problem of clustering data in a non-stationary environment. We already experimented in this context in chapter \ref{ch:evclustering}, where the MKSC algorithm was proposed. However, in that case the focus was more on evolving networks and the model was based on a temporal smoothness assumption. Here an adaptive clustering model based on a different principle is proposed. The new method, called Incremental Kernel Spectral Clustering (IKSC), takes advantage of the out-of-sample property of kernel spectral clustering (KSC) to adjust the initial model over time. Thus, in contrast with other existing incremental spectral clustering techniques, we propose a model-based eigen-update, which guarantees high accuracy. On some toy data we will show the effectiveness of IKSC in modelling the cluster evolution over time via drifting, merging, splitting of clusters and so on. Then we analyse a real-world data-set consisting of PM$_{10}$ concentrations registered during a heavy pollution episode that took place in Northern Europe in the end of January $2010$. We will see how also in this case IKSC is able to recognize some interesting patterns and track their evolution across time, in spite of dealing with the complex dynamics of PM$_{10}$ concentration.

\section{General Overview}
In many real-life applications, from industrial processes \cite{bart_SCP} to the analysis of the blogosphere \cite{Ning_PR} we face the ambitious challenge of dealing with non-stationary data. Therefore researchers perceived the need of developing clustering methods that can model the complex dynamics of evolving patterns in a real-time fashion. This means that a clustering of the data during time-step $t$ must be provided before seeing any data for time-step $t+1$, with the data-distribution changing across time. Some adaptive clustering models with different inspiration are present in the literature: evolutionary spectral clustering techniques discussed in the previous chapter \cite{evol_clustering_chakrabarti,evol_spec_clustering,Langone_physicA,rocco_ICMSQUARE2012}, self-organizing time map (SOTM), where an adaptation of the standard SOM is used to discover the occurrence and explore the properties of temporal structural changes in data \cite{Sarlin_SOTM}, dynamic clustering via multiple kernel learning \cite{diego_IJCNN2013}, incremental $K$-means \cite{incr_kmeans}, where the algorithm at time $t$ is initialized with the centroids found at time $t-1$, data stream clustering algorithms aiming at analysing massive datasets by using limited memory and a single scanning of the data \cite{Guha_2003,Aggarwal_2003}, incremental clustering algorithms, whose main objective is to apply dynamic updates to the cluster prototypes when new data points arrive \cite{Can_1997,Gupta_2004,Ning_2007}, and many others. Here we focus our attention on the family of the Spectral Clustering (SC) approaches briefly discussed in chapter \ref{ch:spectralclustering} \cite{ngjordan,luxburg_tutorial,chung}, which has shown its practical success in many application domains. SC is an off-line algorithm, and the above-cited attempts to make it applicable to dynamic data-sets, although quite appealing, are at the moment not very computationally efficient. In \cite{Ning_PR} and more recently in \cite{Dhanjal_arXiv}, the authors propose some incremental eigenvalue solutions to continuously update the initial eigenvectors found by SC. In the present study, we follow this direction, but with an important difference. The incremental eigen-update we introduce is based on the out-of-sample extension property of Kernel Spectral Clustering (KSC), introduced in section \ref{sc:OOSE_KSC}. The out-of-sample extension alone, without the need of ad-hoc eigen-approximation techniques like the ones proposed in \cite{Ning_PR} and \cite{Dhanjal_arXiv}, can be used to accurately cluster stationary data-streams. However, if the data are generated according to some distribution which changes over time (i.e. non-stationary), the initial KSC model must be updated. In order to solve this issue we introduce the Incremental Kernel Spectral Clustering algorithm (IKSC). The IKSC method exploits the research carried out in \cite{carlos_IJCNN2011} to continuously adjust the initial KSC model over time, in order to learn the complex dynamics characterizing the non-stationary data.

\section{Incremental Kernel Spectral Clustering}
In contrast with other techniques that compute approximate eigenvectors of large matrices like the Nystr\"om method \cite{nystrom}, the work presented in \cite{update_kpca} or the above-mentioned algorithms \cite{Dhanjal_arXiv} and \cite{Ning_PR}, the eigen-approximation we use in the IKSC method is model-based \cite{carlos_IJCNN2011}. This means that based on a training set (in our case the cluster centroids) out-of-sample eigenvectors are calculated using eq. (\ref{eq_outofsample_eig}). These approximate eigenvectors are then used to adapt the initial clustering model over time. In principle, if the training model has been properly constructed, this guarantees high accuracy of the approximated eigenvectors due to the good generalization ability of KSC and LS-SVM in general \cite{multiway_pami,bench_lssvm}.

\subsection{Algorithm}
\label{IKSC_algorithm}
\begin{figure}[htbp]
\centering
\begin{tabular}{c}
\includegraphics[width=4in]{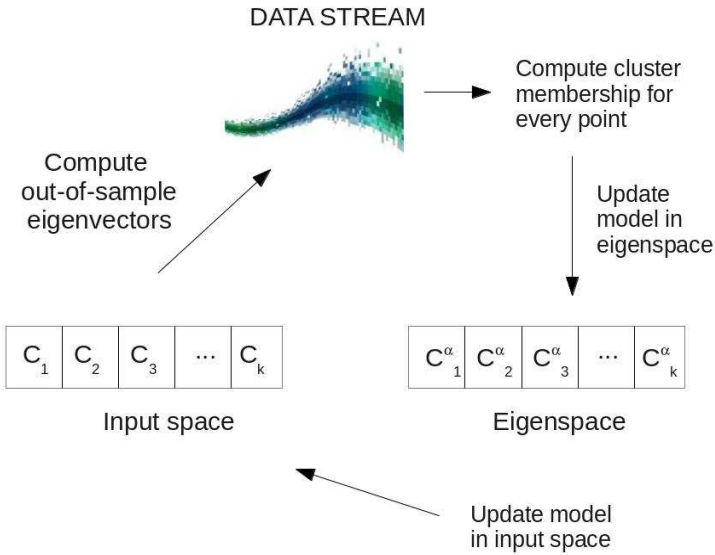}
\end{tabular}
\caption{\textbf{IKSC update scheme} After the initialization phase, when a new point arrives, both the training set and the model (i.e. the cluster centers in the eigenspace) are adapted accordingly.} 
\label{fig_IKSC}
\end{figure}

One big advantage of a model-based clustering tool like KSC is that we can use it online in a straightforward way. Indeed, once we built-up our optimal model during the training phase, we can estimate the cluster membership for every new test point by simply applying eq. (\ref{OOSE}) and the ECOC decoding procedure. However, if the data source is non-stationary, this scheme fails since the initial model is not representative any more of the new data distribution. Therefore to cope with non-stationary data the starting code-book must be adjusted accordingly.              
Here, instead of using the code-book and the ECOC procedure, we propose to express our model in terms of the centroids in the eigenspace and to compute the cluster memberships as measured by the euclidean distance from these centers. In this way it is possible to continuously update the model in response to the new data-stream. In order to calculate the projection in the eigenspace for every new point, we can exploit the second KKT condition for optimality related to the KSC optimization problem which links the eigenvectors and the score variables for training data:         
\begin{equation}
\label{eq_outofsample_eig}
\alpha_{\textrm{test}}^{(l)} = \frac{1}{\lambda_l}D_{\textrm{test}}^{-1}e_{\textrm{test}}^{(l)} 
\end{equation}
with $D_{\textrm{test}}^{-1} = \textrm{diag}(1/ \textrm{deg}(x_1^{\textrm{test}}),\hdots,1/ \textrm{deg}(x_{\textrm{Ntest}}^{\textrm{test}})) \in \mathbb{R}^{\textrm{Ntest}} \times \mathbb{R}^{\textrm{Ntest}}$ indicating the inverse degree matrix for test data. The out-of-sample eigenvectors $\alpha_{\textrm{test}}^{(l)}$ represent the model-based eigen-approximation with the same properties as the original eigenvectors $\alpha^{(l)}$ for training data. With the term eigen-approximation we mean that these eigenvectors are not the solution of an eigenvalue problem, but they are estimated by means of a model built during the training phase of KSC \cite{carlos_IJCNN2011}.       
To summarize, once one or more new points belonging to a data-stream are collected, we update the IKSC model as follows:
\begin{itemize}
\item calculate the out-of-sample extension using eq.(\ref{OOSE}), where the training points $x_i$ are the centroids in the input space $C_1,\hdots,C_k$, and the $\alpha^{(l)}$ are the centroids in the eigenspace $C_1^{\alpha},\hdots,C_k^{\alpha}$  
\item calculate the out-of-sample eigenvectors by means of eq (\ref{eq_outofsample_eig})
\item assign the new points to the closest centroids in the eigenspace
\item update the centroids in the eigenspace
\item update the centroids in the input space
\end{itemize}
To update online a centroid $C_{\textrm{old}}$ given a new sample $x_{\textrm{new}}$, we can use the following formula \cite{Knuth1997}:
\begin{equation}
\label{eq_update_centers}
C_{\textrm{new}} = C_{\textrm{old}} + \frac{x_{\textrm{new}} - C_{\textrm{old}}}{n_{\textrm{old}}} 
\end{equation}
where $n_{\textrm{old}}$ is the number of samples previously assigned to the cluster center $C_{\textrm{old}}$. The same procedure can be used to update the cluster centers in the eigenspace: in this way the initial $\alpha^{(l)}$ provided by KSC are changed over time to model the non-stationary behaviour of the system. A schematic visualization of this procedure is depicted in figure \ref{fig_IKSC}. Finally, the complete IKSC technique is sketched in algorithm \ref{alg:IKSC}. 
\begin{algorithm}[t]
\small
\LinesNumbered
\KwData{Training set $\mathcal{D}_{\textrm{Tr}} = \{x_i\}_{i=1}^{N_{\textrm{Tr}}}$, kernel function $K: \mathbb{R}^d \times \mathbb{R}^d \rightarrow \mathbb{R}$ positive definite and localized ($K(x_i,x_j) \rightarrow 0$ if $x_i$ and $x_j$ belong to different clusters), kernel parameters (if any), number of clusters $k$.}
\KwResult{Clusters $\{\mathcal{A}_1,\hdots,\mathcal{A}_p\}$, cluster centroids in the input space $C_1,\hdots,C_k$, cluster centroids in the eigenspace $C_1^{\alpha},\hdots,C_k^{\alpha}$.} 
Acquire $N_{\textrm{Tr}}$ points.\\ 
Train the KSC model by solving eq. (\ref{dual_KSC}).\\  
Obtain the initial centroids in the input space $C_1,\hdots,C_k$ and the initial centroids in the eigenspace  $C_1^{\alpha},\hdots,C_k^{\alpha}$.\\  
\For{$i\leftarrow N_{\textrm{Tr}}+1$ \KwTo $N$}{
compute the out-of-sample eigenvectors using eq. (\ref{eq_outofsample_eig})\\ 
calculate cluster membership for the new point (or the new batch of points) according to the distance between the out-of-sample eigenvectors and the centroids $C_1^{\alpha},\hdots,C_k^{\alpha}$\\
update centroids in the eigenspace $C_1^{\alpha},\hdots,C_k^{\alpha}$ using eq. (\ref{eq_update_centers})\\
update centroids in the input space $C_1,\hdots,C_k$ according to eq. (\ref{eq_update_centers})\\
new cluster check\\
merge check\\
cluster death\\
}
Outlier elimination.
\caption{IKSC algorithm \cite{IKSC_paper}}
\label{alg:IKSC}
\end{algorithm}
The adaptation to non-stationarities relates to identifying changes in the number of clusters occurring over time by means of some inspections:
\begin{itemize}
\item the new cluster check allows to dynamically create a new cluster if necessary. For every new point the related degree $d_i^{\textrm{test}}$ is calculated. If $d_i^{\textrm{test}} < \epsilon$ where $\epsilon$ is a user-defined threshold, it means that the point is dissimilar to the actual centroids. Therefore it becomes the centroid of a new cluster and it is added to the model. The threshold $\epsilon$ is data-dependent, and can be chosen before processing the data stream based on the degree distribution of the test kernel matrix, when considering as training set the cluster prototypes in the input space.   
\item throughout the merge check, if two centroids become too similar they are merged into one center, and the number of clusters is decreased. In particular, the similarity between two centroids is computed as the cosine similarity in the eigenspace, and two centroids are merged if this similarity is greater than $0.5$.     
\item if the centroid of a cluster is not updated any more the algorithm considers that cluster as disappeared (cluster death). 
\end{itemize}
Finally, if one cluster has size smaller than a user-defined threshold it is considered as outlier and it is eliminated in the end of the data-stream acquisition.

\subsection{Computational Complexity}
\label{cpu_time}
Let us assume that at the beginning of the data-stream acquisition the training set $\mathcal{D}=\{x_i\}_{i=1}^k$, $x_i\in \mathbb{R}^d$ is formed by $k$ centroids in the input space $C_1,\hdots,C_k$ and the initial clustering model is expressed by the centroids in the eigenspace 
$C_1^{\alpha},\hdots,C_k^{\alpha}$, with $C_i^{\alpha} \in \mathbb{R}^{k-1}$. For every new point of the data-stream, as explained in the previous section, we have to compute the out-of-sample extension, the corresponding out-of-sample eigenvectors by means of eq. (\ref{eq_outofsample_eig}) and the update of both the model and the training set, that are the clusters centers in the input space. Since the training set can be kept small enough\footnote{In this work we assume that the training set consists of $k$ points, where $k$ is the number of clusters. In some situations it could happen that such a small number of training points is not enough to define a proper mapping. Nevertheless, by considering more training points $N$ such that $N \ll N_{\textrm{test}}$ the overall complexity of the algorithm does not change.}, the main contribution to the computational complexity is due to the out-of-sample extension part:
\begin{equation}
e_{\textrm{test}}^{(l)} = \Omega_{\textrm{test}}\alpha^{(l)} + b_l1_{\textrm{Ntest}}, l=1, \hdots ,k-1. 
\end{equation}        
The evaluation of the kernel matrix $\Omega_{\textrm{test}}$ needs $O(k^2d)$ operations to be performed. The calculation of the score variables $e_{\textrm{test}}^{(l)}$ takes then $O(k^2d + k^2 + k)$ time. This operation has to be repeated for the $N_{\textrm{test}}$ data-points of the data-stream, so the overall time complexity is $O(N_{\textrm{test}}(k^2d + k^2 + k))$. This can become linear with respect to the number of data-points ($O(N_{\textrm{test}})$) when $k \ll N_{\textrm{test}}$ and $d \ll N_{\textrm{test}}$, which is the case in many applications. This is comparable with other eigen-updating algorithms for spectral clustering like \cite{Ning_PR} and \cite{Dhanjal_arXiv}.

\section{Synthetic Experiments}
\label{artificial_data}

\subsection{Description of the data}
\begin{figure}[htbp]
\centering
\begin{tabular}{cc}
\includegraphics[width=2in]{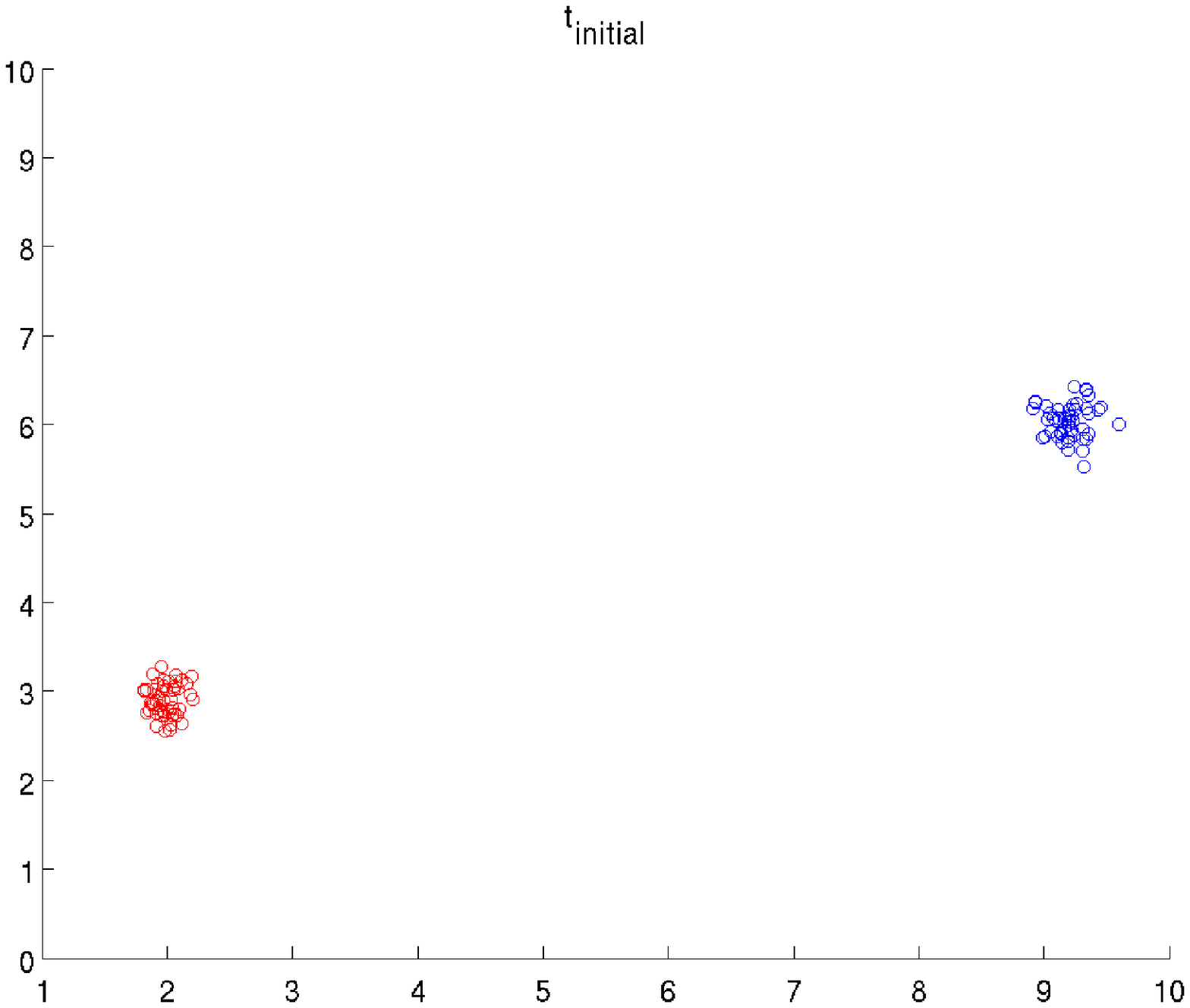}&
\includegraphics[width=2in]{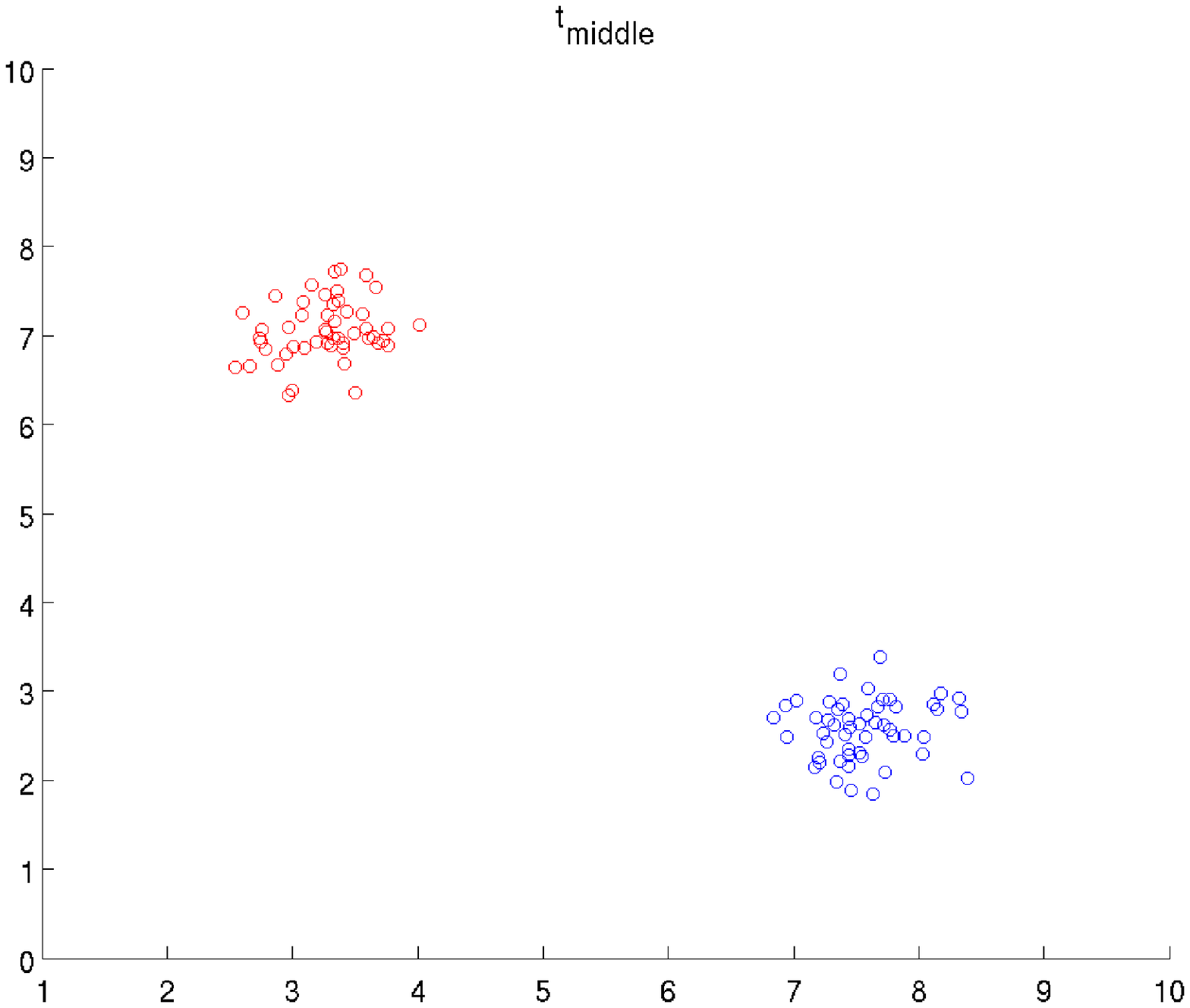}\\
\includegraphics[width=2in]{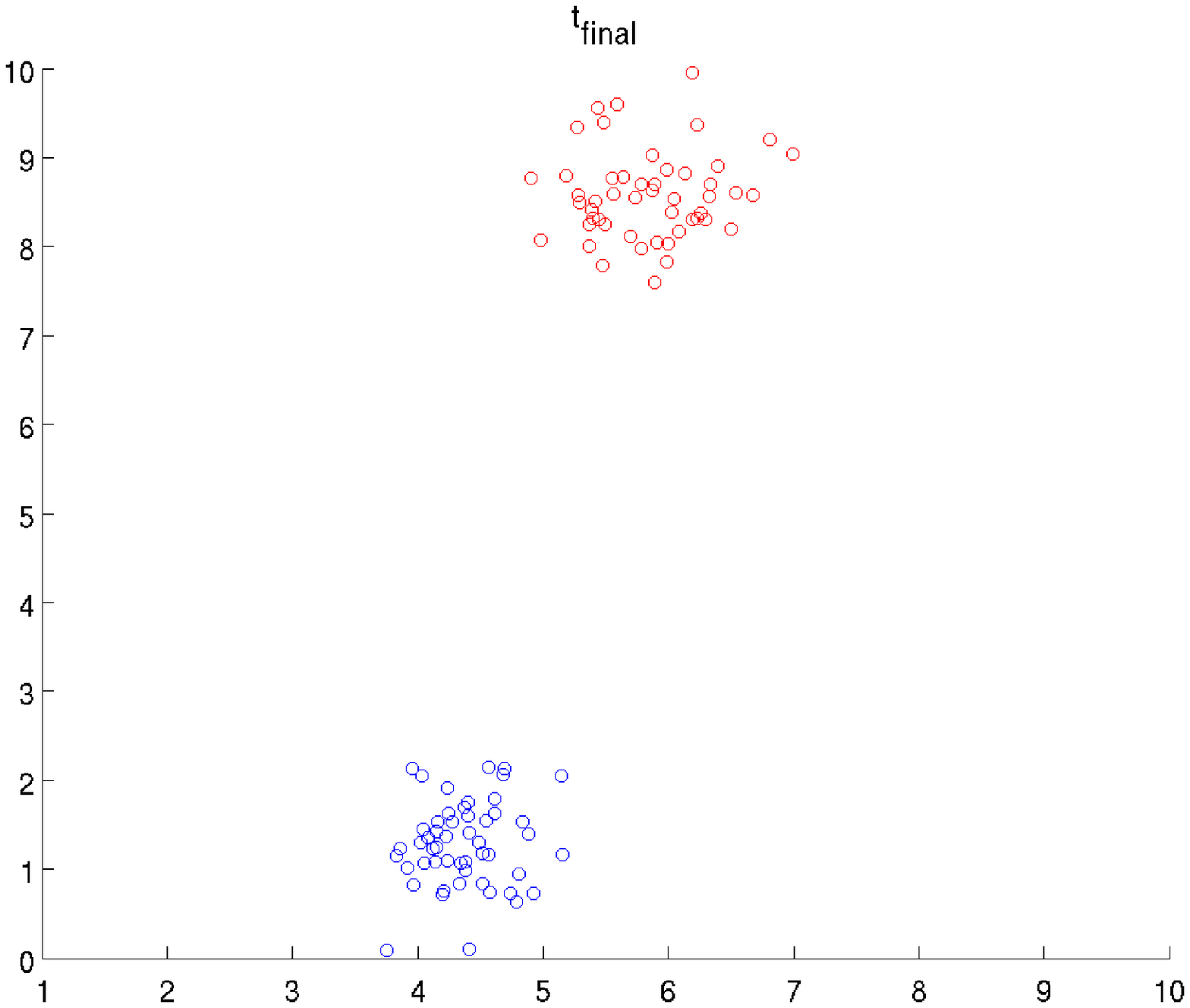}&
\includegraphics[width=2in]{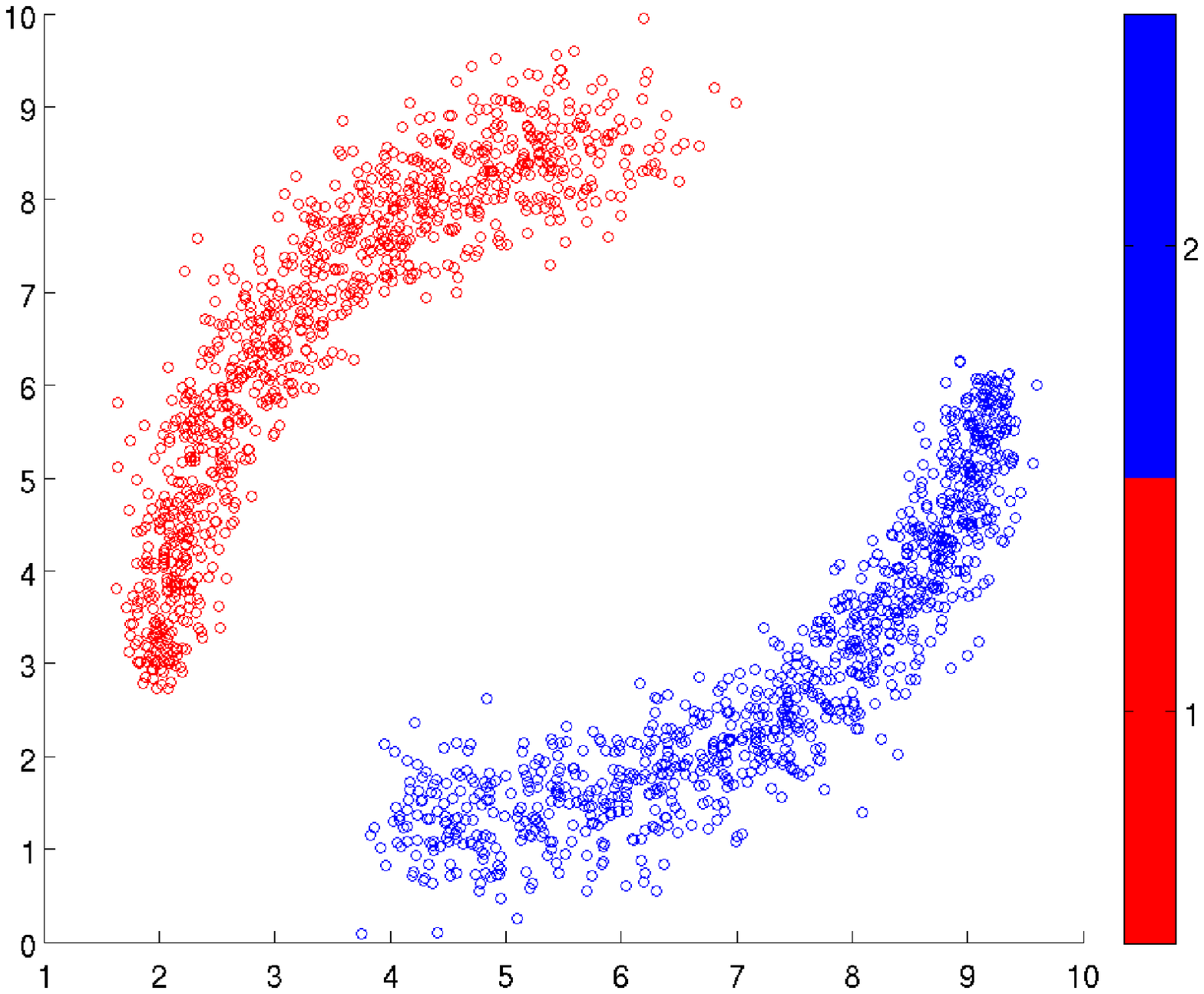}
\end{tabular}
\caption{\textbf{Drifting Gaussian distributions dataset.} Some snapshots of the evolution of the distributions (top and bottom left), and the whole data all at once (bottom right). The axes represent the two dimensions of the $N$ input vectors $x_i \in \mathbb{R}^2, i=1,\hdots,N$.}
\label{gauss_drift1}
\end{figure}

\begin{figure}[htbp]
\centering
\begin{tabular}{cc}
\includegraphics[width=2in]{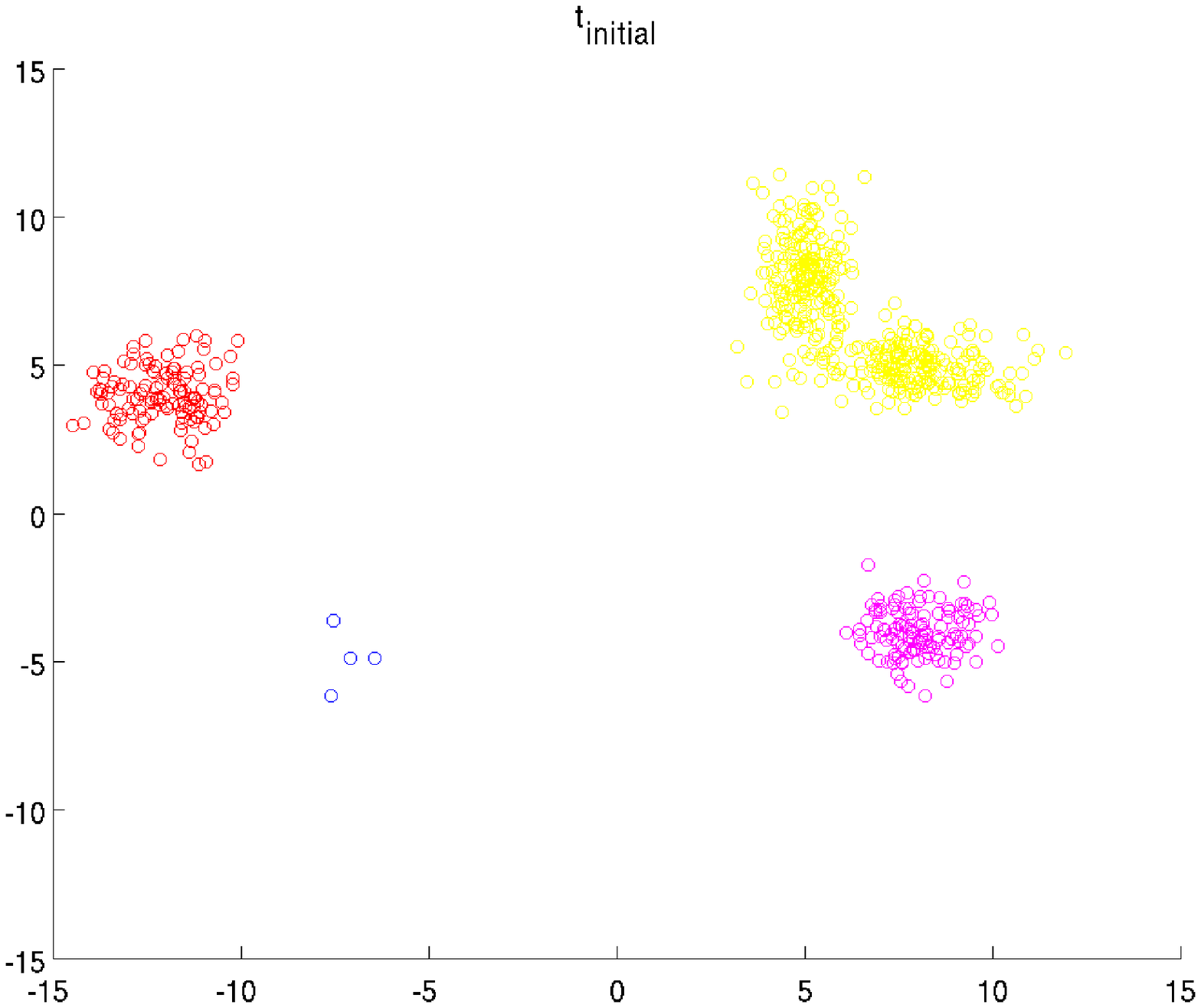}&
\includegraphics[width=2in]{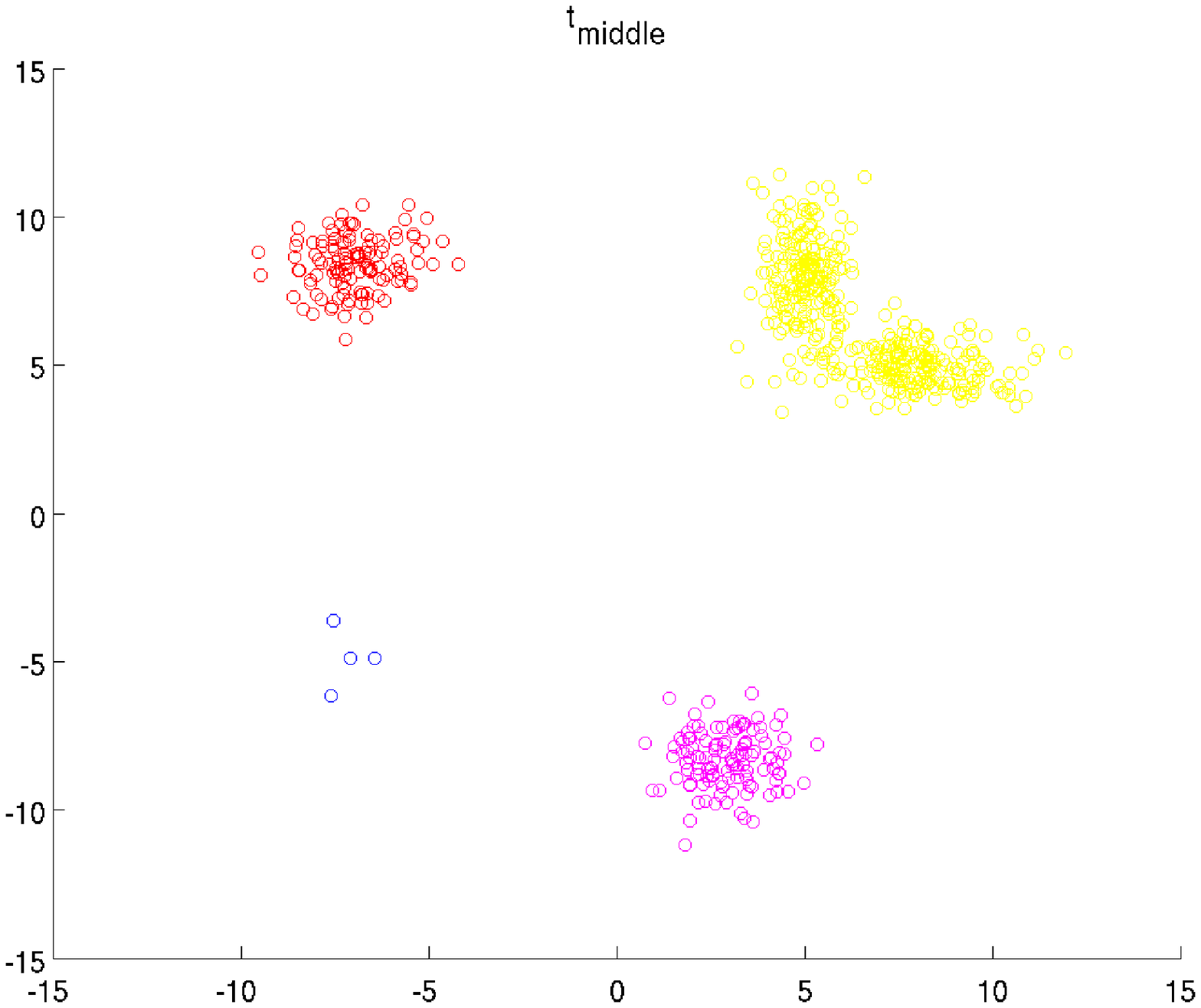}\\
\includegraphics[width=2in]{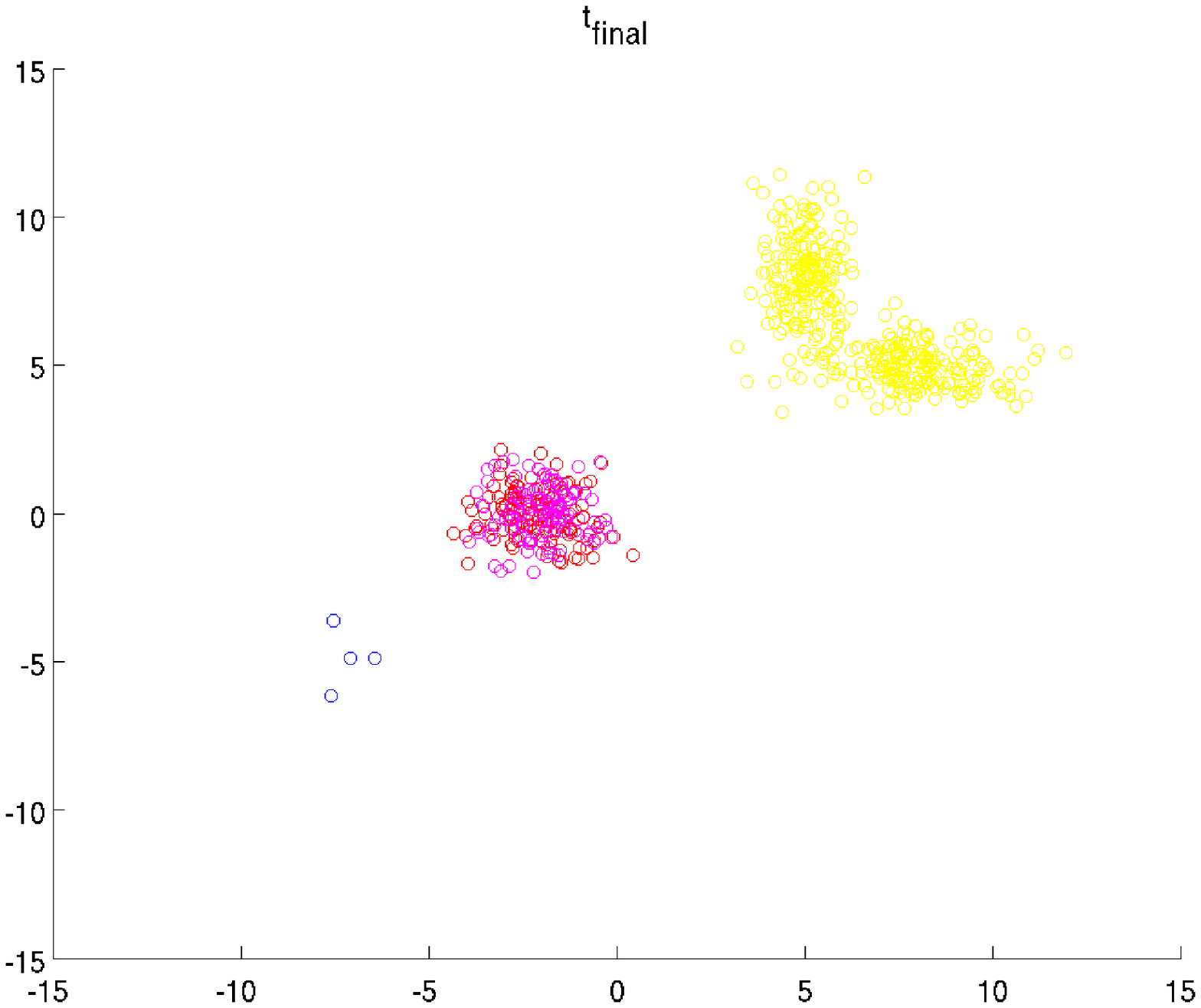}&
\includegraphics[width=2in]{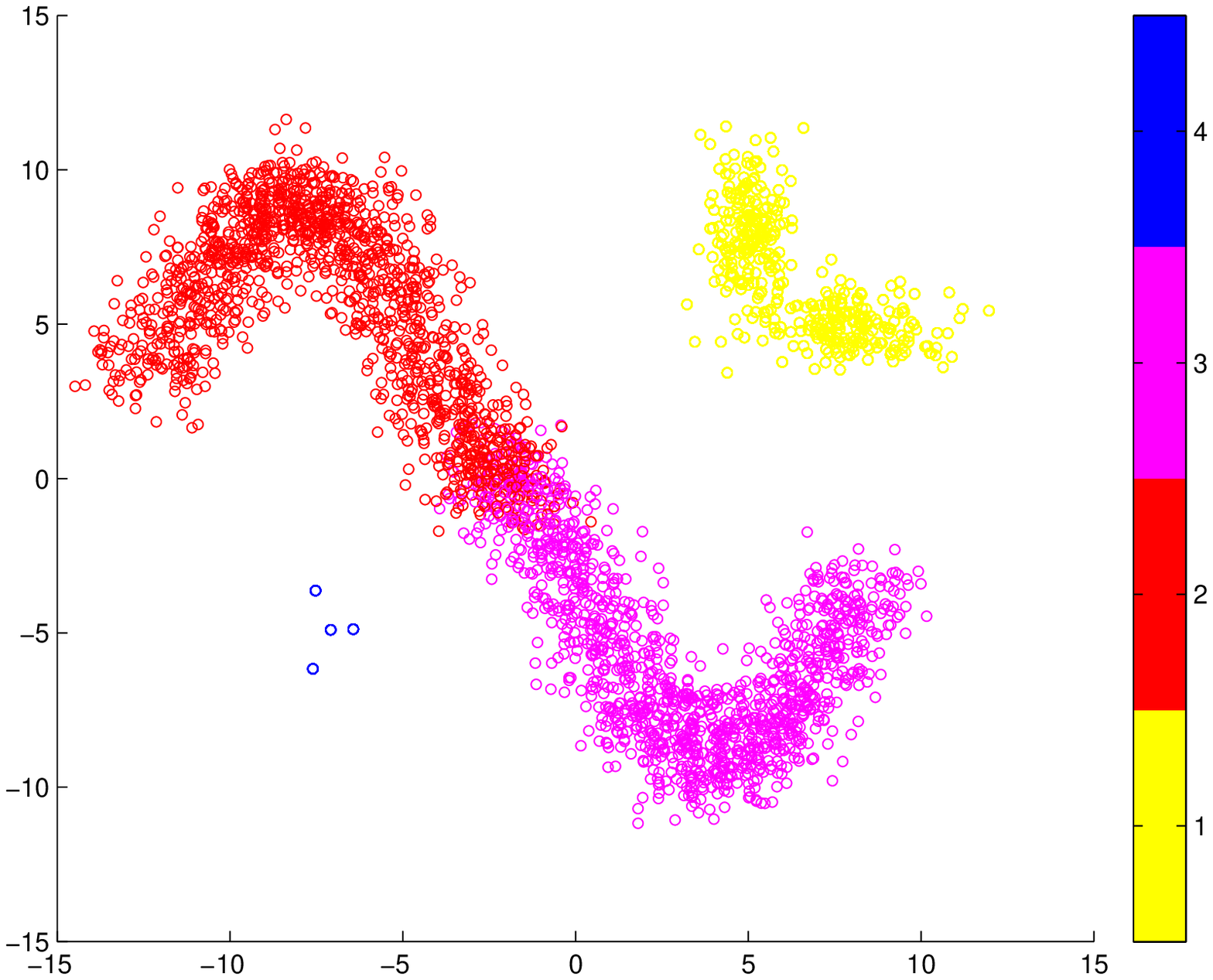}
\end{tabular}
\caption{\textbf{Merging Gaussian distributions dataset.} Some snapshots of the evolution of the distributions (top and bottom left), and the whole data all at once (bottom right). The axes represent the two dimensions of the $N$ input vectors $x_i \in \mathbb{R}^2, i=1,\hdots,N$.}
\label{gauss_drift2}
\end{figure}

\begin{figure}[htbp]
\centering
\begin{tabular}{c}
\includegraphics[width=4.5in]{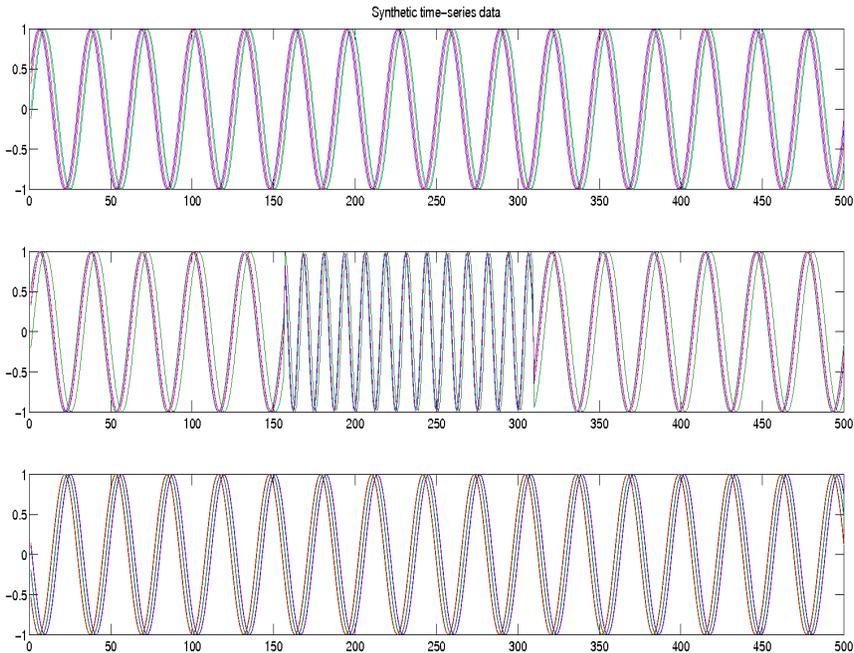}
\end{tabular}
\caption{\textbf{Synthetic time-series.} The $X$ axis indicates the time-step and the $Y$ axis the amplitude of the signals. At $t_1=150$ and $t_2=300$ two change points (change in frequency) can be observed.}
\label{ts_example}
\end{figure}

Three simulations are performed: the first and the second by reproducing the experiments described in \cite{SAKM_boubacar}, and the third with some computer-generated time-series. In the first simulation two Gaussian distributions evolving over time are created. These two clouds of points drift toward each other with increasing dispersal, as illustrated in figure \ref{gauss_drift1}. In the second virtual experiment a multi-cluster non-stationary environment is created. In particular, there are two drifting Gaussian clouds that come to merge, some isolated data forming an outlier cluster of $4$ points and a static cluster consisting of a bi-modal distribution. This second data-set is depicted in figure \ref{gauss_drift2}. In order to test the ability of IKSC to dynamically cluster time-series rather than data-points, we generated $20$ time-series of three types as depicted in figure \ref{ts_example}. The idea behind this experiment is that if we cluster in an online fashion the time-series with a moving window approach, we should be able to detect the appearance of a new cluster given the increase in frequency of the signals of the second type at time step $t_1=150$. Moreover, when these signals get back to their original frequency at time step $t_2 = 300$, the clustering algorithm must detect this change.

In the first two synthetic experiments (Drifting Gaussians and Merging Gaussians) we use the RBF kernel function defined by $K(x_i,x_j) = \textrm{exp}(-||x_i-x_j||^2_2/\sigma^2)$. The symbol $\sigma$ indicates the bandwidth parameter and $x_i$ is the $i$-th data point. In the analysis of the third synthetic data (Synthetic time-series) $x_i$ represents the $i$-th time-series. In this case to better capture the similarity between the time-series we use the RBF kernel with the correlation distance \cite{timeseries_clustering}. Thus $K(x_i,x_j) = \textrm{exp}(-||x_i-x_j||^2_{\textrm{cd}}/\sigma^2)$, where  $||x_i-x_j||_{\textrm{cd}} = \sqrt{\frac{1}{2}(1-R_{ij})}$, with $R_{ij}$ indicating the Pearson correlation coefficient between time-series  $x_i$ and $x_j$.   
By means of extensive experiments we empirically observed that this kernel is positive definite. Moreover the RBF kernel with Euclidean distance has been mathematically proven to fulfil the positive definitiveness property.

\subsection{Simulation results}
Here we show how the proposed IKSC model, thanks to its capacity of adapting to a changing environment, is able to model the complex behaviour of evolving patterns of non-stationary data. To evaluate the outcomes of the model, two cluster quality measures are computed \cite{Halkidi01onclustering}: the average cumulative adjusted rand index (ARI) error defined as $1-\textrm{ARI}$, like in \cite{Dhanjal_arXiv} and the instantaneous mean silhouette value (MSV). 

\begin{figure}[htbp]
\centering
\begin{tabular}{c}
\begin{psfrags}
\psfrag{x}[c][c]{{$x^{(1)}$}}
\psfrag{y}[c][c]{{$x^{(2)}$}}
\psfrag{t}[c][c]{{Input space}}
\includegraphics[width=3in]{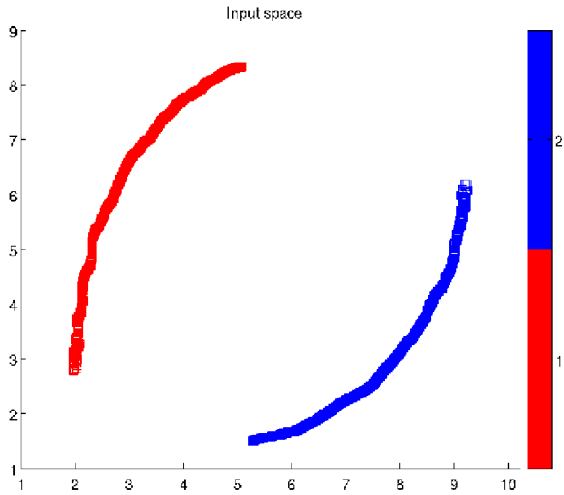}
\end{psfrags}
\end{tabular}
\caption{\textbf{Results of IKSC on the drifting Gaussian distributions.} Evolution of the centroids in the 2D input space. We can notice that the IKSC model can recognize the drifting targets without errors.}
\label{results_gaussdrift1}
\end{figure}

\begin{figure}[htbp]
\centering
\begin{tabular}{c}
\begin{psfrags}
\psfrag{x}[t][t]{{Time instant t}}
\psfrag{y}[b][b]{{MSV}}
\psfrag{t}[b][b]{{}}
\includegraphics[width=3in]{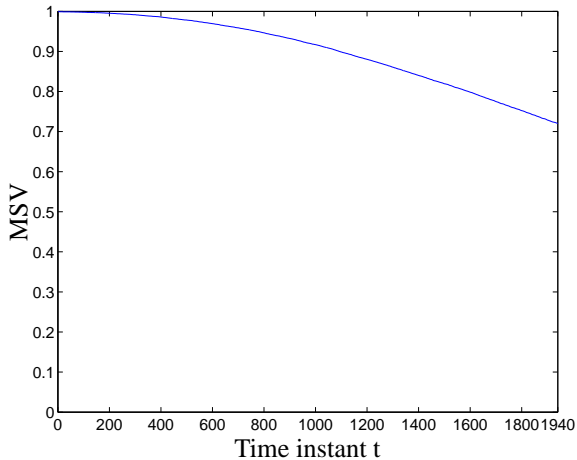}
\end{psfrags}
\end{tabular}
\caption{\textbf{MSV for drifting Gaussian distributions.} The mean silhouette value related to the clusters detected by IKSC stays high over time, meaning that our method is able to model the drift of the distributions.}
\label{sil_gaussdrift1}	
\end{figure}

\begin{figure}[htbp]
\centering
\begin{tabular}{c}
\begin{psfrags}
\psfrag{x}[t][t]{{$x^{(1)}$}}
\psfrag{y}[b][b]{{$x^{(2)}$}}
\psfrag{t}[b][b]{{Input space}}
\includegraphics[width=3in]{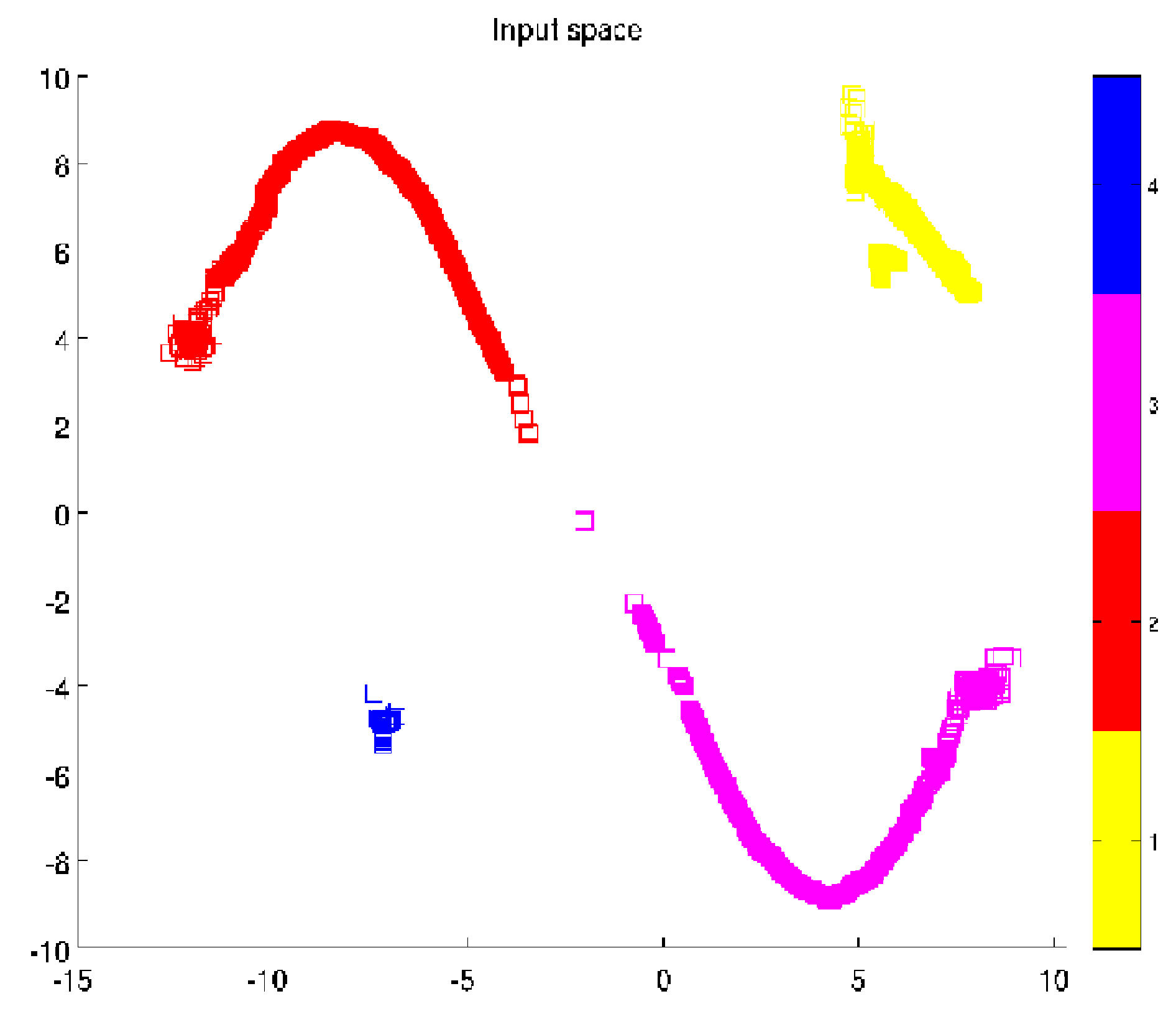}
\end{psfrags}\\
\begin{psfrags}
\psfrag{x}[][]{{$\alpha^{(1)}$}}
\psfrag{y}[][]{{$\alpha^{(2)}$}}
\psfrag{z}[][]{{$\alpha^{(3)}$}}
\psfrag{t}[][]{{Eigenspace}}
\includegraphics[width=3in]{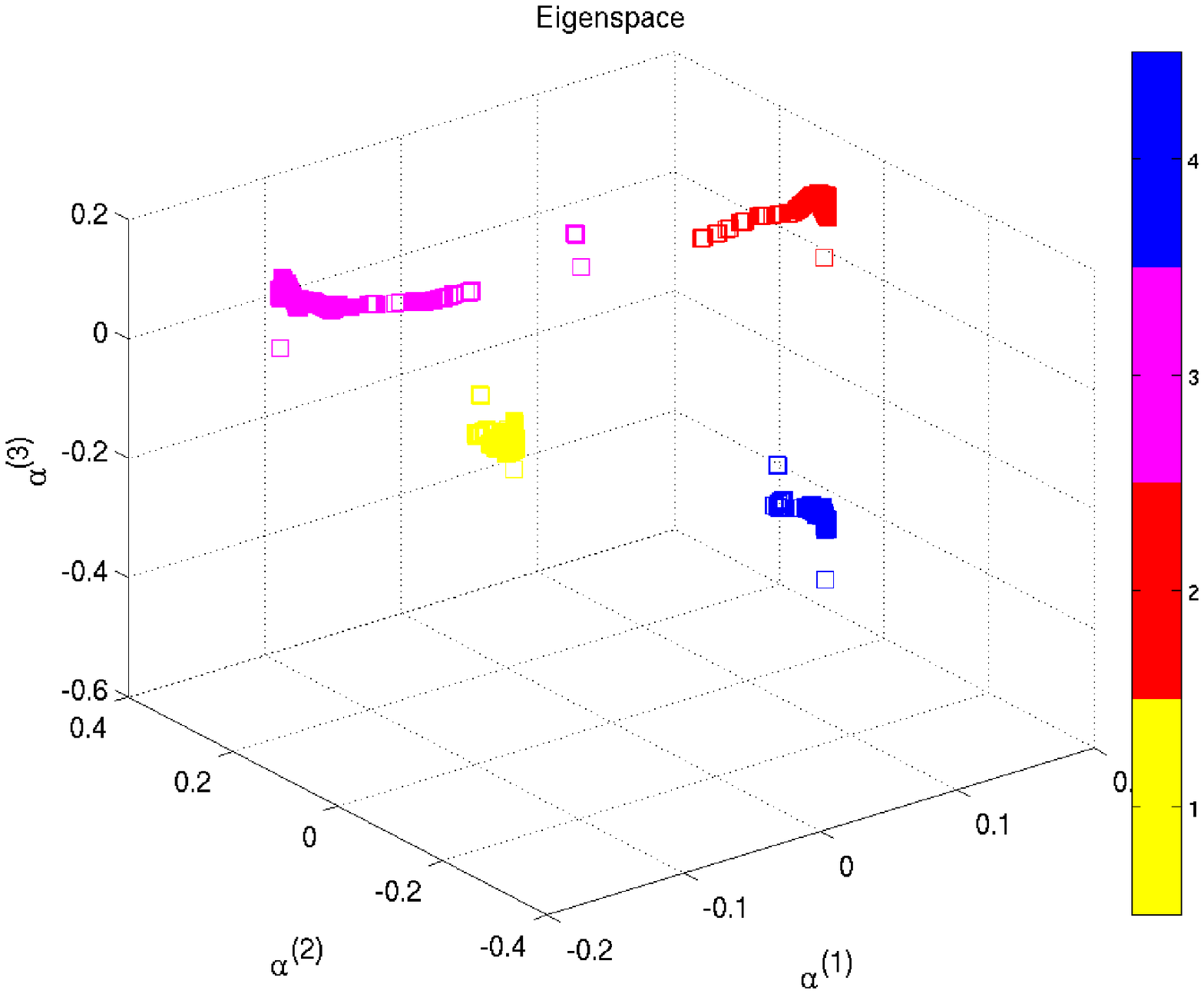}
\end{psfrags}
\end{tabular}
\caption{\textbf{Results of IKSC on the merging Gaussian distributions.} \textbf{Top:} Evolution of the centroids in the 2D input space. \textbf{Bottom:} Model evolution in the eigenspace.}
\label{results_gaussdrift2}
\end{figure}

\begin{figure}[htbp]
\centering
\begin{tabular}{c}
\begin{psfrags}
\psfrag{x}[t][t]{{time instant t}}
\psfrag{y}[b][b]{{MSV}}
\psfrag{t}[b][b]{{}}
\includegraphics[width=3in]{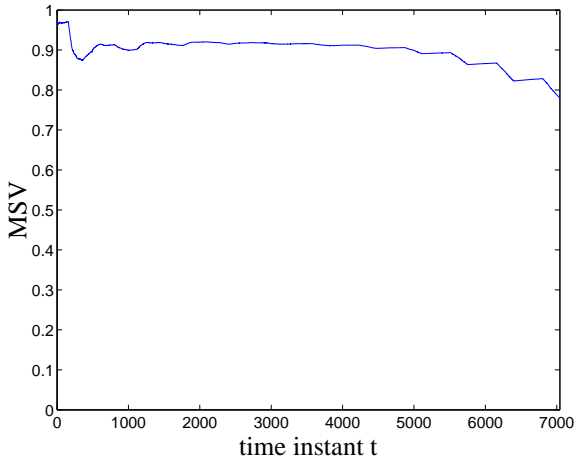}
\end{psfrags}
\end{tabular}
\caption{\textbf{MSV for the merging Gaussian distributions experiment.} The mean silhouette value related to the clusters detected by IKSC remains high over time. Thus, also in this case IKSC manages to properly follow the non-stationary behaviour of the clusters for the whole duration of the experiment.}
\label{sil_gaussdrift2}	
\end{figure}

\begin{figure}[htbp]
\centering
\begin{tabular}{c}
\begin{psfrags}
\psfrag{x}[t][t]{{time instant t}}
\psfrag{y}[b][b]{{Cumulative ARI error}}
\psfrag{t}[b][b]{{}}
\includegraphics[width=3in]{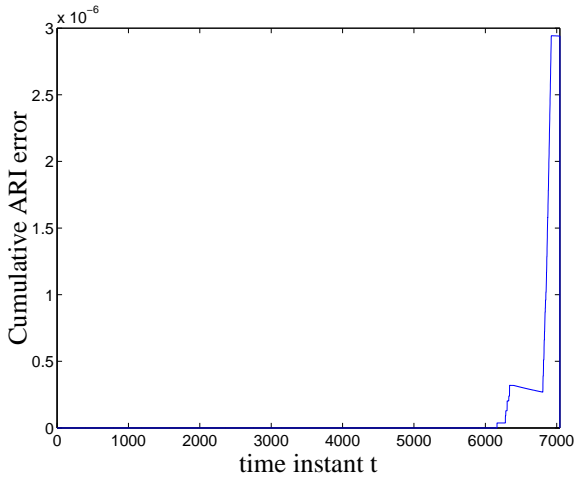}
\end{psfrags}
\end{tabular}
\caption{\textbf{ARI error for Merging Gaussian distributions.} The average cumulative ARI error related to the clusters detected by IKSC is very small over time, with a peak around the merging step at time $t=6926$, in agreement with what was observed also in \cite{SAKM_boubacar}.}
\label{ARIerror_gaussdrift2}	
\end{figure}

\begin{figure}[htbp]
%\centering
\hspace{-5mm}
\begin{tabular}{ll}
\includegraphics[width=2.4in]{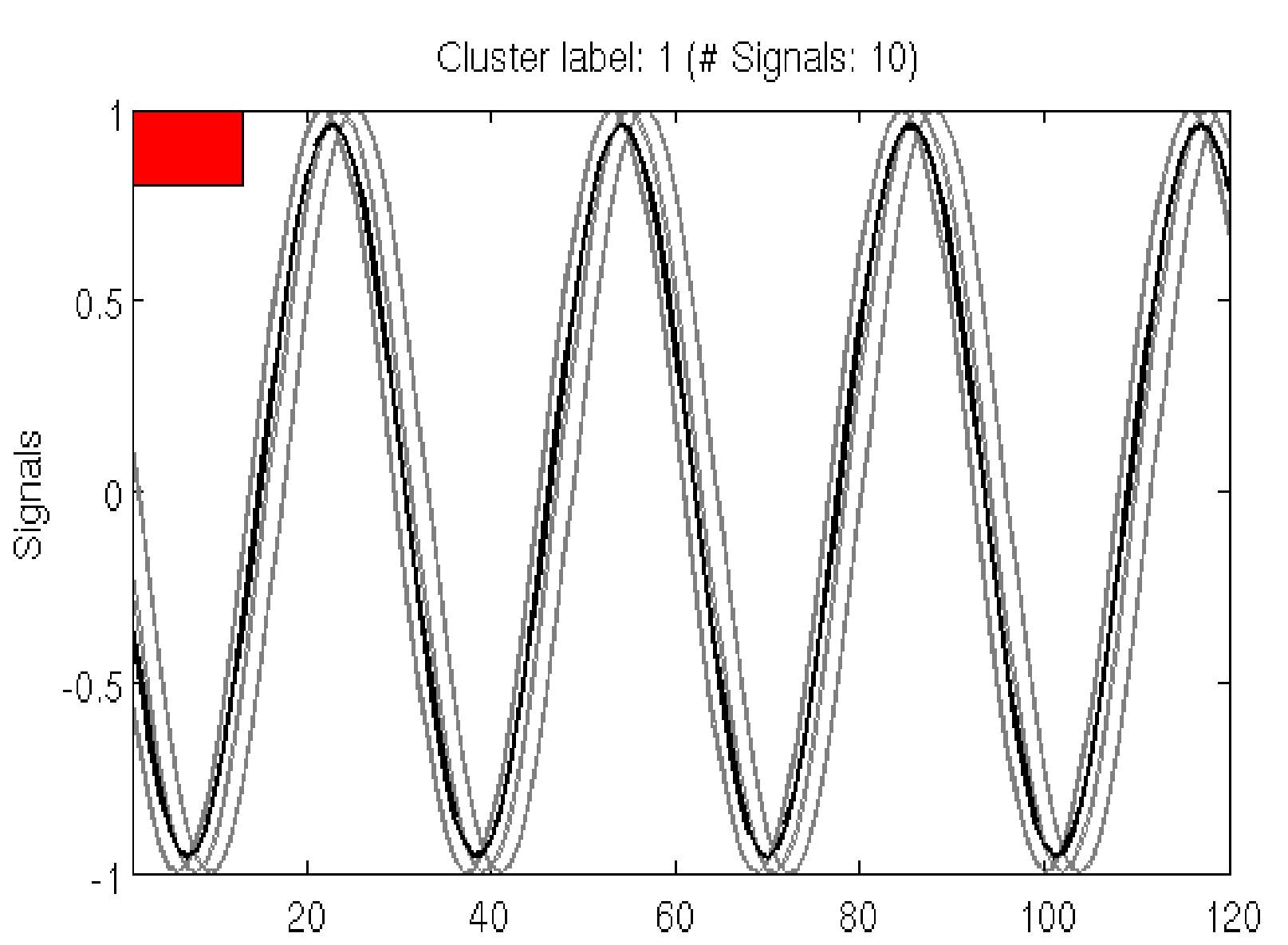}&\includegraphics[width=2.25in]{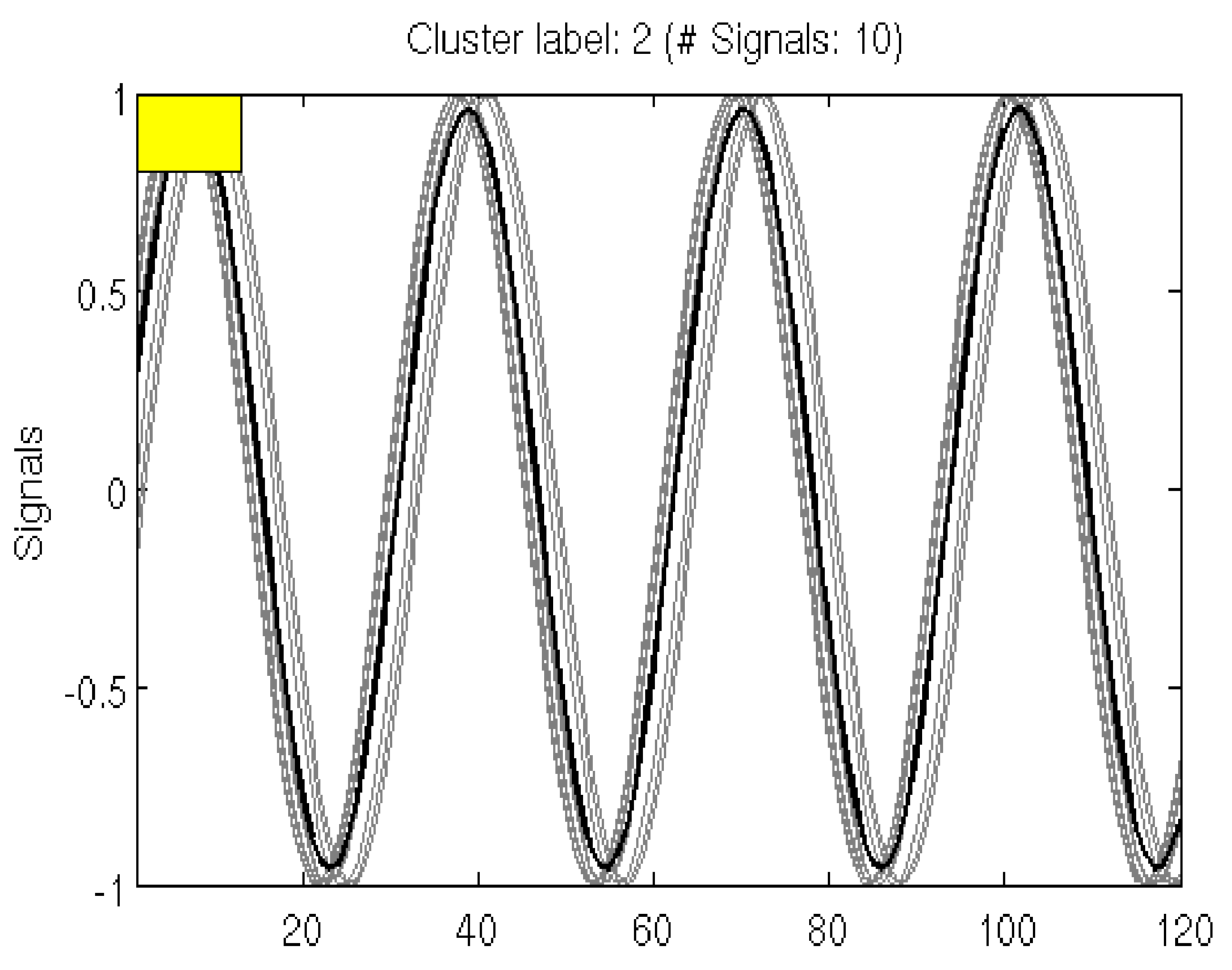}\\
\includegraphics[width=2.5in]{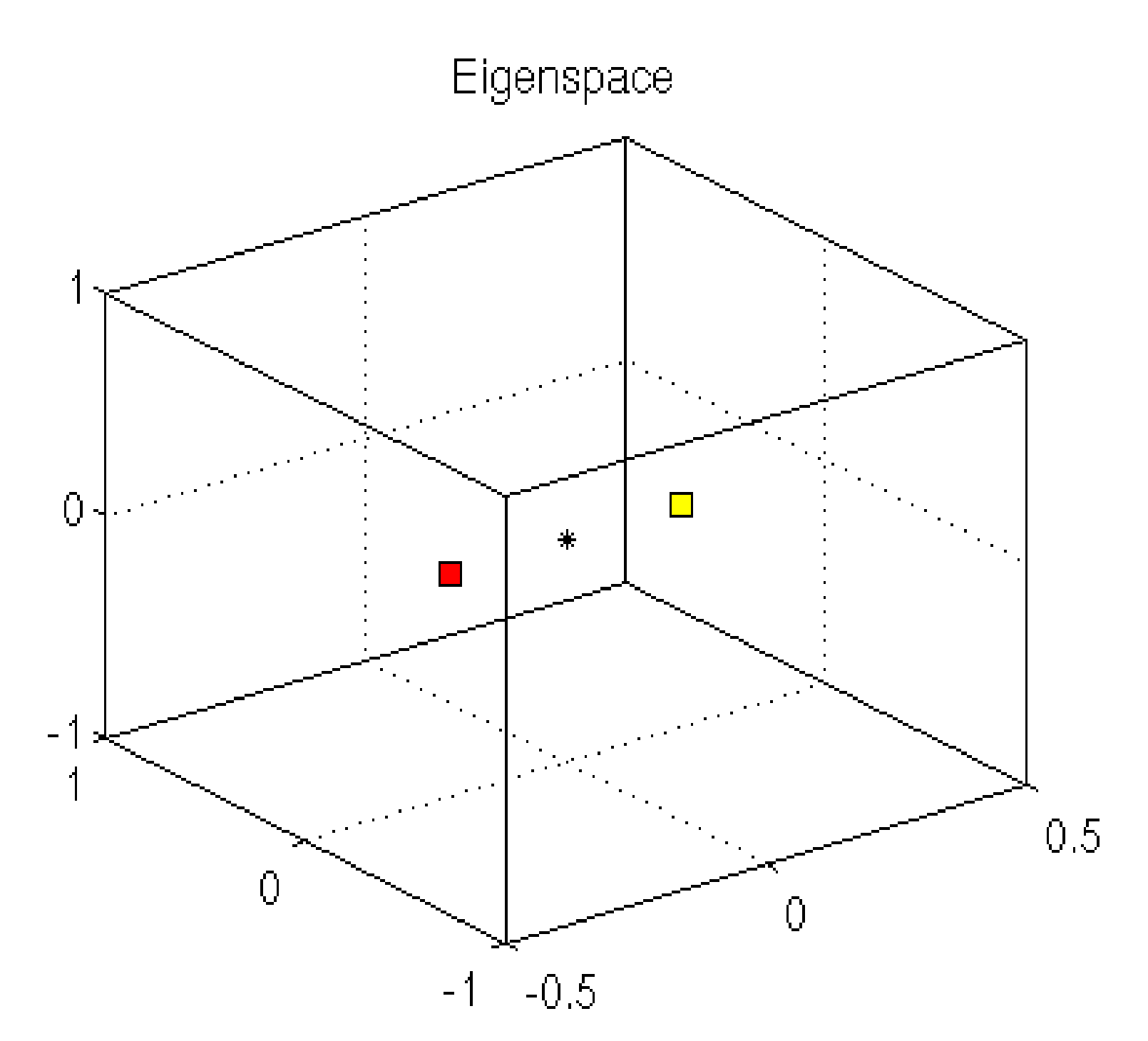}&\includegraphics[width=2in]{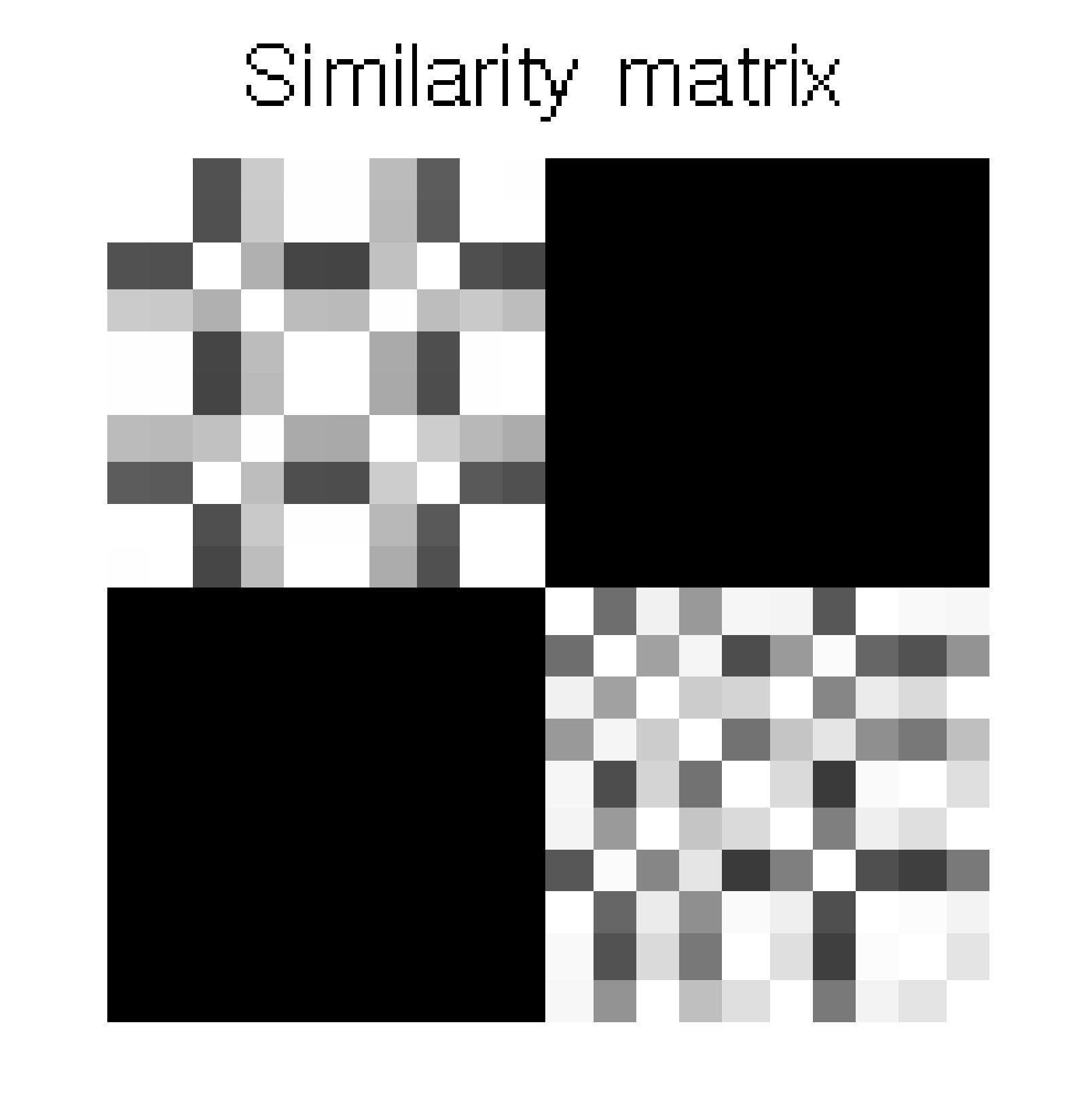}
\end{tabular}
\caption{\textbf{Synthetic time-series initial clustering model.} \textbf{Top:} signals of the two starting clusters. \textbf{Bottom left}: data in the eigenspace spanned by $\alpha^{(1)}$ (the points are mapped in the same location as the related centroids, since the eigenvectors are perfectly piece-wise constant). \textbf{Bottom right}: kernel matrix with a clear block diagonal structure.}
\label{results_ts1}	
\end{figure}

\begin{figure}[htbp]
%\centering
\hspace{-5mm}
\begin{tabular}{ll}
\includegraphics[width=2.25in]{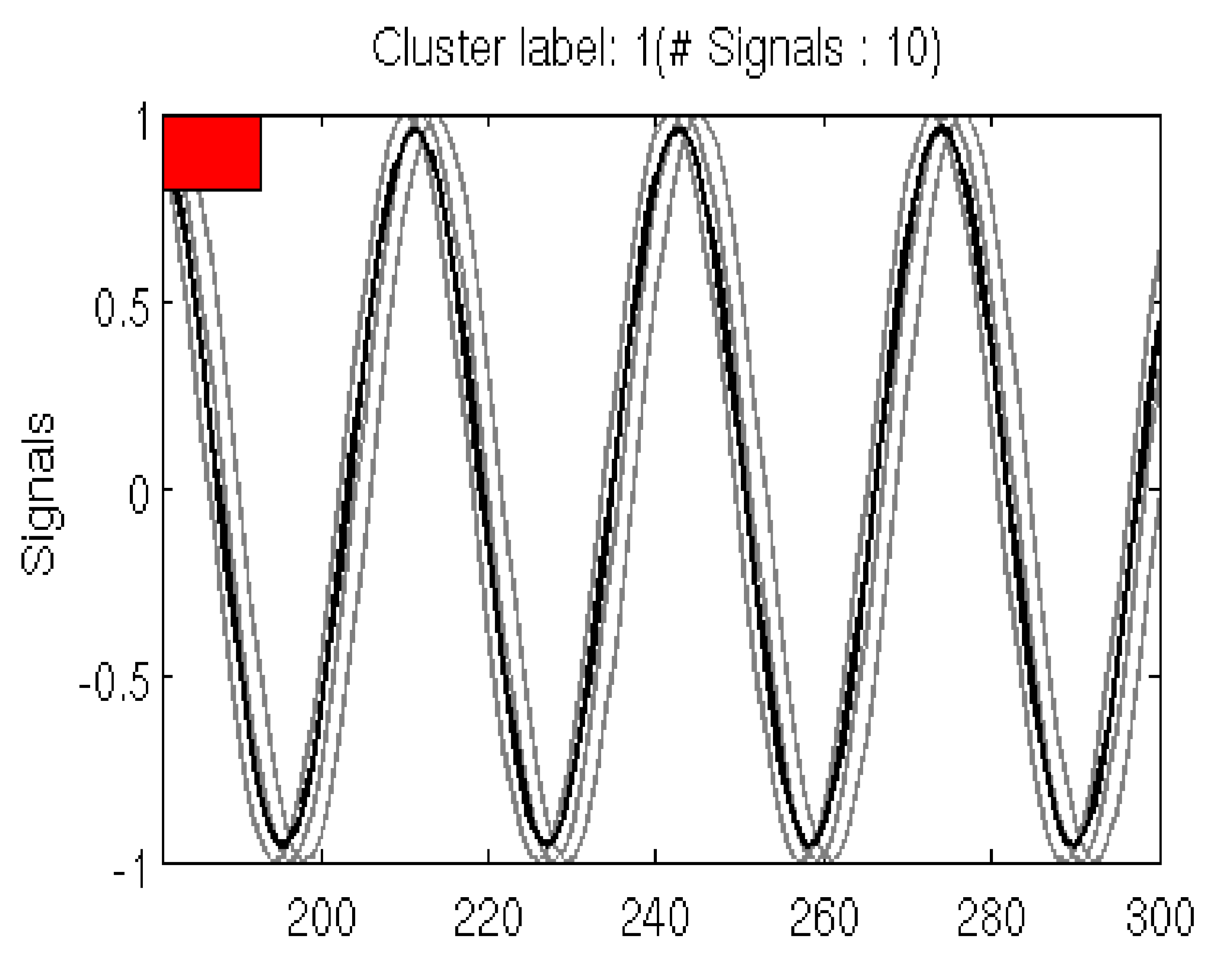}&\includegraphics[width=2.25in]{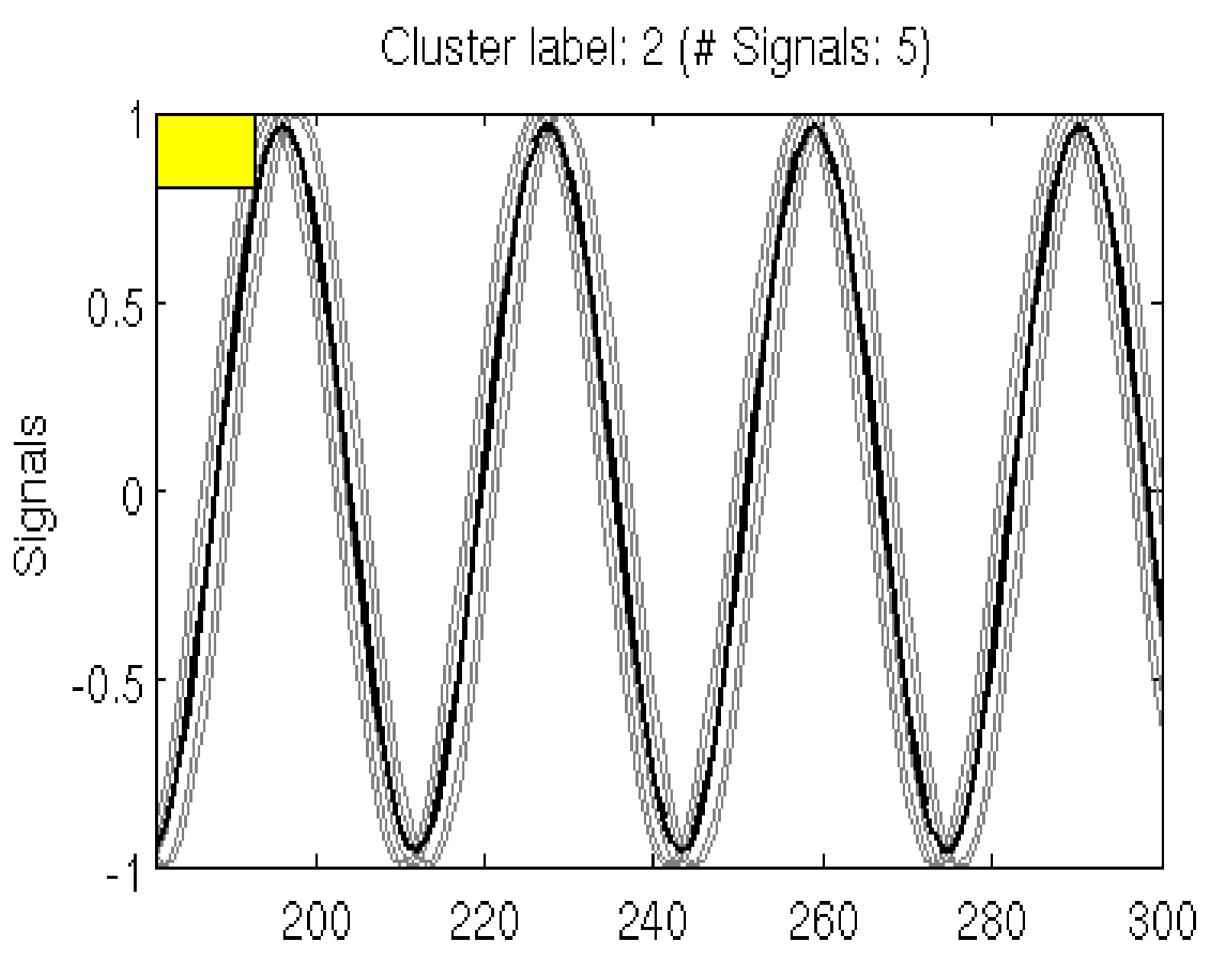}\\
\multicolumn{2}{c}{\includegraphics[width=2.5in]{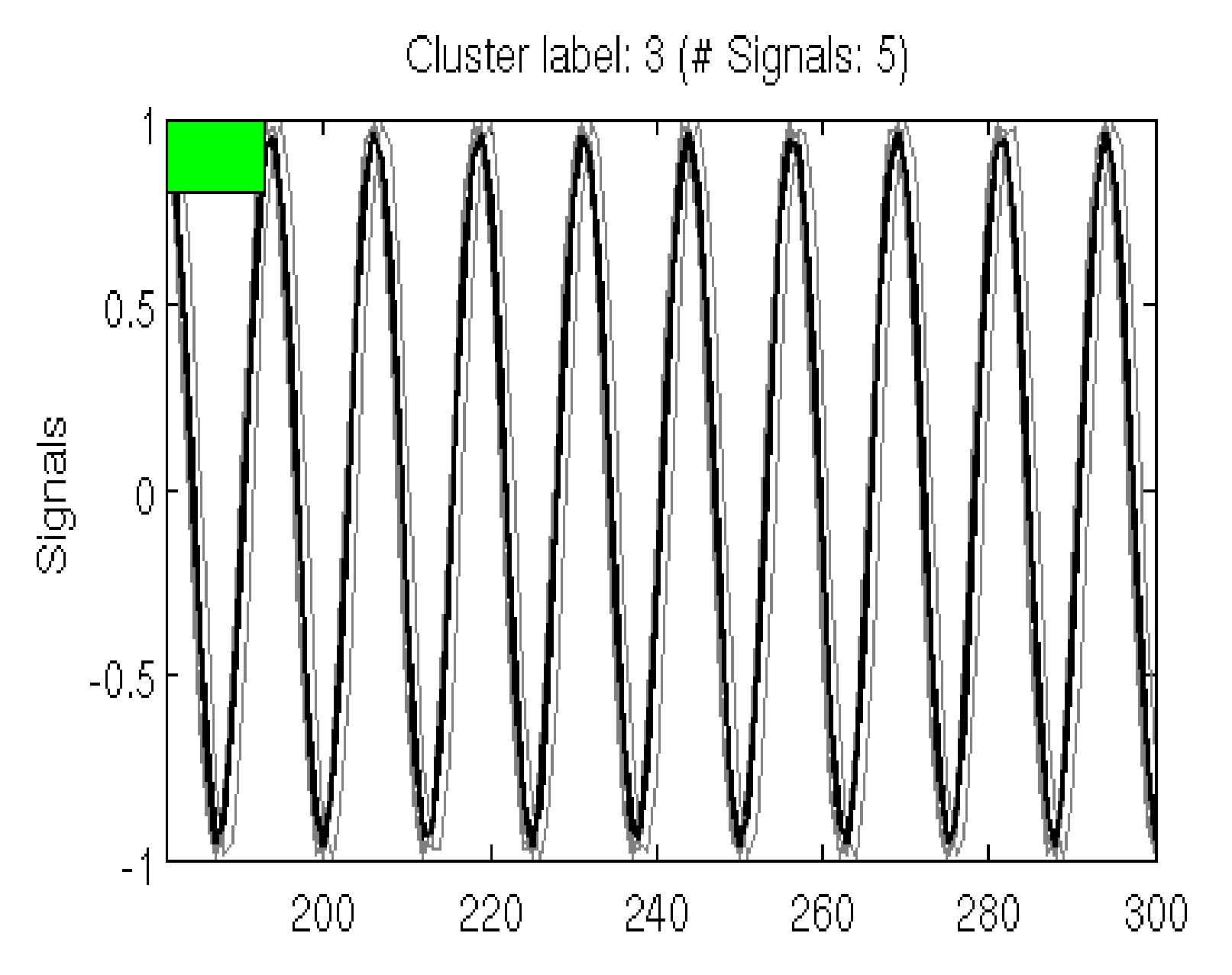}}\\ 
\includegraphics[width=2.5in]{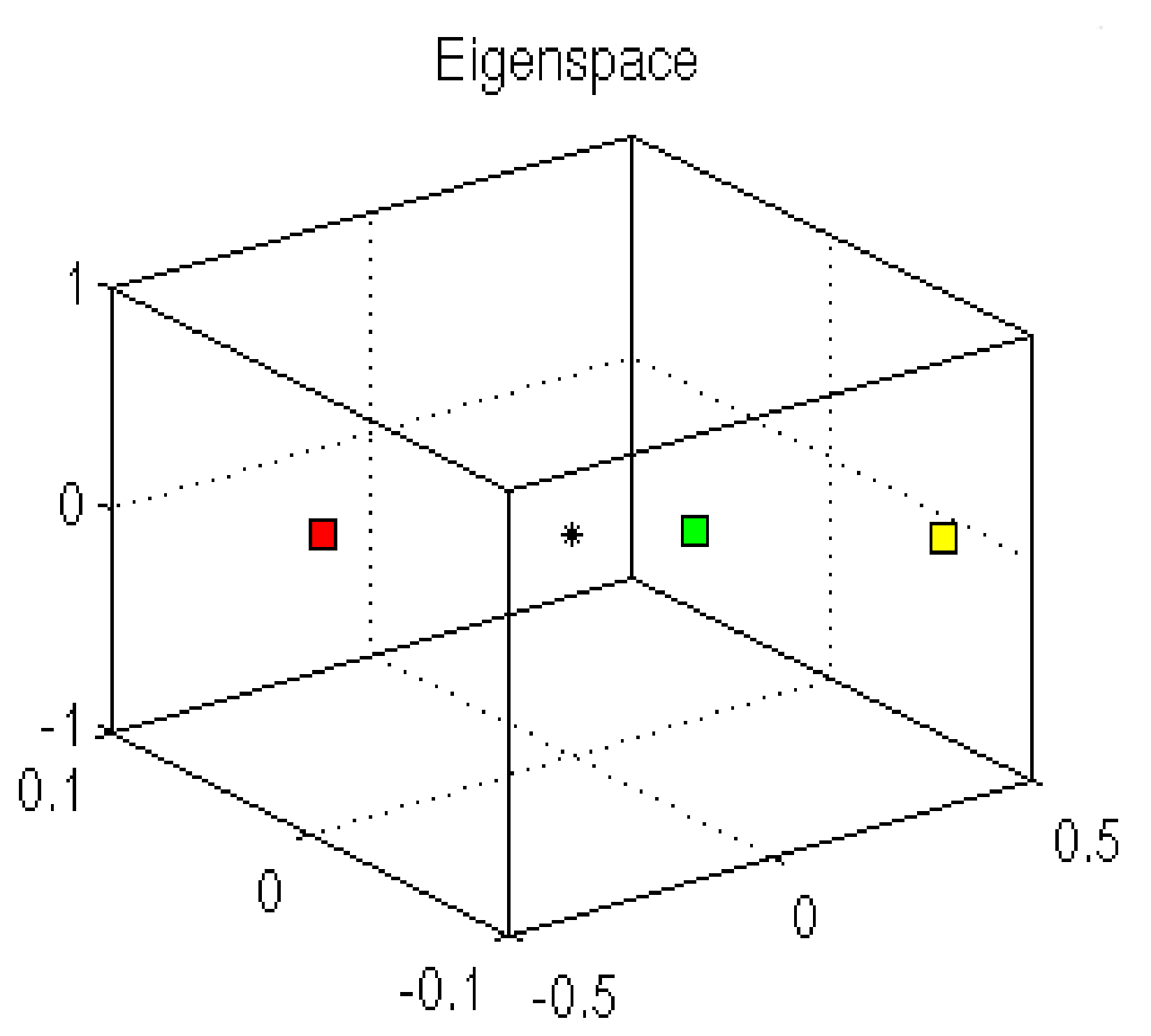}&\includegraphics[width=2in]{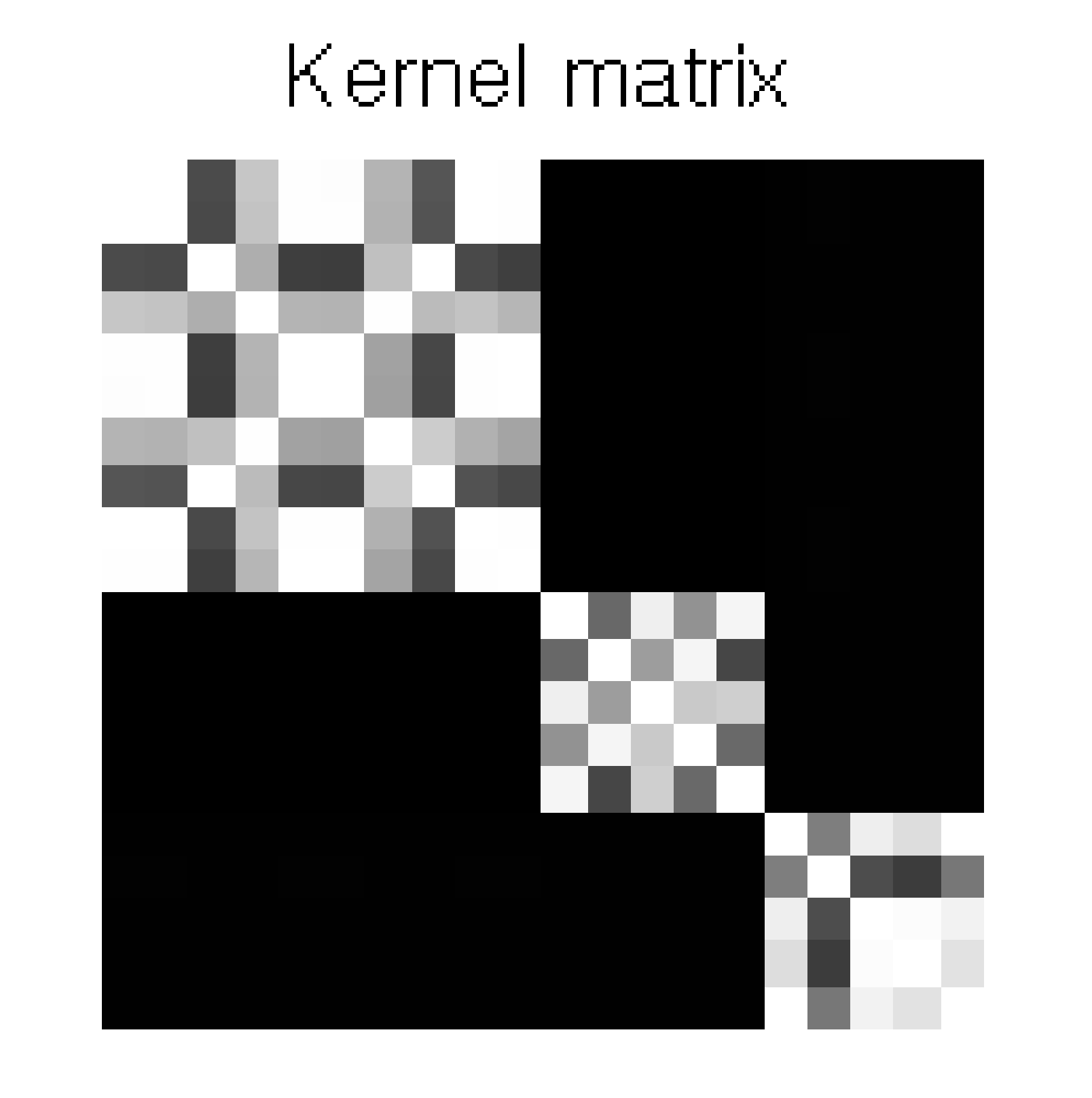}
\end{tabular}
\caption{\textbf{Synthetic time-series clusters after creation.} \textbf{Top and center:} signals of the three clusters after the creation event. \textbf{Bottom left} data in the eigenspace spanned by $\alpha^{(1)}$ and $\alpha^{(2)}$ (the points are mapped in the same location as the related centroids, since the eigenvectors are perfectly piece-wise constant). \textbf{Bottom right:} kernel matrix.}
\label{results_ts2}	
\end{figure}

\begin{figure}[htbp]
%\centering
\hspace{-5mm}
\begin{tabular}{ll}
\includegraphics[width=2.25in]{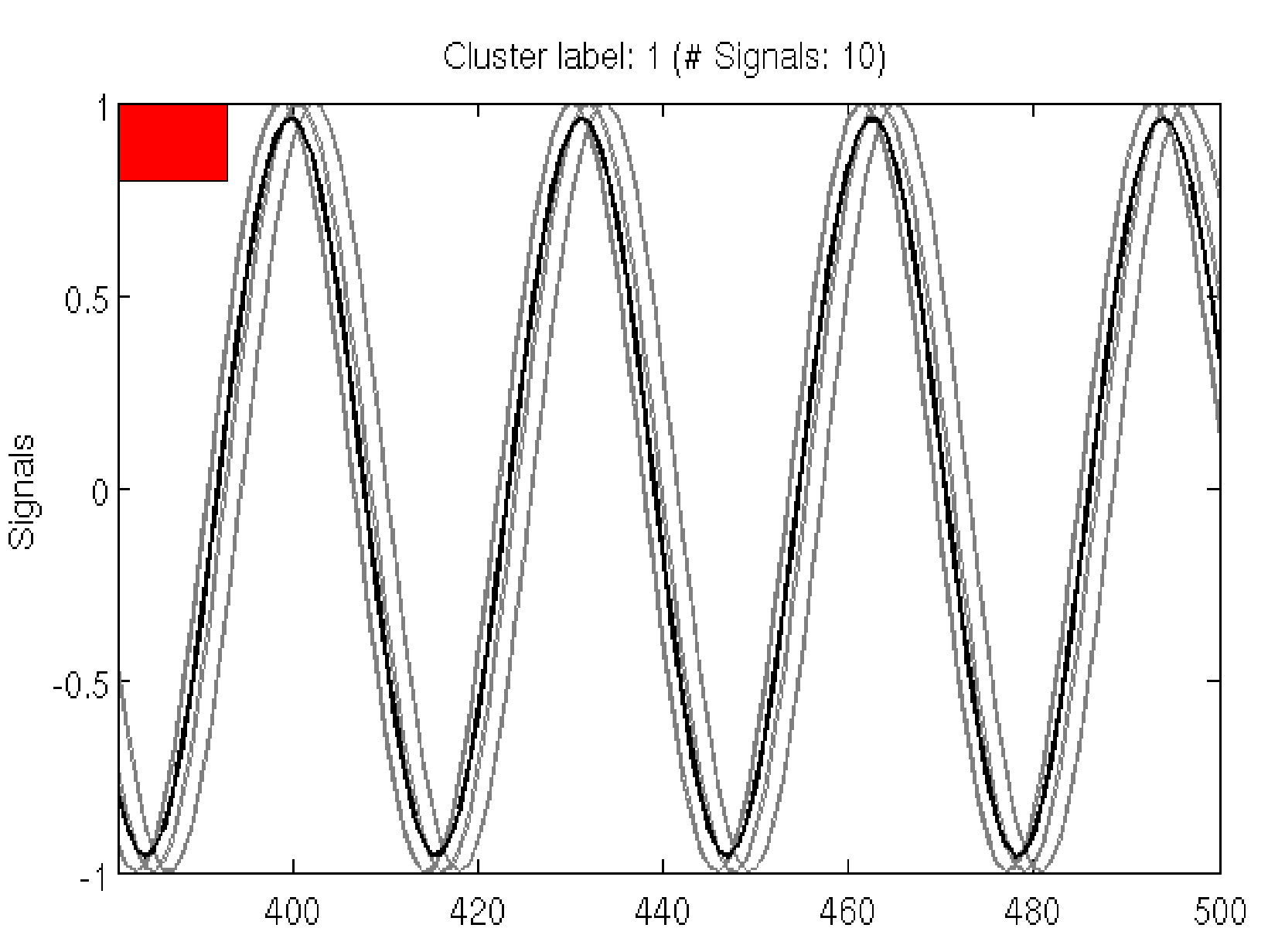}&\includegraphics[width=2.2in]{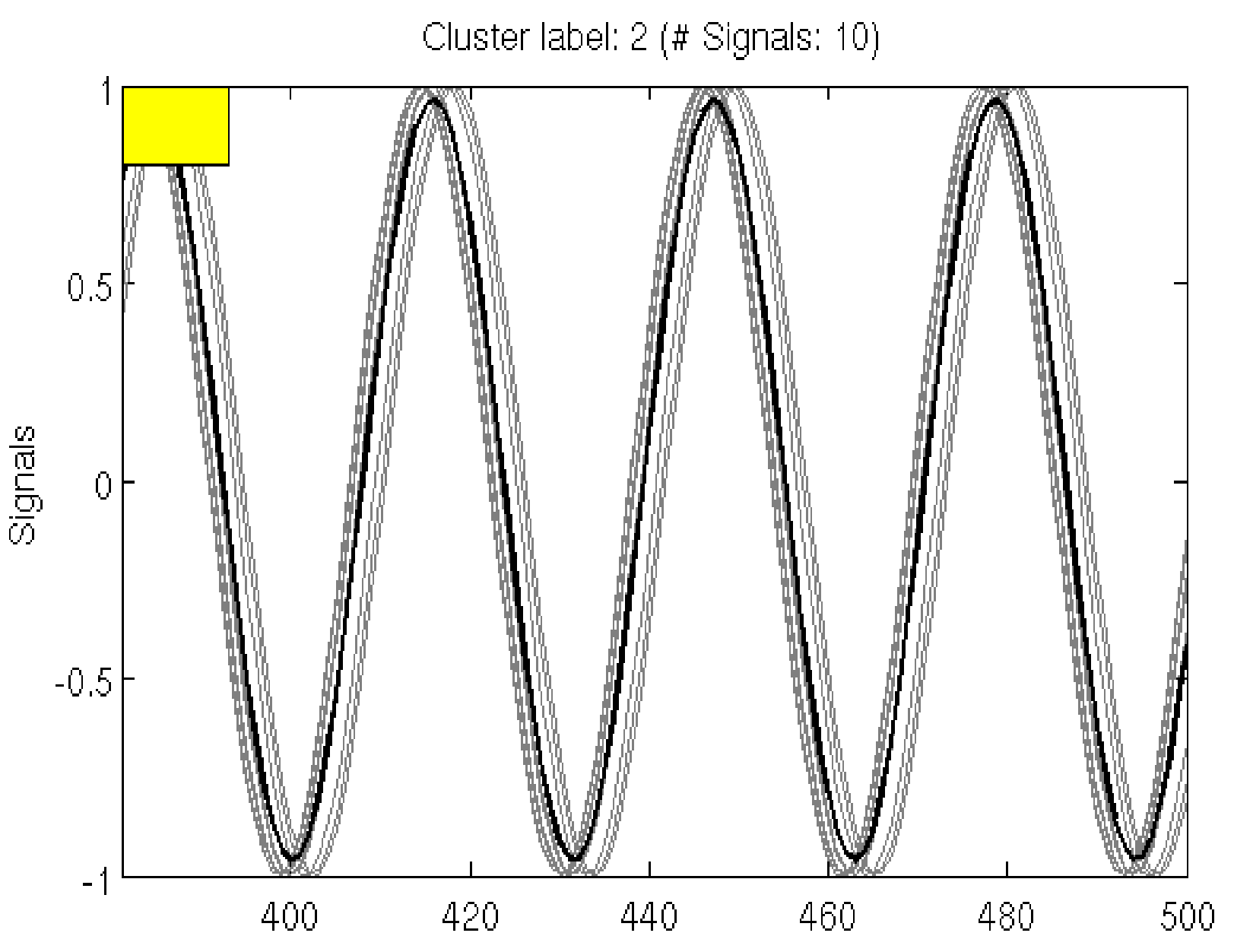}\\
\includegraphics[width=2.5in]{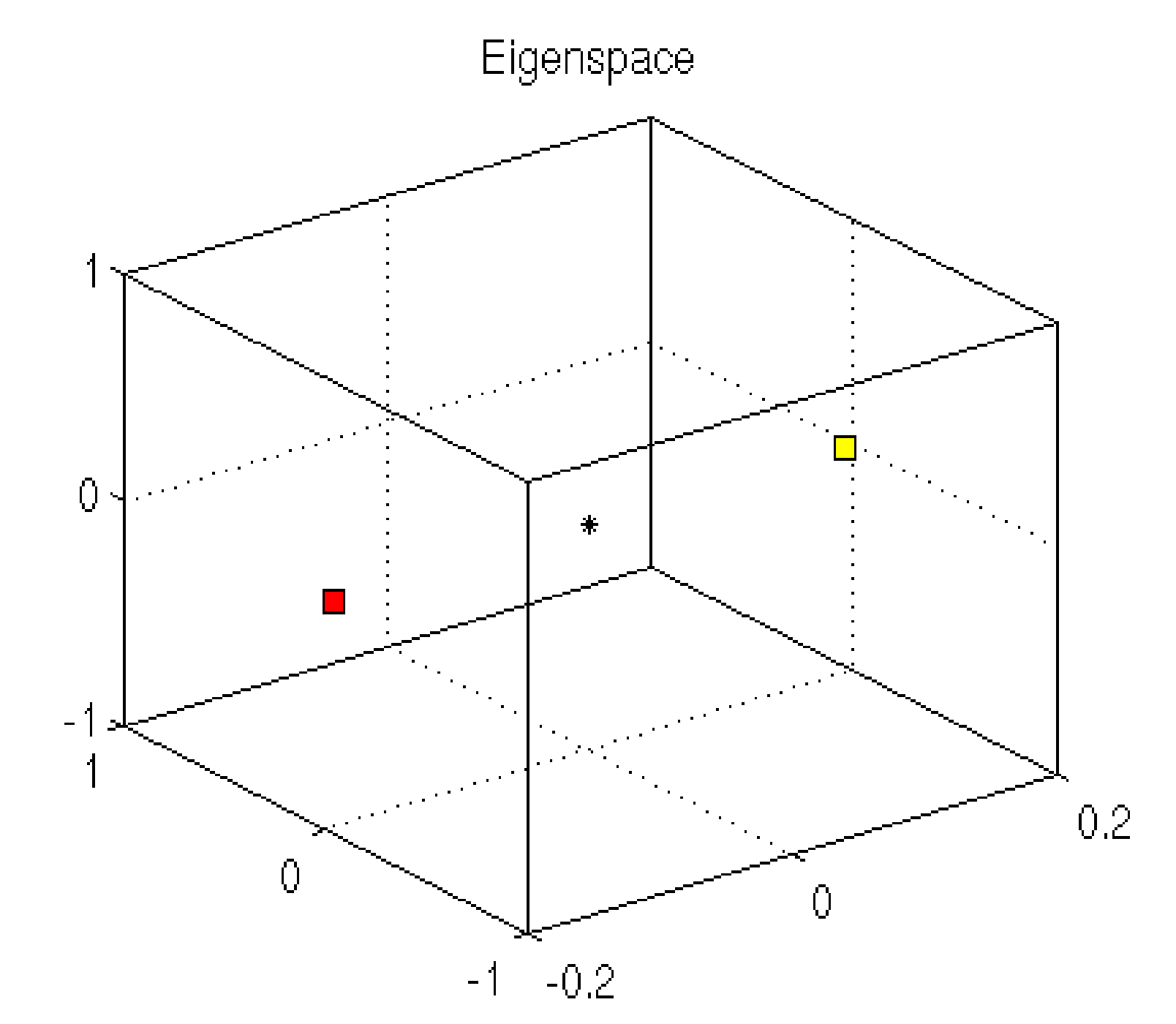}&\includegraphics[width=2in]{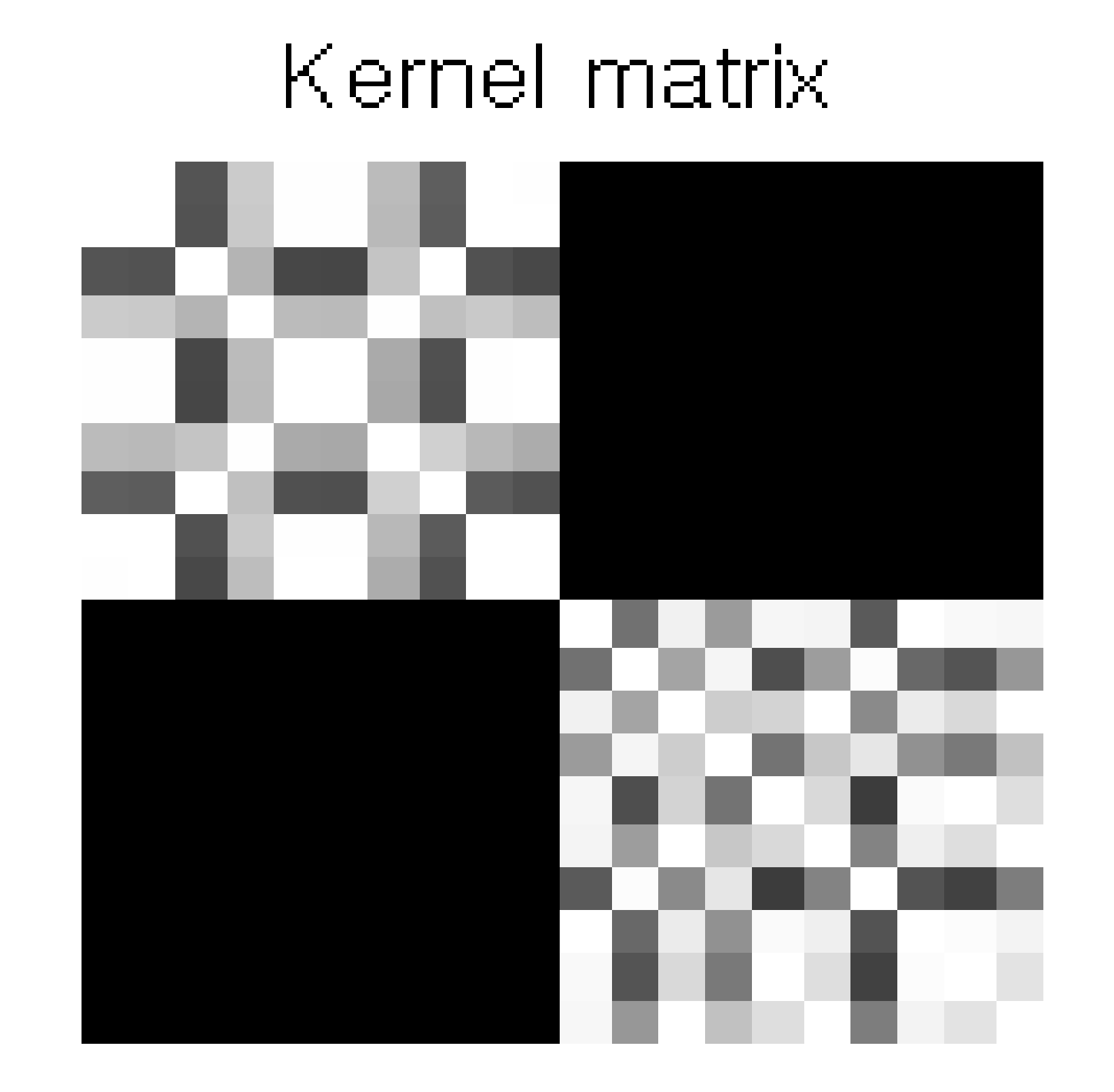}
\end{tabular}
\caption{\textbf{Synthetic time-series final clustering model.} \textbf{Top:} two final clusters after the merging event. \textbf{Bottom left:} clustered data in the eigenspace spanned by $\alpha^{(1)}$ (the points are mapped in the same location as the related centroids, since the eigenvectors are perfectly piece-wise constant). \textbf{Bottom right:} kernel matrix.}
\label{results_ts3}	
\end{figure}

The results of testing the IKSC algorithm on the first synthetic example is presented in figure \ref{results_gaussdrift1}. In the initialization phase $30$ points are used to construct the model. The IKSC algorithm can perfectly model the two drifting distributions: the average cumulative ARI error is equal to $0$. Moreover the quality of the predicted clusters remains very high over time, as demonstrated by the trend of the instantaneous MSV depicted in figure \ref{sil_gaussdrift1}. The results of the simulation related to the second artificial data-set are depicted in figure \ref{results_gaussdrift2}. Similarly to the first artificial experiment, the cluster quality stays high over time as shown in figure \ref{sil_gaussdrift2}, and the partitions found by IKSC are in almost perfect agreement with the ground truth (small ARI error) for the whole duration of the simulation (see figure \ref{ARIerror_gaussdrift2}). Moreover at time-step $t=6926$ the two moving Gaussian clouds are merged, as expected. Only in this case, as observed also in \cite{SAKM_boubacar}, there is a small increase in the average cumulative ARI error. The small cluster at the bottom left side of figure \ref{gauss_drift2} is detected as outlier after the data acquisition. Finally, we discuss the results of IKSC on the synthetic time-series experiment. In the initialization phase the algorithm recognizes $2$ clusters, which are shown in figure \ref{results_ts1}. After some time, we notice that IKSC successfully detects the first change in frequency of the signals of the second type by creating a new cluster at time step $t=223$, as depicted in figure \ref{results_ts2}. Moreover the second change point is detected at time step $t=382$, when a merging of two clusters is performed, as illustrated in figure \ref{results_ts3}.

%\begin{figure}[htbp]
%\centering
%\begin{tabular}{c}
%\includegraphics[width=4.5in]{initial_synthetic_IKSC.eps}
%\end{tabular}
%\caption{\textbf{Synthetic time-series initial clustering model.} \textbf{Top:} signals of the two starting clusters. \textbf{Bottom left}: data in the eigenspace (the points are mapped in the same location as the related centroids, since the eigenvectors are perfectly piece-wise constant). \textbf{Bottom right}: kernel matrix with a clear block diagonal structure.}
%\label{results_ts1}	
%\end{figure}

%\begin{figure}[htbp]
%\centering
%\begin{tabular}{c}
%\includegraphics[width=4.5in]{new_clu_synthetic_IKSC.eps}
%\end{tabular}
%\caption{\textbf{Synthetic time-series clusters after creation.} \textbf{Top and center:} signals of the three clusters after the creation event. \textbf{Bottom left} data in the eigenspace (the points are mapped in the same location as the related centroids, since the eigenvectors are perfectly piece-wise constant). \textbf{Bottom right:} kernel matrix.}
%\label{results_ts2}	
%\end{figure}

%\begin{figure}[htbp]
%\centering
%\begin{tabular}{c}
%\includegraphics[width=4.5in]{merging_synthetic_IKSC.eps}
%\end{tabular}
%\caption{\textbf{Synthetic time-series final clustering model.} \textbf{Top:} two final clusters after the merging event. \textbf{Bottom left:} clustered data in the eigenspace (the points are mapped in the same location as the related centroids, since the eigenvectors are perfectly piece-wise constant). \textbf{Bottom right:} kernel matrix.}
%\label{results_ts3}	
%\end{figure}

\subsection{Analysis of the eigenvectors}
\label{approx_eig}
\begin{figure}[htbp]
\centering
\begin{tabular}{c}
\begin{psfrags}
\psfrag{x}[t][t]{{time instant t}}
\psfrag{y}[b][b]{{$\alpha^{(1)}$}}
\psfrag{t}[b][b]{{}}
\includegraphics[width=4in]{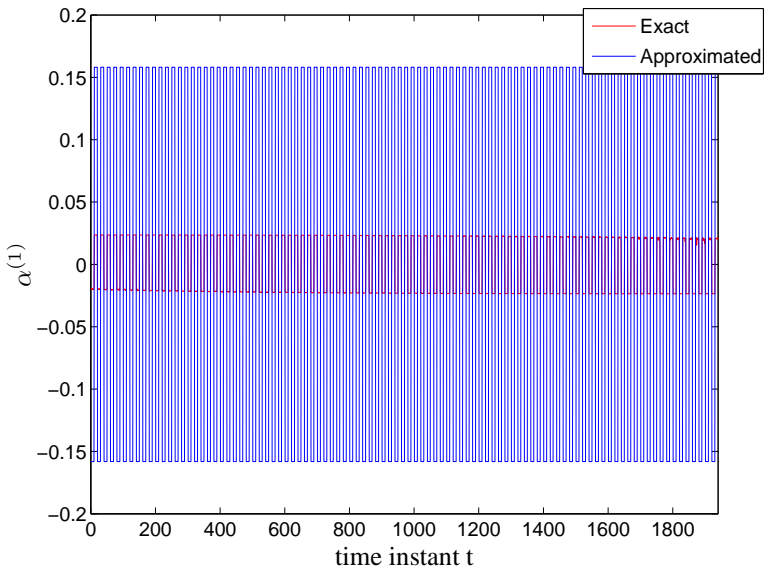}
\end{psfrags}
\end{tabular}
\caption{\textbf{Eigenvector-Drifting Gaussian distributions.} Exact and approximated eigenvector corresponding to the largest eigenvalue of the problem (\ref{dual_KSC}), for the first synthetic example.}
\label{eig_gaussdrift1}	
\end{figure}

\begin{figure}[htbp]
\centering
\begin{tabular}{c}
\begin{psfrags}
\psfrag{x}[t][t]{{time instant t}}
\psfrag{y}[b][b]{{$\alpha^{(1)}$}}
\psfrag{t}[b][b]{{}}
\includegraphics[width=2.5in]{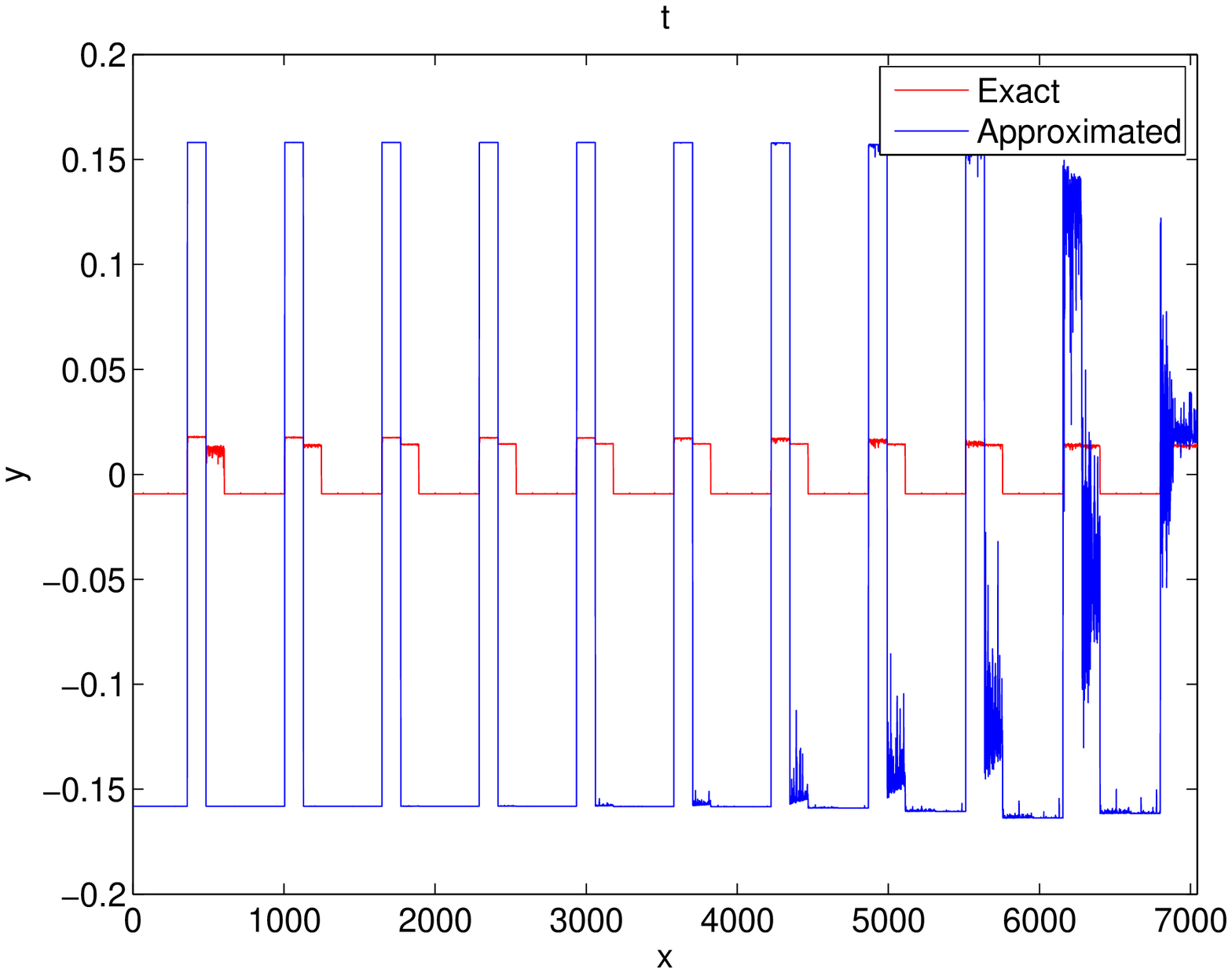}
\end{psfrags}\\\\
\begin{psfrags}
\psfrag{x}[t][t]{{time instant t}}
\psfrag{y}[b][b]{{$\alpha^{(2)}$}}
\psfrag{t}[b][b]{{}}
\includegraphics[width=2.5in]{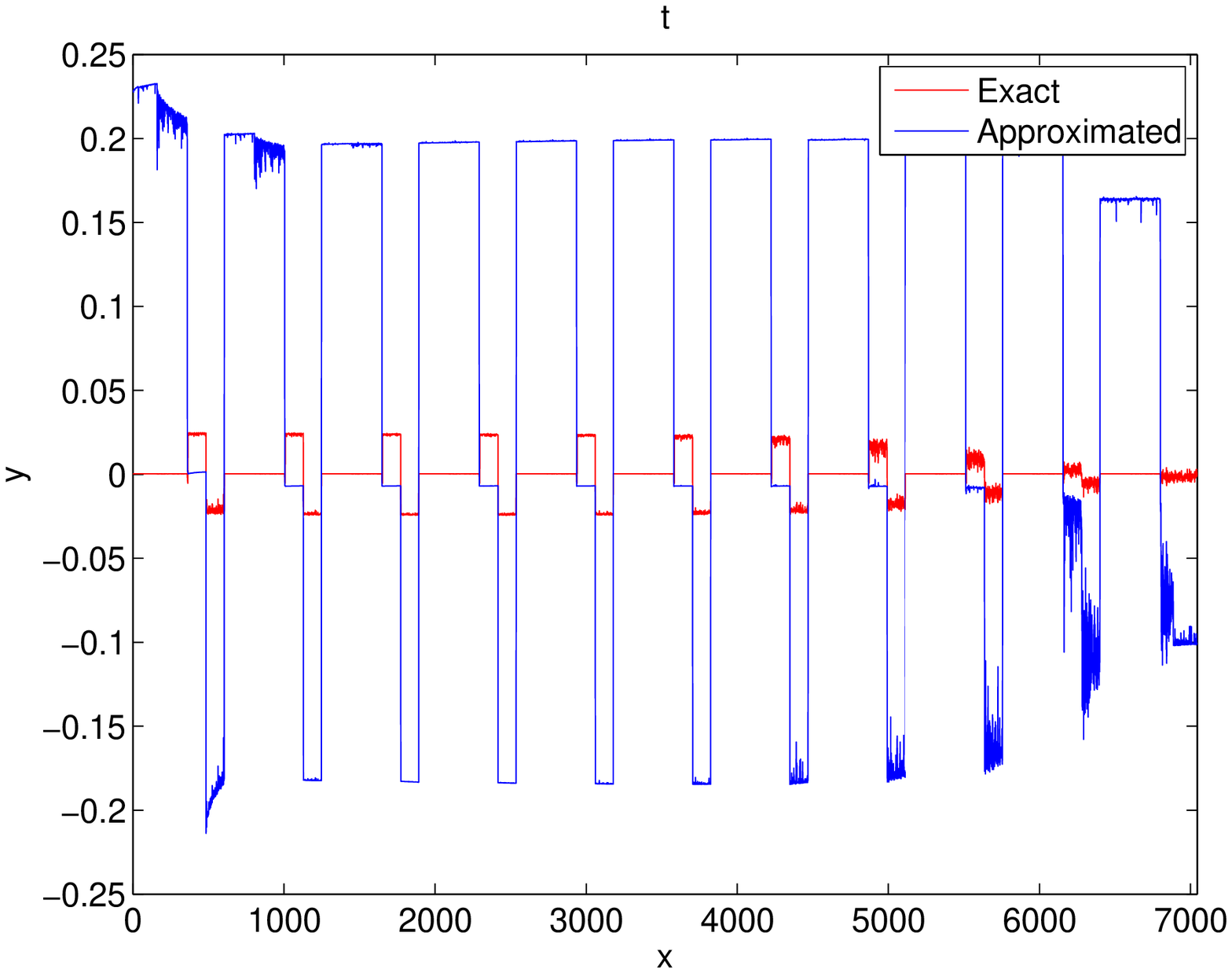}
\end{psfrags}\\\\
\begin{psfrags}
\psfrag{x}[t][t]{{time instant t}}
\psfrag{y}[b][b]{{$\alpha^{(3)}$}}
\psfrag{t}[b][b]{{}}
\includegraphics[width=2.5in]{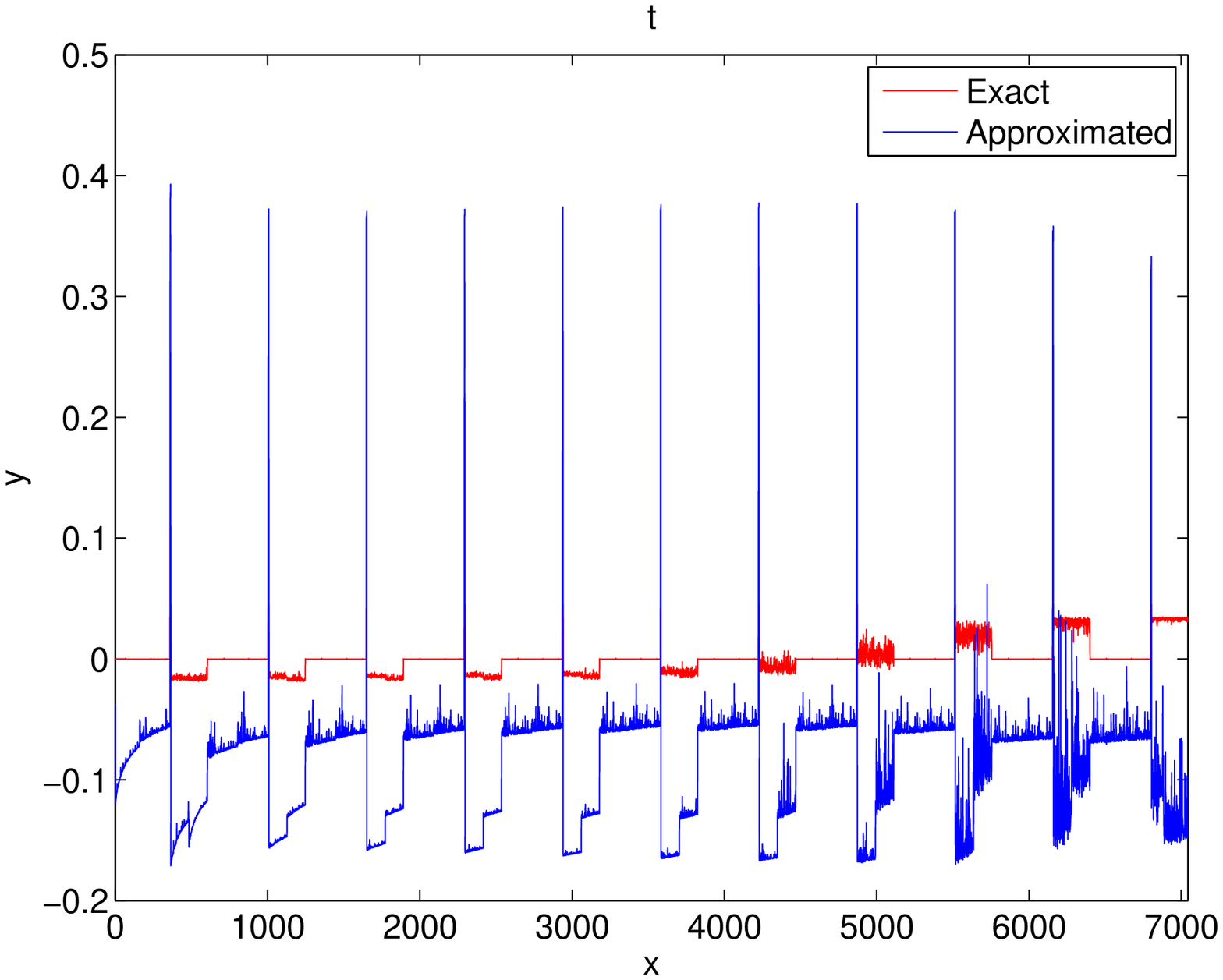}
\end{psfrags}
\end{tabular}
\caption{\textbf{Eigenvectors related to the Merging Gaussian distributions data.} Exact and approximated eigenvectors corresponding to the $3$ largest eigenvalues of the problem (\ref{dual_KSC}), for the second synthetic experiment.}
\label{eigs_gaussdrift2}	
\end{figure}

Here we discuss on the quality of our model-based eigen-updating for kernel spectral clustering. In figure \ref{eig_gaussdrift1} the exact and the approximated eigenvector related to the largest eigenvalue of (\ref{dual_KSC}) for the drifting Gaussians example are shown. We notice that the model-based eigenvectors are less noisy with respect to the exact eigenvectors and a multiplicative bias is present. The first property is quite surprising: basically we are able to recover the perfect separation between the two clusters even when this is somehow masked by the data. This occurs mainly in the end of the simulation when the two Gaussian clouds approach each other. In this case the exact eigenvector is not exactly piece-wise constant due to a small overlap, while the model-based eigenvector is much less perturbed. The multiplicative bias is probably due to the fact that the out-of-sample eigenvectors are computed using an ultra-sparse training set (only the two cluster centroids). The latter allows to process the data-stream very quickly, but lacks of the information related to the spread of the data-points, which may cause the bias. Similar considerations can be done for the second synthetic experiment, i.e. the merging Gaussians. The three eigenvectors corresponding to the largest eigenvalues of (\ref{dual_KSC}) are represented in figure \ref{eigs_gaussdrift2}. In the third approximated eigenvector we can notice $4$ levels, which are not present in the exact eigenvector. Once again this testifies the tight relation between the clustering model of IKSC (the $4$ centroids) and the approximated eigenvectors, which is a unique property of our framework.

\section{Real-Life Example}
\subsection{The PM$_{10}$ data-set}
\label{PM10_data}
\begin{figure}[htbp]
%\centering
\hspace{-6mm}
\begin{tabular}{l}
\includegraphics[width=5in]{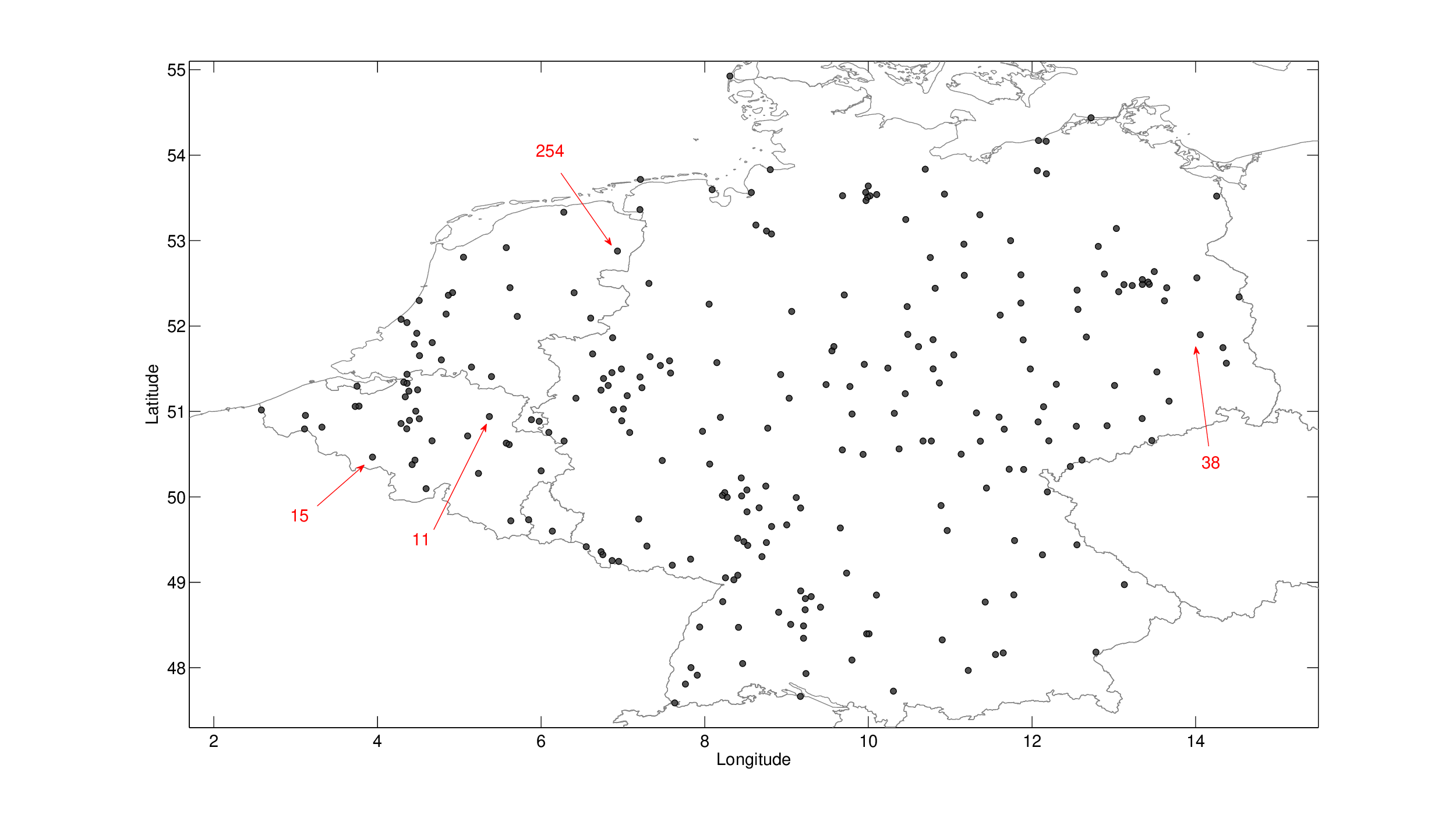}\\
\includegraphics[width=5in]{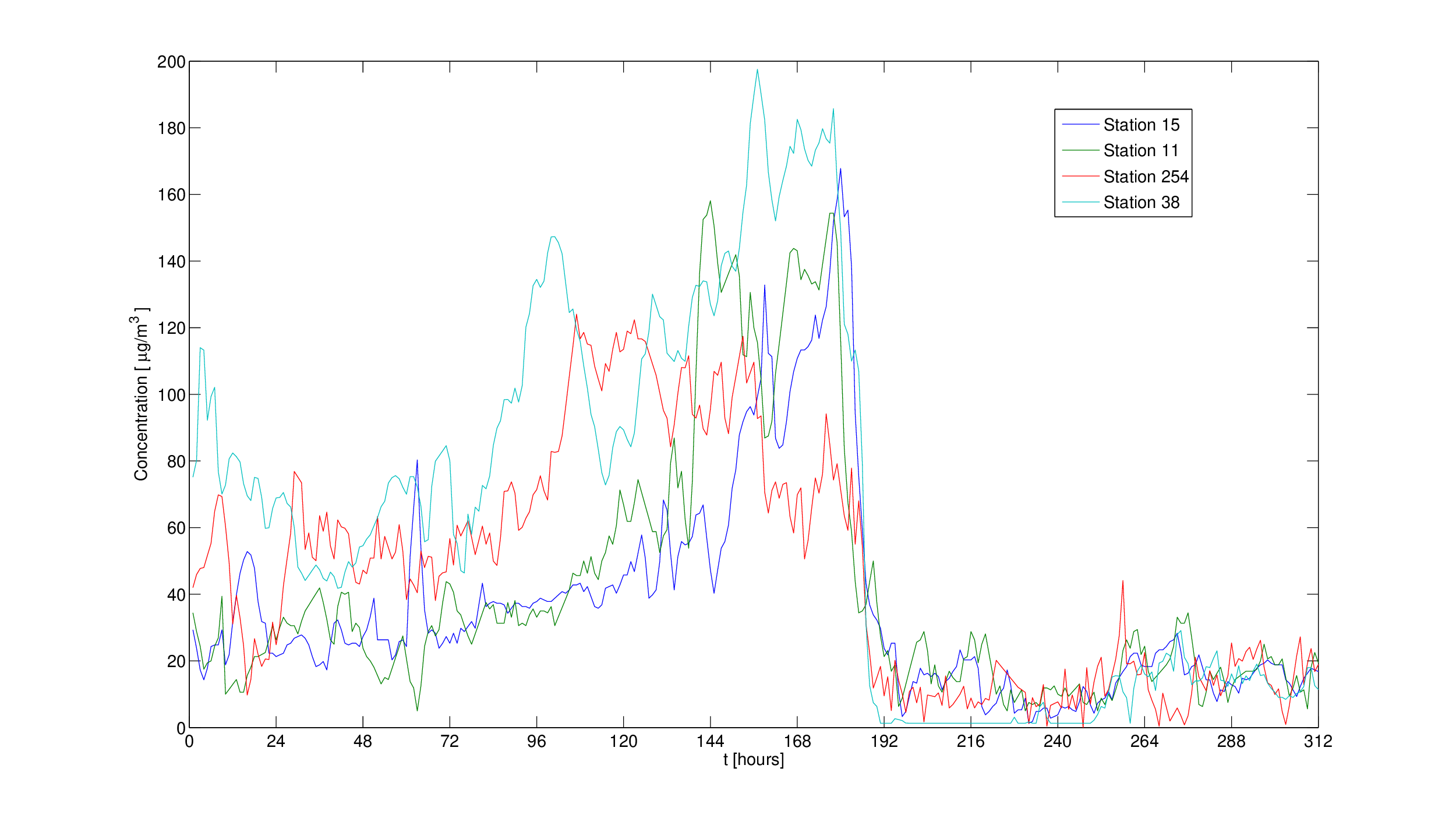}
\end{tabular}
\caption{$\mathbf{PM_{10}}$ \textbf{data.} \textbf{Top:} AirBase monitoring stations. \textbf{Bottom:} Examples of PM$_{10}$ concentrations time-series in Belgium, Netherlands, Luxembourg and Germany, for the whole period under investigation.}
\label{PM10_map}
\end{figure}

Particulate Matter (PM) is the term used for solid or liquid particles found in the air. In particular PM$_{10}$ refers to those particles whose size is up to $10$ micrometers in aerodynamic diameter. The inhalation of
these particles is dangerous for human health since it can cause asthma, lung cancer, cardiovascular issues, etc. Accurate measurements and estimation of PM is then of vital importance by the health care point of view. To this aim the European Environmental Agency manages a publicly available database called AirBase \cite{airbase_url}. This air-quality database contains validated air quality monitoring information of several pollutants for more than $30$ participating countries throughout Europe. In this paper we analyse the PM$_{10}$ data registered by $259$ background stations during a heavy pollution episode that took place between January 20th, 2010 and February 1st, 2010. We focus on an area comprising four countries: Belgium, Netherlands, Germany and Luxembourg (see figure \ref{PM10_map}). The experts attributed this episode to the import of PM originating in Eastern Europe \cite{paper_Mauro}.

\subsection{Results of the simulations}
\begin{figure}[htbp]
\centering
\begin{tabular}{c}
\includegraphics[width=4.5in]{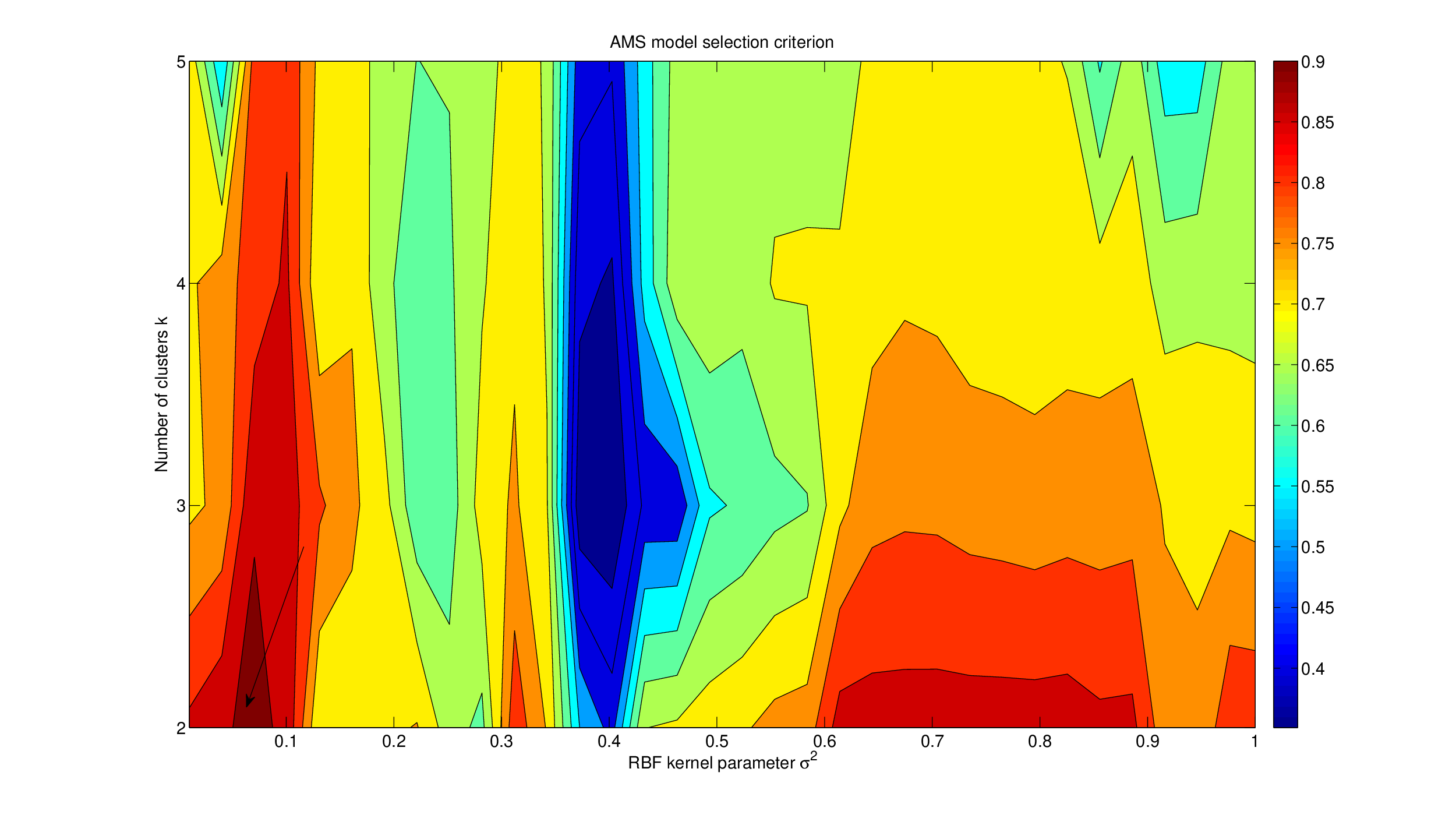}
\end{tabular}
\caption{\textbf{Model selection.} Tuning of the number of clusters and the bandwidth of the RBF kernel in the initialization phase of IKSC for the analysis of the PM$_{10}$ data.} 
\label{PM10_tuning}
\end{figure}

\begin{figure}[htbp]
%\centering
\hspace{-5mm}
\begin{tabular}{l}
\includegraphics[height=2in, width= 5in]{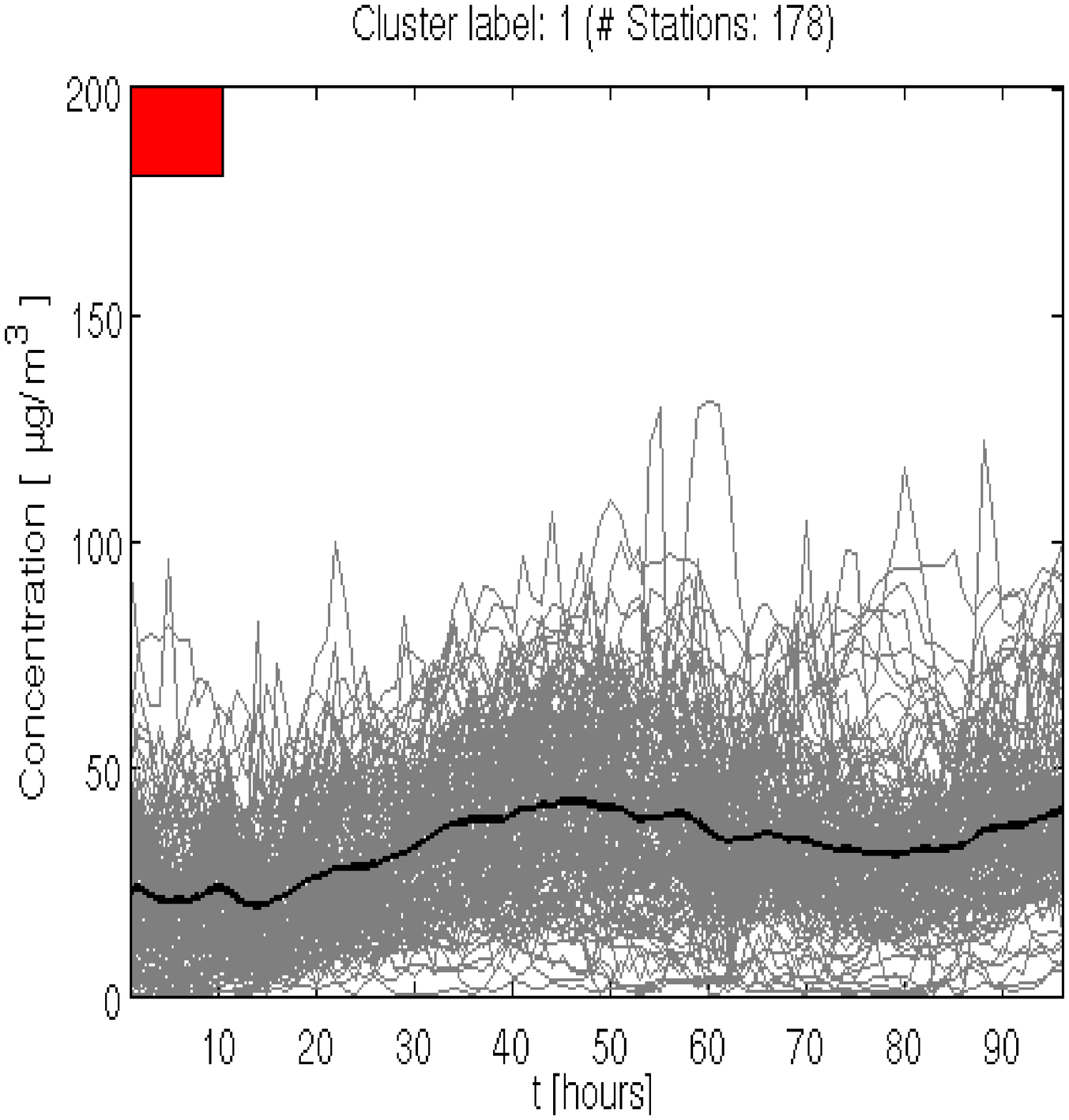}\\
\includegraphics[width= 5in]{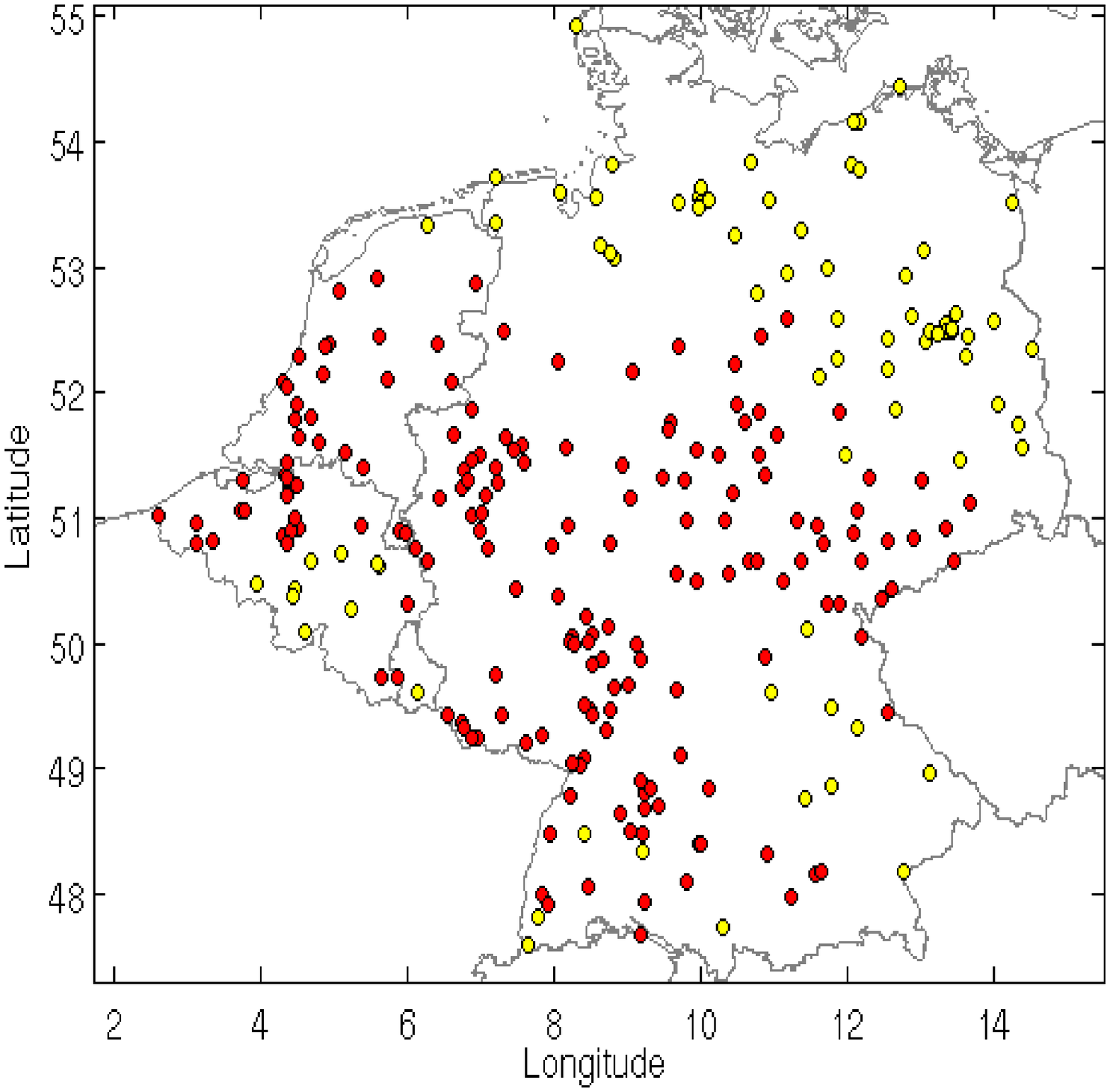}
\end{tabular}
\caption{\textbf{Initial clustering model for the PM$_{10}$ monitoring stations.} \textbf{Top:} signals for the two starting clusters. \textbf{Bottom left:} Spatial distribution of the clusters over Belgium, Netherlands, Luxembourg and Germany. \textbf{Bottom right:} data mapped in the eigenspace (in this case the space spanned by the eigenvector $\alpha^{(1)}$ related to the highest eigenvalue).}
\label{init_model_PM10}	
\end{figure}

In the initialization phase our data-set consists of a time-series of $96$ time steps (i.e. four days) for each station. In order to build-up an initial clustering model we tune the number of clusters $k$ and the proper $\sigma$ for the RBF kernel by using the AMS (Average Membership Strength) model selection criterion explained in chapter \ref{ch:spectralclustering}. After tuning we find $k=2$ and $\sigma^2 = 0.05$ as optimal parameters, as depicted in figure \ref{PM10_tuning}. The initial model, based on these parameters, is illustrated in figure \ref{init_model_PM10}. In this case the $2$ centroids in the input space are the time-series representing the two clusters, while in the eigenspace they are points of dimension $k-1$ (anyway for visualization purposes we always use a 3D plot).

\begin{figure}[htbp]
%\centering
\hspace{-5mm}
\begin{tabular}{l}
\includegraphics[height=1.75in,width= 5in]{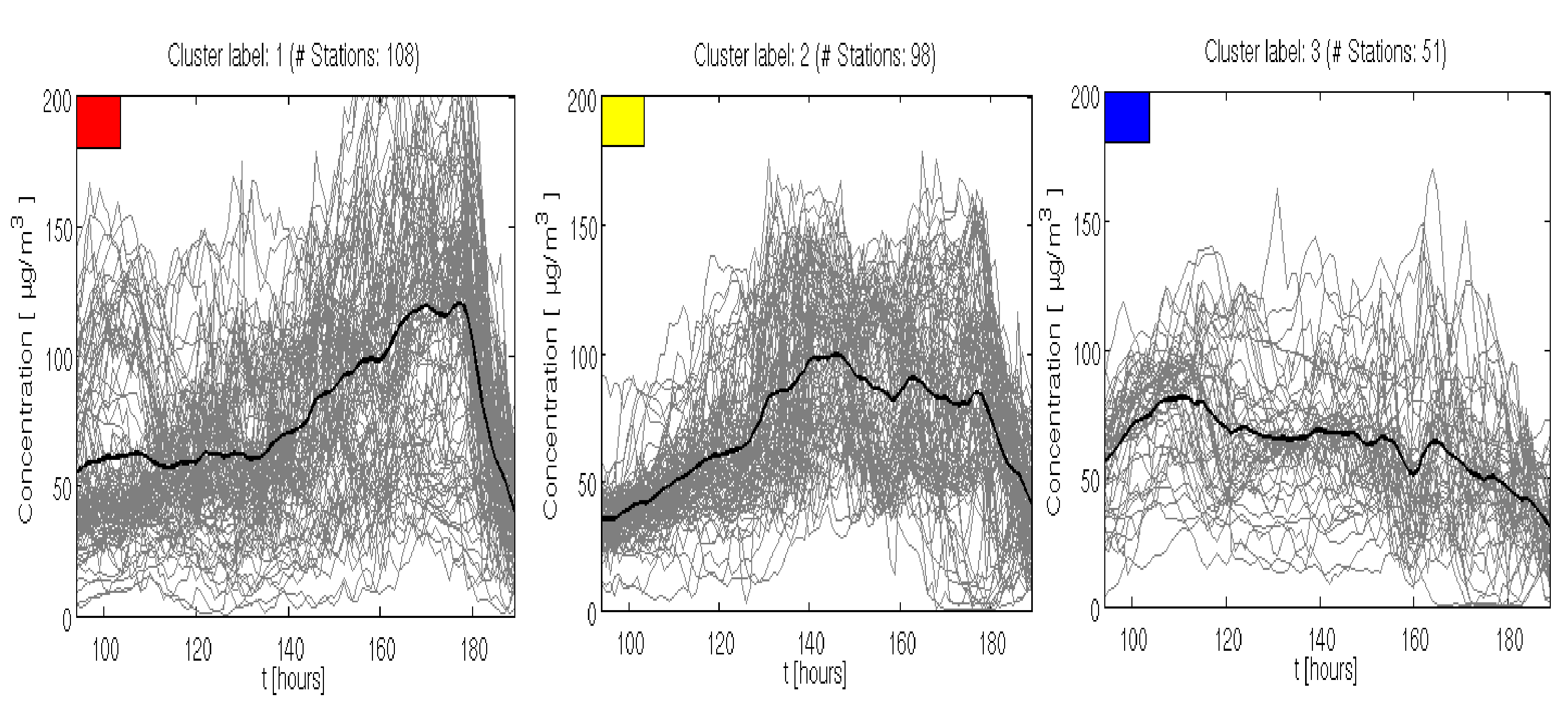}\\
\includegraphics[width= 5in]{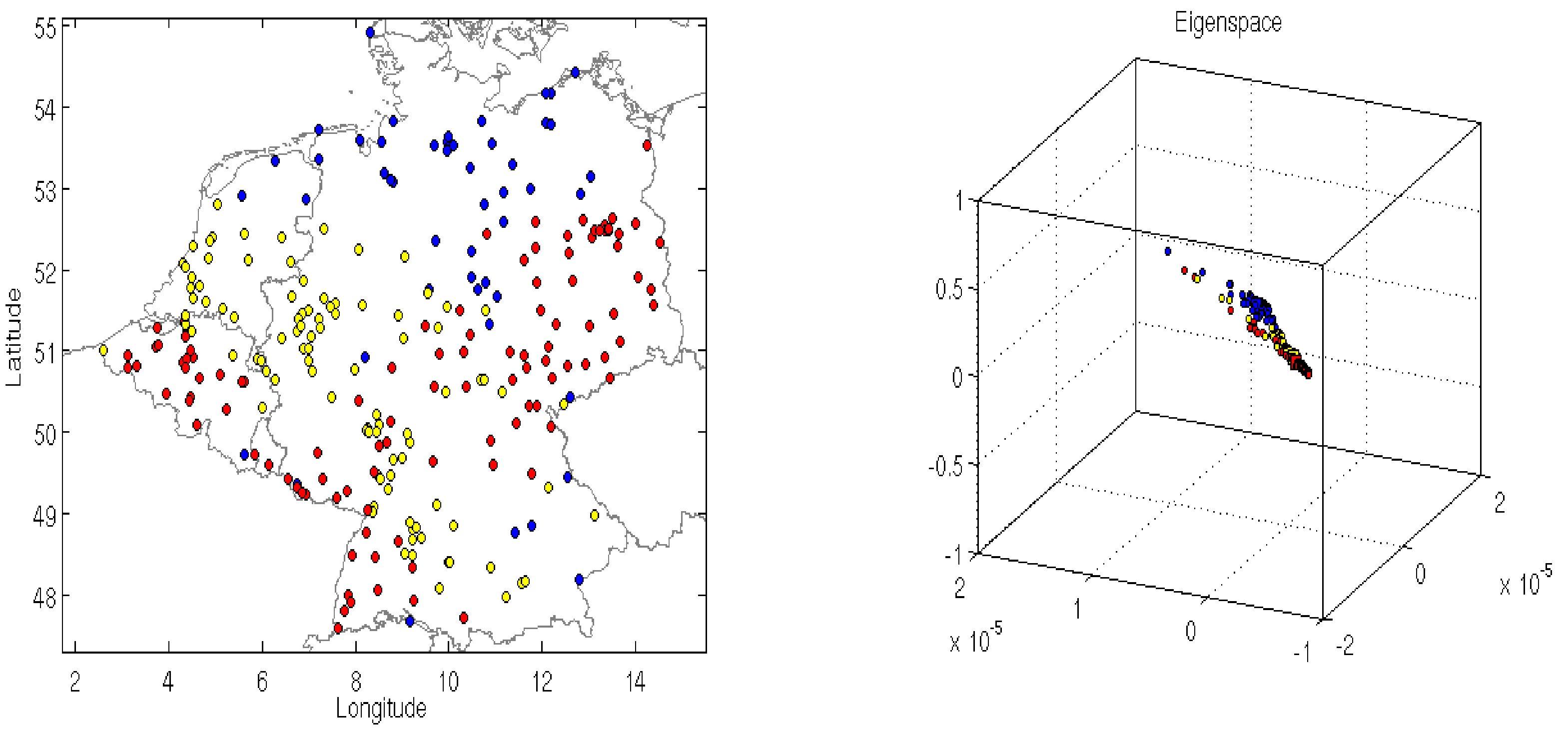}
\end{tabular}
\caption{\textbf{PM$_{10}$ clusters after creation.} \textbf{Top:}  signals for the three clusters after the creation event. \textbf{Bottom left:} Spatial distribution of the clusters over Belgium, Netherlands, Luxembourg and Germany. Interestingly, the new cluster comprises stations located in the North-East part of Germany, which is the area where the pollutants coming from Eastern Europe started to spread during the heavy pollution episode of January $2010$.  \textbf{Bottom right:} data in the space spanned by the eigenvectors $\alpha^{(1)}$ and $\alpha^{(2)}$.}
\label{newclu_PM10}	
\end{figure}

\begin{figure}[htbp]
%\centering
\hspace{-5mm}
\begin{tabular}{l}
\includegraphics[height=2in,width= 5in]{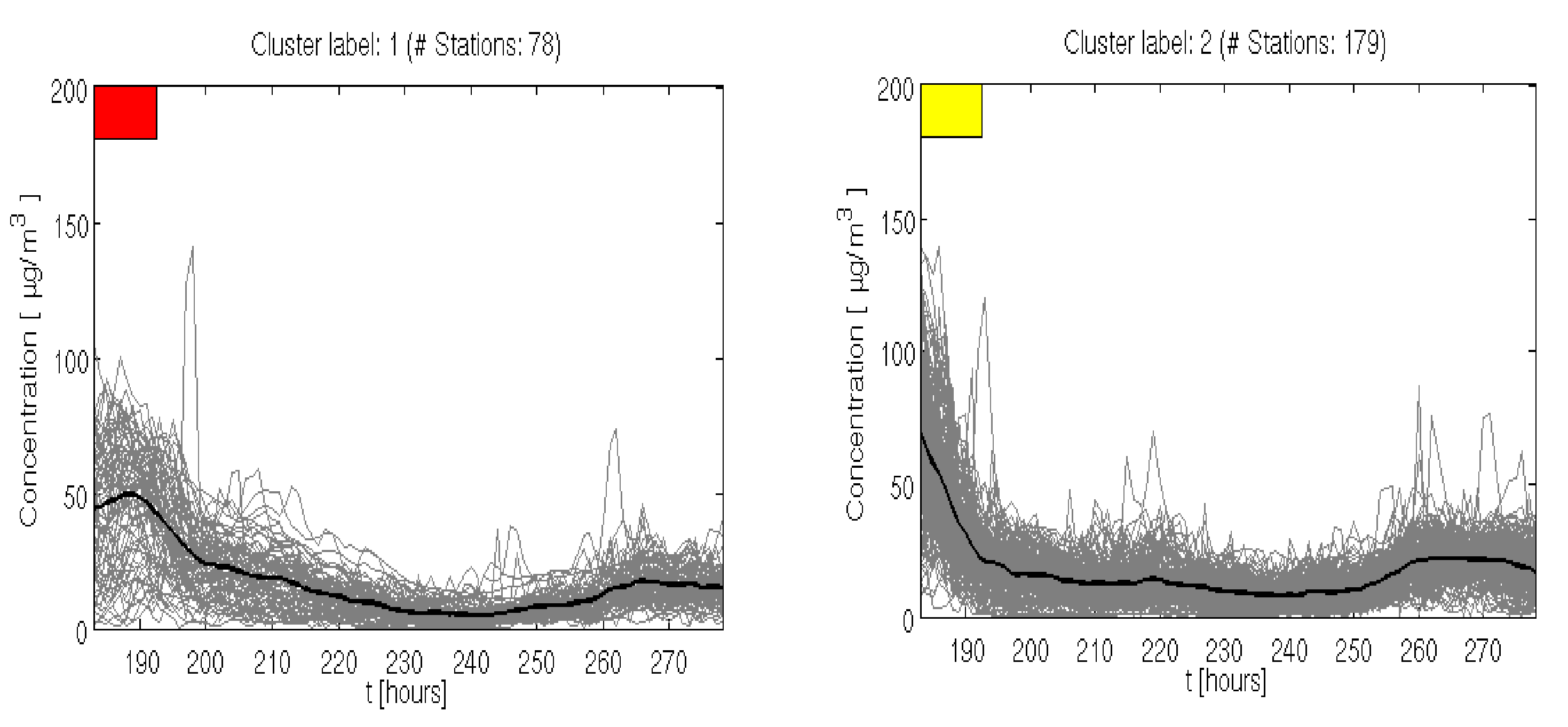}\\
\includegraphics[width= 5in]{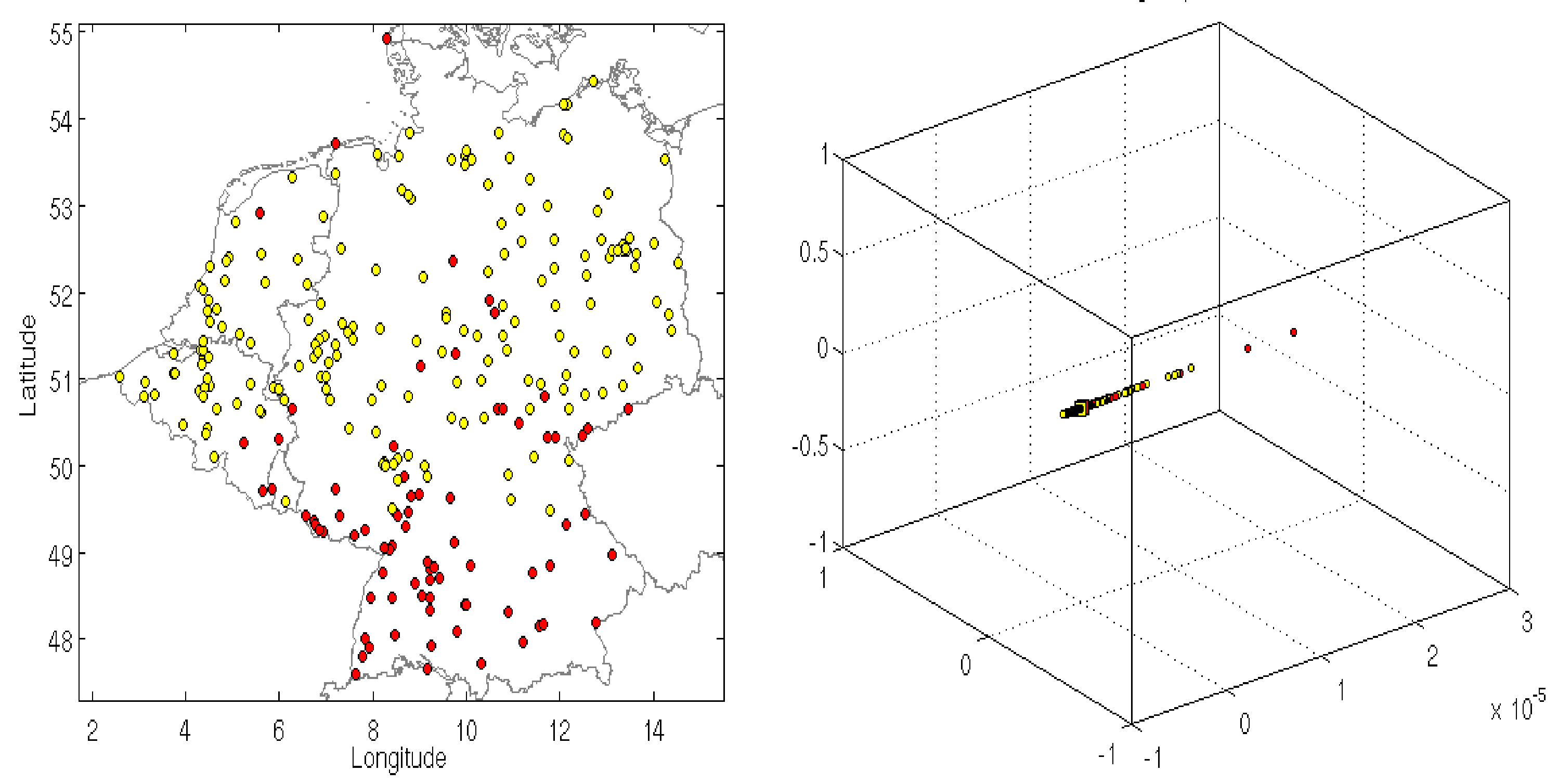}
\end{tabular}
\caption{\textbf{Clustering model of PM$_{10}$ stations after merging.} \textbf{Top:} two clusters left after the merging event occurred at time step $t=251$. \textbf{Bottom left:} Spatial distribution of the clusters over Belgium, Netherlands, Luxembourg and Germany. \textbf{Bottom right:} data in the space spanned by the eigenvector $\alpha^{(1)}$.}
\label{merge_PM10}	
\end{figure}

During the online stage, by adopting a moving window approach, our data-set at time $t$ corresponds to the PM$_{10}$ concentrations measured from time $t-96$ to time $t$. In this way we are able to track the evolution of the pollutants over time. In fact, after some time the IKSC model creates a new cluster, as depicted in figure \ref{newclu_PM10}. Later on these three clusters evolve until a merge of two of them occurs at time step $t = 251$ (see figure \ref{merge_PM10}).
If we analyse more in details the clustering results, we can notice how the new cluster (represented in blue) is concentrated mainly in the Northern region of Germany. Moreover the creation occurs at time step $t=143$, when the window describes the start of the pollution episode in Germany. Afterwards, the new cluster starts expanding in direction South-West. Basically, IKSC is detecting the arrival of the pollution episode originated in Eastern Europe and driven by the wind toward the West. This ability of our clustering model of detecting the dynamics of the pollution cloud at this level of accuracy is rather unexpected. Indeed, IKSC does not have any information about the spatial localization of the stations and the meteorological conditions. At time step $t=251$ two clusters are merged. This can be explained by the fact that the window covers the unusually high PM$_{10}$ concentrations as well as the end of the episode, registered by many of the stations.

%\begin{figure}[htbp]
%\centering
%\begin{tabular}{c}
%%\includegraphics[width=5.5in]{initial_zebrafish_IKSC.eps}
%\includegraphics[width= 4.5in]{initial_PM10.eps}
%\end{tabular}
%\caption{\textbf{Initial clustering model for the PM$_{10}$ monitoring stations.} \textbf{Top:} signals for the two starting clusters. \textbf{Bottom left:} Spatial distribution of the clusters. \textbf{Bottom right:} data mapped in the eigenspace.}
%\label{init_model_PM10}	
%\end{figure}

%\begin{figure}[htbp]
%\centering
%\begin{tabular}{c}
%%\includegraphics[width=5in]{merging_zebrafish_IKSC.eps}
%\includegraphics[width = 4.5in]{newclu_PM10_IKSC.eps}
%\end{tabular}
%\caption{\textbf{PM$_{10}$ clusters after creation.} \textbf{Top:}  signals for the three clusters after the creation event. \textbf{Bottom left:} Spatial distribution of the clusters. Interestingly, the new cluster comprises stations located in the North-East part of Germany, which is the area where the pollutants coming from Eastern Europe started to spread during the heavy pollution episode of January $2010$.  \textbf{Bottom right:} data in the eigenspace.}
%\label{newclu_PM10}	
%\end{figure}

%\begin{figure}[htbp]
%\centering
%\begin{tabular}{c}
%%\includegraphics[width=5in]{newclu_zebrafish_IKSC.eps}
%\includegraphics[width=4.5in]{merge_PM10_IKSC.eps}
%\end{tabular}
%\caption{\textbf{Clustering model of PM$_{10}$ stations after merging.} \textbf{Top:} two clusters left after the merging event occurred at time step $t=251$. \textbf{Bottom left:} Spatial distribution of the clusters. \textbf{Bottom right:} data in the eigenspace.}
%\label{merge_PM10}	
%\end{figure}

\section{Incremental $K$-means Clustering}
\label{Ikmeans_sec}
As mentioned in the previous chapter, one of the most popular data clustering methods in many scientific domains is $K$-means clustering because of its simplicity and computational efficiency.  
In its incremental variant, the $K$-means clustering algorithm is applied online to a data stream. At each time-step Incremental $K$-means (IKM) uses the previous centroids to find the new cluster centers, instead of rerunning the $K$-means algorithm from scratch \cite{incr_kmeans}.
 
In table \ref{table_IKSC_results} a summary of the results regarding all the experiments is presented. The performance of IKSC and IKM are compared in terms of mean ARI and MSV over time. Concerning the experiments with the Gaussian clouds IKSC achieves better cluster accuracy (higher ARI), with a slightly worse MSV with respect to IKM. In the case of the synthetic time-series and the PM$_{10}$ data IKSC outperforms IKM in terms of MSV.                

\section{Conclusions}
In this chapter an adaptive clustering model called Incremental Kernel Spectral Clustering (IKSC) has been introduced. IKSC takes advantage of the out-of-sample property of KSC to adjust the initial model over time. Thus, in contrast with other existing incremental spectral clustering techniques, we proposed a model-based eigen-update, which guarantees high accuracy. On some toy-data we have shown the effectiveness of IKSC in modelling the cluster evolution over-time (drifting, merging, cluster appearance etc.) and in recognizing change points. Then we analysed a real-world data-set consisting of PM$_{10}$ concentrations registered during a heavy pollution episode that took place in Northern Europe in January $2010$. Also in this case IKSC was able to recognize some interesting patterns and track their evolution over-time, in spite of dealing with the complex dynamics of PM$_{10}$ concentration.

\begin{table}[htpb]
\begin{center}
\begin{tabular}{|c|c|c|c|} 
\hline 
\textbf{Experiment} & \textbf{Algorithm} &\textbf{MSV}  & \textbf{ARI}\\
\hline
\multirow{2}{*}{\textbf{Drifting Gaussians}} & {\textbf{IKM}} & $\mathbf{0.89}$ & $1$ \\
& {\textbf{IKSC}} & $0.88$ & $1$\\
\hline
\multirow{2}{*} {\textbf{Merging Gaussians}} & {\textbf{IKM}} & $\mathbf{0.91}$ & $0.95$ \\
& {\textbf{IKSC}} & $0.90$ & $\mathbf{0.99}$\\
\hline
\multirow{2}{*}{\textbf{Synthetic time-series}} & {\textbf{IKM}} & $0.90$ & $-$ \\
& {\textbf{IKSC}} & $\mathbf{0.92}$ & $-$\\
\hline
\multirow{2}{*}{$\mathbf{PM_{10}}$ \textbf{data}} & {\textbf{IKM}} & $0.27$ & $-$ \\
& {\textbf{IKSC}} & $\mathbf{0.32}$ & $-$\\
\hline
\end{tabular}
\end{center}
\caption{\textbf{Cluster quality evaluation.} Average ARI and/or MSV over time for all the experiments that have been performed. In the case of the synthetic time-series and the PM$_{10}$ only MSV is computed since the true partition is unknown.}
\label{table_IKSC_results}
\end{table}

%%%%%%%%%%%%%%%%%%%%%%%%%%%%%%%%%%%%%%%%%%%%%%%%%%
% Keep the following \cleardoublepage at the end of this file, 
% otherwise \includeonly includes empty pages.
\cleardoublepage

% vim: tw=70 nocindent expandtab foldmethod=marker foldmarker={{{}{,}{}}}
%

%
  \graphicspath{{\chaptersdir/conclusion/image/}}%
  %\cleardoublepage%
  \chapter{Conclusions and Future Challenges}

\section{General Conclusions}
\label{ch:conclusion}
This thesis has addressed the issue of clustering in a changing scenario by means of a variety of techniques cast in the LS-SVM primal-dual optimization framework. In chapter \ref{ch:spectralclustering} we introduced the basics of spectral clustering (SC), LS-SVM and kernel spectral clustering (KSC), and we discussed a new algorithm called soft kernel spectral clustering (SKSC). In this method, in contrast with KSC, a fuzzy assignment based on the cosine distance from the cluster prototypes in the projection space is suggested, together with a novel model selection technique called average membership strength criterion (AMS). The soft assignment and the AMS are found to produce more interpretable results and an improved clustering performance with respect to KSC, mainly in cases of large overlap between the clusters. 

In chapter \ref{ch:community_detection}, a whole kernel-based framework for community detection in complex networks is discussed. This methodology is based on five cornerstones, corresponding to the extraction  of a small training sub-graph representative of the community structure of the entire network, the model selection issue, the choice of an appropriate kernel function, the out-of-sample extension and the applicability to large-scale data. In particular, we have introduced a model selection criterion based on the Modularity quality function which is more suited for network data. Moreover, the sampling method to select the training set is based on the greedy maximization of the expansion factor. We showed how our approach produces high-quality partitioning both in terms of Modularity and Conductance, and it outperforms spectral clustering using the Nystr\"om method for out-of-sample extension. As mentioned, we also explained how our technique can be used for large-scale applications and in a distributed computing environment. 

The experiments performed for dealing with the static clustering and community detection problems laid the foundations for the development of models for the analysis of evolving data. In chapter \ref{ch:evclustering} a novel technique called kernel spectral clustering with memory effect (MKSC) is proposed in order to cluster evolving networks. An evolving network can be represented as a sequence of snapshots over time, where each snapshot describes the topology of the network at a particular time instant. When community detection is performed at time $t$, the clustering should be similar to the clustering at the previous time-step $t-1$, and at the same time should accurately incorporate the actual data. In the MKSC algorithm, this sort of temporal smoothness is included in the objective function of the primal problem of KSC through the maximization of the correlation between the actual and the previous models. Moreover, new smoothed measures describing the quality of a partitioning produced at a given time are proposed, and they are used both for model selection and for evaluating the test results. The MKSC method is utilized in two main frameworks: in the first one the smoothness is imposed by fixing the regularization constant $\nu = 1$ and the amount of smoothness is tuned by changing the memory $M$. In the second framework the smoothness gets activated automatically only when it is required, that is when a big change in the data structure occurs. As a consequence, $\nu$ must be properly tuned and it can also be used as a change detection measure. We have also proposed a tracking mechanism to detect important events characterizing the evolution of the communities over time. Moreover, two visualization tools based on a 3D embedding in the space of the dual solution vectors and on the network constructed by the tracking algorithm, respectively, are devised. On a number of synthetic and real-life networks we have shown how MKSC is able to produce high-quality partitioning in terms of classical measures such as Modularity and Conductance, and their smoothed version. Finally, comparisons with state-of-the-art methods like evolutionary spectral clustering (ESC), adaptive evolutionary clustering (AFFECT) and the Louvain method (LOUV) have shown that MKSC has a competitive performance.                     

In chapter \ref{ch:POMclustering}, an application of KSC to an industrial use case is considered. In particular, accelerometer data are collected from a packing machine to monitor its conditions. In order to describe the ongoing degradation process due to the dirt accumulation in the sealing jaws, a windowing operation on the data has been applied. Then an optimal KSC model is trained offline to identify two main regimes, interpretable as normal and faulty state, respectively. Thanks to the out-of-sample extension property and a stationarity assumption, this model is used online to raise a warning when the machine is entering critical conditions. Moreover, we proposed also a soft clustering output that can be interpreted as "probability" to maintenance. In this way KSC could help to optimize the maintenances schedule and to maximize the production capacity.

Finally, in chapter \ref{ch:IKSC} an adaptive model named incremental kernel spectral clustering (IKSC) is introduced in order to work in a non-stationary environment. In the initialization stage a KSC model is constructed, then the latter is updated online according to the acquisition of new points. In particular the clustering model, expressed in terms of cluster prototypes in the eigenspace, is continuously adapted through the out-of-sample eigenvectors calculation. Moreover, the training set is formed only by the cluster centers in the input space, which are also updated in response to new data. In this way, the method has linear complexity and allows to properly track the evolution of complex patterns across time.     
 
\section{Perspectives} 
\label{ch:perspective}
Some possible future research directions may consider the adaptation of the proposed algorithms or the development of new methods dealing with evolving data, in order to cope with challenging applications like video tracking, economic crisis prediction, sentiment analysis etc. Video tracking is the process of locating one or more moving objects over time, by properly associating the target objects in consecutive video frames. It has several uses, like security and surveillance, traffic control, medical imaging and many others. The prediction of an economic crisis is an ambitious challenge, mainly due to the complexity of the economic and financial markets and their non-stationary nature. Nevertheless, unsupervised kernel-based methods could help to catch interesting patterns in the data over time, which may be associated to a forthcoming catastrophe. In this way proper preventive actions can be undertaken to limit the damage. Sentiment analysis aims to determine the attitude of a speaker or a writer with respect to some topic or the overall contextual polarity of a document as positive, negative or neutral. The application to social networks data could allow to track the opinion of millions of people across time on several topics. For instance, this may be a useful tool for the ruling class to understand the impact of some policies and eventually to produce corrective actions. 

Also, a new generation of kernel-based methods cast in a deep architecture, which can learn several levels of nonlinearities, can be developed. Deep learning was first proposed in the artificial neural networks (ANN) domain \cite{Bengio_2009,smale_2009,lequoc_2012} and attracted much attention in the research community. Deep learning allows to learn different levels of abstraction or representations of the original data by stacking together several layers of neurons. Each layer performs unsupervised learning taking as inputs the outputs of the previous layer. In this way a rich understanding of the patterns at different scales is achieved and meaningful features are extracted. As a consequence, if an additional layer is added to the network to perform supervised learning, a breakthrough accuracy in the learning task can be reached. These recent successful experiments are triggering the interest in exploring more powerful models also in other domains, like kernel methods \cite{Bengio_pami2013}. In relation to the methods devised in this thesis, a novel hierarchical clustering model based on a multilayer architecture, inspired by \cite{salakhutdinov_deephierarchical}, could for instance be developed. In particular, the specific models built at each layer can be combined such that the final model produces a result which takes into account the outcomes of the models in the previous layers. This kind of memory effect which propagates the information from layer to layer represents the same principle behind the MKSC algorithm described in chapter \ref{ch:evclustering} in the context of clustering networks over time. Thus, in this case an analogous model where the time is replaced by a scale parameter could be designed.                      
%%%%%%%%%%%%%%%%%%%%%%%%%%%%%%%%%%%%%%%%%%%%%%%%%%
% Keep the following \cleardoublepage at the end of this file, 
% otherwise \includeonly includes empty pages.
\cleardoublepage

% vim: tw=70 nocindent expandtab foldmethod=marker foldmarker={{{}{,}{}}}
%

%%%%%%%%%%%%%%%%%%%%%%%%%%%%%%%%%%%%%%%%%%%%%%%%%%%%%%%%%%%%%%%%%%%%%%

\appendix

\chapter{Appendix}
\label{ch:myappendix}

\section{Cluster Quality Evaluation}
In this Appendix we give a description of the main quality measures used in this thesis to evaluate and compare clustering results.    

\subsection{Data Clustering}
One of the most important issues in cluster analysis is the evaluation of clustering results to find the partitioning that best captures the underlying structure of the data. Intuitively, any good algorithm should search for clusters whose members are close to each other and well separated. Many validity criteria can be found in the literature \cite{milligan_cooper,Halkidi01onclustering}. Here we focus on the Silhouette \cite{silhouette} and the Davies-Bouldin Index (DBI).

Silhouette ranges from -$1$ to $1$ and measures how similar data points in the same clusters are to each other compared
to data points in other clusters. For each data point the more the Silhouette value is positive the more coherent this data point is to points in the same cluster. Negative values of the Silhouette indicates that this data point should be rather assigned to a different cluster. The definition of the Silhouette for the $i$-th datum is as follows:    
\begin{equation*}
\centering
\label{sil}
\textrm{Sil}(i) = \frac{b(i)-a(i)}{\textrm{max}\{a(i),b(i)\}}
\end{equation*}             
where $a(i)$ is the average dissimilarity of data point $i$ with all other data within the same cluster and $b(i)$ indicates the lowest average dissimilarity of $i$ with the data belonging to a different cluster. Finally, the mean Silhouette value (MSV) gives an average measure of the coherence of the groups.
      
In the case of the DBI a measure of similarity $R_{ij}$ between each pair of clusters $A_i$ and $A_j$ is defined based on the ratio between the dispersion of the clusters $s_i$ and $s_j$ and their dissimilarity $d_{ij}$: $R_{ij} = \frac{(s_i+s_j)}{d_{ij}}$. Then DBI for $k$ clusters becomes: 
\begin{equation*}
\centering
\label{DBI}
\textrm{DBI} = \frac{1}{k}\sum_{i=1}^{k}R_i
\end{equation*}             
with $R_i = \textrm{max}_{i \neq j}R_{ij}, i = 1,\hdots,k$.
Roughly speaking DBI quantifies the average similarity between each cluster. Since it is desirable for the clusters to have the minimum possible similarity to each other good partitions correspond to a low value of DBI. 
   
\subsection{Community Detection}
As we mentioned in Chapter \ref{ch:community_detection} clustering network data is mainly referred in the literature as community detection. Given a network, we aim to find a partition such that the nodes within a community are densely connected and the connections are sparse between the communities. Also in this case there has been an extensive work related to assessing the quality of a division of a network into communities \cite{emp_comparison,danon-2005,Lancichinetti2009}. In this dissertation, we utilized two well known criteria: Modularity and Conductance.      
    
Modularity is the most popular quality function of a graph introduced in \cite{newmanmod1,newmanmod2}. It is very well considered due to the fact that closely agrees with intuition on a wide range of real world networks. It is based on the idea that a random graph is not expected to have a cluster structure, so the possible existence of clusters can be revealed by the comparison between the actual density of inter-community edges and the density one would expect to have in the graph if the vertices were attached randomly, which defines the null model. Positive high values indicate the possible presence of a strong community structure. Moreover it has been shown that finding the Modularity of a network is analogous to finding the ground-state energy of a spin system \cite{guimera2004Modularity}. In formulae the definition for an unweighted network\footnote{For weighted graphs an analogue definition has been provided in \cite{newmanmod3}.} is as follows: 
\begin{equation*}
\textrm{Mod} = \frac{1}{2m}\sum_{ij}(S_{ij} -F_{ij})\delta_{ij}
\end{equation*}
with $i,j$ indicating a pair of nodes. The sum runs over all pairs of vertices, $S$ is the similarity matrix, $m$ indicates the total number of edges, and $F_{ij}$ represents the expected number of edges between vertices $i$ and $j$ in the null model. The $\delta_{ij}$ function yields $1$ if vertices $i$ and $j$ are in the same community and $0$ otherwise. Since the standard null model of Modularity imposes that the expected degree sequence matches the actual degree sequence of the graph, the Modularity can be written as: $\textrm{Mod} = \frac{1}{2m}\sum_{ij}(S_{ij} -\frac{d_id_j}{2m})\delta_{ij}$, where we indicate with $d_i=\sum_jS_{ij}$ the degree of the vertex $i$. Thus, a high value of $\textrm{Mod}$ says that the number of inter-community edges of the graph are few in a statistically significant sense, meaning the presence of a strong community-like structure. Moreover, after some linear algebra calculations \cite{jiangmod}, we can write Modularity as: \begin{equation*}
\label{Mod_eq}
\textrm{Mod}=\frac{1}{2m}\textrm{tr}(X^TMX)
\end{equation*}
where $M = S-\frac{1}{2m}dd^T$ is the Modularity matrix, $d = [d_1,\hdots,d_N]$ indicates the vector of the degrees of each node and $X$ represents the cluster indicator matrix. Finally, as it has been pointed out in \cite{good2010performance} and \cite{Fortunato02012007}, Modularity suffers from some drawbacks:
\begin{itemize}
\item resolution limit: it contains an intrinsic scale that depends on the total number of links in the network. Modules that are smaller than this scale may not be resolved 
\item exhibits degeneracies: it typically admits an exponential number of distinct high-scoring solutions and often lacks a clear global maximum
\item it does not capture overlaps among communities in real networks. 
\end{itemize}
Many of the aforementioned problems of Modularity have been solved by properly modifying its definition, as for example in \cite{arenas_fernandez_gomez} where a multi-resolution Modularity has been introduced. In this case, however, as pointed out in \cite{limits_Modularity}, multi-resolution Modularity has a double simultaneous bias: it leads to a splitting of large clusters and a merging of small clusters, and both problems cannot be usually handled at the same time for any value of the resolution parameter. 
 
Another meaningful quality function is the Conductance. For every community in a graph $\mathcal{G} = (\mathcal{V},\mathcal{E})$, the Conductance is defined as the ratio between the number of edges leaving the cluster and the number of edges inside the cluster \cite{Kannan00onclusterings}, \cite{emp_comparison}. In particular, the Conductance denoted by $\textrm{Cond}(S)$ of a set of nodes $\mathcal{A} \subset \mathcal{V}$ is: 
\begin{equation*}
\textrm{Cond}(S)=c_S/ \textrm{min}(Vol(S),Vol(V \setminus S))
\end{equation*}
where $c_S$ denotes the size of the edge boundary, $c_S = \{(u,v) : u \in S, v \notin S \}$, and $Vol(S) = \sum_{u \in S} d_u$, with $d_u$ representing the degree of node $u$. In other words the Conductance describes the concept of a good network community as a set of nodes that has a better internal than external connectivity. Therefore the lower the Conductance the better the community structure. 

\subsection{Comparing Partitions}%ARI,NMI
In order to compare two clustering results two measures of agreement between partitions have been utilized in this thesis: the Adjusted Rand Index (ARI) and the Normalized Mutual Information (NMI). 

Given two partitions $\mathcal{U}$ and $\mathcal{V}$, the ARI \cite{arindex} equals $1$ when the related cluster memberships agree completely, $0$ in case of complete mismatch with equal number of positive and negative mistakes, and $-1$ for a complete disagreement with all negative mistakes\footnote{A positive mistake means that if a partition groups a pair of points in the same cluster, the other does not. A negative mistake means that a pair of points is grouped in different clusters according to an algorithm while the points are assigned to the same cluster by the other algorithm.}. Its most common use is as measure of agreement between the clustering outcomes of an algorithm and a known grouping which acts like a ground-truth.

The NMI \cite{nmijmlr} measures the statistical information shared between the partitions $\mathcal{U}$ and $\mathcal{V}$ it also takes values in the range $[0,1]$. It tells us how much knowing one of these clusterings reduces our uncertainty about the other. The higher the NMI, the more useful the information in $\mathcal{V}$ helps us to predict the cluster memberships in $\mathcal{U}$ and vice versa.

%%%%%%%%%%%%%%%%%%%%%%%%%%%%%%%%%%%%%%%%%%%%%%%%%%
% Keep the following \cleardoublepage at the end of this file, 
% otherwise \includeonly includes empty pages.
\cleardoublepage

% vim: tw=70 nocindent expandtab foldmethod=marker foldmarker={{{}{,}{}}}

%%%%%%%%%%%%%%%%%%%%%%%%%%%%%%%%%%%%%%%%%%%%%%%%%%%%%%%%%%%%%%%%%%%%%%
\backmatter

\includebibliography
\includepublications{publications}
\includecv{biography}

\makebackcoverXII

\end{document}